\documentclass[11pt]{book}

\usepackage{geometry}
\usepackage{mathtools}
\usepackage[boxruled,linesnumbered]{algorithm2e}
\usepackage{blindtext, graphicx}
\usepackage{url}
\usepackage{relsize}
\usepackage{cite}
\usepackage{bm}
\usepackage{changepage}
\usepackage{amssymb}
\usepackage{array}
\usepackage{longtable}
\usepackage[long,nocomma]{optidef}
\usepackage{mdframed}
\usepackage{subfigure}
\usepackage[table,xcdraw]{xcolor}
\usepackage{tabularx, colortbl}
\usepackage{multirow}
\usepackage[nottoc]{tocbibind}
\usepackage{setspace}

\let\oldnl\nl
\newcommand{\nonl}{\renewcommand{\nl}{\let\nl\oldnl}}

\makeatletter
\DeclareOldFontCommand{\rm}{\normalfont\rmfamily}{\mathrm}
\DeclareOldFontCommand{\sf}{\normalfont\sffamily}{\mathsf}
\DeclareOldFontCommand{\tt}{\normalfont\ttfamily}{\mathtt}
\DeclareOldFontCommand{\bf}{\normalfont\bfseries}{\mathbf}
\DeclareOldFontCommand{\it}{\normalfont\itshape}{\mathit}
\DeclareOldFontCommand{\sl}{\normalfont\slshape}{\@nomath\sl}
\DeclareOldFontCommand{\sc}{\normalfont\scshape}{\@nomath\sc}
\makeatother

\newcommand{\name}{{\sf{BWRH}}}
\newcommand{\namefast}{{\sf{BWRHF}}}
\newcommand{\nameiris}{{\sf Iris}}

\begin{document}

\title{Traffic Allocation for Efficient Data Transfers Across Geographically Distributed Datacenters}

\author{\normalsize
    Mohammad Noormohammadpour\\
    \textit{\normalsize Ming Hsieh Department of Electrical Engineering, University of Southern California}
}

\let\cleardoublepage\clearpage

\date{\normalsize   \today}

\begin{titlepage}
    \begin{center}
    {\setstretch{1.0}
        \vspace*{1cm}
        
        {\huge \textbf{On Efficient Data Transfers Across}\\
               \vspace{0.8em}
               \textbf{Geographically Dispersed Datacenters}}
        
        
        \vspace{1cm}
        
        by\\
        
        \vspace{1cm}
        \textbf{\large Mohammad Noormohammadpour}
        
        \vspace{1.5cm}
        
        
        \noindent\makebox[\linewidth]{\rule{\textwidth}{0.4pt}}
        
        \vspace{1.5cm}
        
        {\large A Dissertation Presented to the\\
        FACULTY OF THE USC GRADUATE SCHOOL\\
        UNIVERSITY OF SOUTHERN CALIFORNIA\\
        In Partial Fulfillment of the\\
        Requirements for the Degree\\
        DOCTOR OF PHILOSOPHY\\
        (ELECTRICAL ENGINEERING)
        
        \vspace{1.5cm}
        
        December 2019
        
        \vfill
        
        Copyright 2019 \hfill Mohammad Noormohammadpour}
        
        
        
        
    }
    \end{center}
\end{titlepage}


\frontmatter

\clearpage
\chapter*{Acknowledgements}
\addcontentsline{toc}{chapter}{Acknowledgements}
I want to thank my Ph.D. advisor, Prof. Cauligi Raghavendra, who provided inordinate help with every step in the preparation and making of this dissertation. I want to thank our collaborators Dr. Sriram Rao from Facebook, Dr. Srikanth Kandula from Microsoft, and Dr. Ajitesh Srivastava from the Ming Hsieh Department of Electrical Engineering, University of Southern California. I would also like to thank Prof. Neal Young from the University of California, Riverside, for the helpful comments on Stack Exchange concerning the NP-Hardness proof of the Best Worst-case Routing presented in Appendix \ref{chapter_bwr_hardness}. I finally would like to thank Long Luo from the University of Electronic Science and Technology of China for helpful discussion and collaboration.

I would also like to thank the following researchers and engineers who provided helpful advice and support throughout the Ph.D. program as part of classes and internships. Prof. Minlan Yu now at Harvard; my internship team from Cisco that worked on Non-Volatile Memory for Distributed Storage especially David Oran, Josh Gahm, Atif Fahim, Praveen Kumar, Marton Sipos, and Spyridon Mastorakis; and my internship team at Google NetInfra working on Inter-Datacenter Traffic Engineering especially Jeffrey Liang, Kirill Mendelev, Brad Morrey, Gilad Avidov, and Warren Chen.

\tableofcontents

\listoffigures

\listoftables

\chapter*{Abstract}
\addcontentsline{toc}{chapter}{Abstract}
As applications become more distributed to improve user experience and offer higher availability, businesses rely on geographically dispersed datacenters that host such applications more than ever. Dedicated inter-datacenter networks have been built that provide high visibility into the network status and flexible control over traffic forwarding to offer quality communication across the instances of applications hosted on many datacenters. These networks are relatively small, with tens to hundreds of nodes and are managed by the same organization that operates the datacenters which make centralized traffic engineering feasible. Using coordinated data transmission from the services and routing over the inter-datacenter network, one can optimize the network performance according to a variety of utility functions that take into account data transfer deadlines, network capacity consumption, and transfer completion times. Such optimization is especially relevant for long-running data transfers that occur across datacenters due to the replication of configuration data, multimedia content, and machine learning models.

In this dissertation, we study techniques and algorithms for fast and efficient data transfers across geographically dispersed datacenters over the inter-datacenter networks. We discuss different forms and properties of inter-datacenter transfers and present a generalized optimization framework to maximize an operator selected utility function. Next, in the several chapters that follow, we study, in detail, the problems of admission control for transfers with deadlines and inter-datacenter multicast transfers. We present a variety of heuristic approaches while carefully considering their running time. For the admission control problem, our solutions offer significant speed up in the admission control process while offering almost identical performance in the total traffic admitted into the network. For the bulk multicasting problem, our techniques enable significant performance gain in receiver completion times with low computational complexity, which makes them highly applicable to inter-datacenter networks. In the end, we summarize our contributions and discuss possible future directions for researchers.

\mainmatter

\chapter{Introduction} \label{chapter_introduction}
Datacenters provide an infrastructure for many online services which include services managed by small companies and individuals who do not want to deal with complexities and difficulties of maintaining physical computers \cite{datacenter_survey_1, datacenter_survey_2}. Examples of these online services are on-demand video delivery, storage and file sharing, cloud computing, financial services, multimedia recommendation systems, online gaming, and interactive online tools that millions of users depend on \cite{services_1, services_2, services_3}. Besides, massively distributed services such as web search, social networks, and scientific analytics that require storage and processing of substantial scientific data take advantage of computing and storage resources of datacenters \cite{b4, social_inside, scientific_data_processing}.

Datacenter services may consist of a variety of applications with instances running on one or more datacenters. They may dynamically scale across a datacenter or across multiple datacenters according to end-user demands which enables cost-savings for service managers. Moreover, considering some degree of statistical multiplexing, better resource utilization can be achieved by allowing many services and applications to share datacenter infrastructure.

To reduce costs of building and maintaining datacenters, numerous businesses rely on infrastructure provided by large cloud infrastructure providers such as Google Cloud, Microsoft Azure, and Amazon Web Services \cite{google, azure, aws} with datacenters consisting of hundreds of thousands of servers. This enables the resources needed to run thousands of distributed applications that span hundreds of servers and scale out dynamically as needed to handle additional user load.





\begin{figure}
	\centering
	\includegraphics[width=0.8\textwidth]{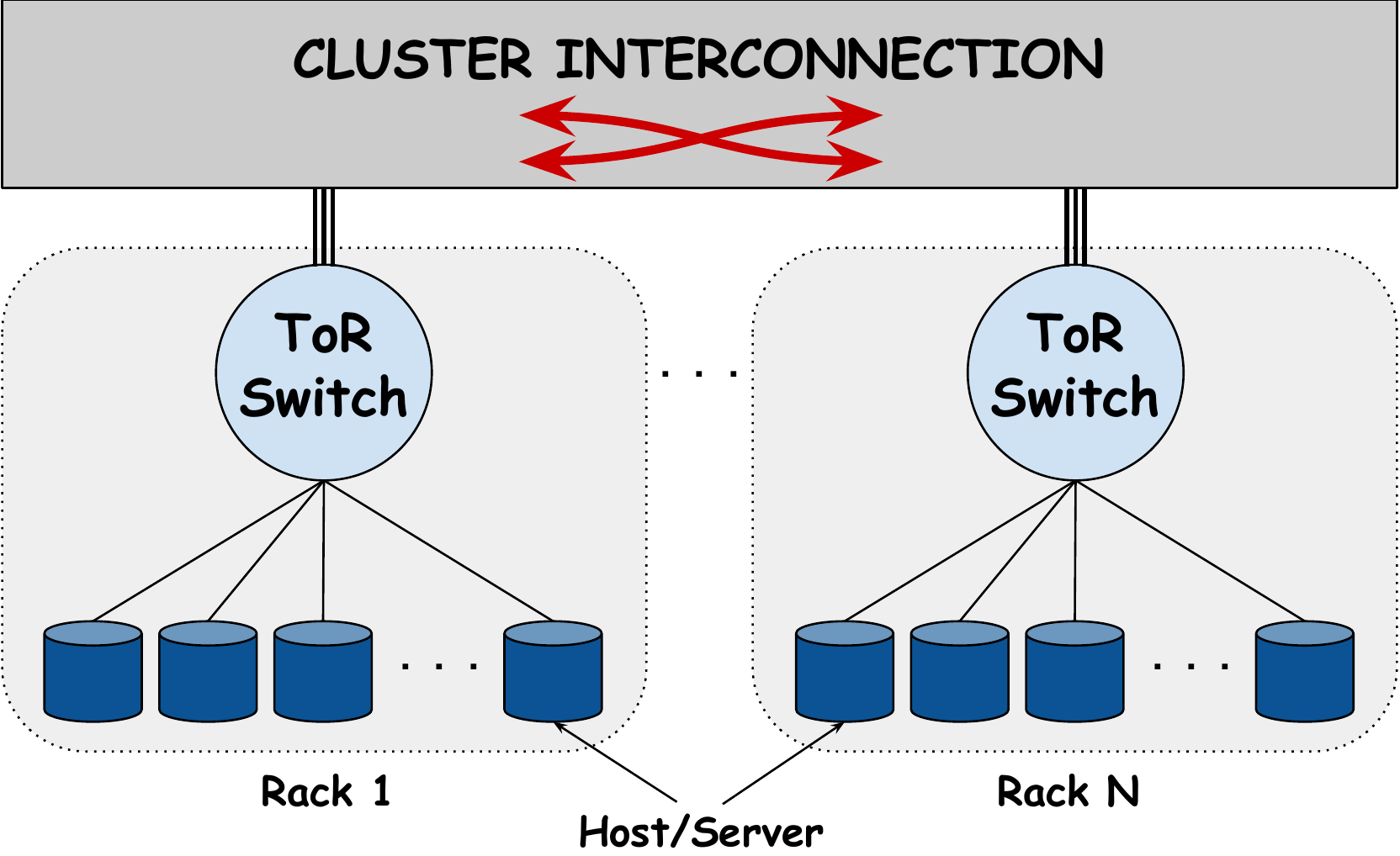}
	\caption{A typical datacenter cluster}
	\label{fig:datacenter-design}
\end{figure}

A datacenter is typically home to multiple server clusters with thousands of machines per cluster that are connected using high capacity networks. Figure \ref{fig:datacenter-design} shows the structure of a typical datacenter cluster network with many racks. A cluster is usually made up of up to hundreds of racks \cite{jupiter, vl2, server-per-rack-facebook}. A rack is essentially a group of machines which can communicate at high speed with minimum latency. All the machines in a rack are connected to a \textbf{Top of Rack (ToR)} switch which provides non-blocking connectivity among them. Rack size is typically limited by maximum number of ports that ToR switches provide and the ratio of downlink to uplink bandwidth. There is usually about tens of machines per rack \cite{jupiter, vl2, server-per-rack-facebook}. ToR switches are then connected via a large interconnection allowing machines to communicate across racks. An ideal network should act as a huge non-blocking switch to which all servers are directly connected allowing them to simultaneously communicate at maximum rate.

Datacenter network topology plays a significant role in determining the level of failure resiliency, ease of incremental expansion, communication bandwidth and latency. The aim is to build a robust network that provides low latency, typically up to hundreds of microseconds \cite{pitfall, dcqcn, hull}, and high bandwidth across servers. Many network designs have been proposed for datacenters \cite{fattree, vl2, hyperx, dcell, leaf-spine, xpander, fbtopology, jellyfish}. These networks often come with a large degree of path redundancy which allows for increased fault tolerance. Also, to reduce deployment costs, some topologies scale into large networks by connecting many inexpensive switches to achieve the desired aggregate capacity and number of machines \cite{clos, jupiter} and the majority of these topologies are symmetrical.

Many services may need to span over multiple racks to access required volume of storage and compute resources. This increases the overall volume of traffic across racks. A high-capacity datacenter network allows for flexible operation and placement of applications across clusters and improves overall resource utilization and on-demand scale out for applications \cite{jupiter, fattree, vl2, fbtopology}. This allows resources of any machine to be used by any application which is essential for hyper-scale cloud providers \cite{google, azure, aws}. However, designing networks that run at very high capacity is costly and unnecessary for smaller companies or enterprises. As a result, many datacenters may not offer full capacity across racks with the underlying assumption that services run mostly within a single rack. To maximize resource utilization across a datacenter, accommodate more services and allow for better scalability, large cloud providers usually build their networks at maximum capacity.

There is growing demand for datacenter network bandwidth. This increase is driven by faster storage devices, rising volume of user and application data, reduced cost of cloud services and ease of access to cloud services. Google reports 100\% increase in their datacenter networking demands every 12 to 15 months \cite{jupiter}. Cisco forecasts a 400\% increase in global datacenter IP traffic and $2.6\times$ growth in global datacenter workloads from 2015 to 2020 \cite{cisco-growth}. This growth in traffic has made network traffic management a necessity for datacenter operators to ensure that services can access the network capacity with minimal interference from other services.

\section{User Experience}
User experience is the cornerstone of online services which have become ubiquitous and are presented to users through a variety of platforms including websites and mobile applications \cite{online_services}. Several factors determine the quality of experience perceived by users while accessing such services the most important of which are latency and availability. It is crucial that users can always access the resources and the faster, the better. For example, a website's load time can affect whether the users will explore the website further. As another example, while watching a video clip on YouTube, users would like the video to start quickly and play smoothly without interruptions or degradation in quality \cite{online_playback_issues}.

To maximize users' quality of experience while interacting with a specific service, operators keep multiple instances of such services up and running at any time and place them closer to local users across regions, countries, and continents \cite{replication_helps_1, replication_helps_2}. This deployment minimizes users' latency while interacting with services and allows for a smooth and responsive experience. Moreover, if an instance is interrupted due to failures or disasters, users will have the option of switching to other running instances of the same service in another datacenter. Doing so will also require services to copy the data based on which they operate across the datacenters on which they run.

\begin{figure}
    \centering
    \includegraphics[width=0.8\textwidth]{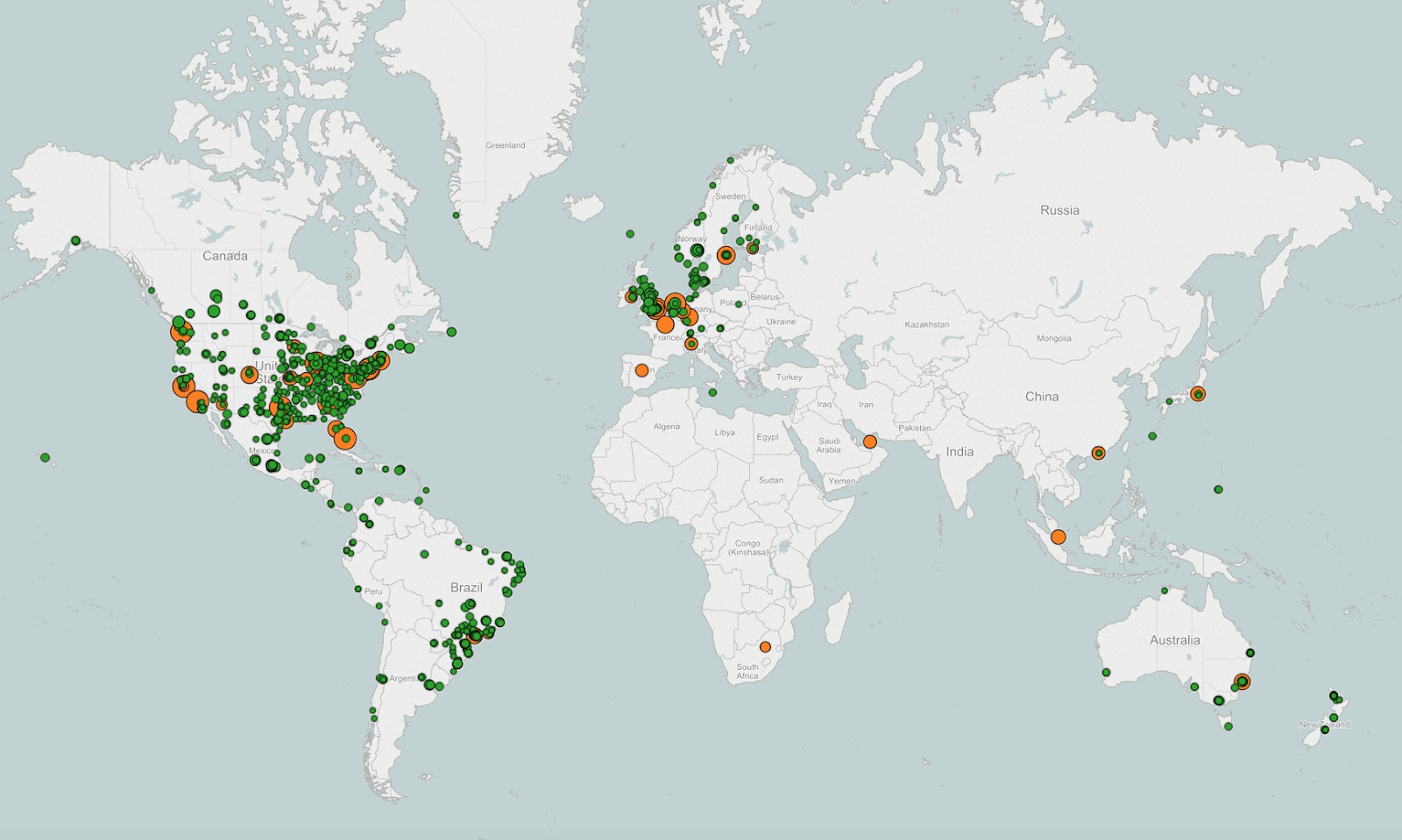}
    \caption{Netflix cache locations as of 2016. Green dots are ISP locations and orange circles are Internet Exchange Points (IXPs) where different network providers connect their networks.\protect\footnotemark}
    \label{fig:netflix_cache_locations}
\end{figure}

An example of such distributed applications is content distribution platforms like Netflix \cite{rep-netflix-locations}. These services copy multimedia content to many locations close to local users for low-latency and high-speed access. Figure \ref{fig:netflix_cache_locations} shows Netflix's cache locations where multimedia content is stored for regional user access \cite{netflix-replication}. Depending on how users are distributed, services can decide how to place copies of data. For example, multimedia content can be distributed to locations where many users are expected to access it. Besides, such copying can be done both proactively and reactively. In the former case, services copy the content to a location before it is accessed by users allowing all users to have fast access to content. In the latter case, services copy the content to a location when a user near that location accesses the content which might lead to first users experiencing less than ideal quality of experience. Although the proactive approach offers a better user experience, it can be more costly for operators.

\addtocounter{footnote}{-1}

\footnotetext{This figure was downloaded from the following URL: \protect\url{https://media.netflix.com/en/company-blog/how-netflix-works-with-isps-around-the-globe-to-deliver-a-great-viewing-experience}}

Another example of distributed services is web search such as Google and Bing \cite{search_1, search_2}. These services crawl billions of web pages and generate significant volumes of search index updates which are distributed across many datacenters for low-latency access by local and regional users \cite{b4, bing}. Search index updates are generated at different frequencies according to how fresh the related results need to be which usually leads to smaller updates at high frequency and larger updates at a low frequency that are pushed from the datacenter that generates them to all other datacenters.

\section{Inter-Datacenter Networks}
\begin{figure}
    \centering
    \includegraphics[width=0.8\columnwidth]{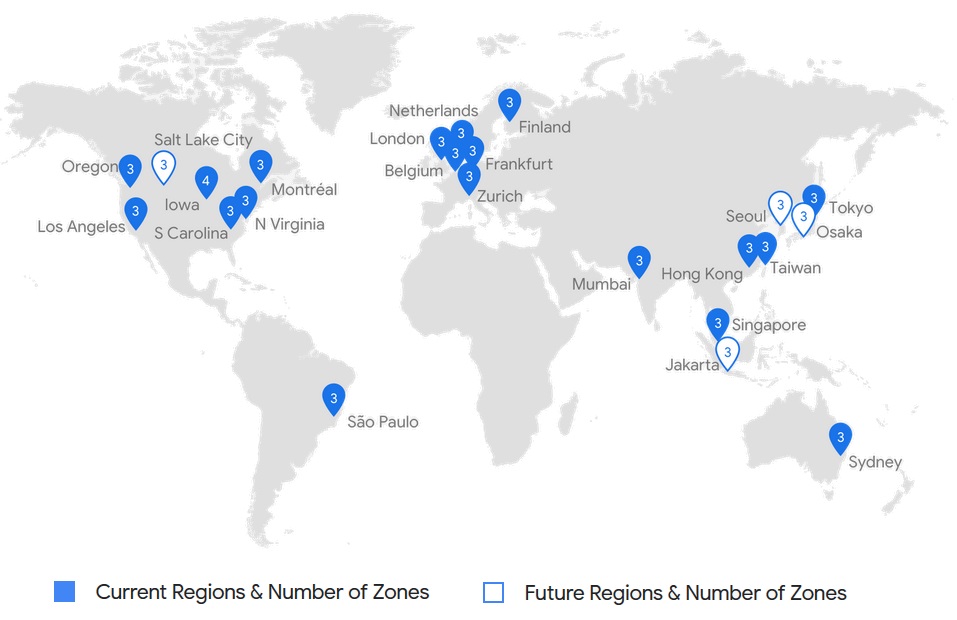}
    \caption{Google, a major cloud services provider, with $19$ functional regions and $4$ currently in progress as of $2019$.\protect\footnotemark}
    \label{fig:datacenter_locations_google}
\end{figure}

There is benefit in providing services using multiple datacenters that are geographically distributed so that required services and data can be brought close to users for low-latency and high-speed access. Accordingly, Google Cloud, Amazon Web Services, and Microsoft Azure operate and maintain multiple geographically distributed datacenters. Google operates across $19$ regions as shown in Figure \ref{fig:datacenter_locations_google} with plans to expand to additional $4$ regions, Microsoft operates across $54$ geographical regions, Amazon runs more than two dozen availability zones each consisting of one or more discrete datacenters, and Facebook employs $7$ datacenters in North America and Europe.

\addtocounter{footnote}{-1}

\footnotetext{This figure was downloaded from the following URL: \protect\url{https://cloud.google.com/about/locations/}}

There is a significant volume of traffic exchanged between datacenters. This traffic is due to frequent copying of large quantities of data and content from one datacenter to one or more datacenters. For this purpose, high bandwidth networks connecting datacenters can be leased or purchased for fast and efficient data transfers \cite{level3, b4, swan-backbone, facebook-express-backbone}. These high-speed wide area networks with dedicated capacity are referred to as inter-datacenter (inter-DC) networks. The resources of these networks may be used by the services that run on the datacenters that they connect. Datacenter operators own the capacity of the inter-DC network and can manage it as needed to maximize the performance of services.

\begin{figure}[t!]
    \centering
    \includegraphics[width=0.7\textwidth]{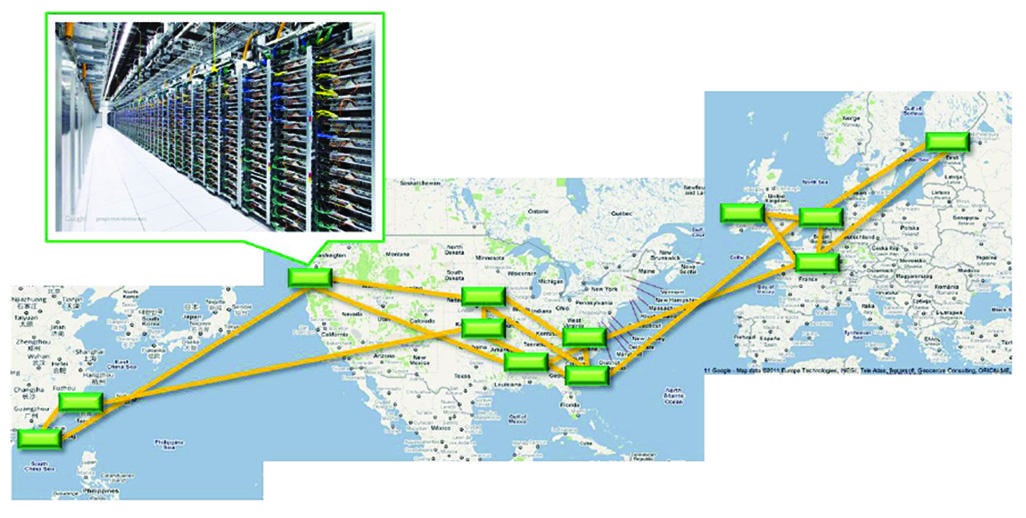}
    \caption{Google's inter-DC network also known as B4.\protect\footnotemark}
    \label{fig:google_wan}
\end{figure}
\begin{figure}[t!]
    \centering
    \includegraphics[width=0.7\textwidth]{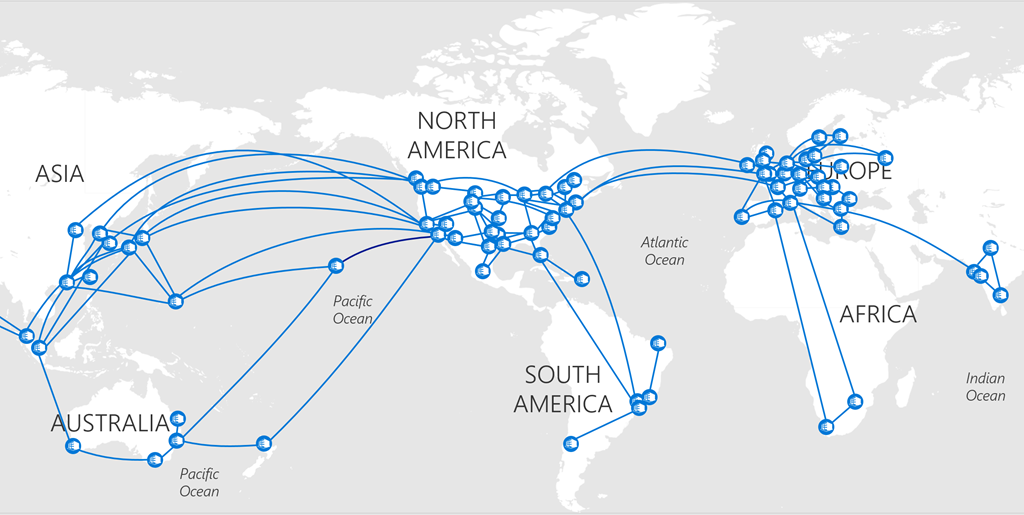}
    \caption{Microsoft Azure's inter-DC network.\protect\footnotemark}
    \label{fig:azure_wan}
\end{figure}

For example, Google B4, shown in Figure \ref{fig:google_wan}, is an inter-DC network that connects Google's datacenters globally.\footnote{This topology is from 2013 and has been well expanded since then.} It hosts the traffic for not only Google but also all the businesses that rely on Google Cloud including thousands of websites, mobile and desktop applications. Another dedicated inter-DC WAN is Microsoft Azure's global backbone \cite{swan, swan-backbone}, shown in Figure \ref{fig:azure_wan}. There are also a variety of third-party companies that offer tools and equipment for medium and small businesses to build their inter-DC networks with dedicated capacity for high performance.

\footnotetext{This figure was downloaded from \cite{google-dc-optical}.}

\footnotetext{This figure was downloaded from the following URL: \protect\url{https://azure.microsoft.com/en-us/blog/how-microsoft-builds-its-fast-and-reliable-global-network/}}

Given that inter-DC networks connect a limited number of locations, usually about tens to hundreds of datacenters, management of their capacity for efficient usage through coordinated resource scheduling is feasible and has been shown to improve utilization and reduce deployment costs \cite{b4, swan, tempus, dynamic_pricing}. Besides, inter-DC networks offer a high level of visibility into network status, and control over network behavior such as routing and forwarding of traffic. These features streamline capacity management which is also the central concept around which this dissertation is shaped.

\section{Inter-DC Transfers}
Datacenter services determine the traffic characteristics and the communication patterns among servers within a datacenter and between different datacenters. Many datacenters, especially cloud providers, run a variety of services that results in a spectrum of workloads. Some popular services include cache followers, file stores, key-value stores, data mining, search indexing, and web search. Some services generate lots of traffic among application instances of the service which is referred to as \textbf{internal traffic}. The reason this traffic is called internal is that they start and end between the instances of the same service without any direct interaction with the users. Examples of communication patterns that generate lots of internal traffic are scatter-gather (also known as partition-aggregate) \cite{nature,d2tcp, dctcp, detail} and batch computing tasks \cite{mapred, dryad}.

Inter-DC transfers occur as a result of geographically distributed services with instances running across various regions and datacenters generating lots of internal traffic across them. For example, multiple instances of services running on different datacenters may need to synchronize by sending periodic or on-demand updates. Besides, in the case of distributed data stores like key-value stores and relational databases, it may be necessary to offer consistency guarantees across multiple instances which requires the constant transmission of replicated data.

\begin{figure}
    \centering
    \includegraphics[width=0.8\textwidth]{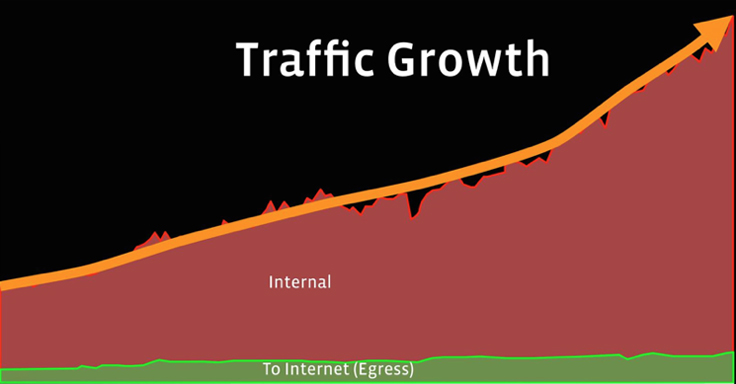}
    \caption{Traffic growth across Facebook's Express Backbone.\protect\footnotemark}
    \label{fig:traffic_growth}
\end{figure}

The volume of internal data transfers across datacenters is growing fast. For instance, Figure \ref{fig:traffic_growth} shows the growth of inter-DC bandwidth across Facebook's datacenters. As can be seen, the amount of internal traffic is a significant portion of the traffic carried by inter-DC network and is growing much faster than user traffic. To support this growing internal traffic, inter-DC network operators, such as Facebook, need to invest in expanding the network capacity which can be expensive. Therefore, efficient utilization of network bandwidth is critical to maximize the support for internal traffic. In this dissertation, we focus on developing efficient algorithms for optimizing internal inter-DC transfers. We consider the multiple research problems around inter-DC networks with a focus on performance, offer several solutions, and perform comprehensive evaluations.

\footnotetext{This figure was downloaded from the following URL: \protect\url{https://code.fb.com/networking-traffic/building-express-backbone-facebook-s-new-long-haul-network/}}



Inter-DC transfers can be classified according to their number of destinations and whether they have completion time requirements. We briefly discuss different types of inter-DC transfers in the following.

\subsection{Point to Point (P2P) Transfers}
Transfers could be generated as a result of data delivery from one datacenter to another datacenter which we refer to as point to point (P2P) transfers \cite{dtb, tempus, b4, amoeba, jetway, owan, social_inside}. Many backup services allow for one geographically distant copy of data in a different region for increased reliability in case of natural disasters or datacenter failures. For example, if a datacenter region on the east coast goes completely off the grid due to a storm, data copied to a datacenter on the west coast can be used to handle user queries. Also, data warehousing services require delivery of data from all datacenters to a datacenter warehouse \cite{fb_warehouse}.

\subsection{Point to Multipoint (P2MP) Transfers}
There are also transfers that deliver an object from one datacenter to multiple datacenters which we refer to as point to multipoint (P2MP) transfers. For example, content delivery networks (CDNs) may push significant video content to regional cache locations \cite{utube, netflix, jetway, ecoflow, social_inside}, cloud storage services may replicate data objects across multiple sites for increased reliability \cite{cassandra, azuresql}, and search engines push substantial updates to their geographically distributed search database on a regular basis \cite{b4}. Data transfers among datacenters for replication of objects from one datacenter to multiple datacenters is referred to as geo-replication \cite{mesa, mdcc, owan, google-dc-optical, mc_flexgrid, mc_icc_overlay, dtb, elastic_optical_networks, b4, yahoo, orchestrating, jetway} and can form a large portion of inter-DC traffic \cite{facebook-express-backbone}.

\subsection{Inter-DC Transfers with Deadlines}
Inter-DC transfers deliver content that may need to become available to applications before specific deadlines. Such deadlines may represent the importance of transfers \cite{tempus, amoeba}. For example, a transfer with a later deadline can be delayed in favor of another transfer with a close deadline. Deadlines are usually due to consumer requirements. For example, the results of some data processing may need to be ready by a specific time. It may also be an internally assigned metric for more efficient scheduling of network transfers. For example, if a data processing task requires two inputs to generate an output, and one of them becomes available sometime in the future, it will not help to deliver the other input data anytime earlier than that time. Assigning a deadline that is in the future, allows the network operators to deliver data that is needed sooner first.

\section{Overview of the Dissertation}
In this dissertation, we develop algorithms and techniques for efficient P2P and P2MP transfers among geographically dispersed datacenters. In Chapter \ref{chapter_background}, we first discuss how a modern inter-DC network manages traffic flow and formally present traffic management problems of interest, specifically online arrival of inter-DC traffic with its requirements. We then discuss performance metrics, such as mean and tail completion times, and finally, give a general optimization formulation for the types of problems we will consider in the rest of the dissertation.


For P2P traffic, path selection for traffic routing is a well-known problem with various existing solutions. However, using  a centralized network architecture and given a dedicated inter-DC network, it is possible to develop routing algorithms that are adaptive to network conditions and therefore more efficient. In Chapter \ref{chapter_adaptive_routing}, we develop a new routing approach referred to as Best Worst-case Routing (BWR) which is capable of considerably reducing inter-DC transfer completion times regardless of the scheduling policy used for transmission of data across the network. We evaluate various heuristics that implement BWR and use them to quickly compute a new path for a newly arriving inter-DC transfer.

In Chapter \ref{chapter_admission_control}, we develop fast admission control algorithms for inter-DC transfers with deadlines. We focus on Point to Point (P2P) transfers to maximize the number of transfers completed before their deadlines. We present a new scheduling policy referred to as the As Late As Possible (ALAP) scheduling and combine it with a load-aware path selection mechanism to perform quick feasibility checks and decide on the admission of new inter-DC transfers. We also perform evaluations across different topologies and using varying network load and show that our approach is scalable and can speed up the admission control by more than two orders of magnitude compared to traditional techniques.

In Chapter \ref{chapter_p2mp_dccast}, we study efficient P2MP transfers where data transfer is needed from one source datacenter to multiple destination datacenters. Although this can be performed as multiple P2P transfers, there is opportunity to do significantly better as all the receiving ends are known apriori and the network traffic forwarding can be centrally controlled. We introduce the concept of load-aware forwarding trees and compute them as weighted Steiner trees.\footnote{A Steiner tree is a tree subgraph of the inter-DC network that connects the sender and all the receivers. The weight of a Steiner tree is the sum of weights of its edges. Selecting a minimum weight Steiner tree over a general graph is NP-Hard \cite{steiner_tree_problem} but fast heuristics exist that offer close to optimal solutions on average \cite{DSTAlgoEvaluation}.} We consider the objective of minimizing the completion time of the slowest transfer and the total bandwidth use of all transfers. We perform extensive evaluations using random and deterministic topologies and show that our tree selection approach can considerably reduce transfer completion times compared to tree selection using other weight assignment techniques. We show that our approach can reduce the completion times of slowest transfers by about $50\%$ compared to performing P2MP using multiple P2P transfers. We also consider deadlines for P2MP transfers and present an admission control solution to maximize the number of P2MP transfers completed before deadlines. Our approach uses load-aware forwarding trees combined with the ALAP scheduling policy to perform fast admission control for P2MP transfers with deadlines. We also perform extensive evaluations and show that compared to state-of-the-art inter-DC admission control solutions our approach admits up to $25\%$ more traffic into the network while saving at least $22\%$ network bandwidth.

For a P2MP transfer, it is in general not required that all receivers get a copy of the data at the same time. In Chapter \ref{chapter_p2mp_quickcast}, we focus on selectively speeding up some datacenters using receiver set partitioning, that is, grouping the receivers of P2MP transfers into multiple partitions and attaching each partition using an independent forwarding tree. That is because a single multicast tree can slow down all receivers to the slowest receiver, although it offers the highest bandwidth savings. We apply our P2MP load-aware tree selection approach per partition to distribute load across the network as well. We also explore different ways of finding the right number of partitions as well as the receivers that are grouped per partition. Using extensive evaluations, we show that our approach can speed up the P2MP receivers by up to $35\times$ when network links have highly varying capacities.

In Chapter \ref{chapter_iris}, we develop a framework to optimize for mixed completion time objectives for P2MP transfers over inter-DC networks. That is, we realize that in general, different applications that distribute copies of objects to many locations, may have different completion time objectives. For example, many applications require one copy of an object to be made quickly while the rest of the replicas can be made slowly. Knowing this requirement, we can select the receiver partitions accordingly to save bandwidth by grouping all the slower receivers into one partition and satisfy the speed requirements by attaching the fastest receiver using an independent path. We present a solution that uses application-specific objectives to optimize the partitioning and tree selection for P2MP transfers. Through simulations and emulations, we show that our approach reduces average receiver completion times by $2\times$ while meeting the requirements specified by applications on completion times.

In Chapter \ref{chapter_p2mp_parallel}, we aim to speed up P2MP transfers using parallel load-aware forwarding trees that are selected as weighted Steiner trees. We attach each partition of receivers using potentially multiple forwarding trees that in parallel deliver data to all its receivers hence increasing their throughput and reducing their completion times. We focus on the selection of edge-disjoint trees to eliminate direct bandwidth contention across the partitions of the same transfer. We perform comprehensive simulations and show that using up to two parallel edge-disjoint trees offers almost all the benefit over various topologies and that by using parallel trees we can speed up P2MP transfers by up to $40\%$.

Finally, in Chapter \ref{chapter_summary}, we provide a summary and set forth several future directions to expand on our work.

\clearpage
\chapter{Inter-DC Network Traffic Engineering} \label{chapter_background}
Inter-DC networks consist of high-capacity links that connect tens to hundreds of datacenters across cities, countries, and continents with dedicated bandwidth \cite{b4, swan, facebook-express-backbone, owan, amoeba, tempus, dynamic_pricing, google-dc-optical}. They can be modeled as a graph with datacenters as nodes and inter-DC links as edges where every edge is associated with the properties of the inter-DC link it represents such as capacity and bandwidth utilization. Given that datacenter operators also manage inter-DC networks, coordination among traffic generation from datacenters and routing of traffic within the inter-DC networks can be used to optimize network utilization and maximize overall utility \cite{tempus, amoeba, luo2018online, dynamic_pricing}.

The context we consider is data transfers that move bulk data across geographically dispersed datacenters over inter-DC networks. Bulk data transfers move the lion share of data across datacenters \cite{social_inside} which makes it highly practical and valuable to optimize their transmission over inter-DC networks. Besides, inter-DC networks are relatively small in terms of the number of edges and nodes which makes it feasible to formulate and solve optimization scenarios to maximize their performance \cite{b4, swan-backbone, facebook-express-backbone}. Finally, inter-DC networks are operated by the same organization that manages the datacenters they connect which makes it possible to control them in a logically centralized fashion as well as apply novel traffic scheduling and routing techniques that cannot be used over the internet.

We consider a centralized traffic management scheme where a logically centralized Traffic Engineering Server (TES) receives traffic requirements from the senders and decides how traffic should be transmitted from the senders and how it should be routed within the inter-DC network across the datacenters. It also communicates with the senders and the network elements to coordinate them. Several inter-DC networks have been built using this principle, and related work has shown that this form of management allows for substantial performance gains \cite{b4, tempus, amoeba, swan, owan, facebook-express-backbone}.

Central traffic allocation offers a variety of benefits: First, it allows for improved performance by minimizing congestion by proactively reserving bandwidth while collectively considering the interplay of many transfers initiated from different datacenters. Second, it offers a highly configurable platform that allows maximizing performance according to various utility functions. Such utility functions can be selected according to an organization's business model. The coordinated routing and scheduling of traffic for maximization of network utility can be formulated as an optimization problem with different constraints as we will show later in this chapter.

The traffic engineering problem we consider is the following. We are given an inter-DC network topology, including the connectivity and link capacities across datacenters, with end-points that generate network traffic located within the datacenters. Data transfers arrive at the network in an online manner at different datacenters, i.e., we assume no prior knowledge of when a future transfer will arrive and what properties it will have. End-points can control the rate at which they transmit traffic. Upon the arrival of a new transfer, the sender communicates with the TES the properties of this transfer and any potential requirements on its transmission. The problem is for TES to compute the best route(s) on which the traffic for this new transfer is forwarded as well as the rate at which the new transfer and all the other existing transfers should transmit their traffic.

The transmission rates need to be updated as new transfers arrive, existing transfers finish, links fail or their capacity changes, or transfers are terminated. To efficiently handle this highly dynamic situation, we assume a slotted timeline and periodically compute end-point transmission rates at the beginning of every timeslot. It is possible to schedule re-computation of rates upon highly critical events in addition to having them run periodically. In this dissertation, we only assume periodic execution of rate calculation for simplicity. Also, the transmission of any new transfer begins as soon as the rates are updated.

We assume that TES makes its optimization decisions given the knowledge of transfers that have already arrived. That is because we do not have deterministic information about transfers that may be created in the future. In general, it may be possible to predict future transfer arrivals and perform further optimizations accordingly, which is out of the scope of this dissertation.

\section{Central Inter-DC Traffic Management Architecture}
\begin{figure}[t!]
    \centering
    \includegraphics[width=0.9\textwidth]{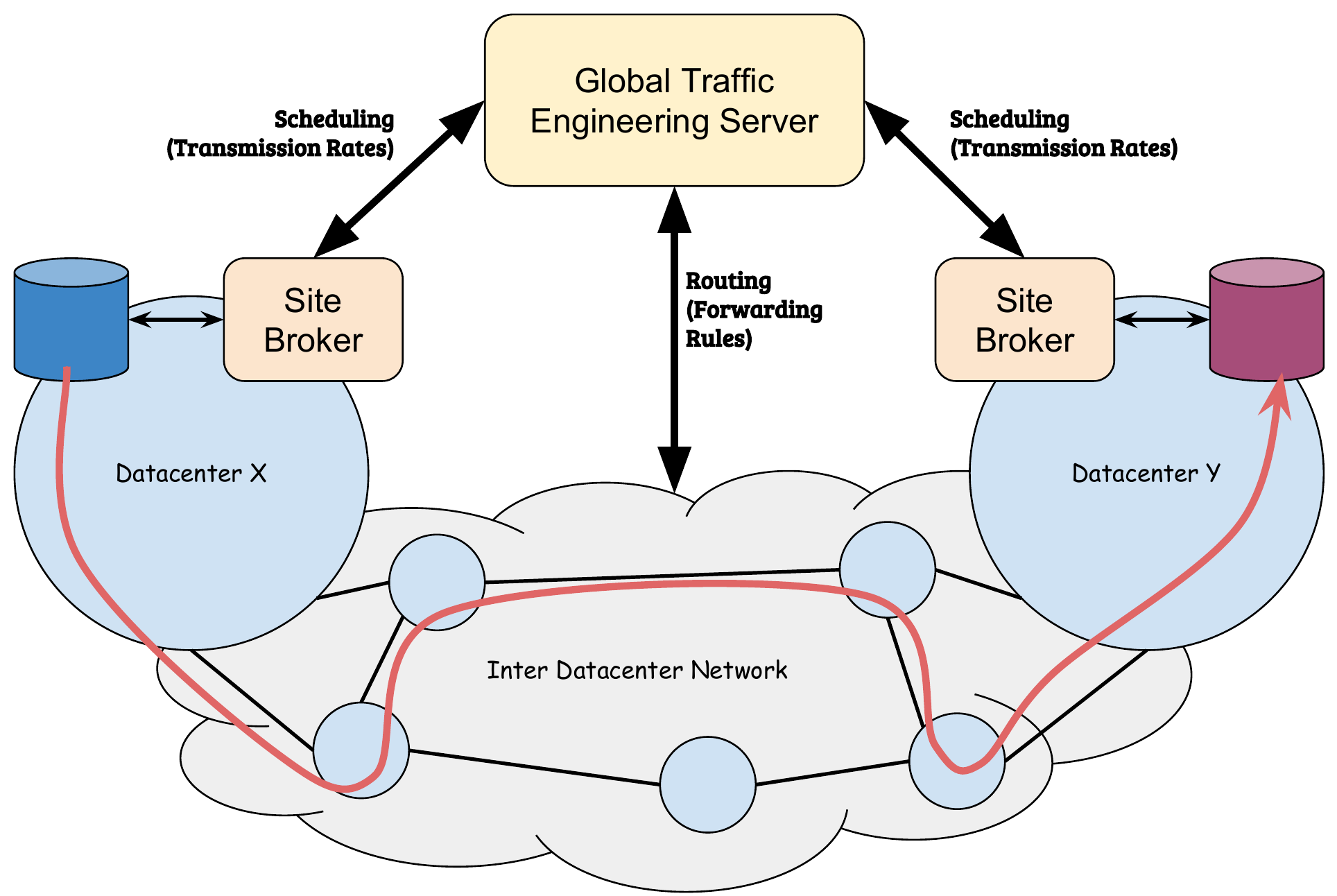}
    \caption{Central traffic management architecture.}
    \label{fig:PROBSETUP}
\end{figure}

\begin{figure}
    \centering
    \includegraphics[width=0.9\textwidth]{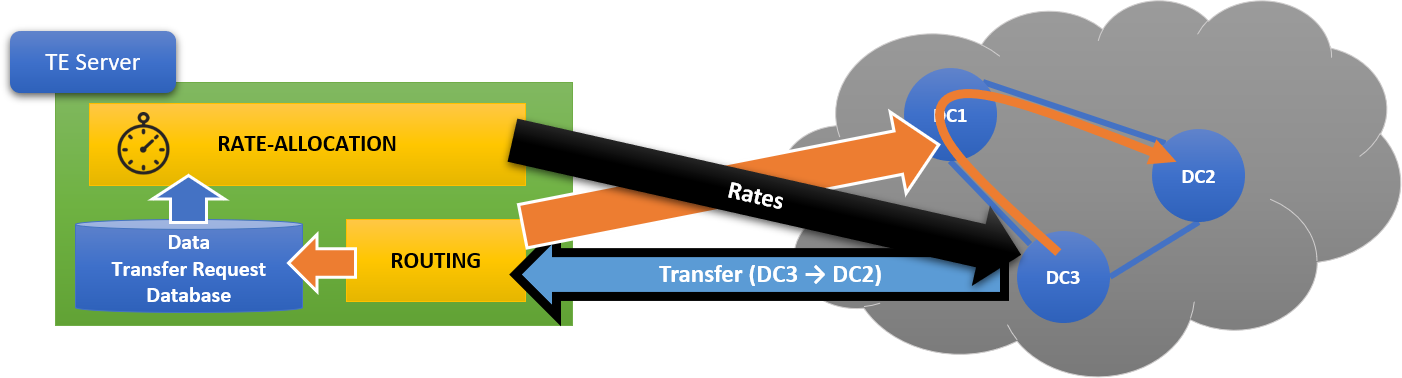}
    \caption{Steps in processing of a new inter-DC transfer.}
    \label{fig:central_process}
\end{figure}

Central network traffic management has two major elements: rate-limiting at the senders and routing/forwarding in the network. Figure \ref{fig:PROBSETUP} shows the overall setup for this purpose adopted by several existing inter-DC networks \cite{swan, bwe}. In this setup, TES calculates transmission rates and routes for submitted transfers as they arrive at the network. Rates are then dispatched to agents that are located at datacenters which are proxies that keep track of local transfers, i.e., transfers initiated within the same datacenters, called site brokers. When TES calculates new routes, they are dispatched to the network by implementing proper forwarding rules on the network's switching elements. Figure \ref{fig:central_process} shows the steps taken by the TES in processing a new inter-DC transfer. The part of the switching elements that does this is referred to as the Forwarding Information Base (FIB).

When a sender wants to initiate a transfer, it first communicates with the site broker in the local datacenter which records the request and forwards it to TES. When TES responds with the transmission rates, site broker records that and forwards it to the sender. The sender then applies rate-limiting at the rate specified by TES. In some setups, the sender should also attach the proper forwarding label to its packets so that its packets are correctly forwarded (like a VLAN tag). Such labeling may also be applied transparently to the sender at a different network entity (hypervisor, border gateway, etc.). This function could also be implemented at the datacenter network edge based on end-point addresses and using real-time packet header modification predicates.

\begin{figure}
    \centering
    \includegraphics[width=0.9\textwidth]{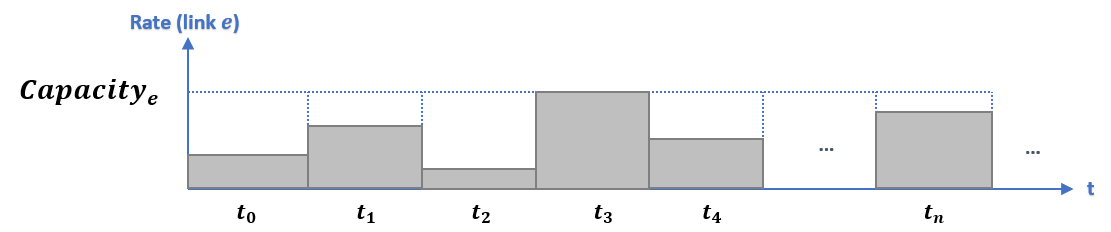}
    \caption{Rate-allocation per link per timeslot.}
    \label{fig:timeslots}
\end{figure}

In order to flexibly allocate traffic with varying rates over time, we break the timeline into small timeslots similar to several current solutions \cite{b4, swan, tempus, amoeba, owan}. Figure \ref{fig:timeslots} shows how this is done for a single link $e$. For a network, capacity is allocated over the whole network per timeslot. We do not assume an exact length for these timeslots as there are trade-offs involved. Having smaller timeslots can lead to inaccurate rate-limiting\footnote{It takes a short amount of time for senders to converge to new rates \cite{carousel}.} and adds the overhead of having to calculate rates for a larger number of timeslots, while having larger timeslots results in a less flexible allocation because the transmission rate is considered constant over a timeslot. Finally, timeslot length depends on transfer sizes. In general, we could select a value based on minimum or average transfer size. Current solutions have used a timeslot duration of $5$ minutes which is long enough to reduce the overhead of rate-computations and short enough to allow the network to adapt to changes in traffic demand \cite{amoeba, owan}.

The purpose of the site broker is manifold by adding one level of indirection between senders and TES. First, it reduces the request-response overhead for TES by maintaining a persistent connection with the server and possibly aggregating many sender requests into a smaller number of messages before sending them off to the server. Second, it allows for the application of hierarchical bandwidth allocation by locally grouping many transfers and presenting them to TES as one.\footnote{This may reduce the accuracy of traffic engineering but makes it significantly scalable in case there is a considerable number of transfers \cite{bwe}.} Finally, site broker can update TES's response according to varying local network conditions, allow senders to switch to a backup TES in case TES goes offline, or even revert to distributed mode.

Centralized traffic management can be realized using Software Defined Networking (SDN) \cite{sdn}. SDN offers many highly configurable features among which is the ability to manage traffic forwarding state centrally and programmatically by installing, updating, or removing forwarding rules in real-time. With a global view of network status and server demands, it is possible to offer globally optimal solutions. WANs operated using SDN have been adopted by an increasing number of companies and organizations over the past few years examples of which include Google \cite{b4}, Microsoft \cite{swan}, and Facebook \cite{facebook-express-backbone}. Of course, there are challenges in such centralized and real-time management of network, for example, routing update inconsistencies or the latency from when forwarding rules are dispatched to when they take effect are two significant issues. Ongoing SDN related research has been addressing these and several other problems \cite{automan, sdn_rule_latency}. In this dissertation, we consider the usage of SDN for controlling dedicated inter-DC networks. We develop algorithms that can be used by TES to compute routes on a per transfer basis as they arrive.

\subsection{Functions of Centralized Traffic Management}
\begin{figure}[t!]
    \centering
    \includegraphics[width=0.7\textwidth]{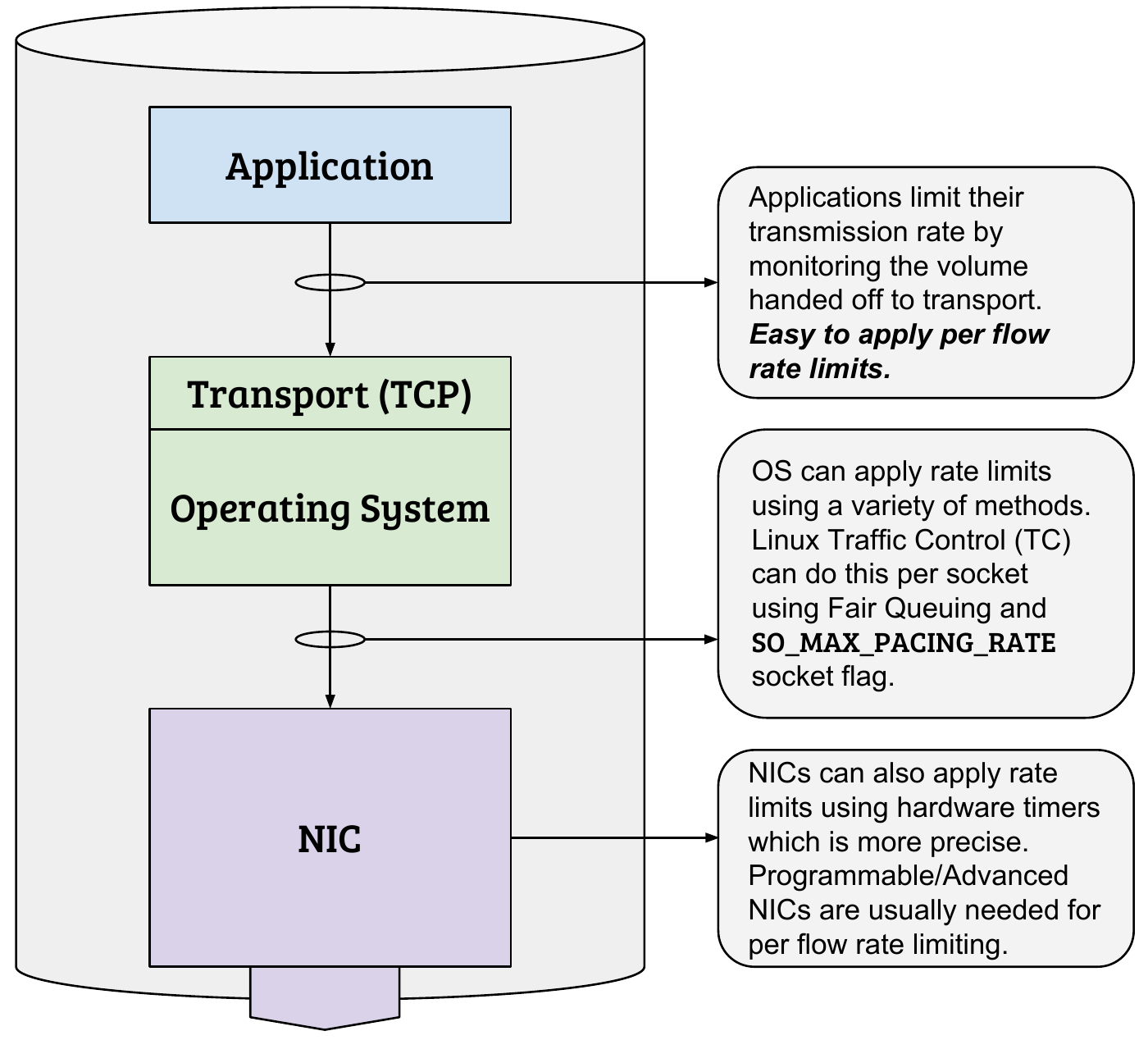}
    \caption{Several end-point rate-limiting techniques.}
    \label{fig:RATE}
\end{figure}

\noindent\textbf{Traffic Rate-limiting:} Figure \ref{fig:RATE} shows how rate-limiting can be applied at the servers before data is transmitted on the wire. The most straightforward approach is for service instances to communicate their demand with the local broker, which in turn makes contact with TES, and only hands off to the transport layer (i.e., socket) as much as specified by TES. This technique requires no changes to the end-points' protocol stack and hardware but requires modifications at the application layer. Another approach is to use the methods supported by the operating system for per-flow rate control. For example, the later versions of Linux allow users to use a socket option along with the Fair Queuing algorithm to specify a pacing rate. Next, it is possible to apply rate limiting in hardware using precise timers. This approach is much more accurate compared to software approaches but requires more sophisticated equipment. There are also hybrid approaches that use a combination of operating system support and hardware rate limiters to apply accurate per transfer rate limiting for a large number of transfers \cite{carousel}.

\vspace{0.5em}
\noindent\textbf{Traffic Routing:} Inter-DC networks are strong candidates for custom routing techniques. Effective routing should take into account the overall load scheduled on links to better use all available capacity while shifting traffic across a variety of paths. Besides, routing should consider the properties of new transfers while assigning routes to them. Conventional routing schemes are incapable of taking into account such parameters to optimize routing with regards to operator-specified utility functions.

\section{Performance Metrics}
A variety of metrics can be used for performance evaluation over inter-DC networks including transfer completion times, total network capacity consumption, and transfers completed before their deadlines. Depending on the services running over inter-DC networks, operators may choose to focus primarily on optimizing one metric or a utility function that generates an aggregate utility value according to all of these metrics. Table \ref{table_metrics} offers an overview of these metrics.

\begin{table}
\begin{center}
\caption{Definition of performance metrics.} \label{table_metrics}
\vspace{0.5em}
\begin{tabular}{ |l|p{8cm}| }
\hline
\textbf{Metric} & \textbf{Description} \\ \hline
Tail completion times & Completion time of the slowest transfer over the evaluation period. In some cases, 99th or 95th percentile may be used instead. \\ \hline
Median completion times & The completion time of the transfer that is slower than 50\% of transfers and faster than the other 50\% over the evaluation period. \\ \hline
Mean completion times & Average of completion times of all transfers over the evaluation period. \\ \hline
Total bandwidth/capacity consumed & Sum of the volume of traffic that was sent on all network edges per edge over the evaluation period. \\ \hline
Ratio of deadline transfers completed & Fraction of transfers the network was able to finish before their deadlines in case a deadline was specified. The network may apply admission control to only accept transfers that it can complete by their deadlines. In this case, we take the fraction of admitted transfers. \\ \hline
Ratio of deadline traffic completed & Ratio of the total volume of transfers the network was able to finish before their deadlines, in case a deadline was specified, by the total volume of transfers. The network may apply admission control to only accept transfers that it can complete by their deadlines. In this case, we take the ratio of admitted traffic by total submitted traffic. \\ \hline
Running time (Network algorithms) & The time to process transfer information, and compute transmission rates and forwarding routes. \\ \hline
\end{tabular}
\end{center}
\end{table}

In general, some of these metrics may be at odds with others, and therefore it may not be possible to optimize all parameters at the same time. The relationship between these metrics also could depend on the operating point of the system. For example, under light traffic load, using more bandwidth usually allows us to reduce the completion times of transfers, while under heavy traffic load, using more bandwidth potentially leads to resource contention and increased completion times.

One can consider two scenarios of transfers with and without deadlines. In the former case, we consider the performance metrics that evaluate the volume of traffic and the total fraction of transfers completed before their deadlines. In the latter case, we pay attention to minimizing tail, median or mean completion times. When deadlines are not present, depending on the services running over the inter-DC networks, we may more strongly consider tail, median or mean completion times.\footnote{It is possible to consider other aggregate metrics as well given the circumstances.} For example, in computing tasks that take multiple inputs from different datacenters, the processing start time depends on when all the inputs become available which increases the importance of reducing tail completion times.

Various data transfer problems considered in this dissertation are all traffic engineering problems over inter-DC networks aiming at optimizing one or more of the metrics stated above. To find efficient solutions to such problems, we can formulate optimization problems using the network and transfer parameters, and consider appropriate performance metrics to optimize. We will develop a general optimization framework in the next section.

\section{General Inter-DC Optimization Formulation}
The inter-DC optimization problem can be formulated in a variety of ways by considering different objective functions and constraints. In each problem, bulk inter-DC transfers will be initiated from one sender to one or more receiving datacenters. In the following, we discuss different types of constraints and objectives that can be combined to form the ultimate framework.

\vspace{0.5em}
\noindent\textbf{Definition of Variables:} Table \ref{table_var_0} shows the list variables used in this section. Data could be transmitted over paths or multicast trees to receivers. Also, in general, data can be transmitted over multiple parallel paths or multicast trees towards the receivers. The notations we defined capture these properties.

\begin{table}[t!]
\begin{center}
\caption{Definition of variables.} \label{table_var_0}
\vspace{0.5em}
\begin{tabular}{ |p{3cm}|p{10.5cm}| }
    \hline
    \textbf{Variable} & \textbf{Definition} \\
    \hline
    \hline
    $t$ and $t_{now}$ & Some timeslot and current timeslot \\
    \hline
    $\omega$ & Width of a timeslot in seconds \\
    \hline
    $e$ & A directed edge \\
    \hline
    $C_e$ & Capacity of $e$ in bytes per second \\
    \hline
    $B_e$ & Current available bandwidth on edge $e$ \\
    \hline
    $G$ & A directed graph representing an inter-DC network \\
    \hline
    $\pmb{\mathrm{E}}_G$ & Set$\langle\rangle$ of edges of directed graph $G$ \\
    \hline
    $\Psi$ & A directed subgraph over which traffic is forwarded to the receivers, could be a path or a multicast tree ($\pmb{\mathrm{E}}_{\Psi} \subset \pmb{\mathrm{E}}_{G}$) \\
    \hline
    $\pmb{\mathrm{R}}$ & Set of all requests (past, current, future) \\
    \hline
    $R_i$ & A transfer request where $R_i \in \pmb{\mathrm{R}},~i \in \pmb{\mathrm{I}}=\{1 \dots I\}$ \\
    \hline
    $S_{R_i}$ & Source datacenter of ${R_i}$ \\
    \hline
    $A_{R_i}$ & Arrival time of ${R_i}$ \\
    \hline
    $\tau_{R_i}$ & Completion time of ${R_i}$ \\
    \hline
    $t_{d_{R_i}}$ & Deadline of ${R_i}$ \\
    \hline
    $\Omega_{R_i}$ & Total network capacity consumed by ${R_i}$ for its completion \\
    \hline
    $\mathcal{V}_{R_i}$ & Original volume of ${R_i}$ in bytes \\
    \hline
    $\pmb{\mathrm{D}}_{R_i}$ & Set$\langle\rangle$ of destinations of $R_i$ \\
    \hline
    $\tau^{i}_{d}$ & Completion time of receiver $d \in \pmb{\mathrm{D}}_{R_i}$ \\
    \hline
    $\pmb{\mathrm{\psi}}^{i}_{d}$ & Directed subgraphs attached to receiver $d \in \pmb{\mathrm{D}}_{R_i}$ from $S_{R_i}$ \\
    \hline
    $f_{\Psi}^{i}(t)$ & Transmission rate of $R_{i}$ on subgraph $\Psi$ at timeslot $t$ \\
    \hline
    $\theta_{\Psi}^{e}$ & Whether edge $e \in \pmb{\mathrm{E}}_G$ is on subgraph $\Psi$ (binary variable) \\
    \hline
    $U$ & A network utility function set by network operators \\
    \hline
\end{tabular}
\end{center}
\end{table}


\vspace{0.5em}
\noindent\textbf{Formal Definition of Completion Times:} We defined a receiver's completion time as the last timeslot with non-zero traffic arriving at that receiver for a specific transfer.

\begin{equation}
\tau^{i}_{d} \triangleq \{t \vert f_{\Psi}^{i}(t) > 0, \exists \Psi \in \pmb{\mathrm{\psi}}^{i}_{d}\}, \forall d \in \pmb{\mathrm{D}}_{R_i}, \forall i \in \{1 \dots I\}
\end{equation}

For a transfer, the completion time is the time at which all receivers of that transfer complete.

\begin{equation}
\tau_{R_i} = \max_{d \in \pmb{\mathrm{D}}_{R_i}} \tau^{i}_{d}, \forall i \in \{1 \dots I\}
\end{equation}

\vspace{0.5em}
\noindent\textbf{Optimization Objective:} A variety of metrics can be considered as part of the optimization objective including transfer completion times (i.e., median, average, tail), total network capacity use, and the number of deadlines missed (or alternatively, number of transfers that could not be admitted to meet their deadlines). In general, a utility function can be defined over these metrics which the optimization problem aims to maximize. This function should be representative of how much profit the business can obtain while using the network.

\begin{equation}
Max ( U((\{\tau_{R_i}\}, \sum_{i} \Omega_{R_i}, \rvert \{i \vert \tau_{R_i} > t_{d_{R_i}}\} \lvert))),~i \in \{1 \dots I\}
\end{equation}

Examples of objective functions include: Minimizing the mean (i.e., average) transfer completion times, i.e., $Min ( \sum_{i \in \{1 \dots I\}} \tau_{R_i})$, minimizing the total network capacity consumption, i.e., $Min ( \sum_{i \in \{1 \dots I\}} \Omega_{R_i})$, minimizing the number of deadline missing transfers, i.e., $Min ( \rvert \{i \in \{1 \dots I\} \vert \tau_{R_i} > t_{d_{R_i}}\} \lvert)$ or a combination of these. For example, we can minimize a weighted sum of completion times and total network capacity consumption, i.e., $Min ( \sum_{i \in \{1 \dots I\}} \tau_{R_i} + \alpha \sum_{i \in \{1 \dots I\}} \Omega_{R_i})$ where $0 < \alpha \ll 1$ is a coefficient used to prioritize minimizing completion times. In all of these cases, $U$ is defined as a negative multiply of these functions. In other words, the network operator profits if these parameters are minimized.

\vspace{0.5em}
\noindent\textbf{Demand Constraints:} The total data transmitted towards a receiver across all the paths or multicast trees connected to it then has to be equal to the total volume of the transfer.

\begin{equation}
\sum_{t} \sum_{\Psi \in \pmb{\mathrm{\psi}}^{i}_{d}} \omega f_{\Psi}^{i}(t) = \mathcal{V}_{R_i}, \forall i \in \{1 \dots I\} \label{demand_const}
\end{equation}

\vspace{0.5em}
\noindent\textbf{Capacity Constraints:} Total transmission rate of all paths and multicast trees sharing an edge must be at most equal to the link's available bandwidth $B_e \le C_e$.

\begin{equation}
\sum_{\substack{i, \Psi\\e \in \pmb{\mathrm{E}}_{\Psi} \vert \Psi \in \{\cup_{d \in \pmb{\mathrm{D}}_{R_i}} \pmb{\mathrm{\psi}}^{i}_{d}\}}} f_{\Psi}^{i}(t) \le B_e, \forall t, e \in \pmb{\mathrm{E}}_{G}
\end{equation}

The available bandwidth on an edge is determined by the volume of traffic used up by short flows (e.g., user-facing, high priority traffic). There is usually a good estimate of how much such traffic is generated as the rate of growth for user traffic is far less than that of business-internal inter-DC transfers \cite{facebook-express-backbone}.

\vspace{0.5em}
\noindent\textbf{Routing Constraints:} To forward traffic from the source to each receiver per transfer, we can use one or more paths or trees. To make sure that each receiver is obtaining a full copy of the data, if any two receivers are connected using the same tree, any tree connected to one of them should also be connected to the other one. In other words, for some request $R_i$, receivers $\pmb{\mathrm{D}}_{R_i}$ can be separated into multiple groups $\pmb{\mathrm{D}}^{j}_{R_i}, j \le \lvert \pmb{\mathrm{D}}_{R_i} \rvert$ each connected using at least one path (i.e., $\lvert \pmb{\mathrm{D}}^{j}_{R_i} \rvert = 1$) or tree (i.e., $\lvert \pmb{\mathrm{D}}^{j}_{R_i} \rvert > 1$).

In general, it is possible to formulate the selection of such paths and trees as part of the optimization framework and create a joint routing and rate computation framework. This however leads to exponential number of constraints and addition of a large number of binary variables to the formulation which in general could take a long time to solve. Another approach would be to compute the paths and trees using some heuristic approach and plug them into the optimization framework which reduces the complexity of the problem allowing it to only focus on computation of the rates.

For the sake of completeness, we briefly discuss how a joint optimization can be formulated by adding constraints to the framework. This can be done by enumerating all possible paths (or trees) from the source to each group of receivers and considering fraction variables that determine how much of the traffic will end up on each path (tree). Also, since we do not know how to group receivers, we need to consider all possibilities and define binary variables that determine which grouping maximizes the utility of the network.

More formally, let us define binary variables $b_k$ as whether we have selected grouping $k \in \{1 \dots K\}$ where $K$ is the total number of ways to partition $\pmb{\mathrm{D}}_{R_i}$ into disjoint sets whose union is equal to $\pmb{\mathrm{D}}_{R_i}$. Also, let us define the groups in $k$\textsuperscript{th} partitioning as $\pmb{\mathrm{D}}^{j,k}_{R_i}, j \in \{1 \dots J\}$. We can write the following constraints:

\begin{align}
    & \sum_{k \in \{1 \dots K\}} b_k = 1 \\
    & \cup_{j \in \{1 \dots J\}} \pmb{\mathrm{D}}^{j,k}_{R_i} = \pmb{\mathrm{D}}_{R_i}, \forall k \in \{1 \dots K\}
\end{align}

Let us define $\pmb{\mathrm{\Psi}}^{j,k}_{i}$ as the set of all paths (trees) that connect $S_{R_i}$ to $\pmb{\mathrm{D}}^{j,k}_{R_i}$ over the inter-DC graph $G$. For every receiver $d \in \pmb{\mathrm{D}}_{R_i}$ we can then define the following constraint to find $\pmb{\mathrm{\psi}}^{i}_{d}$:

\begin{equation}
    \pmb{\mathrm{\psi}}^{i}_{d} = \cup_{\substack{k \in \{1 \dots K\}~\vert~b_k = 1 \\ ~~j \in \{1 \dots J\}~\vert~ d \in \pmb{\mathrm{D}}^{j,k}_{R_i}}} \pmb{\mathrm{\Psi}}^{j,k}_{i}
\end{equation}

The demand constraint of Eq. \ref{demand_const} will then automatically take into account the distribution of traffic across all the paths (trees) that connect to any group of receivers.






\vspace{0.5em}
\noindent\textbf{Hard Deadline Constraint:} A transfer $R_i$ with a hard deadline must complete before its deadline. We can formulate this as an equality of demand over the timeslots prior to the transfer's deadline.

\begin{equation}
\sum_{t \le t_{d_{R_i}}} \sum_{\Psi \in \pmb{\mathrm{\psi}}^{i}_{d}} \omega f_{\Psi}^{i}(t) = \mathcal{V}_{R_i}, \forall d \in \pmb{\mathrm{D}}_{R_i}
\end{equation}

The optimization problem with this constraint may become infeasible. That means the current parameters make it impossible to meet the given deadline. This process is referred to as admission control by performing feasibility checks. In general, fast heuristics exists that allow quick infeasibility checks, however, if a problem is not deemed infeasible by such heuristics, it does not guarantee feasibility.

\begin{figure}
    \centering
    \includegraphics[width=0.5\textwidth]{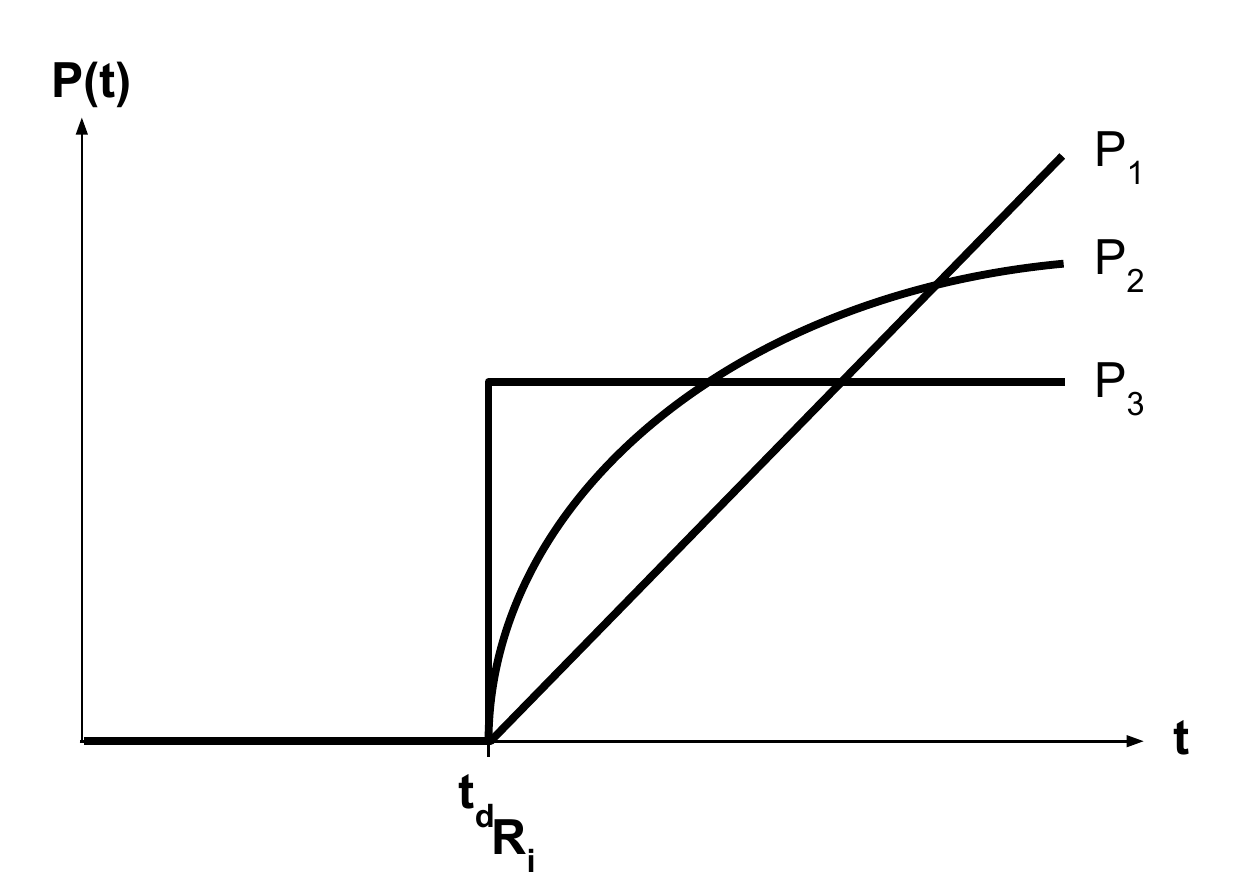}
    \caption{Some penalty functions}
    \label{fig:penalty}
\end{figure}

\vspace{0.5em}
\noindent\textbf{Soft Deadlines:} A soft deadline can be formulated as part of the objective function. Although soft deadlines are not the focus of this dissertation, we provide a short overview of how they can be modeled here. In general, we can use a penalty function that determines the benefit obtained from completing the transfer. In case the transfer is finished way too late, its value could be zero (or even negative as it wastes bandwidth). Here, we define two different penalty functions as shown in Figure \ref{fig:penalty}. These functions are specified according to how the system should handle a deadline miss. A step function, for example, determines that we highly value meeting the deadline, but as soon as a deadline is missed, it does not matter how late we complete the transfer. We define a variable $\gamma_{t}$ that determines how much traffic is delivered per timeslot for a transfer to a specific receiver.

\begin{equation}
\gamma^{i}_{t} \triangleq \sum_{\Psi \in \pmb{\mathrm{\psi}}^{i}_{d}} \omega f_{\Psi}^{i}(t), \forall t, i \in \{1 \dots I\}
\end{equation}

Using this new variable, we can define a system-wide penalty function that can be combined with the objective function in the optimization formulation.

\begin{equation}
\pmb{\mathrm{P}} \triangleq \sum_{i} \sum_{t} \gamma^{i}_{t} P(t)
\end{equation}

And the new objective function can be formulated as follows.

\begin{equation}
Max(U-\pmb{\mathrm{P}})
\end{equation}

\vspace{0.5em}
\noindent\textbf{Other Constraints:} There are many basic constraints such as the valid range of values for variables. In this case, we have the following two basic constraints.

\begin{align}
& \theta_{\Psi}^{e} \in \{0, 1\}, \forall \Psi, e \in \pmb{\mathrm{E}}_{G} \\
& 0 \le f_{\Psi}^{i}(t) \le \min_{e \in \pmb{\mathrm{E}}_{G} ~\vert~ \theta_{\Psi}^{e} = 1} B_e, \forall \Psi
\end{align}

Depending on transfer arrival rate and patterns, this optimization model can become more complex with many variables.\footnote{That is, due to the presence of binary or integer variables and non-linear constraints and objectives.} Solving this optimization framework may be computationally expensive and slow given that it needs to be solved as new transfers arrive. In case transfers have hard deadlines, it may be necessary only to admit new transfers when their deadlines can be met, which essentially requires performing feasibility checks before finding an optimal solution. To address the issue of complexity, throughout this dissertation, we present, implement, and evaluate heuristics that help find quick solutions to different versions of this optimization framework.

\clearpage
\chapter{Adaptive Routing of Transfers over Inter-Datacenter Networks} \label{chapter_adaptive_routing}
Inter-DC networks carry traffic flows with highly variable sizes and different priority classes: long throughput-oriented flows and short latency-sensitive flows. While latency-sensitive flows are almost always scheduled on shortest paths to minimize end-to-end latency, long flows can be assigned to paths according to usage to maximize average network throughput. Long flows contribute huge volumes of traffic over inter-DC WAN. The Flow Completion Time (FCT) is a vital network performance metric that affects the running time of distributed applications and users' quality of experience. Adaptive flow routing can improve efficiency and performance of networks by assigning paths to new long flows according to network status and flow properties. We focus on single path routing while aiming at minimizing completion times and bandwidth usage of internal flows. 

In this chapter, we first discuss a popular adaptive approach widely used for traffic engineering that is based on current bandwidth utilization of links. We propose an alternative that reduces bandwidth usage by up to at least $50\%$ and flow completion times by up to at least $40\%$ across various scheduling policies and flow size distributions. Next, we propose a routing approach that uses the remaining sizes and paths of all ongoing flows to minimize the worst-case completion time of incoming flows assuming no knowledge of future flow arrivals. Our approach can be formulated as an NP-Hard graph optimization problem. We propose \name, a heuristic to quickly generate an approximate solution. We evaluate \name~against several real WAN topologies and two different traffic patterns. We see that \name~provides solutions with an average optimality gap of less than $0.25\%$. Furthermore, we show that compared to other popular routing heuristics, \name~reduces the mean and tail FCT by up to $3.5\times$ and $2\times$, respectively. We then present and evaluate an even faster heuristic called \namefast~which is based on Dijkstra's shortest path algorithm. We perform extensive evaluations to compare \name~and \namefast~to show that they offer relatively similar performance over multiple topologies, scheduling policies, and flow size distributions despite \namefast~being considerably faster and more straightforward.

\section{Background and Related Work}
Although adaptive path selection can be formulated as an online optimization problem, such problems cannot be solved optimally due to no knowledge about future flow arrivals. Alternatively, heuristic schemes can be used by considering a cost (distance) metric and selecting the minimum cost (shortest) path. A variety of metrics have been used for path selection over WAN including static metrics such as hop count and interface bandwidth, and dynamic metrics such as end-to-end latency which is a function of propagation and queuing latency, and current link bandwidth utilization \cite{tvlakshman, routing-metric}. Especially, bandwidth utilization has been extensively used by prior work over inter-DC networks \cite{ospf-is_is, texcp, tempus}.

Our understanding is that while these metrics are effective for routing of short flows, they are insufficient for improving the completion times of long flows as we will demonstrate. Over inter-DC WAN where end-points are managed by the organization that also controls the routing \cite{b4, swan-backbone, facebook-express-backbone}, one can use routing techniques that differentiate long flows from short flows and use flow properties obtained from applications, including flow size information, to reduce the completion times of long flows.

\subsection{A Novel Metric for Adaptive Routing over WAN}
We argue that while assigning paths to new flows, instead of focusing on current bandwidth utilization, \textit{one should consider utilization temporally and into the future,} i.e., by counting total outstanding bytes to be sent per link according to paths assigned to flows and total outstanding bytes per flow. We refer to this total number of remaining bytes per link as its \textit{load} and use it as the cost metric. Compared to utilization, load offers more information about future usage of a link's bandwidth which can help us perform more effective load balancing. Every time a flow is assigned to a path, load variables associated with all edges of that path increase by its demand. Also, a link's load variable decreases continuously as flows on that link make progress.

In addition, we evaluate two heuristics of selecting the path with minimum value of maximum link cost and minimum value of sum of link costs which we refer to as \texttt{MINMAX()} and \texttt{MINSUM()}, respectively. Although the former is frequently used in the literature \cite{ospf-is_is, texcp, tempus}, we find that the latter offers considerably better performance for the majority of traffic patterns and scheduling policies.


\begin{figure*}[t]
\centering

\includegraphics[width=\textwidth]{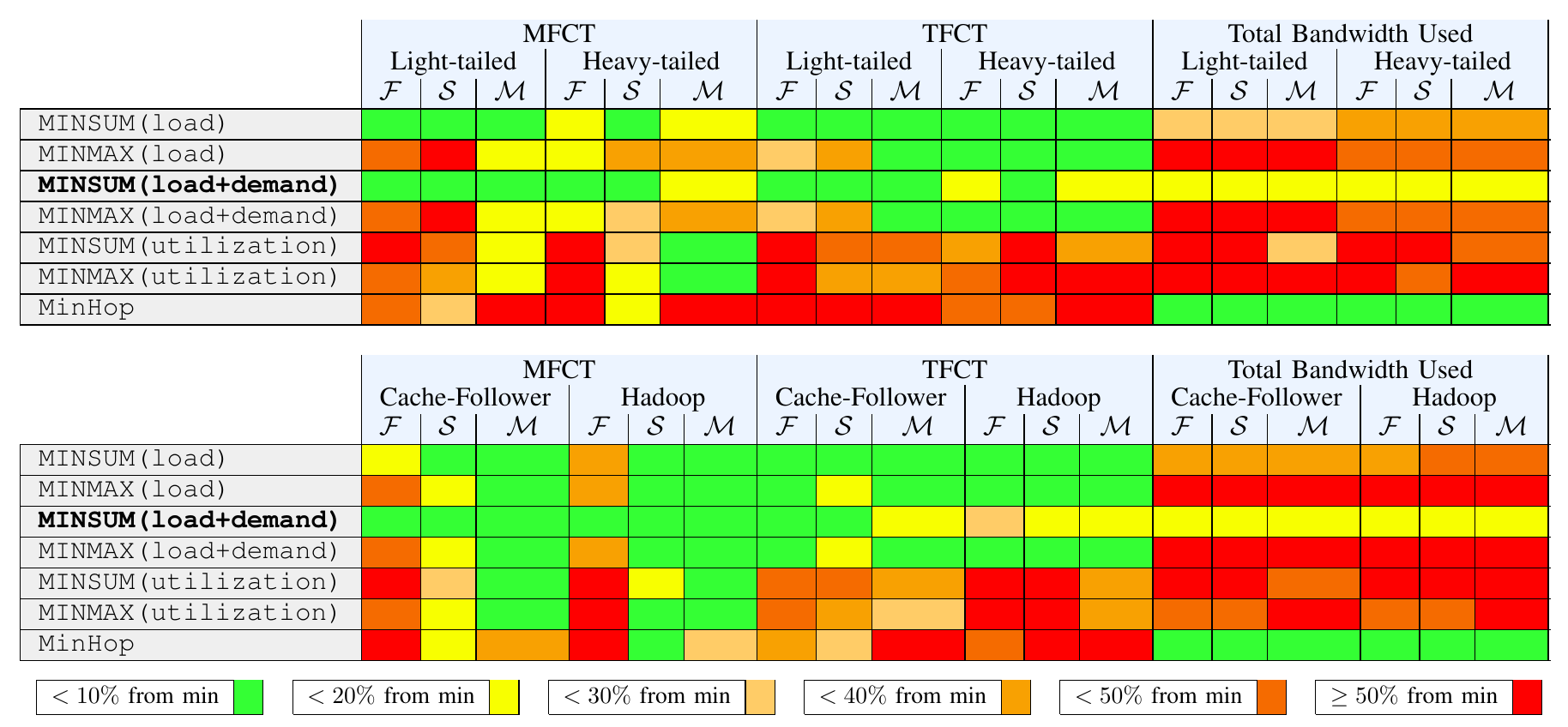}


\caption{Performance of various cost metrics for path selection over Cogent WAN \cite{cogent}, with uniform capacity of $1$ and $\lambda = 1.0$ ($\mathcal{F}$, $\mathcal{S}$ and $\mathcal{M}$ represent the FCFS, SRPT and MMF scheduling policies, respectively), simulation was repeated many times and average was computed. The minimum was computed per column and per metric across all schemes in the column. MFCT and TFCT represent the mean and tail flow completion times, respectively.} \label{fig:cogent}
\end{figure*}

\section{Evaluation of Different Cost Metrics}
We considered a large WAN called Cogent \cite{cogent} with $197$ nodes and $243$ links, four flow demand distributions of light-tailed (Exponential distribution), heavy-tailed (Pareto distribution), Cache-Follower \cite{social_inside} and Hadoop \cite{social_inside} (the last two happen across Facebook datacenters), and a uniform capacity of $1.0$ for all links. A Poisson distribution with rate $\lambda$ was used for flow arrivals. For all flow demand distributions, we assumed an average of $20$ units and a maximum of $500$ units. For heavy-tailed, we used a minimum demand of $2$ units. We considered scheduling policies of First Come First Serve (FCFS), Shortest Remaining Processing Time (SRPT) and Fair Sharing using Max-Min Fairness (MMF). We considered three different cost metrics of ``utilization", ``load", and ``load+demand" per link where demand represents the new flow's size in bytes. To measure a path's cost, we considered two cost functions of \textit{maximum} which assigns any path the cost of its highest cost link (used by \texttt{MINMAX()} heuristic), and \textit{sum} which computes a path's cost by summing up costs of its links (used by \texttt{MINSUM()} heuristic). Combining these path cost functions with the three link cost metrics mentioned above, we obtain six different path selection schemes that select the path with minimum cost for a newly arriving flow. We also considered \texttt{MinHop} which selects a path with minimum hops per flow to compute lower bound of bandwidth usage. For minimum cost path selection, we used Dijkstra's algorithm in JGraphT library. We measured Mean and Tail Flow Completion Times (MFCT/TFCT) and total bandwidth as shown in Figure \ref{fig:cogent}.

\vspace{0.5em}
\noindent\textbf{Flow Completion Times (FCT):} \texttt{MINSUM(load)} and \texttt{MINSUM(load+demand)} perform almost identically in completion times. The rest of schemes offer highly varying performance dictated by scheduling policy or traffic pattern. Schemes based on utilization are at least $40\%$ above the minimum for the majority of scenarios. Also, \texttt{MINMAX(load)} and \texttt{MINMAX(load+demand)} are more than $50\%$ above the minimum in mean completion times for multiple scenarios. Overall, it can be seen that schemes based on ``load" as link cost offer much better tail completion times (less than $10\%$ away from minimum for majority of cases). Also, \texttt{MINSUM(load+demand)} offers the best mean completion times considering all scenarios.

\vspace{0.5em}
\noindent\textbf{Total Bandwidth Usage:} \texttt{MINSUM(load+demand)} offers the minimum extra bandwidth usage compared to \texttt{MinHop} which is below $20\%$ at all times. Schemes based on \texttt{MINMAX()} consume at least $40\%$ extra bandwidth. \texttt{MINSUM(load)} and \texttt{MINSUM(utilization)} use at least $10\%$ more bandwidth at all times compared to \texttt{MINSUM(load+demand)} and at least $20\%$ more bandwidth for the majority of scenarios.

\section{Discussion and Analysis}
We see that \texttt{MINSUM(load+demand)} stays within $20\%$ of minimum for all completion times and within $10\%$ of minimum in the majority of cases. It offers the minimum bandwidth usage across all adaptive approaches (\texttt{MinHop} is static). With this cost metric, larger flows are most likely assigned shorter paths which allows for higher bandwidth savings (due to presence of ``demand" as part of link cost) while shorter flows are assigned to paths with smaller total load which reduces completion times via load balancing. We believe \texttt{MINSUM(load+demand)} performs better than techniques based on \texttt{MINMAX()} since it considers total number of bytes that will eventually be scheduled on a path taking into account all edges and not just the highest loaded/utilized link. Our experiments have shown that \texttt{MINSUM(load+demand)} is also an effective metric for selection of multicast forwarding trees that reduce completion times via load balancing \cite{dccast, quickcast}. It is also interesting to note that \texttt{MINMAX(utilization)}, which is frequently used in traffic engineering research, is far from the best solution for the majority of evaluated scenarios. Centralized frameworks, such as SDN \cite{sdn}, are good candidates for realization of this scheme since they offer access to global view of network status and flow demands.

\section{Best Worst-case Routing (BWR)}
Given the results of the experiments we performed above, it is obvious that current routing heuristics can be far from the optimal over different evaluation scenarios and for various performance metrics. Therefore, we revisit the well-known flow routing problem over inter-DC networks. As mentioned earlier, we focus on long flows which carry tremendous volumes of data over inter-DC networks \cite{b4, tempus}. They are usually generated as a result of replicating large objects such as search index files, virtual machine migration, and multimedia content. For instance, over Facebook's Express Backbone, about $80\%$ of flows for cache applications take at least 10 seconds to complete \cite{social_inside}. Besides, the volume of inter-DC traffic for replication of content and data, which generates many long flows, has been growing at a fast pace \cite{facebook-express-backbone}.

In general, flows are generated by different applications at unknown times to move data across the datacenters. Therefore, we assume that flows can arrive at the inter-DC network at any time and no knowledge of future flow arrivals. Every flow is specified with a source, a destination, an arrival time, and its total volume of data. The Flow Completion Time (FCT) of a flow is the time from its arrival until its completion.

We focus on minimizing the completion times of long flows which is a critical performance metric as it can significantly affect the overall application performance or considerably improve users' quality of experience. For example, in cloud applications such as Hadoop, moving data faster across datacenters can reduce the overall data processing time. As another example, moving popular multimedia content quickly to a regional datacenter via replication allows improved user experience for many local users. To attain this goal, routing and scheduling need to be considered together which can lead to a complex discrete optimization problem. Here, we only address the routing problem, that is, choosing a fixed path for an incoming flow given the network topology and the currently ongoing flows while making no assumptions on the traffic scheduling policy. We focus on single path routing which mitigates the undesirable effects of packet reordering.

Assuming no knowledge of future flow arrivals and no constraints on the network traffic scheduling policy, we propose to minimize the worst-case completion time of every incoming flow given the network topology, the currently ongoing flows' paths, and their remaining number of data units. For any given scheduling policy, we route the flows to minimize the worst-case flow completion time. We refer to this routing approach as the Best Worst-case Routing (BWR).
    
    

\subsection{System Model} \label{model}
We consider a general network topology with bidirectional links and equal capacity of one for all edges and assume an online scenario where flows arrive at unknown times in the future and are assigned a fixed path as they arrive. Each flow is divided into many equal size pieces (e.g., IP datagrams) which we refer to as data units. We also assume knowledge of the flow size (i.e., number of a flow's data units) for the new flow and the remaining flow size for all ongoing flows. Given an index $i$, every flow $F_i$ is defined with a source $s_i$, a destination $t_i$, an arrival time $\alpha_i$, and a total volume of data $\mathcal{V}_i$. In addition, each flow is associated with a path $P_i$, a finish time $\beta_i$ which is the time of delivery of its last data unit, and a completion time $c_i = \beta_i - \alpha_i$. Finally, at any moment, the total number of remaining data units of $F_i$ is $\mathcal{V}^r_i \le \mathcal{V}_i$.

Similar to multiple existing inter-DC networks \cite{b4, facebook-express-backbone, swan-backbone}, we assume the availability of logically centralized control over the network routing. A controller can maintain information on the currently ongoing long flows with their remaining data units and perform routing decisions for an incoming long flow upon arrival.

We employ a slotted timeline model where at each timeslot a single data unit can traverse any path in the network. In other words, we assume a zero propagation and queuing latency which we justify by focusing only on long flows. Given this model, if multiple flows have a shared edge, only one of them can transmit during a timeslot. We say two data units are competing if they belong to flows that share a common edge. Depending on the scheduling policy that is used, these data units may be sent in different orders but never at the same time. Also, if two flows with pending data units use non-overlapping paths, they can transmit their data units at the same time if no other flow with a common edge with either one of these flows is transmitting at the same timeslot.

\subsection{Definition of Best Worst-case Routing}
We aim to reduce long flows' completion times with no assumption on the scheduling policy for transmission of data units. To achieve this goal, we propose the following routing technique referred to as Best Worst-case Routing (BWR):

\vspace{0.5em}
\textbf{Problem 1.} \textit{Given a network topology $G(V,E)$ and the set of ongoing flows $\pmb{\mathrm{F}}=\{F_i, 1 \le i \le N\}$, we want to assign a path $P_{N+1}$ to the new flow $F_{N+1}$ so that the worst-case completion time of $F_{N+1}$, i.e., $\max(c_{N+1})$ is minimized.}

\vspace{0.5em}
Assuming no knowledge of future flows and given the described network model, since only a single data unit can get through any edge per timeslot, the worst-case completion time of a flow happens when the data units of all the flows that share at least one edge with the new flow's path go sequentially and before the last data unit of the new flow is transmitted. Therefore, Problem 1 can be reduced to the following graph optimization problem which aims to minimize the number of competing data units with $F_{N+1}$. 

\vspace{0.5em}
\textbf{Problem 2.} \textit{Given a network topology $G(V,E)$ where every edge $e \in E$ is associated with a set of flows $\pmb{\mathcal{\mathrm{F}}}_e$ (that is, $e \in P_{i}, \forall F_i \in \pmb{\mathcal{\mathrm{F}}}_e$), the set of ongoing flows $\pmb{\mathrm{F}}=\{F_i, 1 \le i \le N\}$, and an incoming flow $F_{N+1}$, we want to find a minimum weight path $P_{N+1}$ where the weight of any path $P$ from $s_{N+1}$ to $t_{N+1}$ is computed as follows:}

\begin{equation}
    W_{P} = \sum_{\{1 \le i \le N ~\vert~ F_i~\in~\{\cup_{e \in P}~\pmb{\mathcal{\mathrm{F}}}_e\}\}} \mathcal{V}^r_i \label{eq1}
\end{equation}

\vspace{0.5em}
\textbf{Proposition 1.} \textit{Assuming no knowledge of future flow arrivals, $P_{N+1}$ selected by solving Problem 2 minimizes the worst-case completion time of $F_{N+1}$ regardless of the scheduling policy used for transmission of data units.}

\vspace{0.5em}
\textbf{\textit{Proof.}} $P_{N+1}$ is chosen to minimize the maximum number of data units ahead of $F_{N+1}$ given the knowledge of ongoing flows' remaining data units which minimizes the worst-case $\beta_{N+1}$, that is, the maximum number of timeslots the last data unit of $F_{N+1}$ has to wait before it can be sent. Since $\alpha_{N+1}$ is fixed, this minimizes $\max(c_{N+1})$.

\begin{figure}[t]
    \centering
    \includegraphics[width=0.7\columnwidth]{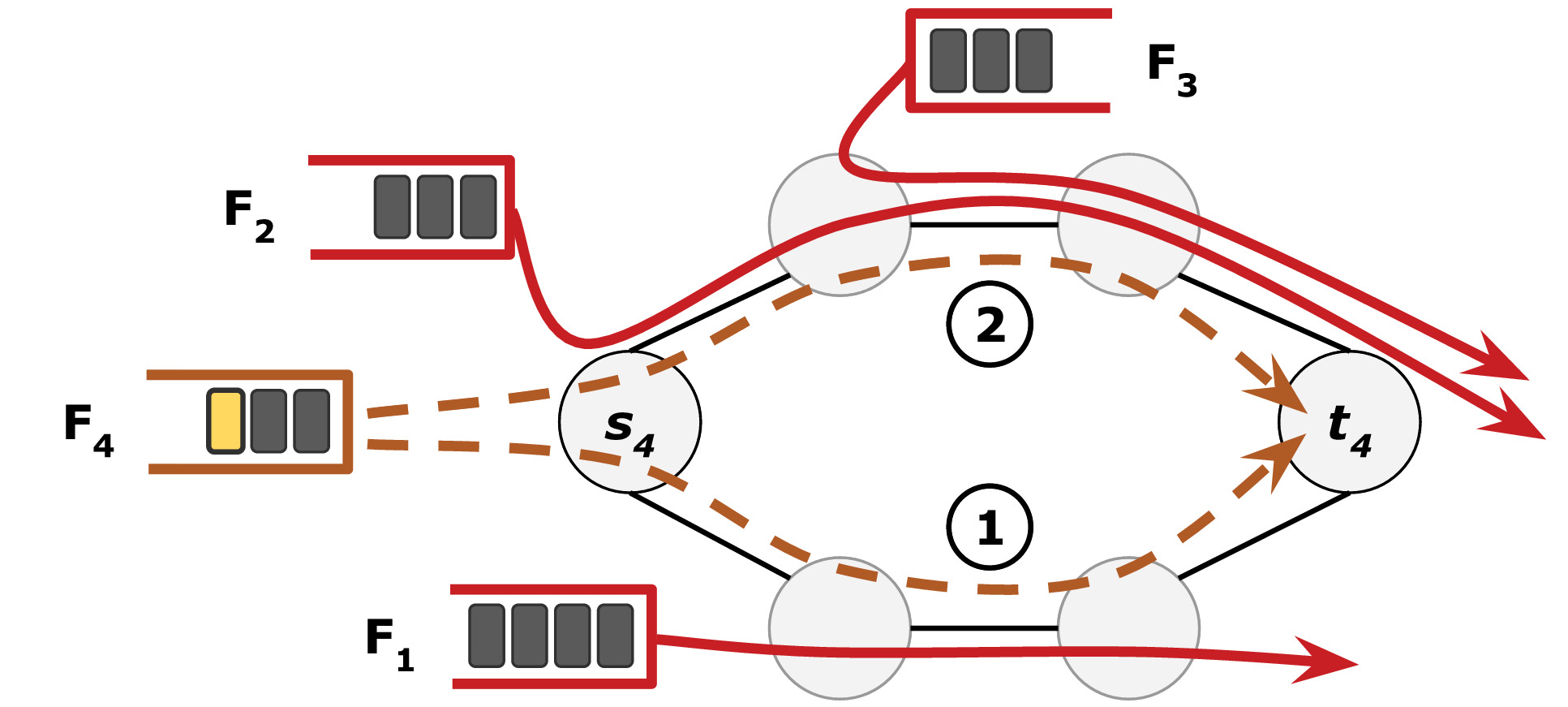}
    \caption{Example of routing a new flow $F_4$} \label{fig:example}
\end{figure}

\vspace{0.5em}
\textbf{Example:} Consider the scenario shown in Figure \ref{fig:example}. A new flow $F_4$ with 3 data units has arrived and has two options of sharing an edge with $F_1$ that has 4 remaining data units (path 1) or sharing edges with $\{F_2, F_3\}$ which have a total of 6 remaining data units (path 2). Our approach tries to minimize the worst-case completion time of $F_4$ given ongoing flows. If path 1 is chosen, the worst case completion time of $F_4$ will be 7 while with path 2 it will be 9 and therefore, the logically centralized network controller will select path 1 for $F_4$. The worst-case completion times are not affected by the scheduling policy and are independent of it. Also, the fact that $F_2$ has three common edges with path 2 and $F_3$ has two common edges with path 2 does not affect the worst-case completion time of $F_4$ on path 2.

\subsection{BWR Heuristic (\name)}
The path weight assignment used in Problem 2 is not edge-decomposable. Finding a minimum weight path for $F_{N+1}$ is NP-Hard and requires examining all paths from $s_{N+1}$ to $t_{N+1}$.\footnote{Please see Appendix \ref{chapter_bwr_hardness} for proof.} We propose a fast heuristic here, called \name, that finds an approximate solution to Problem 2. Algorithm \ref{bwrh} shows our proposed approach to finding a path $P_{N+1}$ for $F_{N+1}$. At every iteration, the algorithm finds the minimum weight path from $s_{N+1}$ to $t_{N+1}$ with at most $K$ hops by computing the weight of every such path according to Eq. \ref{eq1}. The algorithm starts by searching all the minimum hop paths from $s_{N+1}$ to $t_{N+1}$ and finding the weight of the minimum weight path among such paths. It then increases the number of maximum hops allowed (i.e., $K$) by one, extending the search space to more paths. This process continues until the weight of the minimum weight path with at most $K$ hops is the same as $K-1$, i.e., there is no gain while increasing the number of hops.

The termination condition used in \name~may prevent us from searching long paths. Therefore, if the optimal path is considerably longer than the minimum hop path, it is possible that the algorithm terminates before it reaches the optimal path. Let us call the optimal path $P_o$ and the path selected with our heuristic $P_h$. The optimality gap, defined as $\frac{W_{P_h} - W_{P_{o}}}{W_{P_{o}}}$, is highly dependant on the number of remaining data units of ongoing flows. We find that the worst-case optimality gap can be generally unbounded. However, it is highly unlikely, in general, for the optimal path to be long as having more edges increases the likelihood of sharing edges with more ongoing flows which increases the weight of the path. We will later confirm this intuition through empirical evaluations and show that \name~provides solutions with an average optimality gap of less than a quarter of percent.

\SetAlgoVlined
\begin{algorithm}[t]
\caption{\name} \label{bwrh}
\SetKw{KwBy}{by}
\SetKwProg{FindPath}{FindPath}{}{}

\vspace{0.4em}
\KwIn{$F_{N+1}$, $G(V,E)$, $P_i,\mathcal{V}^r_i,1 \le i \le N$}

\vspace{0.4em}
\KwOut{$P_{N+1}$}

\nonl\hrulefill

\vspace{0.4em}
$K \gets$ $\#$hops on the minimum hop path from $s_{N+1}$ to $t_{N+1}$\;

\vspace{0.4em}
$W_{min}^K \gets$ Weight of the minimum weight path from $s_{N+1}$ to $t_{N+1}$ with at most $K$ hops by examining all such paths\;

\vspace{0.4em}
\Repeat{$W_{min}^{K} \ge W_{min}^{K-1}$}{
    \vspace{0.4em}
    $K \gets K+1$\;
    
    \vspace{0.4em}
    Compute $W_{min}^K$\;
    
    \vspace{0.4em}
}

\vspace{0.4em}
$P_{N+1} \gets$ The minimum weight path from $s_{N+1}$ to $t_{N+1}$ with at most $K-1$ hops (if multiple minimum weight paths exist, choose the one with minimum hops)\;
\end{algorithm}

\subsection{Application to Real Network Scenarios}
We discuss how \name~can be used to find a path for an incoming flow on a real network assuming a uniform link capacity. We can use the same topology as the actual topology as input to \name. Since we focus on long flows for which the transmission time is significantly larger than both propagation and queuing latency along existing paths, it is reasonable to ignore their effect in routing (hence the assumption that these values are zero in \S \ref{model}). Next, assuming that all data units are of the same size, we can use the total number of remaining bytes per ongoing flow in place of the number of remaining data units as it does not affect the selected path. In practice, some data units may be smaller than the underlying network's MTU, which for the long flows with many data units, has minimal effect on the selected path. Once \name~selects a path, the network's forwarding state is updated accordingly to route the new flow's traffic, for example, using SDN \cite{b4, tempus}.

In general, network traffic is a mix of short and long flows. Since our dissertation targets the long flows, routing of short flows will not be affected and could be done considering the propagation and queuing latency. Incoming long flows can be routed according to the knowledge of current long flows while ignoring the effect of short flows. 

\subsection{Evaluations} \label{bwrh_eval}
We considered two flow size distributions of light-tailed (Exponential) and heavy-tailed (Pareto) and considered Poisson flow arrivals with the rate of $\lambda$. We also assumed an average flow size of $\mu$ data units with a maximum of 500 data units along with a minimum size of 2 data units for the heavy-tailed distribution. We considered the scheduling policies of First Come First Serve (FCFS), Shortest Remaining Processing Time (SRPT) and Fair Sharing based on max-min fairness \cite{max-min-fairness}.

\vspace{0.5em}
\noindent\textbf{Topologies:} We used GScale \cite{b4} with 12 nodes and 19 edges, AGIS \cite{agis} with 25 nodes and 30 edges, ANS \cite{ans} with 18 nodes and 25 edges, AT\&T North America \cite{att} with 25 nodes and 56 edges, and Cogent \cite{cogent} with 197 nodes and 243 edges. We assumed bidirectional edges with a uniform capacity of 1 data unit per time unit for all of these topologies. 

\vspace{0.5em}
\noindent\textbf{Schemes:} We considered three schemes besides \name. The \textit{Shortest Path (Min-Hop)} approach simply selects a fixed shortest hop path from the source to destination per flow. The \textit{Min-Max Utilization} approach selects a path that has the minimum value of maximum utilization across all paths going from the source to the destination. This approach has been extensively used in the traffic engineering literature \cite{tvlakshman, tempus}. The \textit{Shortest Path (Random-Uniform)} selects a path randomly with equal probability across all existing paths which are at most one hop longer than the shortest hop path.

\begin{figure}[b]
    \centering
    \includegraphics[width=0.8\columnwidth]{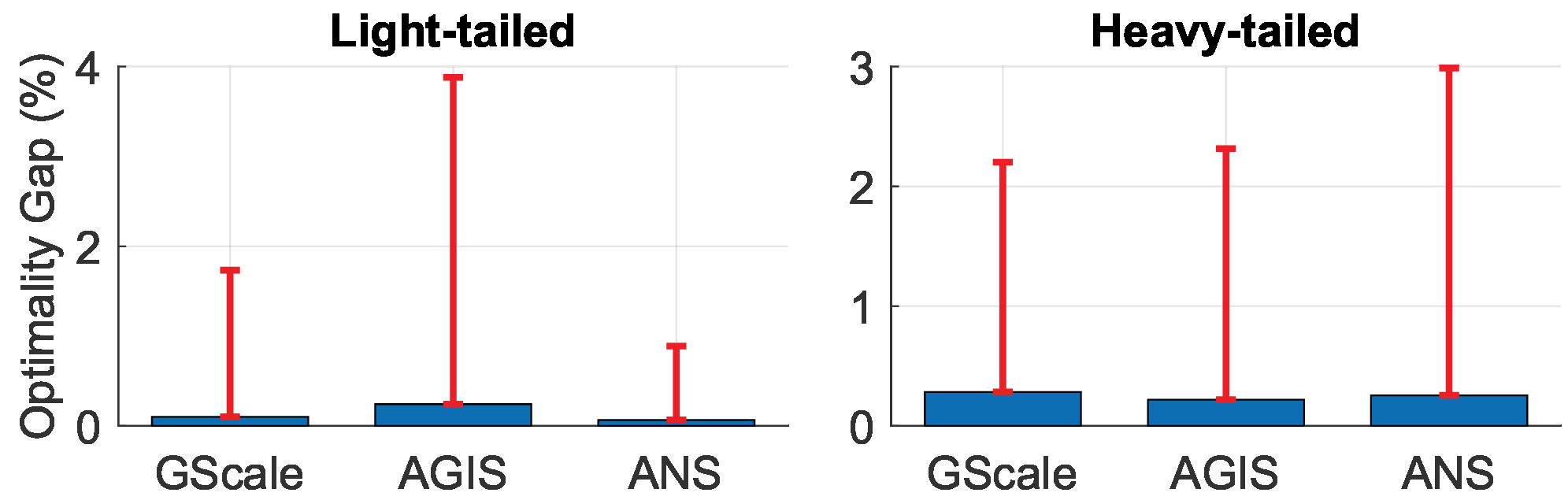}
    \caption{\name's optimality gap for $\lambda = 10$ and $\mu = 50$ computed for 1000 flow arrivals.} \label{fig:exp0}
\end{figure}


\begin{figure}[p]
    \centering
    \subfigure[AT\&T Topology \cite{att}]
    {
        \includegraphics[width=0.83\textwidth]{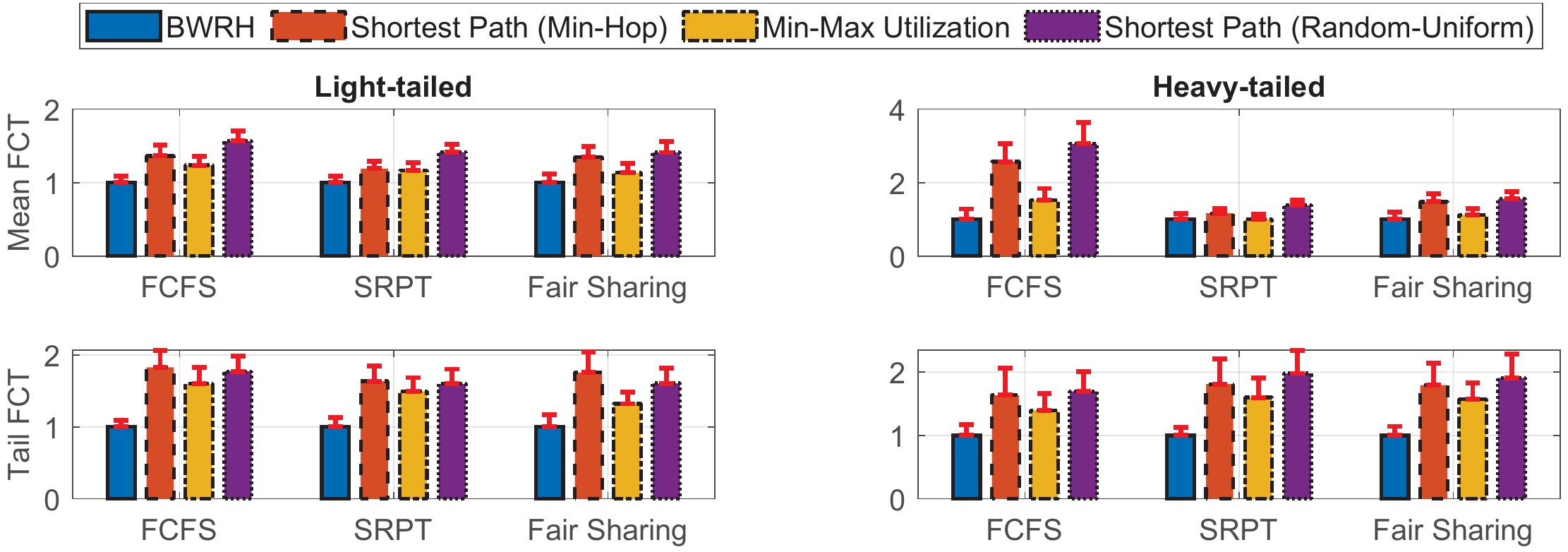}
    }
    \\
    \subfigure[Cogent Topology \cite{cogent}]
    {
        \includegraphics[width=0.83\textwidth]{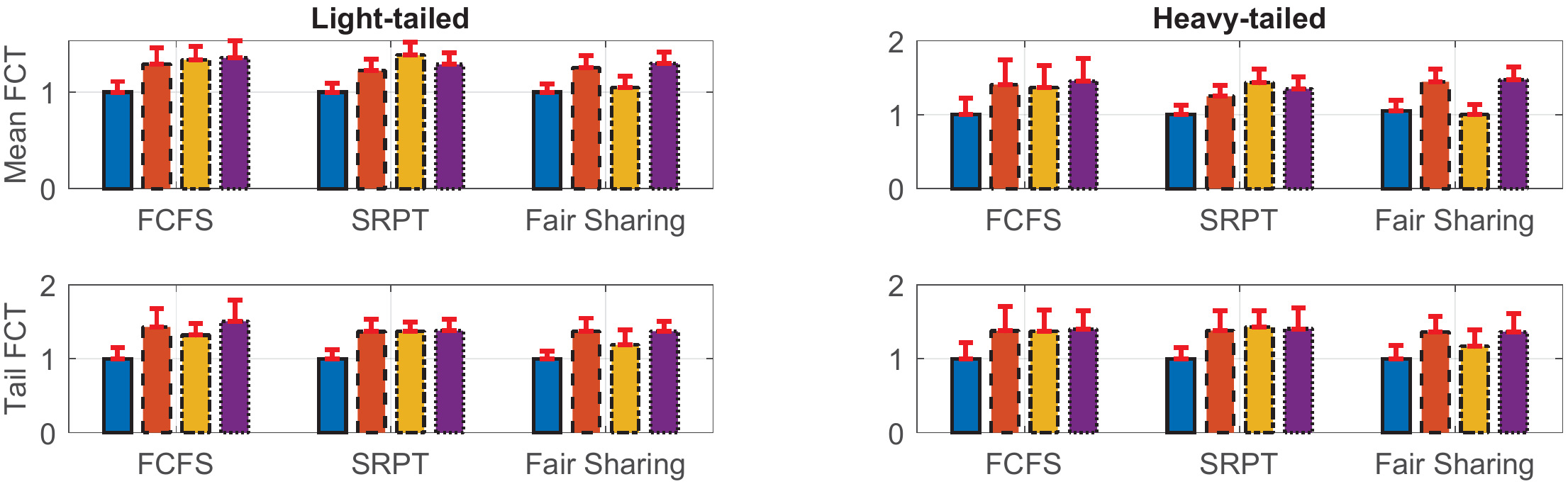}
    }
    \\
    \subfigure[GScale Topology \cite{b4}]
    {
        \includegraphics[width=0.83\textwidth]{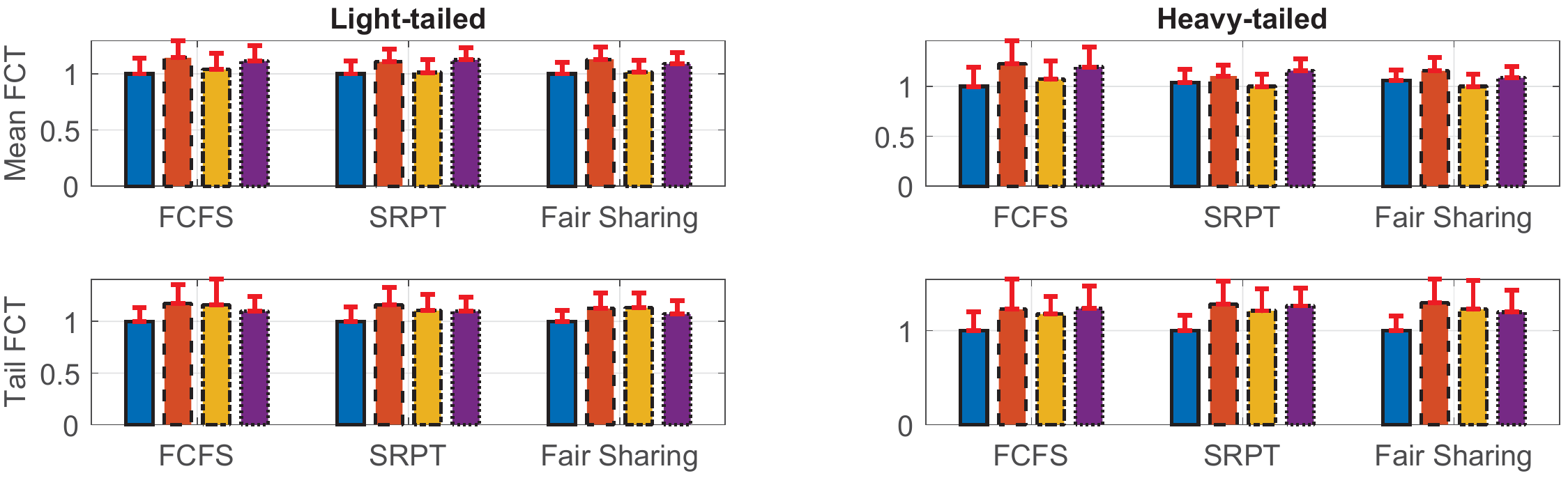}
    }
    \\
    \subfigure[ANS Topology \cite{ans}]
    {
        \includegraphics[width=0.83\textwidth]{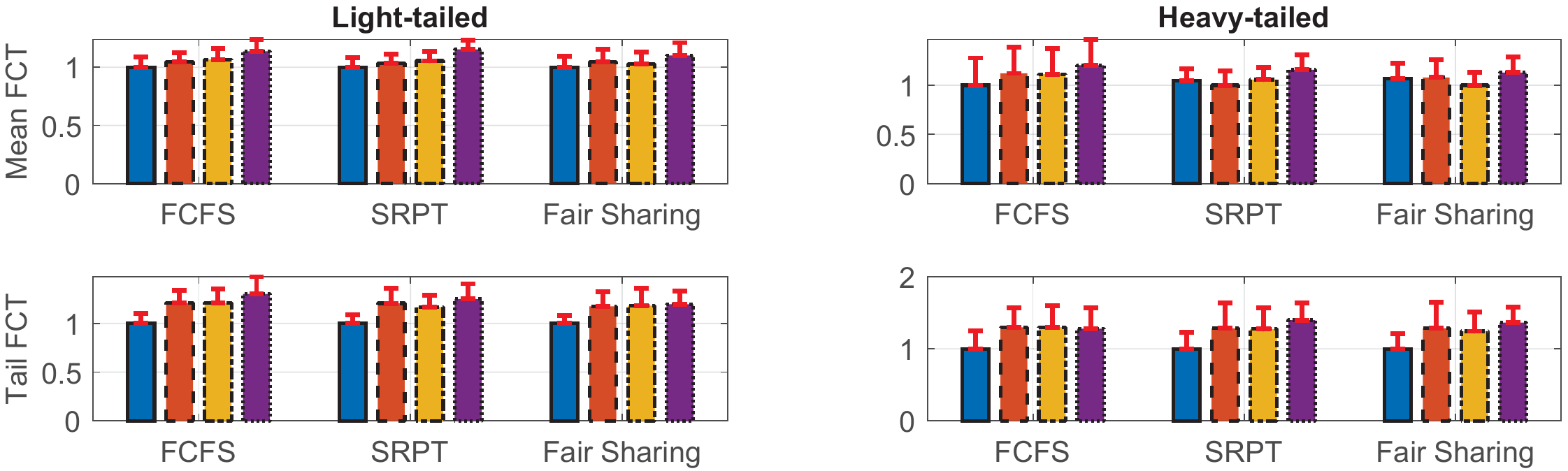}
    }
    \caption{Online routing techniques by flow scheduling policy assuming $\lambda=1$, $\mu=50$, and various topologies over 500 time units. All simulations were repeated $20$ times and the average results have been reported along with standard deviations.} \label{fig:exp2}
\end{figure}

\vspace{0.5em}
\noindent\textbf{\name's Optimality Gap:} In Figure \ref{fig:exp0} we compute the optimality gap of solutions found by \name~over three different topologies and under two traffic patterns. The optimal solution was computed by taking into account all existing paths and finding the minimum weight path on topologies of GScale, AGIS, and ANS. We also implemented a custom branch and bound approach which would require less computation time with a small number of ongoing flows (i.e., $< 20$ in our setting) and an intractable amount of time for a large number of ongoing flows (i.e., $> 30$ in our setting). According to the results, the average gap is less than $0.25\%$ over all experiments. We could not perform this experiment on larger topologies as computing the optimal solution would take an intractable amount of time.

\vspace{0.5em}
\noindent\textbf{Effect of Scheduling Policies:} In Figure \ref{fig:exp2}, we fixed the flow arrival rate to 1 and mean flow size to 50 and tried various scheduling policies under the four topologies of AT\&T North America, Cogent, GScale, and ANS. All simulations were repeated 20 times and the standard deviation for each instance has been reported. The minimum value normalizes each group of bars. We see that \name~is consistently better than other schemes regardless of the scheduling policy used. We can also see that compared to each other, the performance of other schemes varies considerably with the scheduling policy applied. To quantify, \name~provides up to $3.5\times$ and $2\times$ better mean and tail completion times than the other schemes across all scenarios on average, respectively.

\vspace{0.5em}
\noindent\textbf{Running Time:} We implemented Algorithm \ref{bwrh} in Java using the JGraphT library. To exhaustively find all paths with at most $K$ hops, we used the class \texttt{AllDirectedPaths} in JGraphT. We performed simulations while varying $\lambda$ from 1 to 10 and $\mu$ from 5 to 50 over 1000 flow arrivals per experiment which covers both lightly and heavily loaded regimes. We also experimented with all the four topologies pointed to earlier, both traffic patterns of light-tailed and heavy-tailed, and all three scheduling policies of FCFS, SRPT, and Fair Sharing. The maximum running time of Algorithm \ref{bwrh} was $222.24$ milliseconds, and the average of maximum running time across all experiments was $27$ milliseconds. This latency can be considered negligible given the time needed to complete long flows once they are routed.

\clearpage
\section{A Faster BWR Heuristic (\namefast)}
In the previous section, we showed that even for large topologies, \name~is a fast heuristic. Even so, the tail latency associated with finding a path can be hundreds of milliseconds. To be able to apply BWR to shorter flows, we propose a heuristic called \namefast~that runs much faster than \name~with the caveat that its solutions are on average farther from the optimal.

\namefast~is based on Dijkstra's algorithm and works by simply assigning weights to edges of the inter-DC graph and selecting a minimum weight path. Despite its simplicity, empirical evaluations show its significant and consistent gains. Algorithm \ref{bwrh2} shows our proposed approach to finding a path $P_{N+1}$ for $F_{N+1}$. The coefficient $\epsilon$ allows us to select the shortest hop path in case there are multiple paths with the same weight.

\SetAlgoVlined
\begin{algorithm}[ht]
\caption{\namefast} \label{bwrh2}
\SetKw{KwBy}{by}
\SetKwProg{FindPath}{FindPath}{}{}

\vspace{0.4em}
\KwIn{$F_{N+1}$, $G(V,E)$, $P_i,\mathcal{V}^r_i,1 \le i \le N$, and $0 < \epsilon \ll 1$}

\vspace{0.4em}
\KwOut{$P_{N+1}$}

\nonl\hrulefill

\vspace{0.4em}
Assign edge weights, $W_e = (\sum_{F_i \in \pmb{\mathcal{\mathrm{F}}}_e} \mathcal{V}^r_i) + \epsilon$, $\forall e \in E$\;

\vspace{0.4em}
$P_{N+1} \gets$ Find a minimum weight (shortest) path from the source to the destination of $F_{N+1}$\;
\end{algorithm}

We will find the worst-case optimality gap for \namefast~based on the number of data units of flows already in the system. Without loss of generality, let us assume that flows $F_i,1 \le i \le N$ have been sorted by their remaining data units from the smallest ($F_1$) to the largest ($F_N$). Let us call the optimal path $P_o$ and the path selected with our heuristic $P_h$.

\vspace{0.5em}
\textbf{Theorem 1.} \textit{$\frac{W_{P_h}}{W_{P_{o}}} \le \frac{\sum_{1 \le i \le N} \mathcal{V}^r_i}{\mathcal{V}^r_1}$.}

\begin{figure}[ht]
    \centering
    \includegraphics[width=0.7\columnwidth]{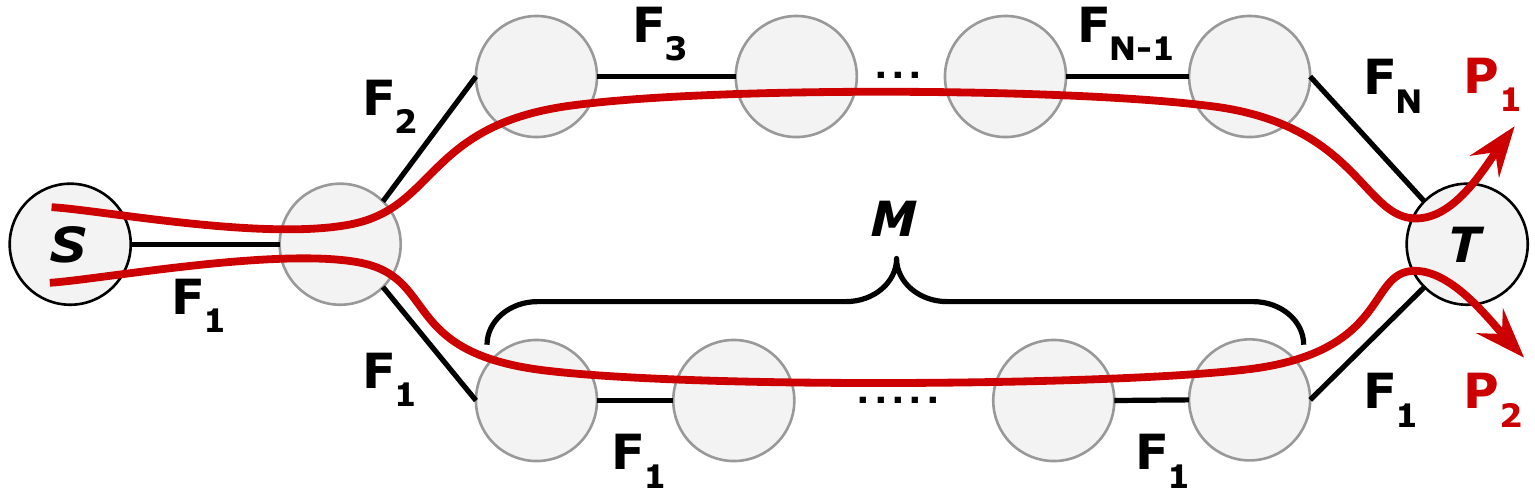}
    \caption{Worst-case routing scenario} \label{fig:worstcase_dijkstra}
\end{figure}

\vspace{0.5em}
\textit{\textbf{Proof.}} In case there exists a path with weight of zero from $s_{N+1}$ to $t_{N+1}$, Algorithm \ref{bwrh2} and the optimal solution will both choose a path with weight of zero. In case the weight of the optimal path is greater than zero, the quality of paths selected by Algorithm \ref{bwrh2} is highly correlated with the existing flows, their remaining data units and paths, and the network topology. We construct a simple example, as shown in Figure \ref{fig:worstcase_dijkstra}, that obtains the worst-case optimality gap. There are two possible paths, $P_1$ and $P_2$, for $F_{N+1}$. Let us choose the number of intermediate nodes $M$ on $P_2$ so that $M > \frac{\sum_{1 \le i \le N} \mathcal{V}^r_i}{\mathcal{V}^r_1}$. Apparently, from $S$ to $T$, the optimal solution for Problem 2 is $P_2$ with a total weight of $\mathcal{V}^r_1$. However, Algorithm \ref{bwrh2} will choose $P_1$ with a total weight of $\sum_{1 \le i \le N} \mathcal{V}^r_i$. This represents the worst-case as the weight of optimal path is the minimum and the weight of the chosen path is the maximum.

The worst-case optimality gap is highly dependant on the remaining flow data units and can potentially be large. However, the worst-case scenario is highly unique. We will show, through experiments, that Algorithm \ref{bwrh2} offers close to optimal solutions under different traffic patterns and network loads.

\subsection{Evaluations} \label{bwrhf_eval}
We performed extensive simulations to compare the two schemes of \name, \namefast and an exact implementation of BWR using exhaustive search by finding and evaluating all existing paths between the source and destination of every incoming flow. We used the same simulation parameters and topologies discusses in \S \ref{bwrh_eval}. We compared the earlier schemes with respect to network load and scheduling policies.

\begin{figure}[p]
    \centering
    \subfigure[Light-tailed Traffic]
    {
        \includegraphics[width=\textwidth]{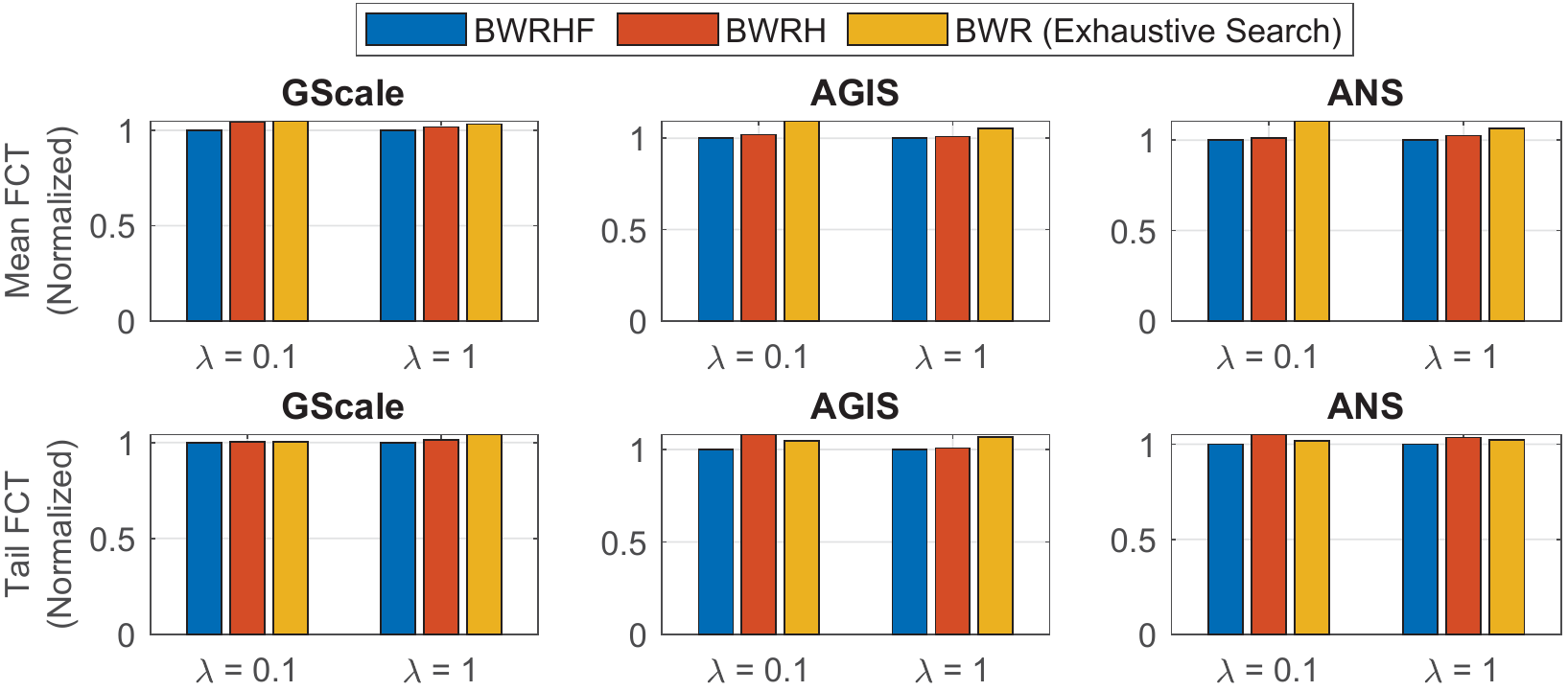}
    }
    \\
    \subfigure[Heavy-tailed Traffic]
    {
        \includegraphics[width=\textwidth]{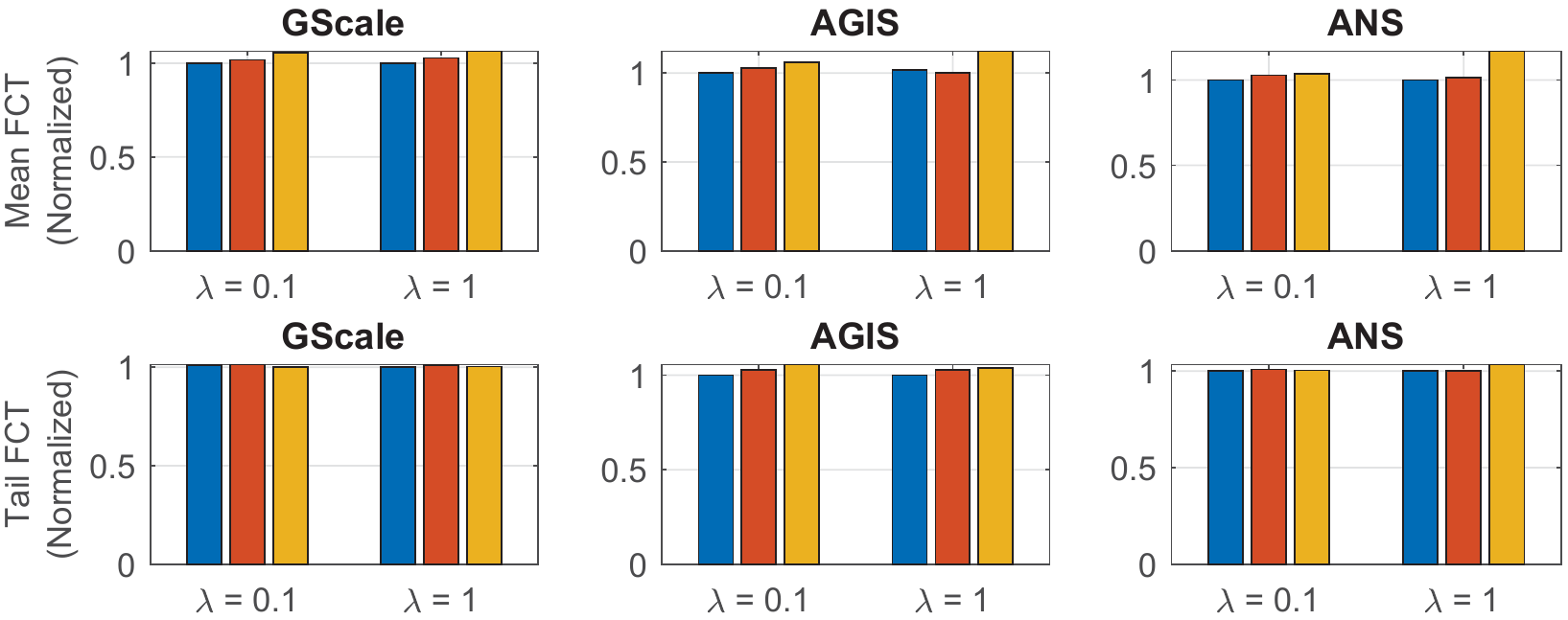}
    }
    \caption{Comparison of mean and tail flow completion times for the three implementations of BWR for the three topologies of GScale \cite{b4}, AGIS \cite{agis} and ANS \cite{ans}. Exhaustive search finds all possible paths between the end-points and then finds a minimum weight path. We considered $\mu = 50$ data units and performed the simulation over 500 time units. All simulations were repeated 20 times and the average results have been reported. We applied the Fair Sharing policy based on max min fairness which is most widely used.} \label{fig:exp_bwrhf_1}
\end{figure}

\begin{figure}[p]
    \centering
    \subfigure[Light-tailed Traffic]
    {
        \includegraphics[width=\textwidth]{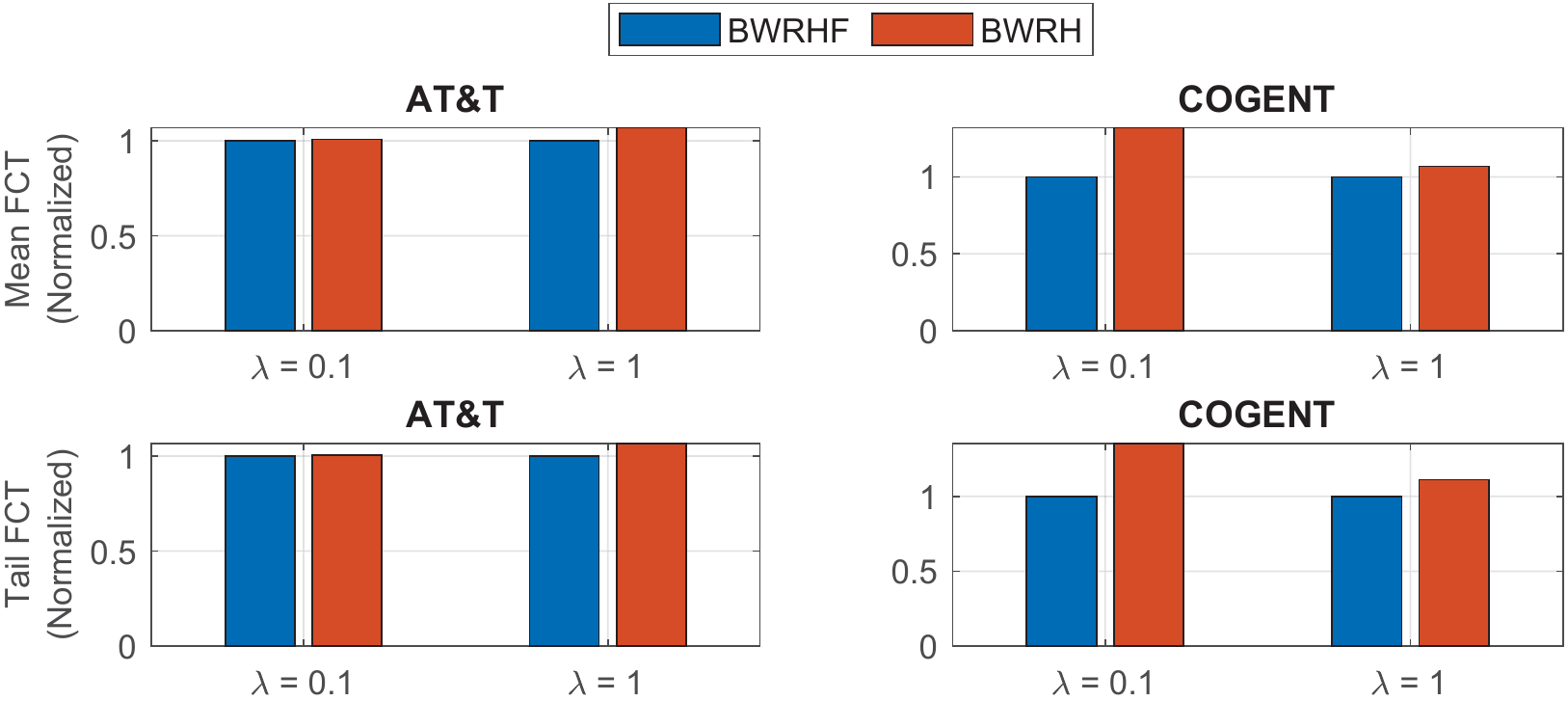}
    }
    \\
    \subfigure[Heavy-tailed Traffic]
    {
        \includegraphics[width=\textwidth]{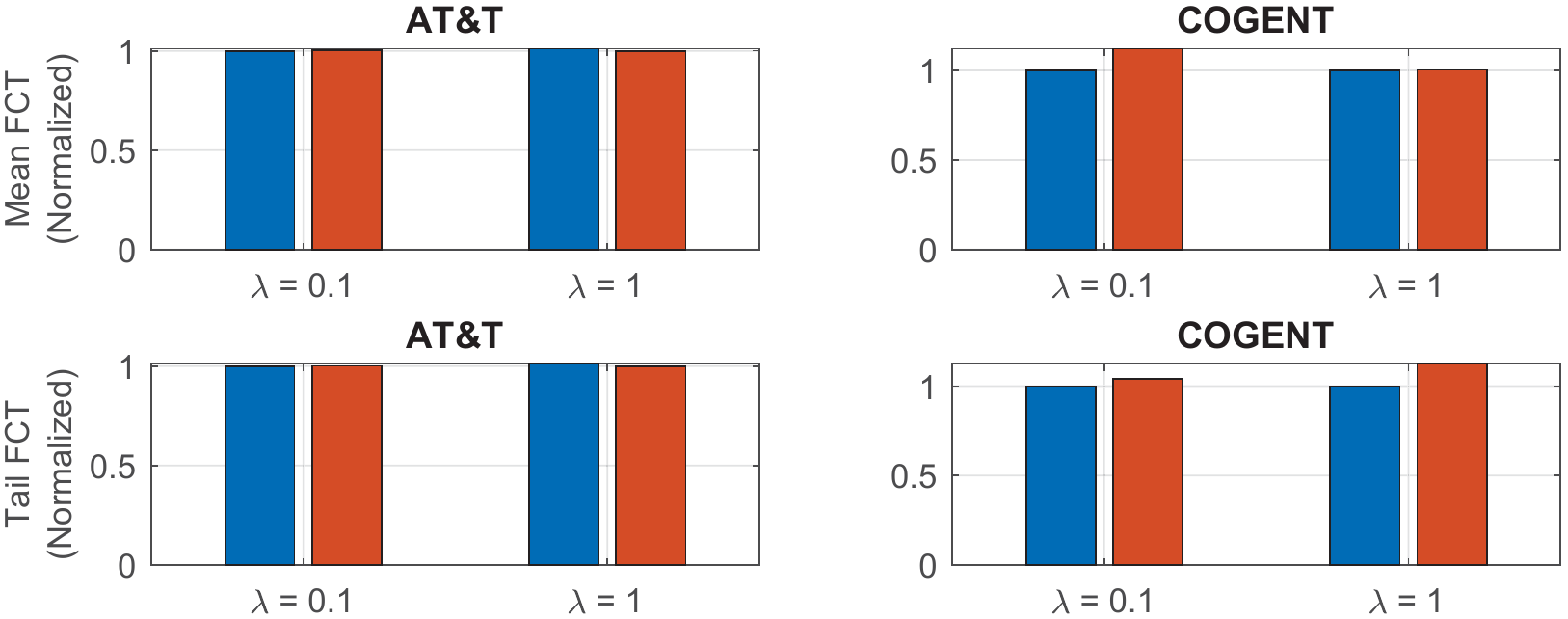}
    }
    \caption{Comparison of mean and tail flow completion times for the three implementations of BWR over two large topologies of AT\&T \cite{att} and Cogent \cite{cogent}. We excluded exhaustive search as it would take intractable amount of time for the topologies considered here. We considered $\mu = 50$ data units and performed the simulation over 500 time units. All simulations were repeated 20 times and the average results have been reported. We also applied the Fair Sharing scheduling policy based on max min fairness which is most widely used.} \label{fig:exp_bwrhf_2}
\end{figure}

\vspace{0.5em}
\noindent\textbf{\namefast's Performance by Network Load:} In Figures \ref{fig:exp_bwrhf_1} and \ref{fig:exp_bwrhf_2}, we explore the effect of load on the mean and tail completion times of various schemes considering the fair sharing policy. We consider multiple topologies with a different number of nodes and multiple degrees of connectivity. We see that regardless of incoming load (i.e., for different values of $\lambda$), all schemes offer close performance values. The performance gap is affected by both topology and load. We see a negligible difference in performance under both GScale and AT\&T topologies. For the topologies of Cogent, AGIS, and ANS, we observe that performance differs by up to 35\% across the schemes in a couple of cases. We also understand that although more straightforward, \namefast~offers better completion times in almost all instances. Knowing that BWR itself is a greedy online approach, this can be explained by noticing that making sub-optimal decisions for new flows as they arrive (i.e., the case for \namefast), can help future flows perform better in many cases. Since we evaluate the performance by looking at system-wide metrics (i.e., mean and tail flow completion times), it is reasonable to make sub-optimal decisions for routing of a new flow upon its arrival if that potentially helps the future flows, which we are unaware of, perform better and hence give us a better system-wide performance. For example, while the exact BWR implementation might choose a long path with minimum outstanding data units for a new flow, doing so might consume considerable network capacity due to many edges. Selecting a shorter path with marginally more data units can save more network bandwidth over extended periods and allow future flows to complete faster. Besides, it should be noted that the approach we took in Eq \ref{eq1} for computing the worst-case completion time of a new flow may overshoot, that is, the worst-case may be larger than necessary. This could happen as edge-disjoint flows that intersect with a path for the new flow may be able to transmit their data units in parallel. Computing tighter bounds on the worst-case, however, requires taking into account the dependencies of current flows and so can be computationally intensive in general.

\begin{figure}[p]
    \centering
    \subfigure[Light-tailed Traffic]
    {
        \includegraphics[width=\textwidth]{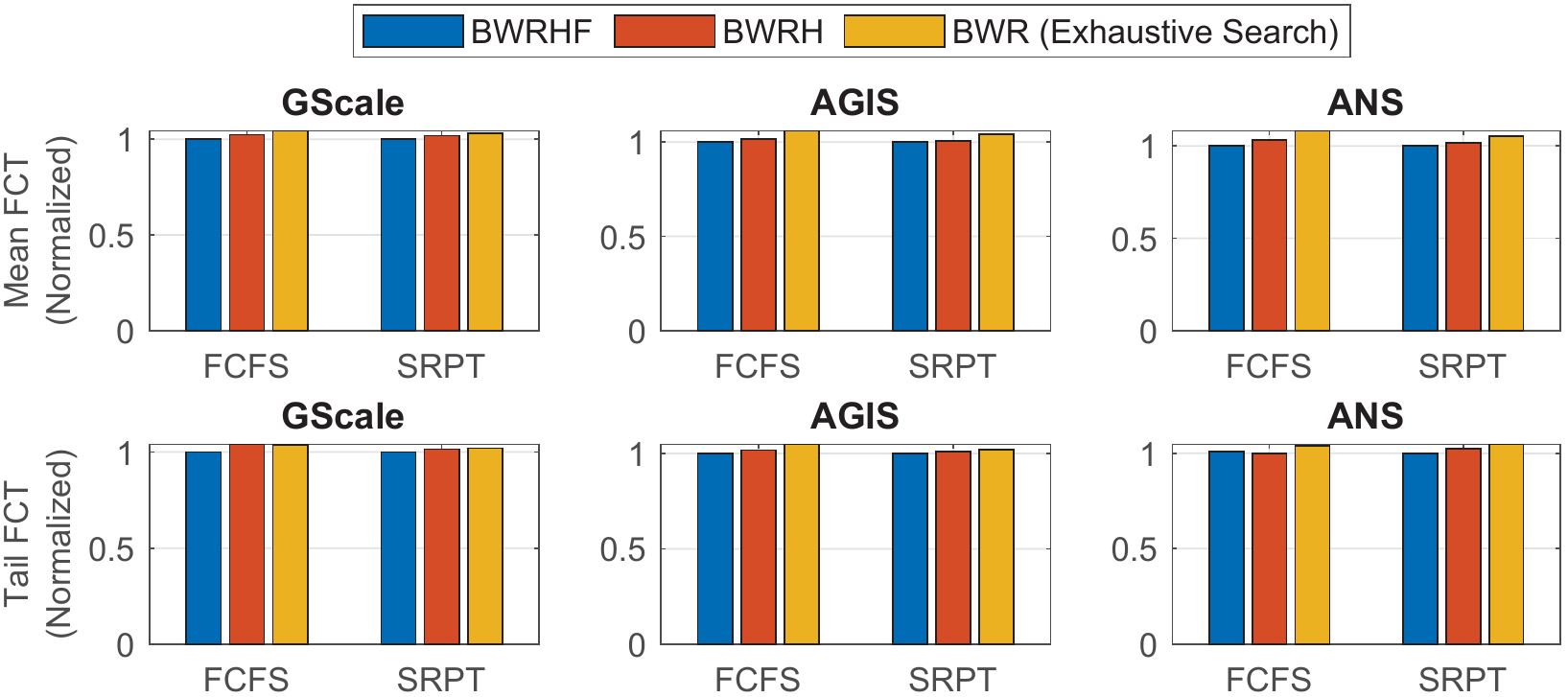}
    }
    \\
    \subfigure[Heavy-tailed Traffic]
    {
        \includegraphics[width=\textwidth]{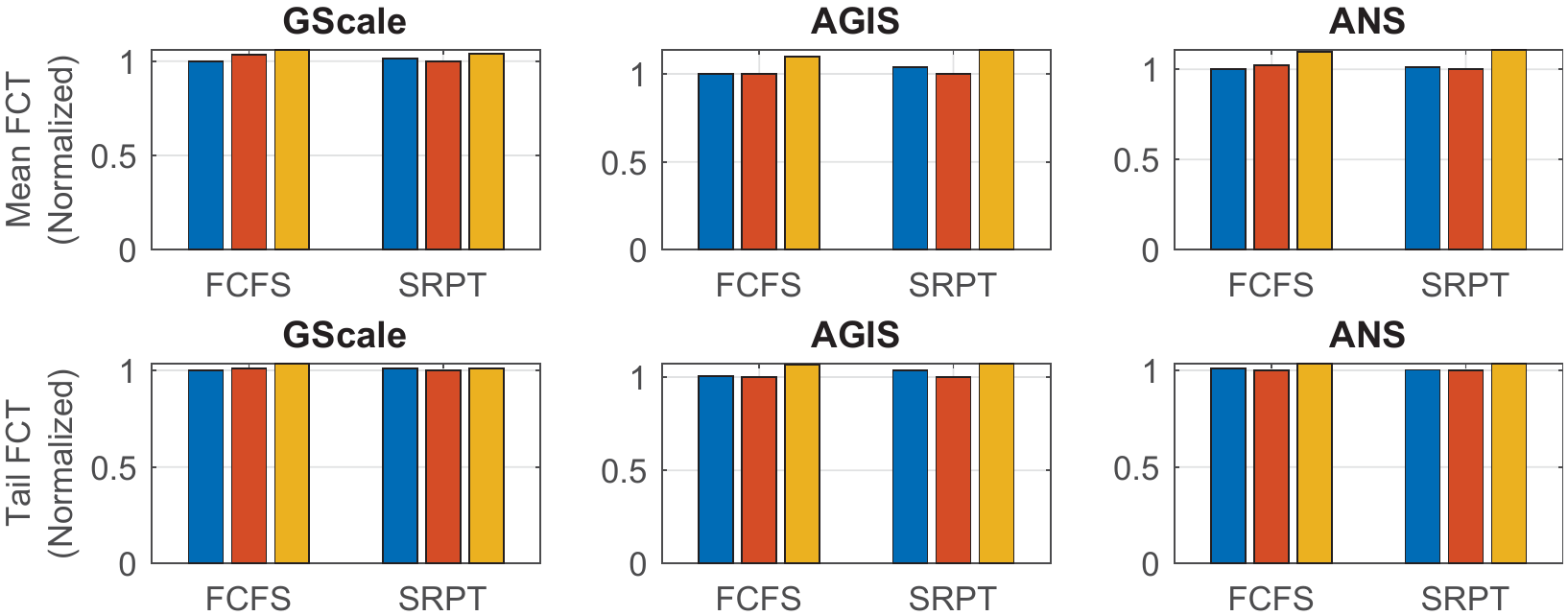}
    }
    \caption{Comparison of mean and tail flow completion times for the three implementations of BWR for the three topologies of GScale \cite{b4}, AGIS \cite{agis} and ANS \cite{ans}. Exhaustive search finds all possible paths between the end-points and then finds a minimum weight path. We considered $\lambda = 1$ and $\mu = 50$ data units and performed the simulation over 500 time units. All simulations were repeated 20 times and the average results are reported.} \label{fig:exp_bwrhf_3}
\end{figure}

\begin{figure}[p]
    \centering
    \subfigure[Light-tailed Traffic]
    {
        \includegraphics[width=\textwidth]{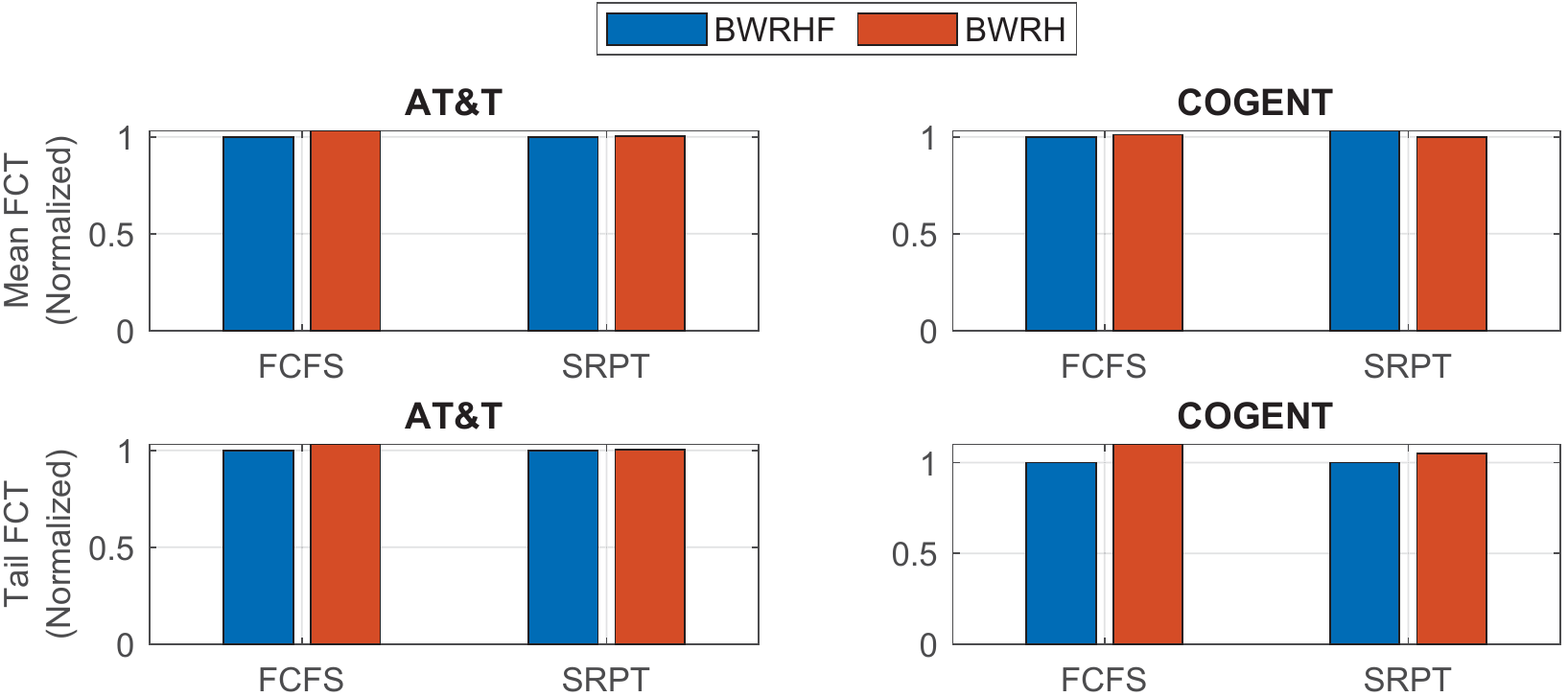}
    }
    \\
    \subfigure[Heavy-tailed Traffic]
    {
        \includegraphics[width=\textwidth]{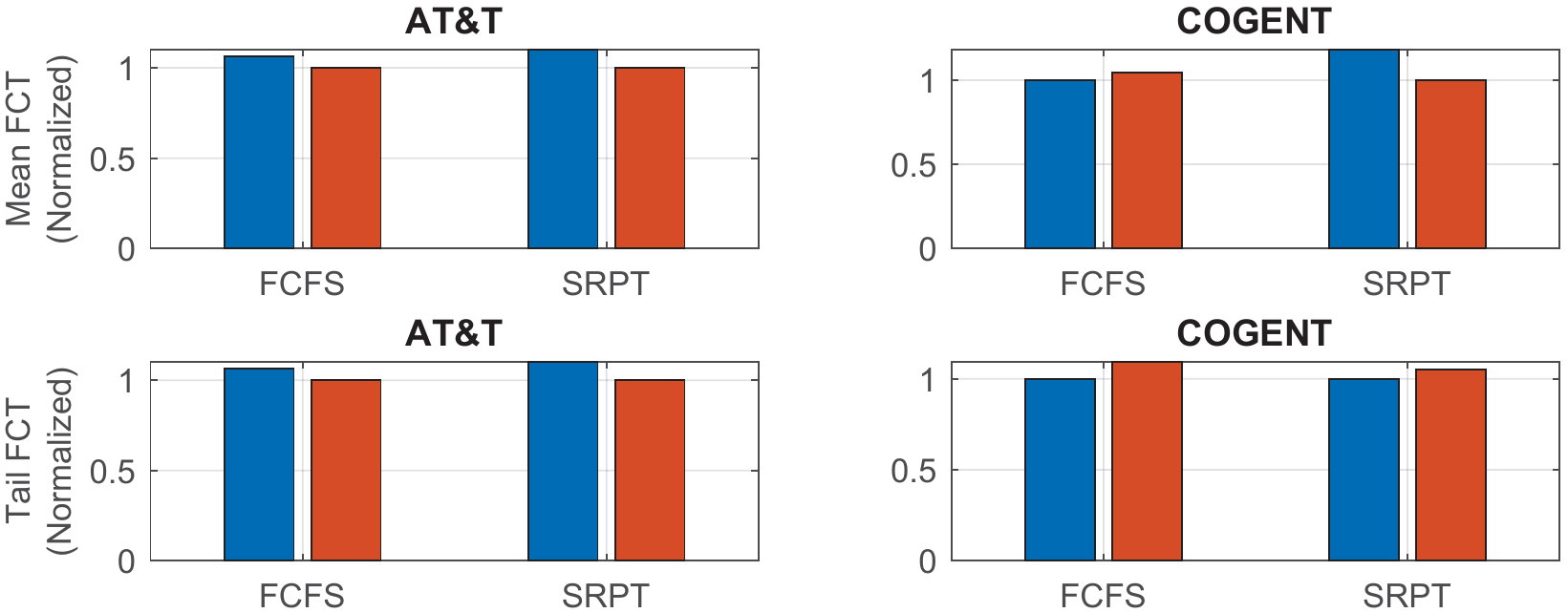}
    }
    \caption{Comparison of mean and tail flow completion times for the three implementations of BWR over two large topologies of AT\&T \cite{att} and Cogent \cite{cogent}. We excluded exhaustive search as it would take intractable amount of time for the topologies considered here. We considered $\lambda = 1$ and $\mu = 50$ data units and performed the simulation over 500 time units. All simulations were repeated 20 times and the average results have been reported.} \label{fig:exp_bwrhf_4}
\end{figure}

\begin{figure}[p]
    \centering
    \subfigure[AT\&T Topology \cite{att}]
    {
        \includegraphics[width=\textwidth]{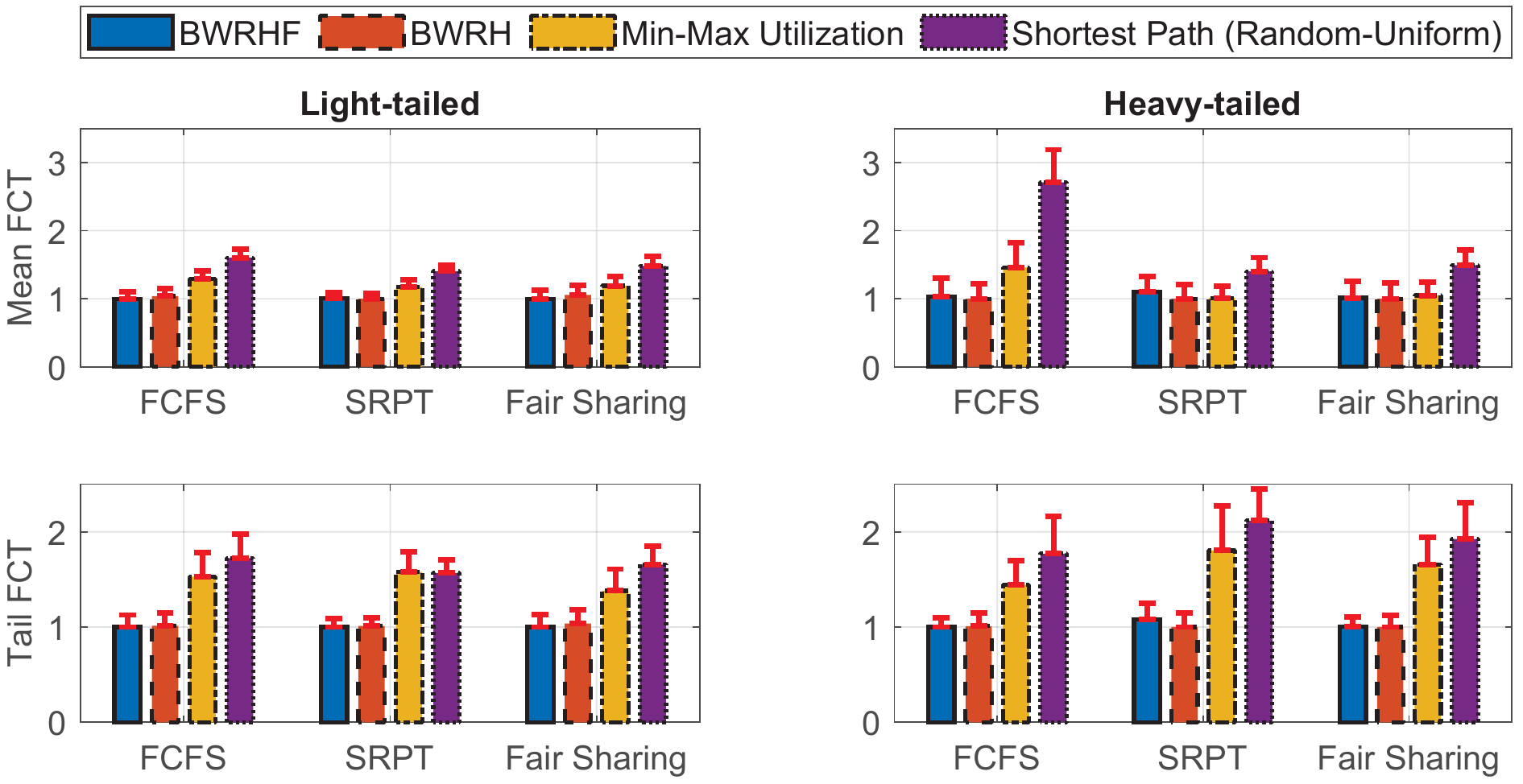}
    }
    \\
    \subfigure[Cogent Topology \cite{cogent}]
    {
        \includegraphics[width=\textwidth]{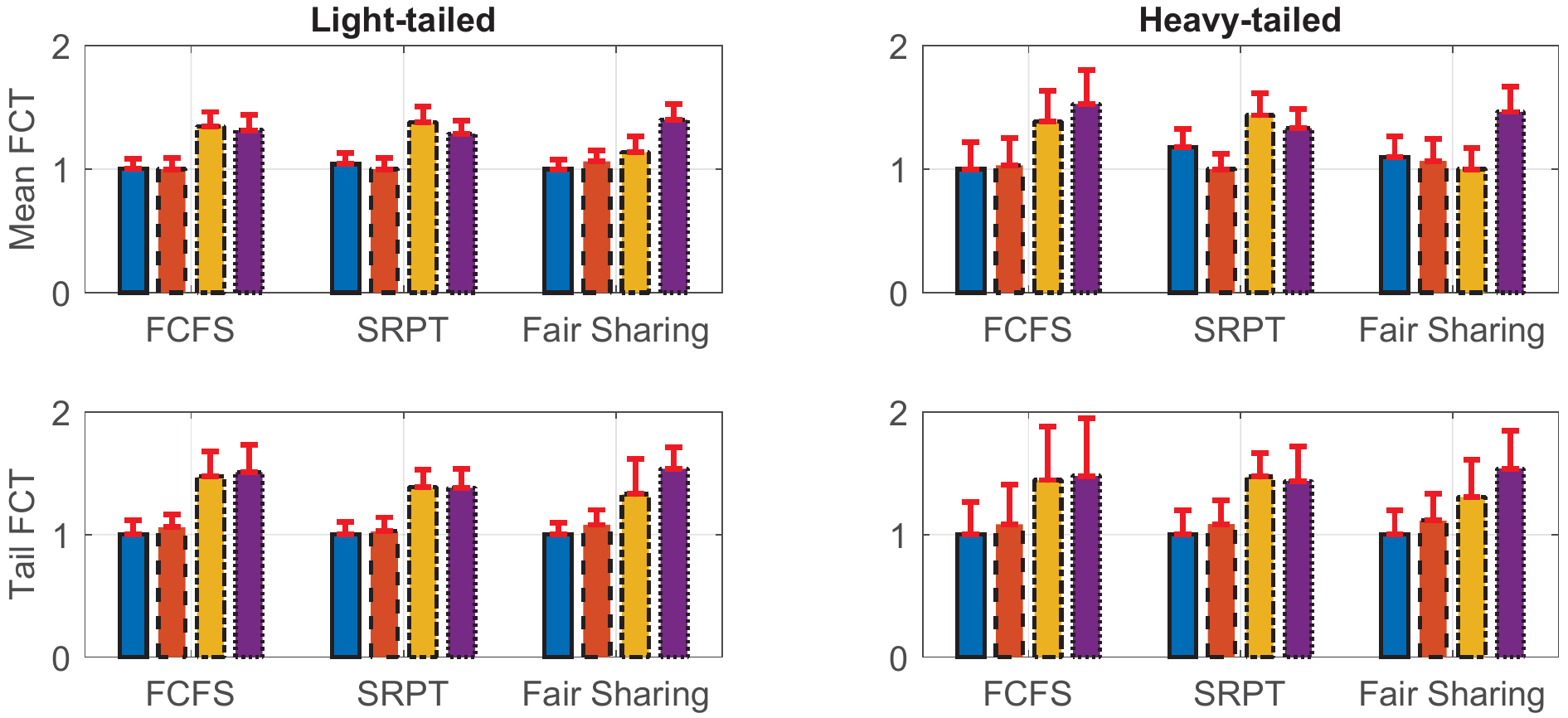}
    }
    \caption{Online routing techniques by flow scheduling policy assuming $\lambda=1$, $\mu=50$, and AT\&T and Cogent topologies over 500 time units. All simulations were repeated $20$ times and the average results have been reported along with standard deviations.} \label{fig:exp_bwrhf_5}
\end{figure}

\begin{figure}[p]
    \centering
    ~~~~~~~\includegraphics[width=0.95\textwidth]{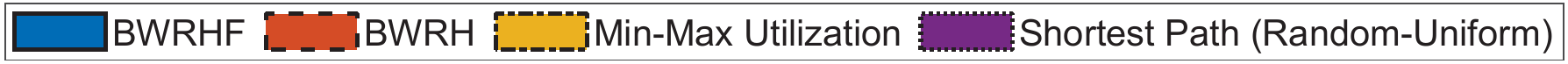}
    \\
    \subfigure[GScale Topology \cite{b4}]
    {
        \includegraphics[width=\textwidth]{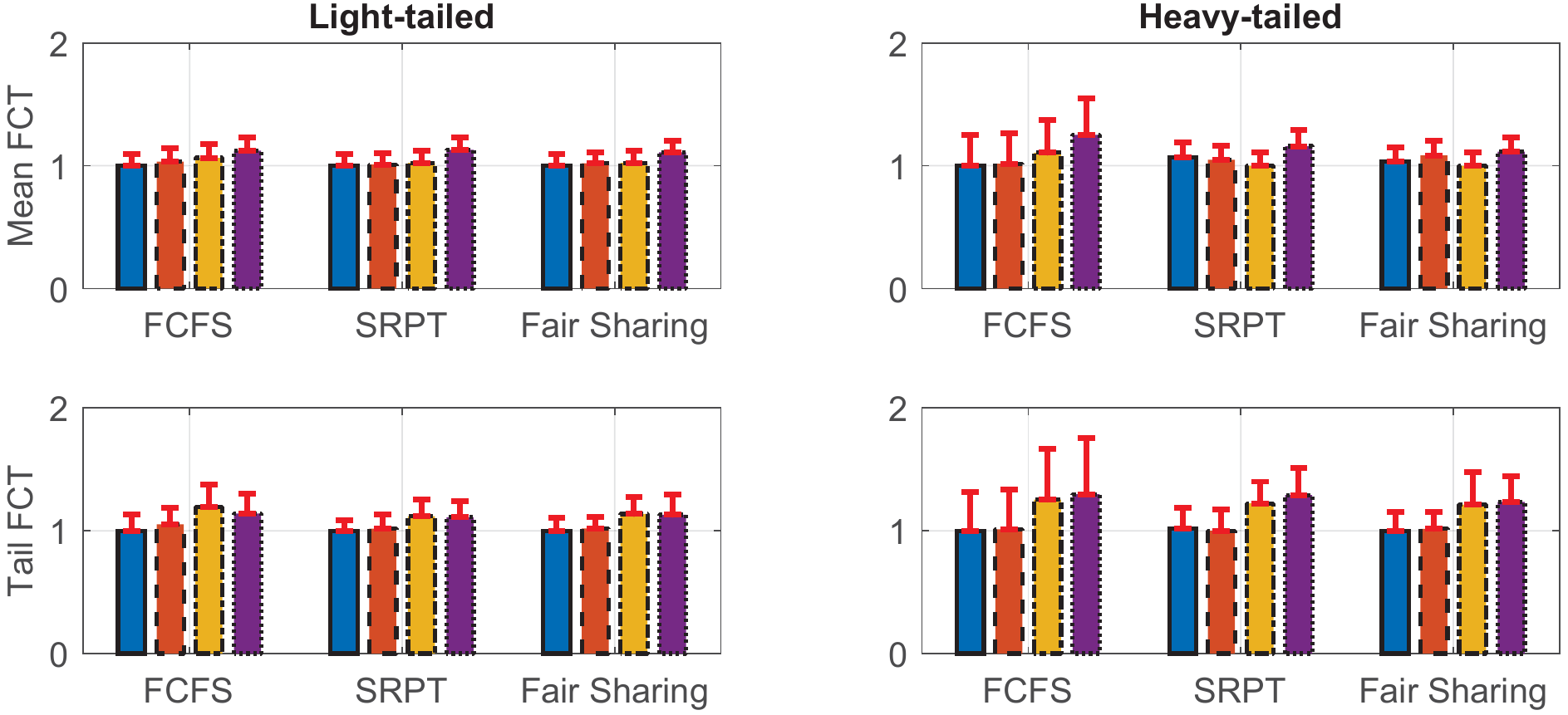}
    }
    \\
    \subfigure[ANS Topology \cite{ans}]
    {
        \includegraphics[width=\textwidth]{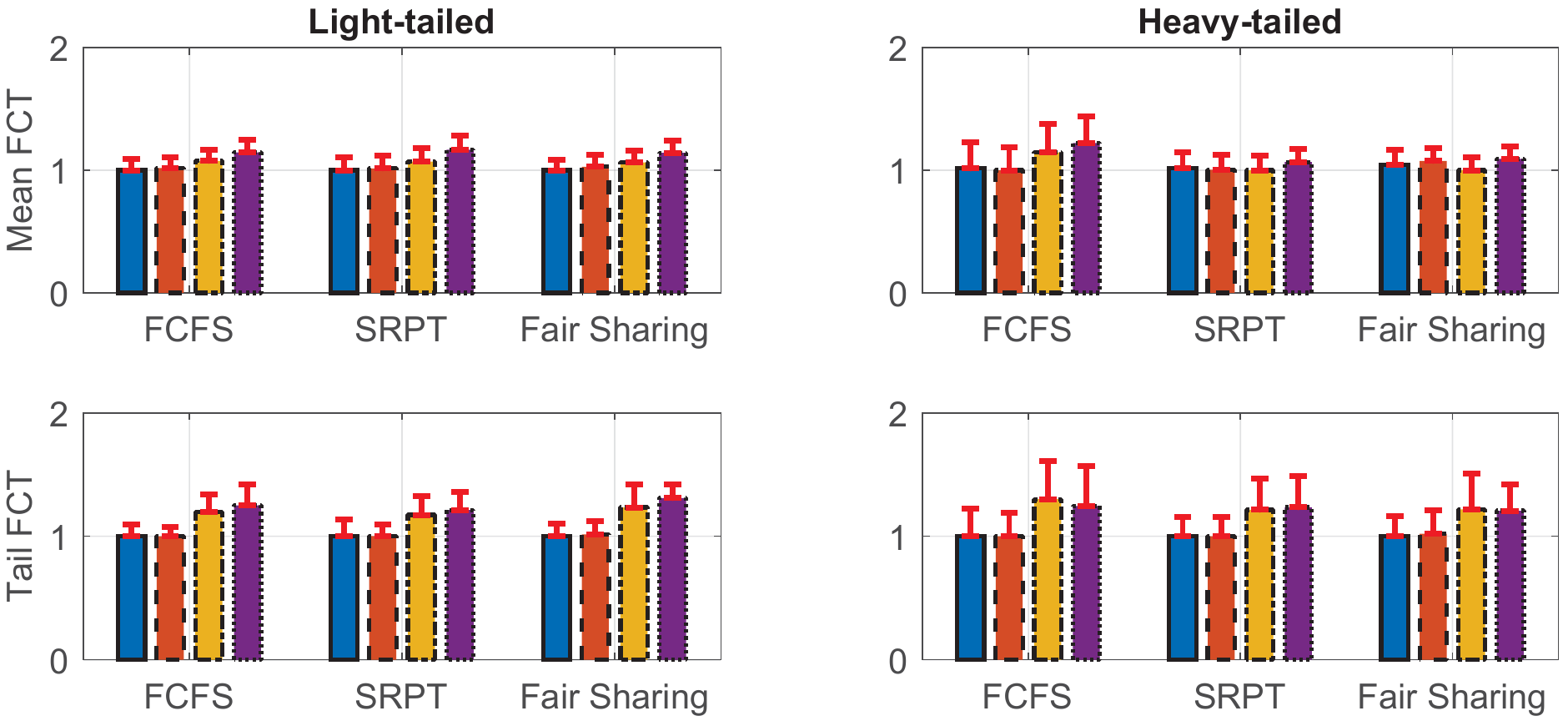}
    }
    \caption{Online routing techniques by flow scheduling policy assuming $\lambda=1$, $\mu=50$, and GScale and ANS topologies over 500 time units. All simulations were repeated $20$ times and the average results have been reported along with standard deviations.} \label{fig:exp_bwrhf_6}
\end{figure}

\begin{figure}
    \centering
    ~~~~~~~\includegraphics[width=0.95\textwidth]{exp_2_extended_legend.pdf}
    \\
    \subfigure[AGIS Topology \cite{agis}]
    {
        \includegraphics[width=\textwidth]{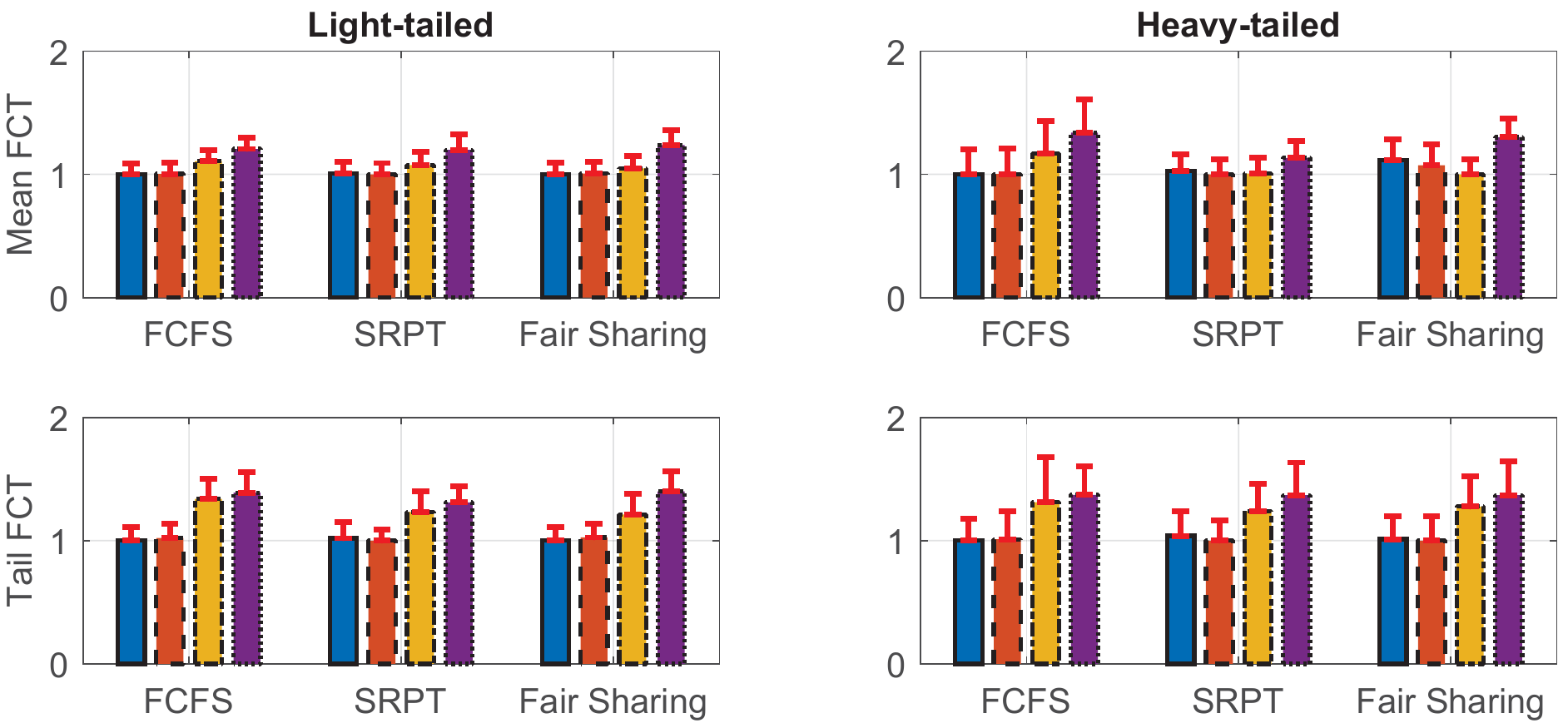}
    }
    \caption{Online routing techniques by flow scheduling policy assuming $\lambda=1$, $\mu=50$, and AGIS topology over 500 time units. All simulations were repeated $20$ times and the average results have been reported along with standard deviations.} \label{fig:exp_bwrhf_7}
\end{figure}

\vspace{0.5em}
\noindent\textbf{\namefast's Performance by Scheduling Policy:} In Figures \ref{fig:exp_bwrhf_3} and \ref{fig:exp_bwrhf_4}, we explore the effect of scheduling policies of SRPT and FCFS on the mean and tail completion times of various schemes.\footnote{The effect of the fair sharing policy was already discussed in Figures \ref{fig:exp_bwrhf_1} and \ref{fig:exp_bwrhf_2}.} Again, we observe that the straightforward heuristic of \namefast~performs well compared to \name~and the exact BWR implementation. We also see that under the heavy-tailed distribution of flow sizes, the effect of scheduling policies is more obvious. We see little difference in the performance of different schemes over all the topologies given different scheduling policies. In most cases, we see that \namefast~performs little better (i.e., up to 10\%) than \name. For a few scenarios, \namefast~performs little worse (i.e., up to 15\%). The same two arguments discussed in the effect of network load above also applies to why this may be the case. In Figures \ref{fig:exp_bwrhf_5}, \ref{fig:exp_bwrhf_6} and \ref{fig:exp_bwrhf_7}, we compare \name~and \namefast~with two other schemes of path selection that we earlier used in \S \ref{bwrh_eval}. We observe that for multiple scheduling policies, flow size distributions, and topologies, the two heuristics of \name~and \namefast~perform almost equally well and better than the other schemes, i.e., up to $2.6\times$ and $2.1\times$ better in mean and tail completion times, respectively.

\begin{figure}[t]
    \centering
    \includegraphics[width=0.7\columnwidth]{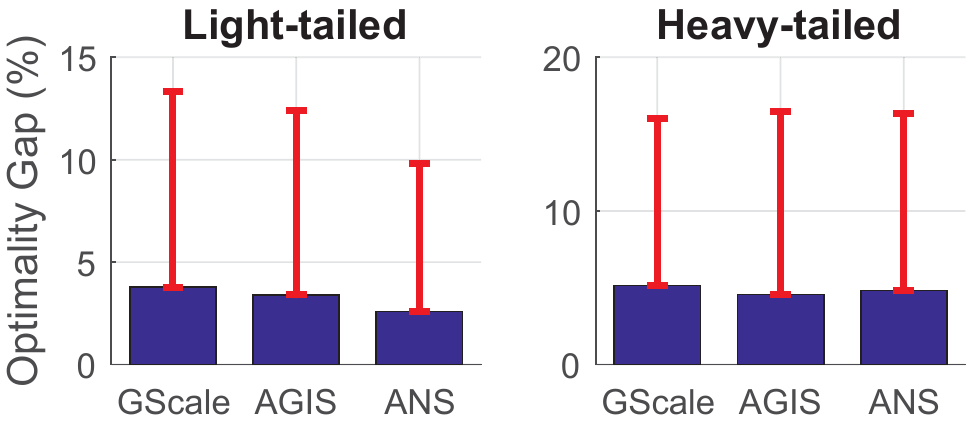}
    \caption{\namefast's optimality gap for $\lambda = 10$ and $\mu = 50$ computed for 1000 flow arrivals.} \label{fig:exp0_fast}
\end{figure}

\vspace{0.5em}
\noindent\textbf{\namefast's Optimality Gap:} In Figure \ref{fig:exp0_fast}, we compute the optimality gap of solutions found by \namefast~over three different topologies and under two traffic patterns. The optimal solution was computed by taking into account all existing paths and finding the minimum weight path on topologies of GScale, AGIS, and ANS. We also implemented a custom branch and bound approach which would require less computation time with a small number of ongoing flows (i.e., $< 20$ in our setting) and an intractable amount of time for a large number of ongoing flows (i.e., $> 30$ in our setting). According to the results, while the optimality gap may be large occasionally (i.e., $>50\%$ for $<2\%$ of the incoming flows, not shown), the average gap is less than $5\%$ over all experiments. We could not perform this experiment on larger topologies as computing the optimal solution would take an intractable amount of time.

\vspace{0.5em}
\noindent\textbf{Running Time:} \namefast~aims to find one minimum weight path using Dijkstra's algorithm which is on average much less computationally intensive compared to \name. We implemented Algorithm \ref{bwrh2} in Java using the JGraphT library. We performed simulations while varying $\lambda$ from 1 to 10 and $\mu$ from 5 to 50 over 1000 flow arrivals per experiment which covers both lightly and heavily loaded regimes. We also experimented with all the four topologies pointed to earlier, both traffic patterns of light-tailed and heavy-tailed, and all three scheduling policies of FCFS, SRPT, and Fair Sharing. The maximum running time of Algorithm \ref{bwrh} was $17.88$ milliseconds, and the average of maximum running time across all experiments was $1.38$ milliseconds. This latency is about $10\times$ less than what was observed from \name~under identical circumstances.

\section{Conclusions}
In this chapter, we explored a variety of routing heuristics and showed that the current routing techniques are insufficient for reducing the completion times of inter-DC transfers, even compared to several simple routing heuristics that we discovered. We then presented a new technique for routing based on flow size information, called the Best Worst-case Routing (BWR), to reduce flow completion times. Accordingly, the online routing problem turns into finding a minimum weight path on the topology from the source to the destination where the weight is computed by summing up the number of remaining data units of all the flows that have a common edge with the path. Since this is a hard problem, we developed two fast heuristics with small average optimality gaps. We also discussed how information from a real network scenario could be used as input to our network model to find a path on an actual inter-DC network for an incoming flow.

\clearpage
\chapter{Fast Deadline-based Admission Control for Inter-DC Transfers} \label{chapter_admission_control}
We consider the problem of admission control for point to point inter-DC transfers with deadlines. As the total capacity of inter-DC networks is limited, the purpose of admission control is to only accept new transfers when we can complete them prior to their deadlines while meeting the deadlines of all other transfers already in the system. To achieve this, traffic scheduling is needed for future timeslots because by focusing only on current timeslot we cannot guarantee that admitted transfers will finish before their deadlines \cite{tempus}. Besides, any algorithm used to perform such inter-DC admission control is desired to maximize the transfer admission rate and make efficient use of existing network resources. Speed in processing new transfer requests is another requirement. That is because for large scale applications that have millions of users, large number of transfers may have to be processed and allocated every minute. In this chapter, we propose and discuss a new scheduling policy called As Late As Possible (ALAP) scheduling and combine it with a novel routing policy to perform fast and effective admission control.

\section{Background and Related Work}
There is considerable work on maximizing the number of deadline meeting flows for traffic inside datacenters. These approaches, however, do not perform admission control which leads to wasted bandwidth. In \cite{d2tcp, pdq}, authors propose deadline aware transport protocols which increase the number of transfers that complete prior to their assigned deadlines by adjusting the transmission rate of such transfers based on their deadlines. Also, multiple previous studies have focused on improving the efficiency and performance of inter-DC communications through proper scheduling of transfers. In \cite{tempus}, authors propose TEMPUS which improves fairness by maximizing the minimum portion of transfers delivered to destination before the transfer deadlines. TEMPUS cannot guarantee that admitted transfers are completed prior to their deadlines. 
In \cite{amoeba}, authors propose Deadline-based Network Abstraction (DNA) which allows tenants to specify deadlines for transfers, and a system called Amoeba which performs admission control for new inter-DC transfers. When a request is submitted, Amoeba formulates an optimization scenario, performs feasibility checks, and decides whether the new request can be satisfied using available resources. If a transfer cannot be completed prior to its deadline, Amoeba tries to reschedule a subset of previously admitted requests to push traffic further away out of the new request's timeline. The admission process is performed on a first-come-first-served (FCFS) basis and requests are not preempted, that is, the system does not drop a previously admitted request as this can lead to thrashing.

\section{Fast Admission Control on A Network Path}
We discuss a new scheduling approach that allows fast admission control over a single network path. We will extend this idea to general networks in the next section.

\subsection{System Model}
In this section, we consider a simple topology where multiple transfers are scheduled over the same path. We will use the same notation as that in Table \ref{table_var_0}. Assume we are allocating traffic for a timeline starting at $t_{now}$ representing current time and ending at $t_{end}$ which corresponds to the latest deadline for all submitted requests. New requests may be submitted to the scheduler at any time. Every request $R_i$ is identified with two parameters $\mathcal{V}_{R_i}$ and $t_{d_{R_i}}$ representing request size and deadline, respectively. Since all requests are scheduled over the same path, they all have the same source and destination. Requests are instantly allocated upon arrival over timeslots for which $t > t_{now}$. We consider a TES that receivers the inter-DC transfer requests and decides whether they can be admitted. If yes, the TES has to also compute a transmission schedule which determines the rate at which the source node should send packets associated with every transfer per timeslot.

\subsection{Currently Used Approach}
To perform admission control, one can formulate and solve a linear program (LP) involving all current transfers and the new transfer with demand and capacity constraints populated based on link capacities (for the links on the path) and request volumes. We can then attempt to solve this LP. If this LP is feasible, then the transfer can be admitted. This LP has to be solved every time a new request is submitted and can result in changing the allocation of already scheduled requests. The problem with this approach is its high complexity (solving possibly large LPs over and over is computationally inefficient) as the frequency of arrivals increases.

\subsection{As Late As Possible (ALAP) Scheduling}
We propose As Late As Possible Scheduling (ALAP) \cite{rcd}, which is a fast traffic allocation technique that minimizes the time required to perform the admission process. It avoids rescheduling already admitted requests to quickly decide whether a new request can be admitted. It also achieves high utilization and can efficiently use network resources. We present the rules based on which ALAP operates:

\vspace{0.5em}
\noindent\textbf{Rule 1:} Similar to previous schemes \cite{amoeba}, preemption is not supported. Preempting a request that is partly transmitted is wasteful. Also, it may result in thrashing if requests are consecutively preempted in favor of future requests. 

\vspace{0.5em}
\noindent\textbf{Rule 2:} To be fast, ALAP does not change the allocation of already allocated traffic unless there is leftover bandwidth in current timeslot ($t_{now}$). In which case, it fetches traffic from the earliest timeslot that is not empty and sends it. This is done until either we fully utilize the current timeslot or there is no more traffic to send.

When a new transfer $R_{new}$ is submitted, ALAP creates a small LP, only involving the new request, to schedule it. The number of variables in this LP is $(t_{d_{R_{new}}} - t_{now})$. Assume the amount of bandwidth allocated to new transfer at time $t$ is $f^{new}_{P}(t)$ and $C - B(t)$ is the residual capacity on the path at timeslot $t$ assuming a path capacity of $C = \min_{e \in P}(C_e)$ and available bandwidth of $B(t) = \min_{e \in P}(B_e(t)), \forall t$ where $P$ is the path on which we perform admission control. We use the LP of equation \ref{eq:eq2} with the objective function of equation \ref{eq:eq1} to do the allocation. If the following LP does not yield a feasible solution, we reject the request.

\begin{align}
& U(R_{new}) \triangleq \sum_{t=t_{now}+1}^{t_{d_{R_{new}}}} t~f^{new}_{P}(t) \label{eq:eq1} \\
& \max (U(R_{new})) \\
& \sum_{t = t_{now}+1}^{t_{d_{R_{new}}}} f^{new}_{P}(t) = \mathcal{V}_{R_{new}} \\
& 0 \le f^{new}_{P}(t) \le C - B(t), ~~~ t_{now} < t \le t_{d_{R_{new}}} \label{eq:eq2}
\end{align}

\vspace{0.5em}
Now consider a scenario where transfers $R_1, R_2, ~... ~ R_K$ arrive at the network in order to be allocated on path $P$. We show that upon arrival of $R_k, ~ 1 \le k \le K$, ALAP allocation for previously admitted requests is so that we cannot increase the chance of admission for $R_k$ by rearranging the allocation of already allocated requests (previous $k - 1$ requests). Recall that the deadline of $R_k$ is shown as $t_{d_{R_k}}$ and at any time $t$, the latest deadline of all admitted requests is $t_{end}$. 

\vspace{0.5em}
\textbf{Theorem 1.} If we draw a vertical line at time $t_{d_{R_k}} \le t_{end}$ in our traffic allocation, it is not possible to increase the free space behind the line by moving traffic from left side of the line ($t \le t_{d_{R_k}}$) to the right side ($t_{end} \ge t > t_{d_{R_k}}$).

\begin{figure*}[t]
\centering
\includegraphics[width=\textwidth]{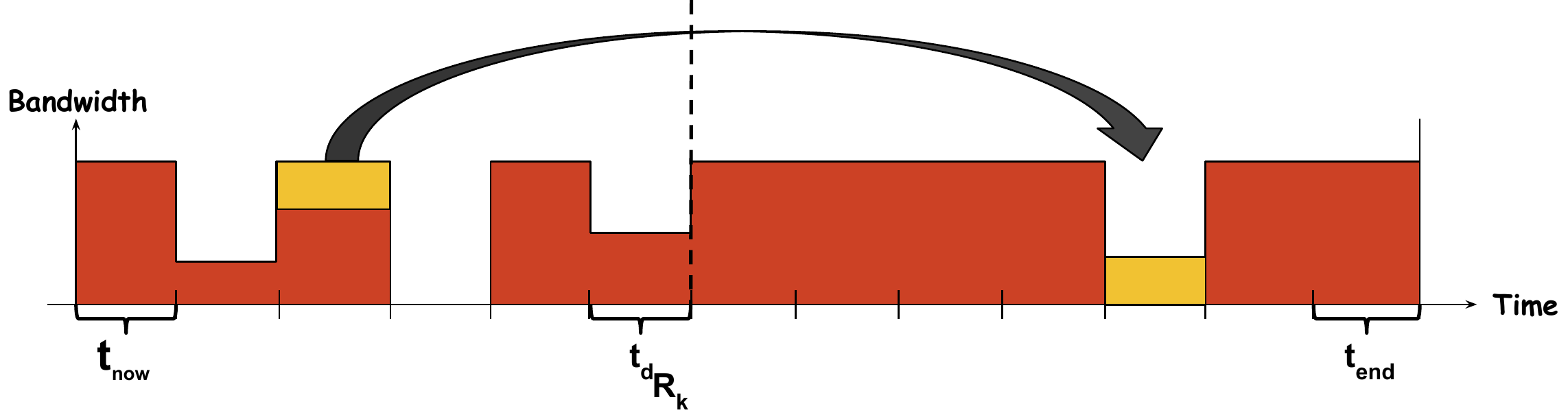}
\caption{A traffic allocation used in proof of Theorem 1.}
\label{fig:theorem1}
\end{figure*}

\vspace{0.5em}
\textbf{\textit{Proof.}} Let us assume we have the allocation shown in Figure \ref{fig:theorem1} for the first $k-1$ requests on path $P$. To schedule all requests, we used the utility function of equation \ref{eq:eq1} which assigns a higher cost to future timeslots. Let us assume that we can move some traffic volume from left side to the right side. If so, this volume belongs to at least one of the admitted requests and that means we are able to increase the utility for that request further. This is not possible because the LP in equation \ref{eq:eq2} gives the maximum utility. That means if we were to move traffic from left side of the line to the right side it would either result in violation of link capacity constraints or violation of deadline constraints.

Now let's assume a new transfer arrives. If it can be allocated using the residual link capacity on all the links of path $P$, then we can admit it. If not, based on Theorem 1, there is no way we can shift already allocated traffic so that we can accommodate the new transfer. Since every new transfer is scheduled closest possible to its deadline, we refer to this policy as As Late As Possible (ALAP) scheduling since traffic cannot be pushed further closer to the deadline.

\begin{figure}
    \centering
    \includegraphics[width=\columnwidth]{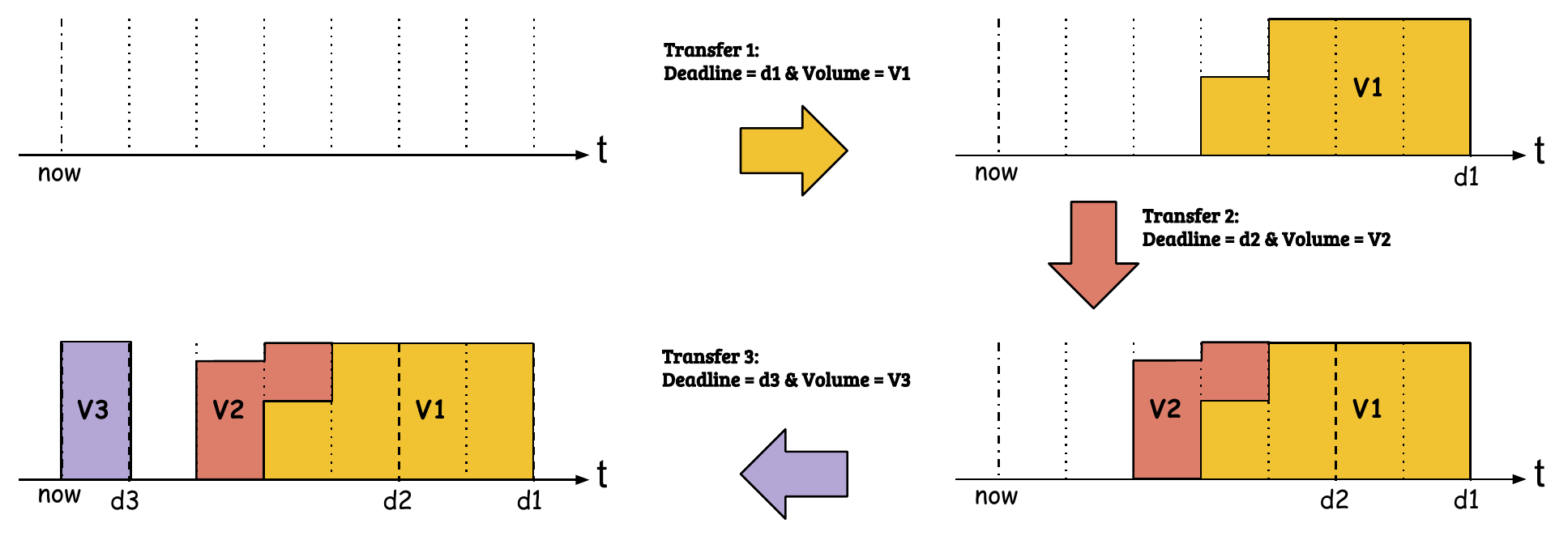}
    \caption{An example of ALAP allocation.}
    \label{ALAP}
\end{figure}

Figure \ref{ALAP} provides an example of the ALAP allocation technique. As can be seen, when the first transfer is received, the timeline is empty and therefore it is allocated adjacent to its deadline. The second transfer is allocated as close as possible to its deadline. The implication of this type of scheduling is that requests do not use resources until it is absolutely necessary. This means resources will be available to other requests that may currently need them. When the third transfer arrives, resources are free and it just grabs as much bandwidth as needed. If we had allocated the first two requests closer to current time we may have had to either reject the third transfer or move the first two transfers ahead freeing resources for the third transfer (which would have required rescheduling).

\subsection{Simulation Results}
We compare the performance and speed of ALAP with Amoeba \cite{amoeba}. Other schemes, such as \cite{swan, tempus}, are deadline-agnostic and have an effective link utilization of less than $50\%$ \cite{amoeba}. Amoeba, on the other hand, only accepts requests when it can guarantee that the deadline can be fully met.

\vspace{0.5em}
\noindent\textbf{Setup:} We consider a topology with multiple equal capacity links with a capacity of $1$ attached in a line and traffic is transmitted from one end to the other. We assume that high priority traffic (e.g., user generated, real-time, etc.) takes a fixed amount of bandwidth and allocate the leftover among inter-DC transfer requests. Simulation is performed for $576$ timeslots each lasting $5$ minutes which is equal to $2$ days. We performed the simulations three times and calculated the average.  

\vspace{0.5em}
\noindent\textbf{Metrics:} Fraction of inter-DC transfer requests that were rejected, average link utilization, and average allocation time, in timeslots, per request are the three metrics measured and presented. 

\vspace{0.5em}
\noindent\textbf{Workload:} We generate inter-DC transfers according to a Poisson distribution of rate $1 \le \lambda \le 8$ request(s) per timeslot. The difference between the arrival time of requests and their deadlines follows an exponential distribution with an average of $12$ timeslots. In addition, the demand of each request also follows an exponential distribution with an average of $0.286$ (a maximum of $1$ unit of traffic can be sent in each timeslot on every link). 

\begin{figure*}
\centering
\includegraphics[width=\textwidth]{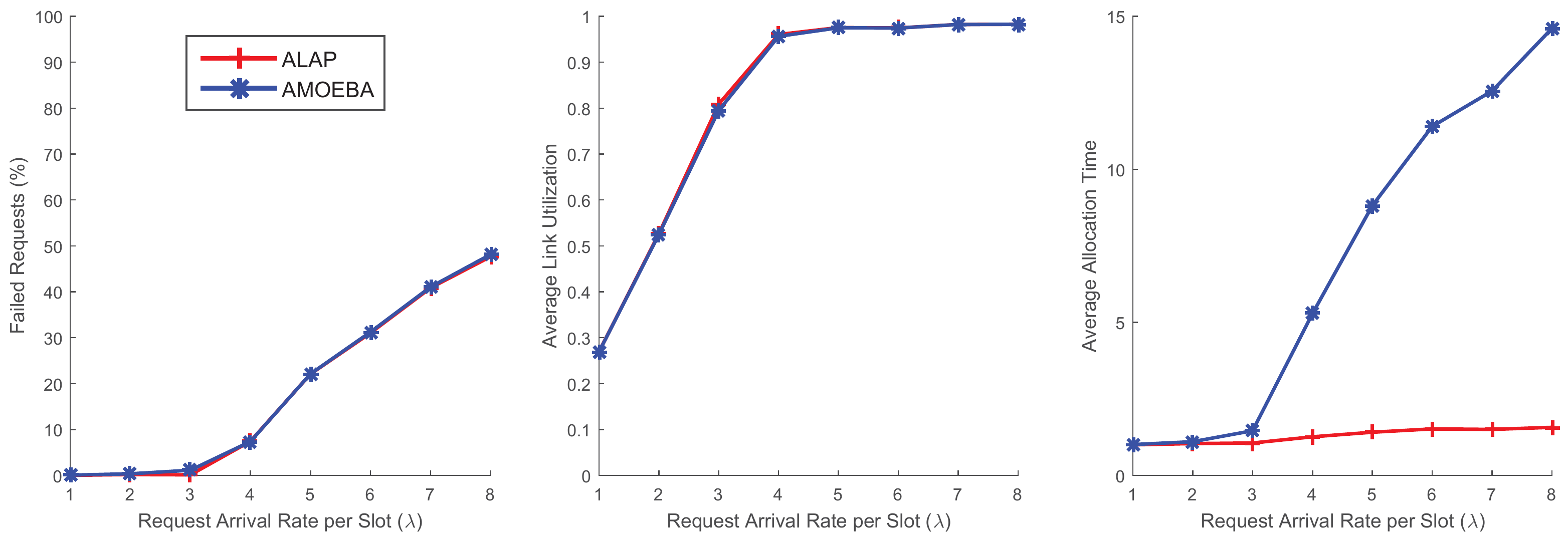}
\caption{Comparison between Amoeba and ALAP scheduling.}
\label{fig:results_1}
\end{figure*}

\vspace{0.5em}
Figure \ref{fig:results_1} shows the aforementioned simulation metrics for both Amoeba and ALAP. As can be seen, both algorithms result in similar rejection rate (and so admission rate) and utilization. However, ALAP achieves the same performance metrics with much less complexity. ALAP is up to $15\times$ faster than Amoeba. Also, the complexity of ALAP grows slowly as the frequency of arrivals increases, i.e., up to $1.6\times$ while arrivals increase by a factor of $8$.

With regards to the trend for time complexity as shown in Fig. \ref{fig:results_1}, when the request arrival rate is small, most of the capacity is left unused. Therefore, Amoeba does not have to move already allocated requests to push in a new one. As the arrival rate increases, we see a higher utilization. Starting the arrival rate of $4$, utilization grows close to $1$ and we can see a huge jump in the time complexity of Amoeba (by a factor of $3.7\times$). That is because Amoeba has to move around multiple already allocated requests to push in the new request. 

For an arrival rate of $8$ requests per timeslot, we see that both algorithms drop almost half of the requests. This can happen as a result of capacity loss in the network. For example, when a datacenter is connected using only two links and one of them fails for a few timeslots. While Amoeba can get really slow, ALAP scheduling is able to handle such situations almost as fast as when there is low link utilization.

\section{Application of ALAP over General Network Topologies} \label{section_dcroute}
We empirically showed that ALAP can speed up the allocation process by allowing new transfers to be scheduled only considering the residual bandwidth on the edges of a path $P$ which results in creation of much smaller LPs. In this section, we consider the routing problem in addition to the ALAP scheduling policy for admission control over a general network. We focus on single path routing and develop a solution called DCRoute \cite{dcroute}.

\vspace{0.5em}
\noindent\textbf{Minimizing Packet Reordering:} Avoiding packet reordering allows data to be instantly delivered to applications upon arrival of packets. In addition, inter-DC networks have characteristics similar to WAN networks (including asymmetric link delays and large delays for links that connect distant locations) for which multiplexing packets over different paths has been shown to considerably degrade TCP performance \cite{wan-reordering}. Putting out of order packets and segments back in order can be expensive in terms of memory and CPU usage, especially when transmitting at high rates.

\vspace{0.5em}
\noindent\textbf{Admission Control over General Networks:} In contrast to routing over a single path, for a network, each request is routed on multiple links and there are many ways to schedule requests ALAP. If some links are used by multiple requests routed on different edges, how traffic is allocated on common links can affect multiple other links which will affect the requests that use those links later on. We propose a routing heuristic that allows us to select a least loaded path for a new request over which we attempt to allocate a new request. We will show that using the ALAP scheduling policy, we can greatly speed up the allocation process while sacrificing negligible performance.

\subsection{System Model}
At any given moment, we have two parameters $t_{now}$ and $t_{end}$ which represent current timeslot and the latest deadline among all current transfers, respectively. A request arriving sometime in timeslot $t$ can be allocated starting timeslot $t+1$ since the schedule and transmission rate for current timeslot is already decided and broadcast into all senders. Also, at any moment $t$, $t_{now}$ is the timeslot that includes $t$ (current timeslot), and $t_{now}+1$ is the next available timeslot for allocation (next timeslot). A request is considered \textbf{active} if it is admitted into the system and its deadline has not passed yet. Some active requests may take many timeslots to complete transmission. The total unsatisfied demand of an active request is called the residual demand of that request. We will use the same notation as that in Table \ref{table_var_0}. We will define some additional variables in this section as shown in Table \ref{table_var_1}.

\begin{table}
\begin{center}
\caption{Variables used in this chapter in addition to those in Table \ref{table_var_0}} \label{table_var_1}
\vspace{0.5em}
\begin{tabular}{ |p{2cm}|p{13cm}| }
    \hline
    \textbf{Variable} & \textbf{Definition} \\
    \hline
    \hline
    $L_e(t)$ & Total load currently scheduled on edge $e$ prior to and including timeslot $t$ \\
    \hline
    $L_e$ & Total load currently scheduled on edge $e$ (same as $L_e(t_{now})$) \\
    \hline
    $\mathcal{V}^{r}_{R_{i}}$ & Current residual demand of request $R_i$ \\
    \hline
\end{tabular}
\end{center}
\end{table}

\vspace{0.5em}
\noindent\textbf{Definition of Edge Load $L_e$ and $L_e(t)$:} We define a new metric called edge load which determines the total remaining volume of traffic per edge for all the transfers that share that edge. This metric provides a measure of how busy a link is expected to be on average over future timeslots. $L_e(t)$ is the total volume of traffic scheduled on an edge prior to and including timeslot $t$. $L_e$ can then be written as $L_e(t_{end})$. For a new request with a deadline of $t_{d_{R_i}}$, it would only make sense for us to consider all the traffic scheduled on edges prior to $t_{d_{R_i}}$, i.e., $L_e(t_{d_{R_i}}), \forall e \in \pmb{\mathrm{E}}_{G}$ is the metric we will use to select a path.

\vspace{0.5em}
Upon arrival of a transfer request, a central controller decides whether it is possible to allocate it considering some criteria that includes the total available bandwidth over future timeslots. If there is not enough room to allocate a request, the request is rejected and can be resubmitted to the system later with a new deadline. 

\vspace{0.5em}
\noindent\textbf{Allocation Problem:} Given active requests $R_1$ through $R_n$ with residual demands $\mathcal{V}^{r}_{R_1}$ to $\mathcal{V}^{r}_{R_n}$ ($0 < \mathcal{V}^{r}_{R_i}, ~ 1 \le i \le n$), is it possible to allocate a new request $R_{n+1}$? If yes, we want to find a valid path over the inter-DC network and a transmission schedule that respects capacity and deadline constraints.

\vspace{0.5em}
There are many ways to formulate this as an optimization problem. We can solve this allocation problem by forming a linear program (LP) considering capacity constraints of the network edges as well as demand constraints of requests while considering a subset of available paths between the source and the destination of new request. We can also formulate an edge-based optimization problem that automatically considers all possible paths. These formulations, however, do not consider the single path routing constraint we have to minimize packet reordering. Adding the single path constraint will turn this into a Mixed Integer Linear Program (MILP) which are in general NP-Hard. If the constructed LP (or MILP) is feasible, the solution will give us a possible allocation. Although this approach maybe straightforward, considering the number of active requests, number of links in network graph, and how far we are planning ahead into the future due to deadlines ($t_{end}$), the resulting LP (or MILP) could be large and may take a long time to solve.

One of the ways to speed up this process is to limit the number of possible paths between every pair of nodes \cite{tempus}, for example, by using only the K-Shortest Paths \cite{amoeba}. Another method to speedup is to limit the number of considered active requests based on some criterion \cite{amoeba} such as having a common edge with the new request on their paths (if we know what path or potential paths we will assign to the new request). It is also possible to use custom iterative methods to solve the resulting LP models faster based on the solutions of previous LP models in a way similar to the water filling process \cite{tempus}.

\subsection{Network-wide ALAP Scheduling}
We do not create an LP model by employing a fast routing heuristic that allows us to select a path according to the total load scheduled on network edges and by trying to allocate new requests only knowing the residual bandwidth on the edges for different timeslots. DCRoute relies on the following three rules.

\vspace{0.5em}
\noindent\textbf{Rule 1:} A path $P_{i}$ is selected for every request $R_{i}$ upon their arrival based on the total outstanding load on the edges of the candidate paths.

\vspace{0.5em}
\noindent\textbf{Rule 2:} $R_{i}$ is initially allocated according to the ALAP policy on $P_{i}$.

\vspace{0.5em}
\noindent\textbf{Rule 3:} If the upcoming timeslot is underutilized, network utilization is maximized by pulling traffic from the closest timeslots into the future.

\vspace{0.5em}
Pulling traffic from closest timeslot into the future to maximize utilization allows the ALAP property of allocation to hold true afterwards. That is, all residual demands will still be allocated as close to their deadlines as possible. Over a network, however, requests with different paths could have common edges which could create complex dependencies that would prevent us from pulling traffic from earliest timeslots with non-zero allocation. This means, to maximize utilization, we may have to pull traffic from later timeslots which might render the ultimate allocation non-ALAP. To fix this, we add a procedure that runs afterwards, scans the timeline, and pushes the allocation forward as much as possible to make it ALAP.

\begin{figure}[t!]
    \centering
    \includegraphics[width=\textwidth]{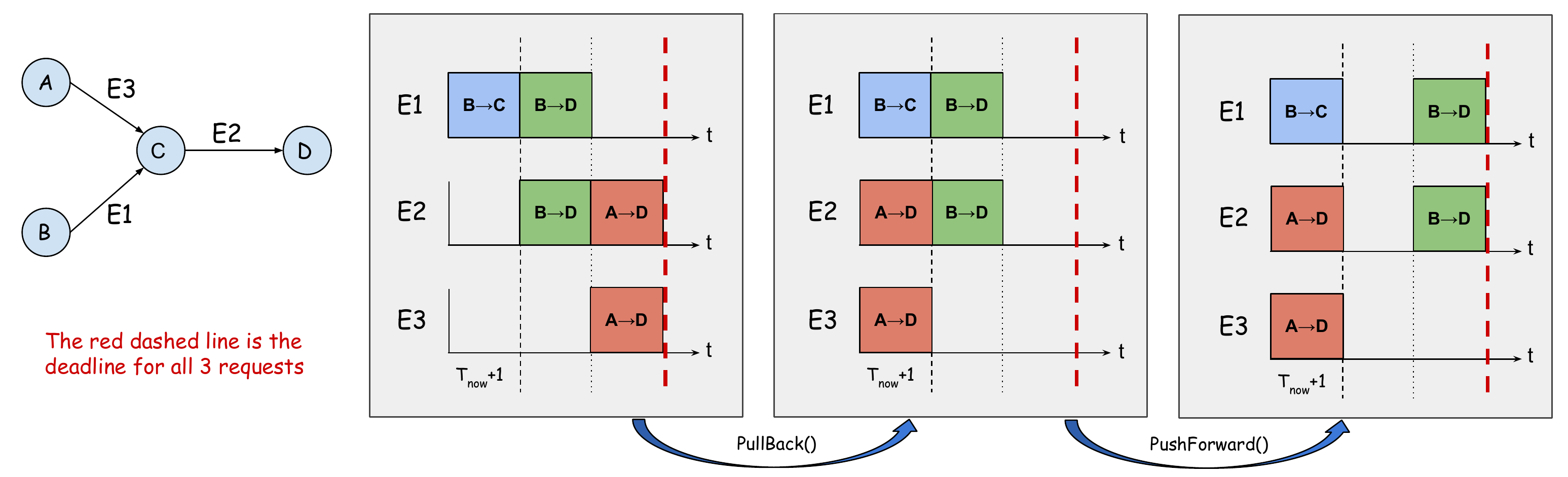}
    \caption{An example of improving utilization (i.e., PullBack phase) while keeping the final allocation ALAP (i.e., PushForward phase).}
    \label{ex}
\end{figure}

Figure \ref{ex} shows an example of this process. There are three different requests all of which having the same deadline. It is not possible to pull back the green request as the link $E1$ is already occupied. Therefore, we have to pull the orange request (PullBack phase). Afterward, the allocation is not ALAP anymore, so we push the green request toward its deadline (PushForward phase). The final assignment is ALAP, and the utilization of upcoming timeslot is maximum.

\subsection{Load-based Dynamic Routing} \label{load_balancing_routing}
\begin{figure}[t!]
    \centering
    \includegraphics[width=0.9\textwidth]{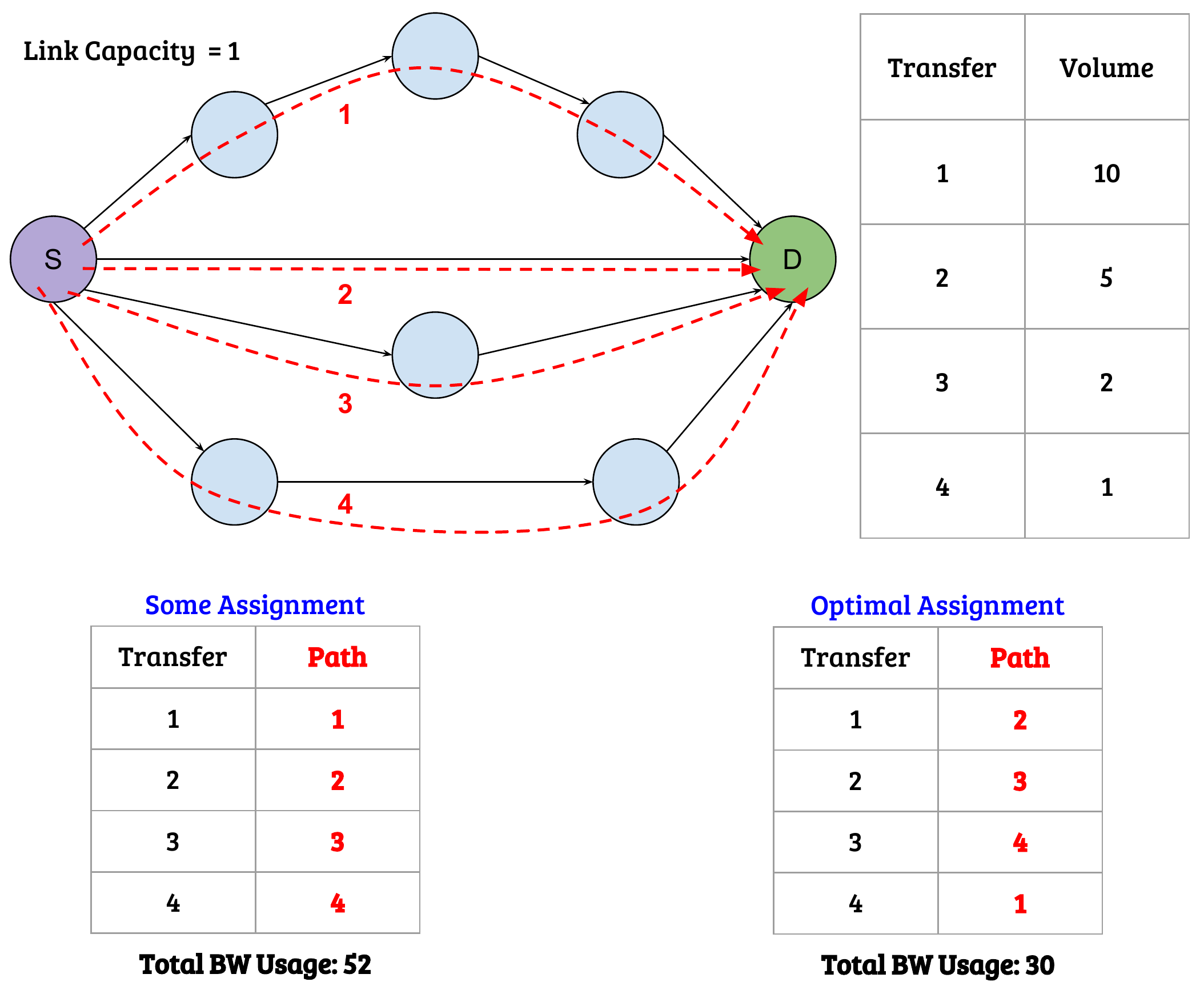}
    \caption{An example of assigning paths to transfers and their total network capacity use assuming that no two transfers should be assigned the same paths for load balancing.}
    \label{fig:ROUTING}
\end{figure}

The next part is assigning paths to new transfers as they arrive. A transfer from any source to any destination can be generally routed over many paths. To avoid packet reordering, we limit the number of paths per transfer to $1$. In general, one can assign static paths to every new request given the source and the destination just like the K-Shortest Paths approach. However, as we will demonstrate, it is better to assign paths to new requests according to their sizes. To understand why this is important, we created the example of Figure \ref{fig:ROUTING}. By assigning shorter paths to larger transfers, the total capacity usage decreases across the network leaving more room for future requests on average. Such savings can pile up as time goes by with arrival of many transfers. This is especially important if transfer sizes are skewed which is what this study from Facebook confirms \cite{social_inside}.

Next, we would like routing to assign different transfers to different paths as much as possible to balance load across the network. If all transfers are assigned to the shortest path, it will be overloaded and slowed down while there is leftover capacity over some longer paths. As a result, we need a path selection technique that takes into account both load balancing and path assignment according to volumes. A well-established technique is to assign available paths some cost that is calculated as a function of transfer size and path properties and select the path with minimum cost. We propose a straight-forward cost assignment scheme that meets the stated criteria and is quick to compute.

\vspace{0.5em}
\noindent\textbf{Routing Cost:} Given a new transfer $R_{new}$ and a set of available paths $\mathrm{\textbf{P}}$ where for every path $P \in \mathrm{\textbf{P}}$ path cost $C_P(t)$ is defined as total outstanding load prior to $t$ which is calculated by summing up the total load scheduled on $P$ prior to $t$ if $R_{new}$ were to be put on $P$ considering the new transfer's size $\mathcal{V}_{R_{new}}$.

\vspace{0.5em}
Let us assume a graph $G(V,E)$ connecting datacenters with bidirectional links with equal capacity for simplicity. We have variables $L_e(t)$ that represent the total sum of traffic volume scheduled over edge $e$ from time $t_{now}+1$ to $t$ (total load that is scheduled but not sent prior to time $t$). The value of $L_e(t)$ depends on transfer arrivals. As new transfers arrive this value increases on some edges and as we send traffic over time, this value decreases. With this notation, the cost assigned to path $P$ with $\lvert P \rvert$ edges given transfer $R_{new}$ with deadline $t_{d_{R_new}}$ and volume $\mathcal{V}_{R_{new}}$ will be:

\begin{equation}
    C_P(t) = \sum_{e \in P} L_e(t) + \mathcal{V}_{R_{new}} ~\lvert P \rvert = \sum_{e \in P}(L_e(t) + \mathcal{V}_{R_{new}})
\end{equation}

\vspace{0.5em}
\noindent\textbf{Routing Objective:} We want to select path $P$ that with minimum value of $C_P(t_{d_{R_{new}}})$ for all valid paths given $R_{new}$. This means selecting the path over which routing $R_{new}$ results in minimum total load (considering $R_{new}$ itself) prior to and including $t_{d_{R_{new}}}$.

\vspace{0.5em}
\noindent\textbf{Implications:} Since this cost assignment is edge decomposable (i.e., cost of a path is sum of costs of its edges), a path with minimum cost can be simply selected using Dijkstra's shortest path algorithm. For small transfers where $\mathcal{V}_{R_{new}}$ is much smaller than $L_e(t_{d_{R_{new}}})$, the total already scheduled load on edges is dominant and as a result the assignment selects paths with minimum total load prior to the new transfer's deadline. If $\mathcal{V}_{R_{new}}$ is considerably larger than $L_e(t_{d_{R_{new}}})$ for candidate paths, the cost function leans toward selecting shorter paths to minimize network capacity usage. This is essentially effective for heavy-tailed transfer size distributions. That is, the few enormous transfers will be scheduled on shortest paths while the rest of transfers are distributed across longer paths for load balancing.

\subsection{DCRoute Algorithm}
Every time a new request is submitted to the system, $t_{end}$ is updated to the latest deadline. We define the \textbf{active window} as the range of timeslots over all edges from time $t_{now}+1$ to $t_{end}$ which are the timeslots DCRoute operates on. DCRoute is made up of four procedures explained in the following.

\SetAlgoVlined
\begin{algorithm}[t]
\small

\vspace{0.4em}
\KwIn{$R_{new}({\cal V}_{R_{new}},S_{R_{new}},D_{R_{new}},t_{d_{R_{new}}})$, $G(V, E)$, $W$, $L_{e}(t)$ and $B_e(t)$, $\forall e \in \pmb{\mathrm{E}}_G, t > t_{now}$}

\vspace{0.4em}
\KwOut{Whether $R_{new}$ should be admitted, and a minimum cost path $P$}

\nonl\hrulefill

\vspace{0.4em}
To every edge $e \in \pmb{\mathrm{E}}_G$, assign cost $L_e(t_{d_{R_{new}}}) + \mathcal{V}_{R_{new}}$\;

\vspace{0.4em}
Find path $P$ by running Dijkstra's algorithm for shortest (minimum cost) path\;


\vspace{0.4em}
$t^{\prime}$ ~$\gets$~ $t_{d_{R_{new}}}$ and $\mathcal{V}^{\prime}$ ~$\gets$~ $\mathcal{V}_{R_{new}}$ \;

\vspace{0.4em}
\While{$\mathcal{V}^{\prime} > 0$ \textnormal{\textbf{and}} $t^{\prime} > t_{now}$} {
 \vspace{0.4em}
 $B_P(t^{\prime})$ ~$\gets$~ $\min_{e \in \pmb{\mathrm{E}}_P}(B_e(t^{\prime}))$ \;
 
 \vspace{0.4em}
 Schedule $R_{new}$ on $P$ with rate $\min(B_P(t^{\prime}),\frac{\mathcal{V}^{\prime}}{\omega})$ at timeslot $t^{\prime}$ \;
 
 \vspace{0.4em}
 $t^{\prime}$ ~$\gets$~ $t^{\prime}+1$ and $\mathcal{V}^{\prime}$ ~$\gets$~ $(\mathcal{V}^{\prime}-\min(B_P(t^{\prime}),\frac{\mathcal{V}^{\prime}}{\omega}) \times \omega)$ \;
}

\vspace{0.4em}
\Return{\textnormal{$P$ if $\mathcal{V}^{\prime} = 0$, otherwise, reject $R_{new}$}}\;

\caption{Allocate($R_{new}$)} \label{algo_1}
\end{algorithm}

\vspace{0.5em}
\noindent\texttt{Allocate($R_{new}$):} Algorithm \ref{algo_1} is executed upon arrival of a new request $R_{new}$ and performs path selection, admission control and bandwidth allocation. To do so, it assigns a cost of $L_e(t_{d_{R_{new}}}) + \mathcal{V}_{R_{new}}$ to every edge $e \in \pmb{\mathrm{E}}_G$ of the graph and then runs Dijkstra's algorithm to select the path $P$ with minimum cost. It then tries to schedule transfer $R_{new}$ on $P$ according to the ALAP policy starting from timeslot $t_{d_{R_{new}}}$ backward until $R_{new}$ is completely satisfied. It rejects the request, if there is not enough capacity on $P$ from $t_{now}$ to $t_{d_{R_{new}}}$.

\vspace{0.5em}
\noindent\texttt{PullBack():} This procedure sweeps the timeslots starting $t_{now}+2$ to $t_{end}$ and pulls back traffic to the next timeslot to be scheduled, i.e., $t_{now}+1$. The objective is to maximize network resource utilization. When pulling back traffic, all edges on a transfer's path have to be checked for available capacity and updated together and atomically as we pull traffic back.

\vspace{0.5em}
\noindent\texttt{PushForward():} After pulling some traffic back, it may be possible for some other traffic to be pushed ahead even further to make the allocation ALAP. This procedure scans all future timeslots starting $t_{now}+2$ and makes sure that all demands are allocated ALAP. If not, it moves as much traffic as possible to the future timeslots until all residual demands are ALAP. Note that there may be many ALAP schedules due to spacial and temporal dependencies across transfers. This procedure finds one of such schedules by scanning through time and edges in a fixed order.

\vspace{0.5em}
\noindent\texttt{Walk():} This procedure is executed when the allocation for next timeslot is final. It broadcasts to all datacenters the allocation that is finalized for the next timeslot and adjusts requests' remaining demands accordingly by deducting what is scheduled to be sent from the total demand.

\subsection{Simulation Results}
In this section, we perform simulations to evaluate the performance of DCRoute. We generate synthetic traffic requests with Poisson arrival and input the traffic to both DCRoute and a few other techniques that can be used for deadline-aware traffic allocation. Two metrics are being measured and compared: \textbf{allocation time} and \textbf{fraction of rejected traffic} both of which are desired to be small.

\vspace{0.5em}
\noindent\textbf{Simulation Parameters:} We used the same traffic distributions as described in \cite{amoeba}. Requests arrive with Poisson distribution of rate $\lambda$. Also, total demand of each request $R$ is distributed exponentially with mean $\frac{1}{8}$ proportional to the maximum transmission volume possible prior to $t_{d_R}$. In addition, the deadline of requests is exponentially distributed for which we assumed a mean of $10$ timeslots. We performed the simulations over $500$ timeslots. We considered a uniform link capacity of $1$ for all edges.

\vspace{0.5em}
We compare DCRoute with the following allocation schemes for all of which we used the same objective function as \cite{amoeba}:

\vspace{0.5em}
\noindent\textbf{Global LP:} This technique is the most general and flexible way of allocation which routes traffic over all possible edges. All active requests are considered for all timeslots on all edges creating a potentially large linear program. The solution here gives us a lower bound on traffic rejection rate.

\vspace{0.5em}
\noindent\textbf{K-Shortest Paths:} Same as Global LP, however, only the K-Shortest Paths between each pair of nodes are considered in routing. The traffic is allocated using a linear program over such paths. We simulated four cases of $K \in \{1, 3, 5, 7\}$. It is obvious that as $K$ increases, the overall rejection rate will decrease as we have higher flexibility for choosing paths and multiplexing traffic.

\vspace{0.5em}
\noindent\textbf{Pseudo-Integer Programming (PIP):} In terms of traffic rejection rate, comparing DCRoute with the previous two techniques is not fair as they allow multiplexing packets on multiple paths. The aim of this technique is to find a lower bound on traffic rejection rate when all packets of each request are sent over a single path. To do so, the general way is to create an integer program involving a list of possible paths (maybe all paths) for the new request and fixed paths for requests already allocated. The resulting model would be a non-linear integer program which cannot be solved using standard optimization libraries available. We instead created a number of linear programs each assigning one of the possible K-Shortest Paths for the newly arriving request. We then compare the objective values manually and choose the best possible path. In our implementation, we chose $K = 20$. This $K$ seems to be more than necessary as we saw negligible improvement in traffic rejection rate even when increasing $K$ from $5$ to $7$. Using PIP, the path over which a request is transferred is decided upon admission and does not change afterwards. We implemented two versions of this scheme:

\begin{itemize}
    \item Pure Minimum Cost (PMC): We choose the path that results in smallest objective value.
    \item Shortest Path, Minimum Cost (SPMC): Amongst all shortest paths that result in a feasible solution and have the least number of hops, we choose the one with smallest objective value.
\end{itemize}

\subsubsection{Experiment 1: Google's GScale Network}
\begin{figure}[p]
    \centering
    \includegraphics[width=\textwidth]{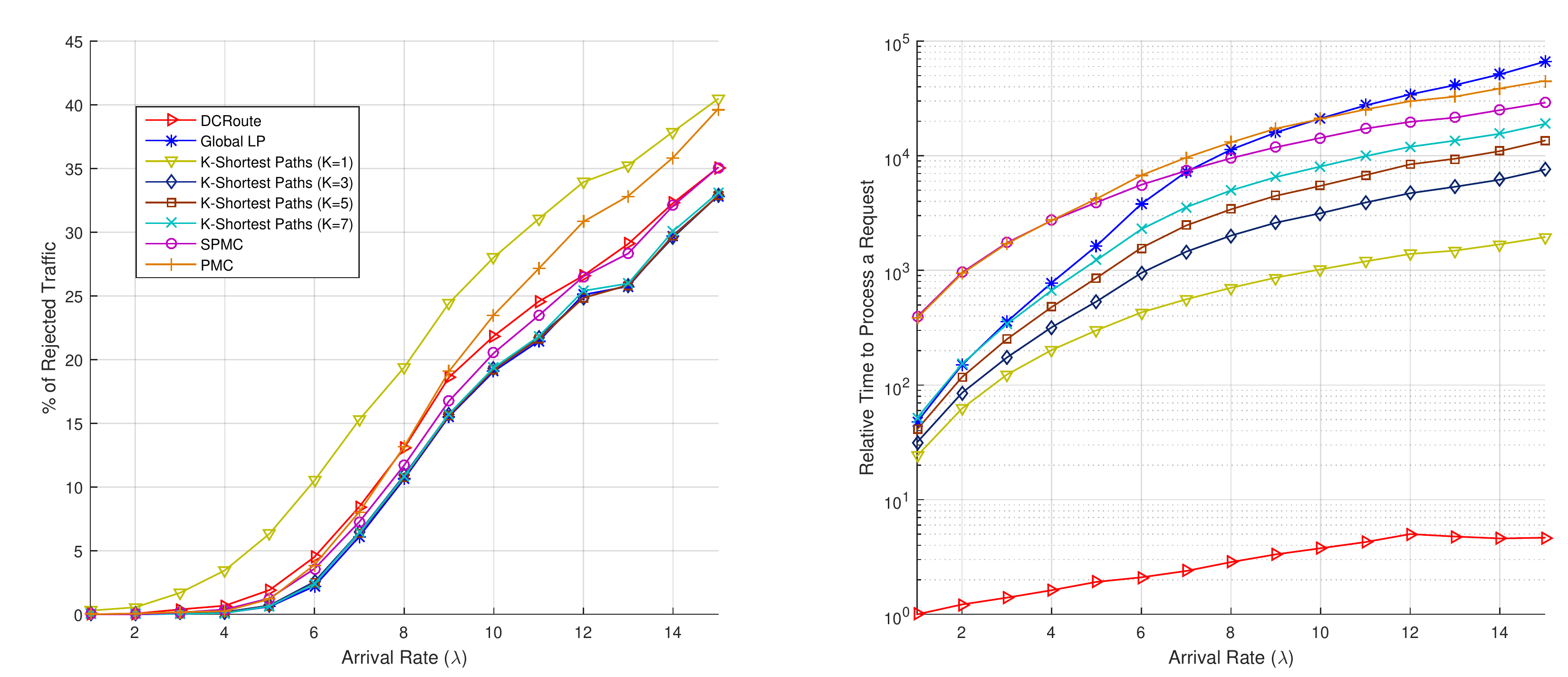}
    \caption{Total \% of rejected traffic and relative request processing time for GScale network with 12 nodes and 19 links.}
    \label{115}
\end{figure}

GScale network \cite{b4} comprised of $12$ nodes and $19$ links connects Google datacenters worldwide. We used the same topology to evaluate DCRoute as well as other allocation schemes. Figure \ref{115} shows the rejection rate of different techniques for different arrival rates from low load ($\lambda=1$) to high load ($\lambda=15$). We have included the schemes that potentially multiplex traffic over multiple paths just to provide a lower bound. Comparing with PMC and SPMC schemes over all arrival rates, DCRoute performs $< 2\%$ worse than the one with minimum rejection rate. Also, compared to all schemes, DCRoute rejects at most $4\%$ more traffic. Figure \ref{115} shows the relative time to process a request using different schemes. This time is calculated dividing the total time to allocate and adjust all requests over all timeslots by the total number of requests. DCRoute is about $3$ orders of magnitude faster than either PMC or SPMC. It should be noted that the rate at which time complexity grows drops as we move toward higher arrival rates since there is less capacity available for new requests and many arriving requests get rejected by failing simple capacity constraint checks.

\begin{figure}[p]
    \centering
    \includegraphics[width=\textwidth]{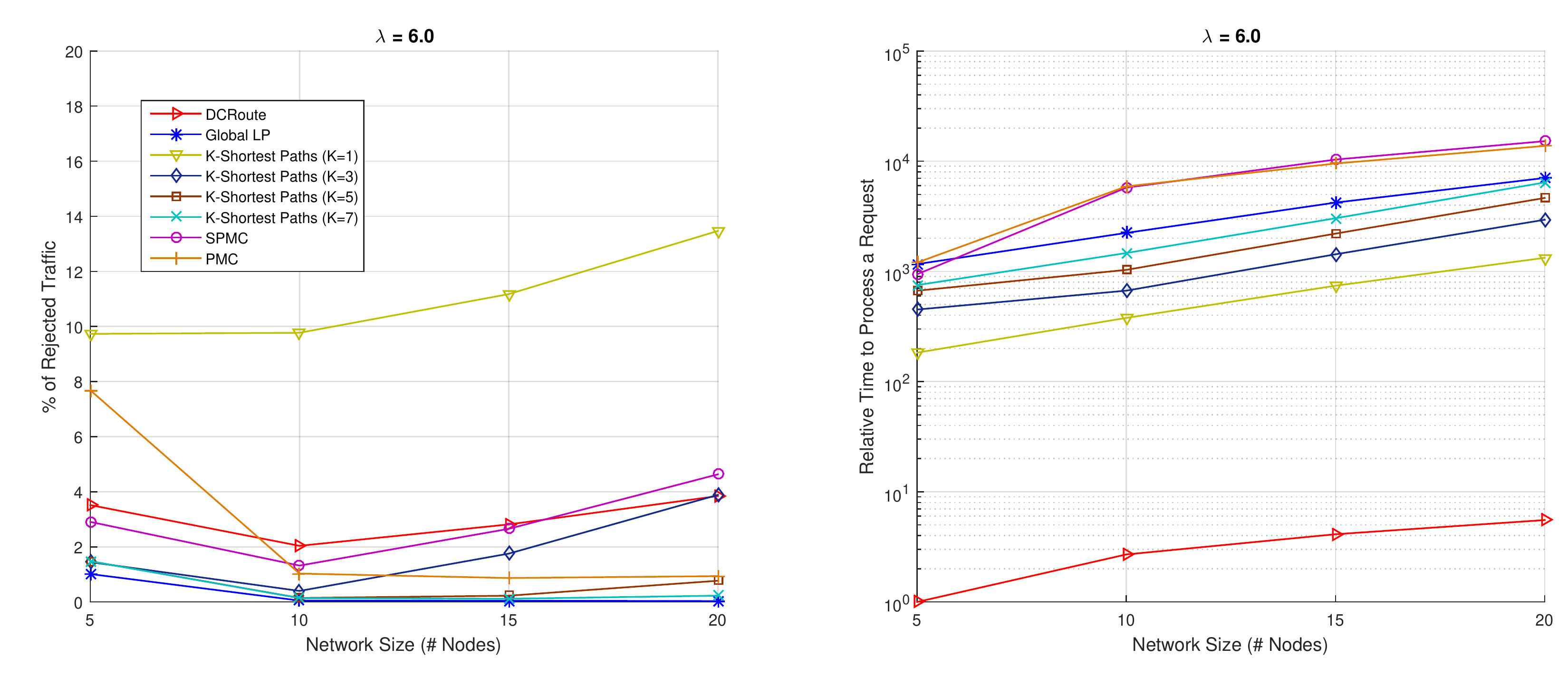}
    \caption{Total \% of rejected traffic and relative request processing time for networks with different sizes.}
    \label{520}
\end{figure}

\subsubsection{Experiment 2: Network with Variable Size}
We simulated different methods against four networks from $5$ to $20$ nodes: $(N, M)$ $\in$ $\{(5, 7), (10, 17),$ $(15, 27), (20, 37)\}$. In our topology, each node was connected to $3$ or $4$ other nodes at most $2$ hops away. The arrival rate was kept constant at $\lambda=6.0$ for all cases. Figure \ref{520} shows the rejection rate of different schemes for different network sizes. As network size increases, since $\lambda$ is kept constant, the total capacity of network increases compared to the total demand of requests. As a result, for a scheme that multiplexes request traffic over different paths, we expect to see a decrease in rejection rate. For the K-Shortest Paths case with $K \in \{1,3\}$, we see an increase in rejection rate which we think is because these schemes cannot multiplex packets that much. Increasing the network size for these cases can cause more requests to have common links as the network is sparsely connected and create more bottlenecks resulting in a higher rejection rate. 

PMC has a high rejection rate for small networks since choosing the minimum cost path might result in selecting longer (more hops) paths that create larger number of bottlenecks due to collision with other requests. Increasing network size, there are more paths to choose from and that results in less bottlenecks and therefore less rejection rate. In contrast, SPMC enforces the selection of paths with smaller number of hops resulting in lower rejection rates for small networks (due to request paths colliding less) and more rejections as network grows due to less diversity of chosen paths.

Compared to these two approaches, DCRoute balances the choice between smaller and longer paths. The assigned path has the least sum of load on the entire path and the least bottleneck load among all such paths. Paths with heavily loaded links and unnecessarily larger number of hops are avoided. As a result, rejection rate compared to $\min($PMC, SPMC$)$ is relatively small ($<3\%$) for all network sizes. Also, as Figure \ref{520} shows, similar to previous simulation, DCRoute is almost three orders of magnitude faster than PIP schemes and more than $200\times$ faster than all considered schemes.

\section{Admission Control with Multipath ALAP Scheduling}
\begin{figure}
    \centering
    \includegraphics[width=0.8\textwidth]{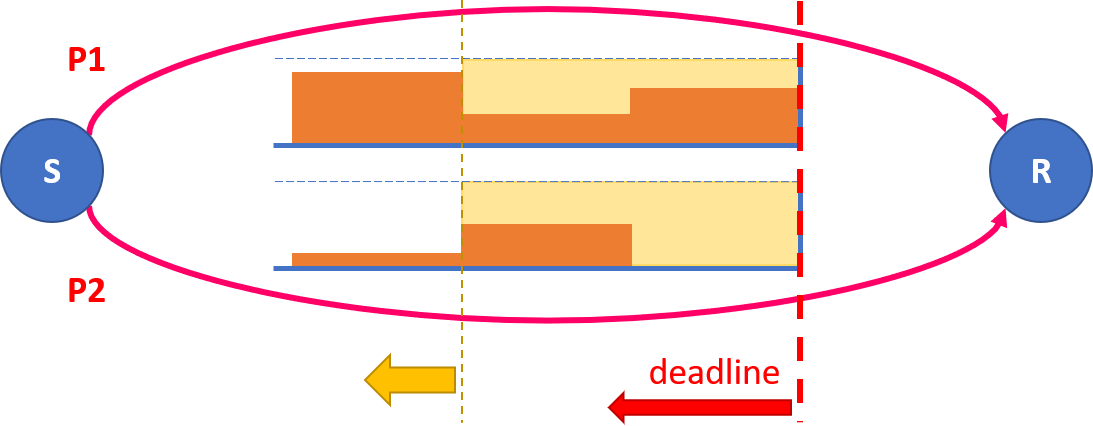}
    \caption{An example of multipath ALAP scheduling: traffic is allocated on edge-disjoint paths from the deadline backward in parallel.}
    \label{fig:mp_dcroute_basics}
\end{figure}

In some scenarios we may be inclined to pay the reordering cost in order to increase the throughput, especially since inter-DC capacity is costly. In case packet reordering is not an issue, we can use multipath routing to increase network throughput and maximize the chances of admission for new transfers with deadlines. According to the ALAP policy, we will need to schedule traffic as close as possible to the deadlines over the multiple paths. That can be done by starting from the deadline on both paths and allocating as much as possible, then moving one timeslot back and allocating as much as possible on both paths, and so on. This has been shown in Figure \ref{fig:mp_dcroute_basics}.

\subsection{Multipath Routing}
There are a variety of ways to select multiple paths for new transfers. We focus on application of parallel edge-disjoint paths to increase throughput. The benefit of using edge-disjoint paths is that the traffic for the same transfer will not have to compete with itself over common edges. We explain how multiple paths are selected and name our technique MP-DCRoute.

We want to select paths in a similar way to the approach presented in \ref{load_balancing_routing} which allowed for quick selection of paths while it balanced the load across network edges. To select more than one path with such properties, after finding the first path, we mark all of its edges as deleted and then search for the next path. This way, we are sure to obtain the same good load balancing properties while guaranteeing that the paths are edge-disjoint. We can keep searching for new paths until there are no more paths remain, or we can terminate the search as soon as we find a given number of paths. We combined these two conditions to allow for up to $K$ load balancing paths per transfer where $K$ is a configuration parameter.

The parameter $K$ needs to be selected carefully as using too many parallel paths per transfer can waste bandwidth and exhaust network capacity. That is because as we select more paths, the paths tend to grow longer, or use edges that are heavily loaded. This means that, under light load, using more paths can improve throughput while under heavy load, doing such can quickly saturate the network and lead to rejection of transfers. In general, $K$ can be selected adaptively according to the network's overall load factor. That is, the operators can monitor incoming traffic load and update $K$ accordingly for the new transfers.

\subsection{Simulation Results}
In this section, we perform simulations to evaluate the performance of MP-DCRoute. We generate synthetic traffic requests with Poisson arrival and input the traffic to both MP-DCRoute and DCRoute presented in the previous section. Three metrics are being measured and compared: \textbf{allocation time}, \textbf{fraction of rejected requests} and \textbf{fraction of rejected traffic} all of which are desired to be small.

\vspace{0.5em}
\noindent\textbf{Simulation Parameters:} We used the same traffic distributions as described in \cite{amoeba}. Requests arrive with Poisson distribution of rate $\lambda$. Also, total demand of each request $R_{new}$ is distributed exponentially with mean $\frac{1}{2}$ proportional to the maximum transmission volume possible prior to $t_{d_{R_{new}}}$. In addition, the deadline of requests is exponentially distributed for which we assumed a mean of $1$ timeslot. We performed the simulations over $1000$ timeslots. We considered a uniform link capacity of $1$ for all edges.

\vspace{0.5em}
We compare the following allocation schemes which are basically single and multipath ALAP scheduling techniques:

\vspace{0.5em}
\noindent\textbf{DCRoute (1 path):} The technique proposed in \S \ref{section_dcroute}. It uses a single path that is selected adaptively according to network load to balance load and minimize packet reordering.

\vspace{0.5em}
\noindent\textbf{MP-DCRoute (up to $K$ paths):} We use the technique proposed in this section to select up to $K$ edge-disjoint adaptively selected paths that balance load across the network.

\begin{figure}[p]
    \centering
    \includegraphics[width=\textwidth]{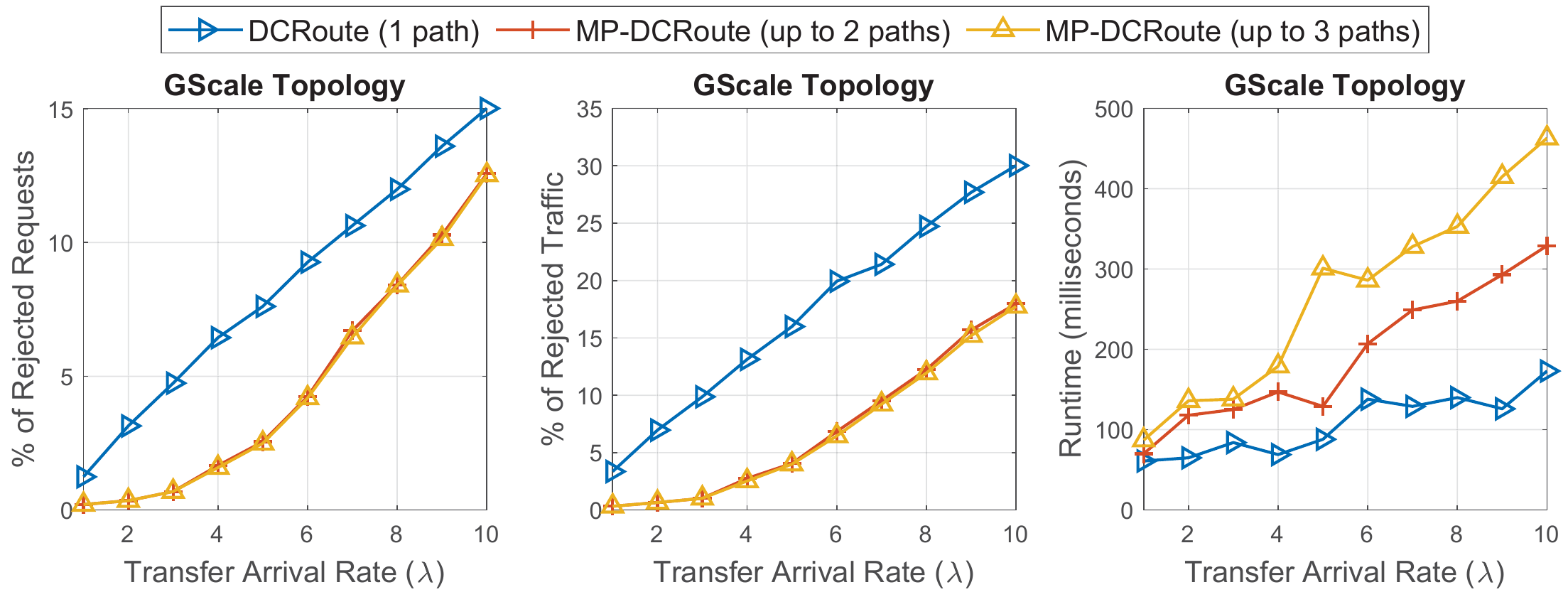}
    \caption{Multipath ALAP scheduling over GScale \cite{b4} topology.} \label{fig:mp_dcroute_gscale}
\end{figure}

\begin{figure}[p]
    \centering
    \includegraphics[width=\textwidth]{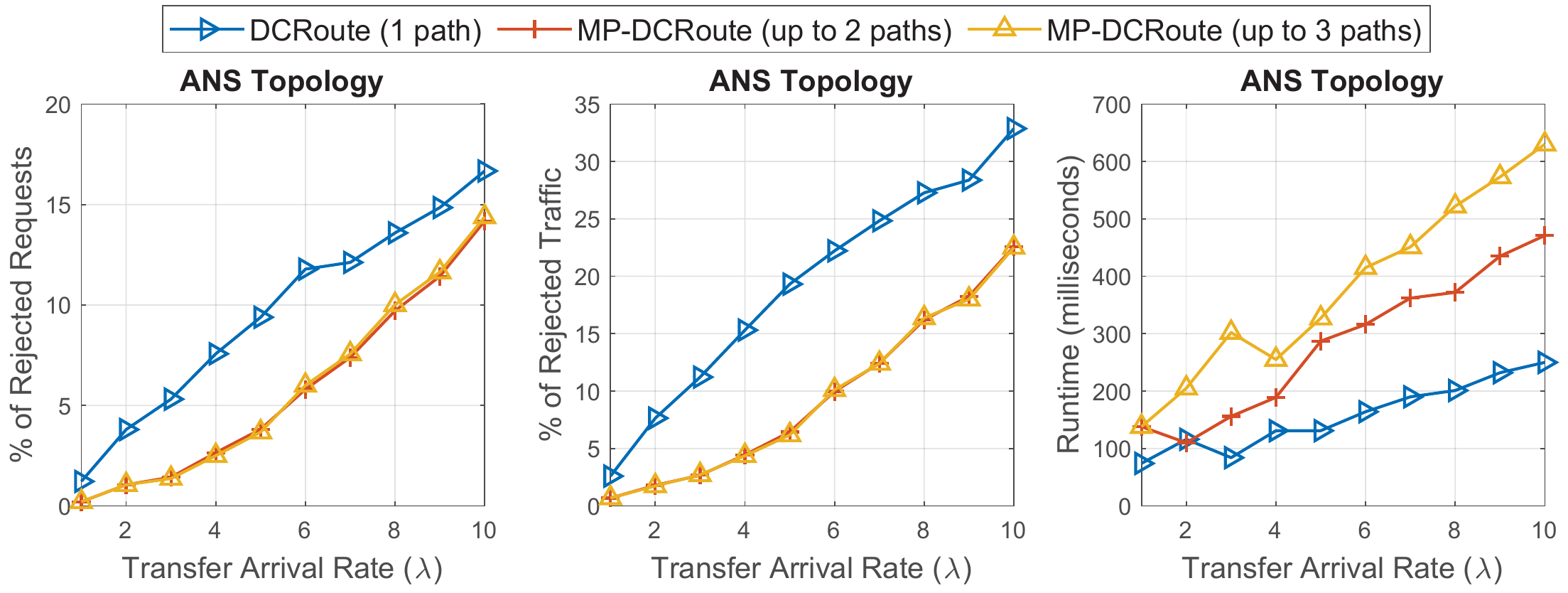}
    \caption{Multipath ALAP scheduling over ANS \cite{ans} topology.} \label{fig:mp_dcroute_ans}
\end{figure}

\begin{figure}[p]
    \centering
    \includegraphics[width=\textwidth]{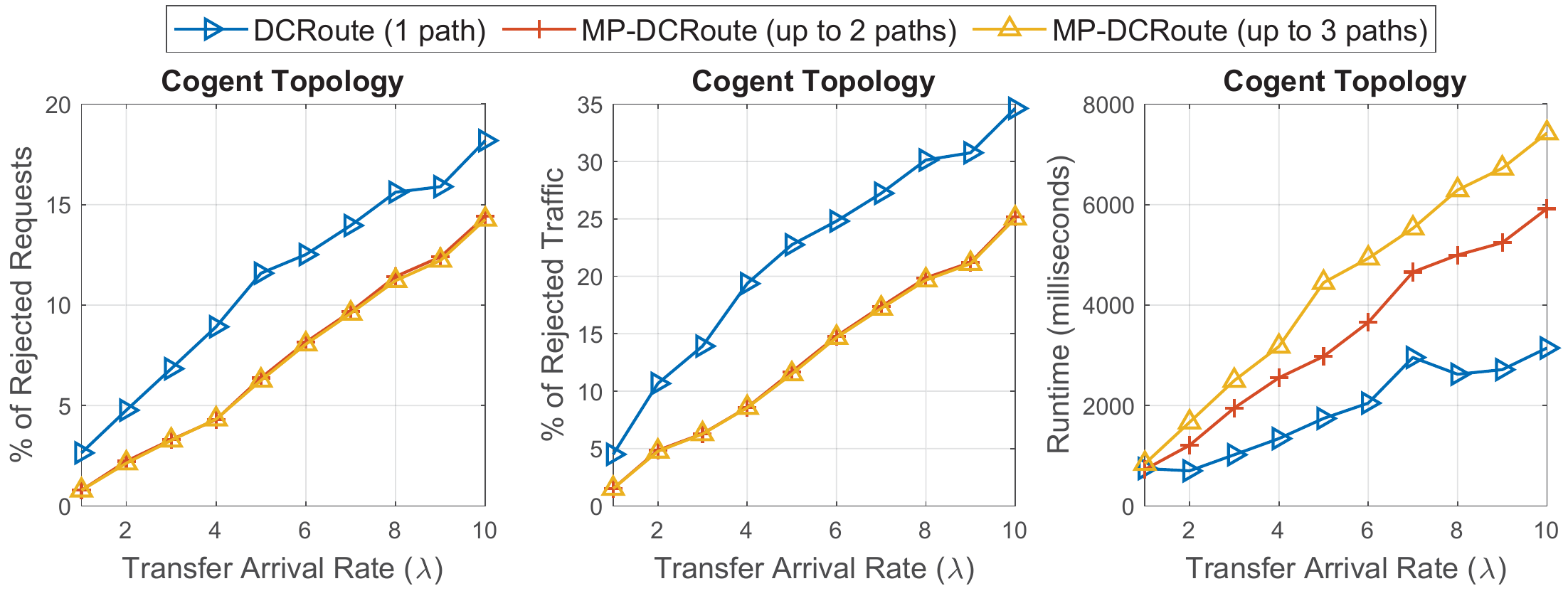}
    \caption{Multipath ALAP scheduling over Cogent \cite{cogent} topology.} \label{fig:mp_dcroute_cogent}
\end{figure}

\vspace{0.5em}
We compare DCRoute and MP-DCRoute over three different topologies as shown in Figures \ref{fig:mp_dcroute_gscale}, \ref{fig:mp_dcroute_ans} and \ref{fig:mp_dcroute_cogent}. In terms of the total traffic admitted and the total number of requests admitted, we see that MP-DCRoute does considerably better, i.e., up to $12\%$ more traffic and up to $5\%$ more transfers are admitted to the inter-DC network. We also see that the gain of using multiple paths reduces as we increase the network load by increasing the arrival rate of transfers. Also, we see that all the benefit of using multiple paths is received with $2$ paths and increasing the number of paths to $3$ has virtually no benefit.\footnote{In most cases, using 3 paths instead of 2 hurts the performance.}

We then evaluate the running time of different techniques which is the total computation time to handle all $1000$ timeslots. We see that MP-DCRoute can be between up to $2\times$ to $3\times$ slower than DCRoute which is due to the time needed to find additional paths and schedule traffic over multiple paths per transfer. However, since the total time to process a single request is small,\footnote{In the order of milliseconds.} this should not cause any practical impediments.

\clearpage
\section{Conclusions}
In this chapter, we discussed the problem of admission control for inter-DC transfers with deadlines which is an essential problem given that inter-DC networks have limited capacity. Sending traffic without paying attention to deadlines could waste bandwidth as the value of completed transfers past their deadlines may be significantly less. We discussed why current approaches based on linear programming or mixed integer linear programming are not effective in general as they could take a long time to solve and require considerable computing resources. We presented a new scheduling technique called As Late As Possible (ALAP) policy that allows the scheduler to quickly decide whether a new transfer can be admitted on a given path. We then developed an adaptive routing approach that balances load across the network and saves network capacity by routing larger transfers over shorter paths. Finally, we realized that, although using a single path per transfer can minimize packet reordering, which is a desired property, it can also limit the obtainable throughput. We then applied an edge-disjoint multipath routing technique that improves the traffic admitted to the network. We performed extensive simulations to confirm the effectiveness of our approaches showing that our methods can reduce the time needed to perform admission control and compute a valid schedule by orders of magnitude at little or no cost to the total traffic admitted to the inter-DC network.

\clearpage
\chapter{Efficient Point to Multipoint Transfers over Inter-DC Networks} \label{chapter_p2mp_dccast}
As discussed in Chapter \ref{chapter_introduction}, a large volume of inter-DC traffic is due to replication of data and content from one datacenter to multiple other datacenters. We refer to such transfers as Point to Multipoint (P2MP) which have a known sender and set of receivers upon arrival. Also, in general, we do not have knowledge of arrival times for these transfers and have to manage them as they arrive at the network, i.e., in an online fashion.

We consider efficient routing and scheduling of P2MP data transfers, with the objective of minimizing transfer completion times and total network capacity consumption. Using centralized scheduling and load-aware multicast tree selection, we can significantly improve the performance. Our approach is different from traditional multicasting in that we select multicast trees atomically given the source and all the destinations whereas traditional multicasting builds multicast trees incrementally as destinations join. With a global view of network topology and edge load status, it is possible to find near optimal weighted Steiner trees that connect any given source datacenter to its destination datacenters per P2MP transfer. We define appropriate edge weights and select minimum weight Steiner trees which lead to efficient bandwidth utilization of all network edges. To our knowledge, the research set forth, at the time of publication,\footnote{This chapter was originally published in \cite{dccast}.} was the first to explore and study efficient P2MP transfers over inter-DC networks.

\section{Background and Related Work}
\begin{figure}[t!]
    \centering
    \includegraphics[width=\textwidth]{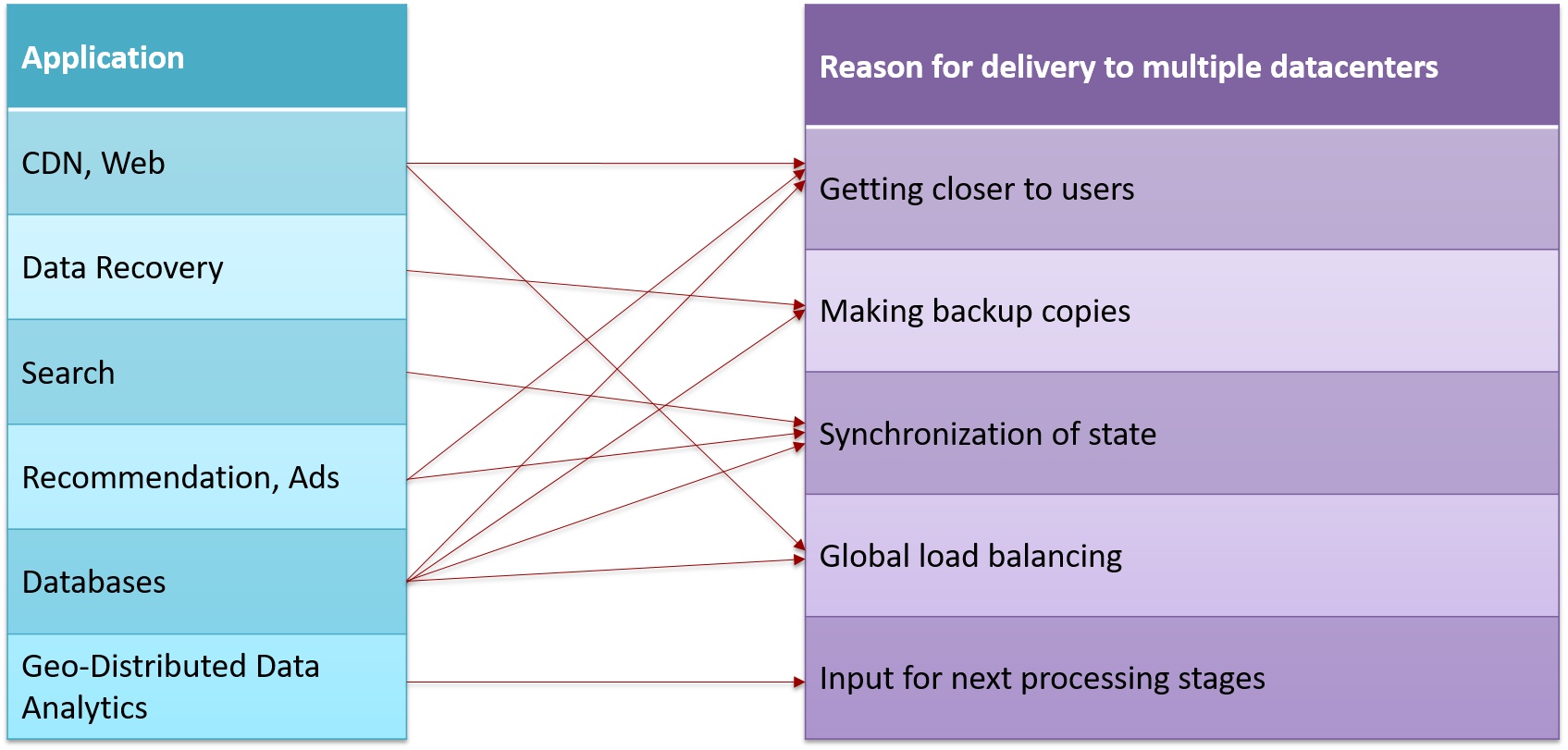}
    \caption{Applications that generate transfers potentially with multiple destinations.}
    \label{fig:p2mp_app}
\end{figure}

A variety of datacenter services replicate content and data from one location to many locations. Table \ref{table_0} provides a brief list of how many replicas are made for some applications. Also, Figure \ref{fig:p2mp_app} offers a list of applications that perform P2MP transfers and gives a short description of why such replication is done.

\begin{table}[t!]
\begin{center}
\caption{Various services that perform data replication.} \label{table_0}
\vspace{0.5em}
\begin{tabular}{ |p{4cm}|p{11cm}| }
    \hline
    \textbf{Service} & \textbf{Replicas} \\
    \hline
    \hline
    Facebook & Across availability regions \cite{rep-facebook}, $\ge 4$ \cite{rep-facebook-2}, for various object types including large machine learning configs \cite{fb-holistic} \\
    \hline
    CloudBasic SQL Server & Up to $4$ secondary databases with active Geo-Replication (asynchronous) \cite{rep-cloudbasic} \\
    \hline
    Azure SQL Database & Up to $4$ secondary databases with active Geo-Replication (asynchronous) \cite{rep-azure} \\
    \hline
    Oracle Directory Server & Up to the number of datacenters owned by an enterprise for regional load balancing of directory servers \cite{rep-oracle-1, rep-oracle-2}\\
    \hline
    AWS Route $53$ GLB & Across multiple regions and availability zones for global load balancing \cite{route-53} \\
    \hline
    Youtube & Function of popularity, content potentially pushed to many locations (could be across $\ge33$ datacenters \cite{rep-youtube}) \\
    \hline
    Netflix & Across $2$ to $4$ availability regions \cite{rep-netflix-regions}, and up to $233$ cache locations distributed globally \cite{rep-netflix-locations} \\
    \hline
\end{tabular}
\end{center}
\end{table}

\begin{figure}
    \centering
    \includegraphics[width=0.5\columnwidth]{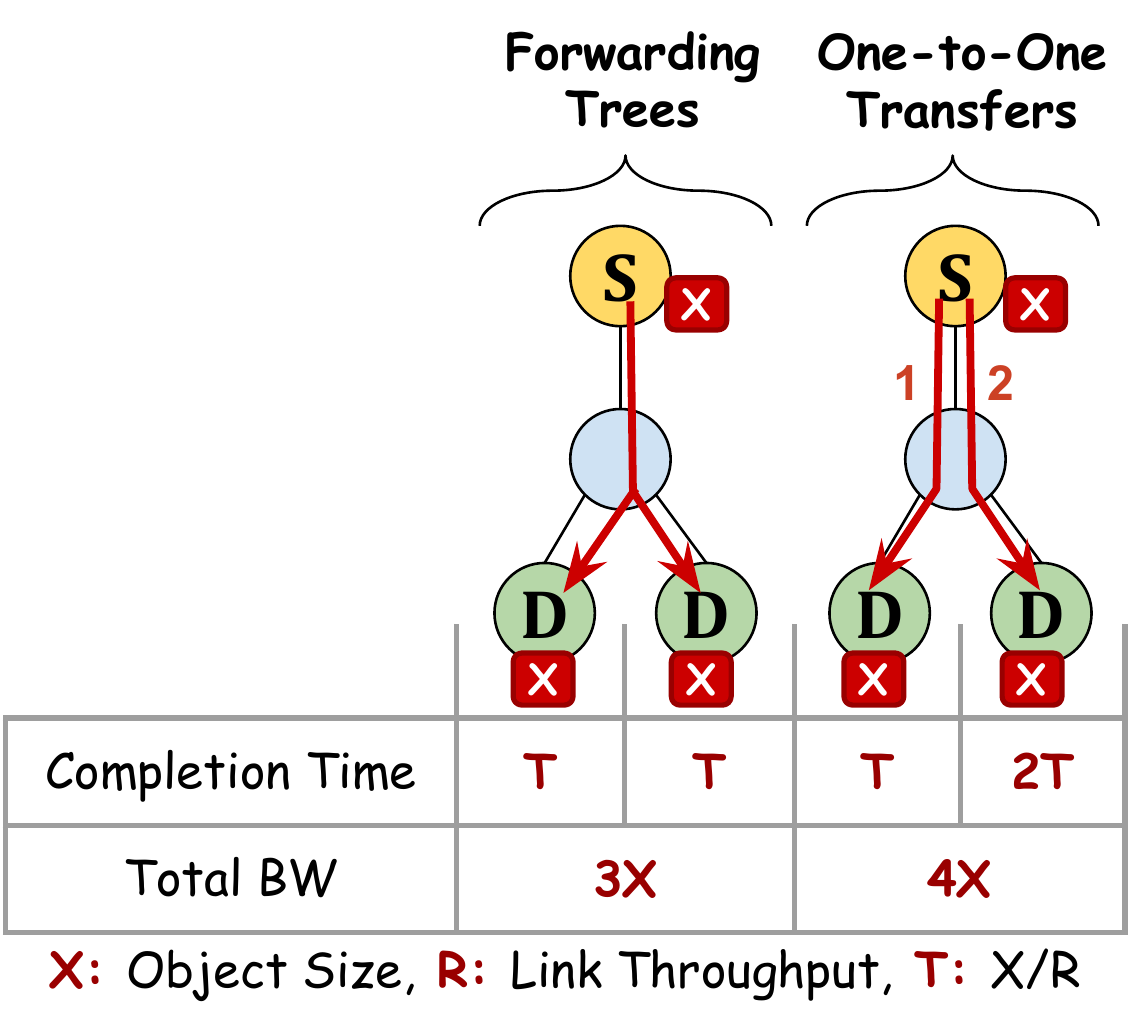}
    \caption{Inter-DC multicasting can reduce total bandwidth consumption as well as completion times of transfers.}
    \label{fig:p2mp_0}
\end{figure}

One solution is to perform P2MP transfers as multiple independent P2P transfers that are scheduled separately \cite{mbdt_initial, ssnf, netstitcher, postcard, dtb,  grease, geo_backup_selection, amoeba, tempus, ecoflow, orchestrating, dcroute}. There may however be more efficient ways, in terms of total bandwidth usage and transfer completion times, to perform P2MP transfers by sending at most one copy of the message across any link given that the source datacenter and destination datacenters are known apriori. In Figure \ref{fig:p2mp_0}, an object $X$ is to be transferred from datacenter $S$ to two $D$ datacenters considering a link throughput of $R$. In order to send $X$ to destinations, one could initiate individual transfers, but that wastes bandwidth and increases delivery time since the link attached to $S$ turns into a bottleneck.

We present an elegant solution using minimum weight Steiner Trees \cite{steiner_tree_problem} (a.k.a., Forwarding Trees, or Multicast Trees) for P2MP transfers that achieves reduced bandwidth usage and tail completion times for receivers. We briefly go over some of the related work in this space and survey their objectives and methods.

\vspace{0.5em}
\noindent\textbf{Internet Multicasting:} 
A large body of general multicasting approaches have been proposed where receivers can join multicast groups anytime to receive required data and multicast trees are incrementally built and pruned as nodes join or leave a multicast session such as IP multicasting \cite{ip_multicast}, TCP-SMO \cite{tcp-smo} and NORM \cite{norm}. These solutions focus on building and maintaining multicast trees, and do not consider link capacity and other ongoing multicast flows while building the trees.

\vspace{0.5em}
\noindent\textbf{Multicast Traffic Engineering:}
An interesting work \cite{online_multicast_bw_guarantees} considers the online arrival of multicast requests with a specified bandwidth requirement. The authors provide an elegant solution to find a minimum weight Steiner tree for an arriving request with all edges having the requested available bandwidth. This work assumes a fixed transmission rate per multicast tree, dynamic multicast receivers, and unknown termination time for multicast sessions whereas we consider variable transmission rates over timeslots, fixed multicast receivers, and deem a multicast tree completed when all its receivers download a specific volume of data. MTRSA \cite{sdn_multicast} considers a similar problem to \cite{online_multicast_bw_guarantees} but in an offline scenario where all multicast requests are known beforehand while taking into account the number of available forwarding rules per switch. MPMC \cite{MPMC_2013, MPMC_2016} maximizes the throughput for a single multicast transfer by using multiple parallel multicast trees and coding techniques. None of these works aims to minimize the completion times of receivers while considering the total bandwidth consumption.

\vspace{0.5em}
\noindent\textbf{Datacenter Multicasting:}
A variety of solutions have been proposed for minimizing congestion across the intra-datacenter network by selecting multicast trees according to link utilization. Datacast \cite{datacast} sends data over edge-disjoint Steiner trees found by pruning spanning trees over various topologies of FatTree, BCube, and Torus. AvRA \cite{avalanche} focuses on tree and FatTree topologies and builds minimum edge Steiner trees that connect the sender to all receivers as they join. MCTCP \cite{mctcp} reactively schedules flows according to link utilization. These works do not aim at minimizing the completion times of receivers and ignore the total bandwidth consumption.

\vspace{0.5em}
\noindent\textbf{Overlay Multicasting:}
With overlay networks, end-hosts can form a multicast forwarding tree in the application layer. RDCM \cite{rdcm} populates backup overlay networks as nodes join and transmits lost packets in a peer-to-peer fashion over them. NICE \cite{nice} creates hierarchical clusters of multicast peers and aims to minimize control traffic overhead. AMMO \cite{AMMO} allows applications to specify performance constraints for selection of multi-metric overlay trees. DC2 \cite{dc2} is a hierarchy-aware group communication technique to minimize cross-hierarchy communication. SplitStream \cite{split-stream} builds forests of multicast trees to distribute load across many machines. BDS \cite{overlay_hkust} generates an application-level multicast overlay network, creates chunks of data, and transmits them in parallel over bottleneck-disjoint overlay paths to the receivers. Due to limited knowledge of underlying physical network topology and condition (e.g., utilization, congestion or even failures), and limited or no control over how the underlying network routes traffic, overlay routing has limited capability in managing the total bandwidth usage and distribution of traffic to minimize completion times of receivers. In case such control and information are provided, for example by using a cross-layer approach, overlay multicasting can be used to realize solutions such as those presented in this dissertation.

\vspace{0.5em}
\noindent\textbf{Reliable Multicasting:}
Various techniques have been proposed to make multicasting reliable including the use of coding and receiver (negative or positive) acknowledgments. Experiments have shown that using positive ACKs does not lead to ACK implosion for medium scale (sub-thousand) receiver groups \cite{tcp-smo}. TCP-XM \cite{tcp-xm} allows reliable delivery by using a combination of IP multicast and unicast for data delivery and re-transmissions. MCTCP \cite{mctcp} applies standard TCP mechanisms for reliability. Another approach is for receivers to send NAKs upon expiration of some inactivity timer \cite{norm}. NAK suppression has been proposed to address implosion which can be applied by routers \cite{arm}. Forward Error Correction (FEC) has been used to reduce re-transmissions \cite{norm} and improve the completion times \cite{avalanche_code} examples of which include Raptor Codes \cite{raptor} and Tornado Codes \cite{tornado}. These techniques can be applied complementary to the algorithms and techniques presented in this dissertation.

\vspace{0.5em}
\noindent\textbf{Multicast Congestion Control:}
Existing approaches track the slowest receiver. PGMCC \cite{pgmcc}, MCTCP \cite{mctcp} and TCP-SMO \cite{tcp-smo} use window-based TCP like congestion control to compete fairly with other flows. NORM \cite{norm} uses an equation-based rate control scheme. With rate allocation and end-host based rate limiting applied over inter-DC networks, need for distributed congestion control becomes minimal; however, such techniques can still be used as a backup in case there is a need to fall back to distributed inter-DC traffic control.

\vspace{0.5em}
\noindent\textbf{Other Related Work:}
CastFlow \cite{castflow} precalculates multicast spanning trees which can then be used at request arrival time for fast rule installation. ODPA \cite{odpa} presents algorithms for dynamic adjustment of multicast spanning trees according to specific metrics. BIER \cite{bier} has been recently proposed to improve the scalability and allow frequent dynamic manipulation of multicast forwarding state in the network and can be applied complementary to our solutions in this dissertation. Peer-to-peer approaches \cite{promise, bittorrent, slurpie} aim to maximize throughput per receiver without considering physical network topology, link capacity, or total network bandwidth consumption. Store-and-Forward (SnF) approaches \cite{netstitcher, mbdt, dtb, mbdt_initial} focus on minimizing transit bandwidth costs which does not apply to dedicated inter-DC networks. However, SnF can still be used to improve overall network utilization in the presence of diurnal link utilization patterns, transient bottleneck links, or for application layer multicasting. BDS \cite{bds} uses many parallel overlay paths from a multicast source to its destinations storing and forwarding data from one destination to the next. Application of SnF for bulk multicast transfers considering the physical topology is complementary to our work in this dissertation. Recent research \cite{ddccast,multicast_deadline,AGE,dartree} also consider bulk multicast transfers with deadlines with the objective of maximizing the number of transfers completed before the deadlines.

\section{Adaptive Forwarding Tree Selection for P2MP Transfers} \label{dccast}
We present an efficient scheme for P2MP transfers called DCCast \cite{dccast} which aims to optimize tail transfer completion times as well as total network capacity consumption. It selects forwarding trees according to a weight assignment that tries to balance load across the network. 

\subsection{System Model} \label{dccast_sys_model}
To allow for flexible bandwidth allocation, we consider a slotted timeline \cite{tempus, amoeba, dcroute} where the transmission rate of senders is constant during each timeslot, but can vary from one timeslot to next. This can be achieved via rate-limiting at end-hosts \cite{swan, bwe}. A central scheduler is assumed that receives transfer requests from end-points, calculates their temporal schedule, and informs the end-points of rate-allocations when a timeslot begins. We focus on scheduling large transfers that take more than a few timeslots to finish and therefore, the time to submit a transfer request, calculate the routes, and install forwarding rules is considered negligible in comparison. We assume equal capacity for all links in an online scenario where requests may arrive anytime. A more advanced solution that considers non-uniform link capacity is discussed in the next chapter. We will use the same notation as that in Table \ref{table_var_0} with some additional variables in this section as shown in Table \ref{table_var_dccast}.

\begin{table}
\begin{center}
\caption{Variables used in this chapter in addition to those in Table \ref{table_var_0}} \label{table_var_dccast}
\vspace{0.5em}
\begin{tabular}{ |p{2cm}|p{13cm}| }
    \hline
    \textbf{Variable} & \textbf{Definition} \\
    \hline
    \hline
    $L_e$ & Total load currently scheduled on edge $e$ (same as $L_e(t_{now})$) \\
    \hline
    $T_{R_{i}}$ & The forwarding tree (i.e., multicast tree) selected for request $R_i$ \\
    \hline
\end{tabular}
\end{center}
\end{table}

\vspace{0.5em}
\noindent\textbf{Definition of Edge Load $L_e$:} We define a new metric called edge load which provides a measure of how busy a link is expected to be on average over future timeslots. $L_e, \forall e \in \pmb{\mathrm{E}}_{G}$ is the total volume of traffic scheduled on an edge $e$ which is computed by summing up the number of remaining bytes for all the transfers that share $e$ at $t_{now}$.

\subsection{Selection of Forwarding Trees}
Our proposed approach is, for each P2MP transfer, to jointly route traffic from source to all destinations over a forwarding tree to save bandwidth. Using a single forwarding tree for every transfer also minimizes packet reordering which is known to waste CPU and memory resources at the receiving ends especially at high rates \cite{juggler, mptcphard}.

To perform a P2MP transfer $R_{new}$ with volume $\mathcal{V}_{R_{new}}$, the source $S_{R_{new}}$ transmits traffic over a Steiner Tree that spans across $\pmb{\mathrm{D}}_{R_{new}}$. At any timeslot, traffic for any transfer flows with the same rate over all links of a forwarding tree to reach all the destinations at the same time. The problem of scheduling a P2MP transfer then translates to finding a forwarding tree and a transmission schedule over such a tree for every arriving transfer in an online manner. A relevant problem is the minimum weight Steiner tree \cite{steiner_tree_problem} that can help minimize total bandwidth usage with proper weight assignment. Although it is a hard problem, heuristic algorithms exist that often provide near optimal solutions \cite{robins2005tighter, Watel2014}.

\subsection{Scheduling Policy}
When forwarding trees are found, we schedule traffic over them according to First Come First Serve (FCFS) policy using all available residual bandwidth on links to minimize the completion times. This allows us to provide guarantees to users on when their transfers will complete upon their arrival. We do not use a preemptive scheme, such as Shortest Remaining Processing Time (SRPT), due to practical concerns: larger transfers might get postponed over and over which might lead to the starvation problem and it is not possible to make promises on exactly when a transfer would complete. Optimal scheduling discipline to minimize tail times rests on transfer size distribution \cite{caltech-tail}.

\subsection{DCCast Algorithms}
DCCast is made up of two algorithms as follows.\footnote{An implementation of DCCast is available on Github: \url{https://github.com/noormoha/DCCast}}

\vspace{0.5em}
\noindent\texttt{Update()}: This procedure is executed upon beginning of every timeslot. It simply dispatches the transmission schedule, that is the rate for each transfer, to all senders to adjust their rates via rate-limiting and adjusts $L_{e}~(e \in \pmb{\mathrm{E_G}})$ by deducting the total traffic that was sent over $e$ during current timeslot.

\vspace{0.5em}
\noindent\texttt{Allocate($R_{new}$)}: This procedure is run upon arrival of every request which finds a forwarding tree and schedules $R_{new}$ to finish as early as possible. Pseudo-code of this function has been shown in Algorithm \ref{algo_1}. Statically calculating forwarding trees can lead to creation of hot-spots, even if there exists one highly loaded edge that is shared by multiple trees. As a result, DCCast dynamically chooses a forwarding tree that reduces the tail transfer completion times while saving considerable bandwidth.

It is possible that larger trees provide higher available bandwidth by using longer paths through least loaded edges, but using which would consume more overall bandwidth since they send same traffic over more edges. To model this behavior, we use a weight assignment that allows balancing these two possibly conflicting objectives. The weights represent traffic load allocated on links. Selecting links with lower weights will improve load balancing that would be better for future requests. The trade off is in avoiding heavier links at the expense of getting larger trees for a more even distribution of load.

The forwarding tree $T_{R_{new}}$ selected by Algorithm \ref{algop2mp_1} will have a total weight of:

\begin{equation}
    \sum_{e \in \pmb{\mathrm{E}}_{T_{R_{new}}}}(L_e + \mathcal{V}_{R_{new}})
\end{equation}

This weight is essentially the total load over $T_{R_{new}}$ if request $R_{new}$ were to be allocated on it. Selecting trees with minimal total weight will most likely avoid highly loaded edges and larger trees. To find an approximate minimum weight Steiner Tree, we used GreedyFLAC \cite{Watel2014, DSTAlgoEvaluation}, which is quite fast and in practice provides results not far from the optimal.

\SetAlgoVlined
\begin{algorithm}[t!]
\small
\vspace{0.4em}
\KwIn{$R(\mathrm{V}_{R_{new}},S_{R_{new}},\pmb{\mathrm{D}}_{R_{new}})$, $G(\pmb{\mathrm{V}}_G, \pmb{\mathrm{E}}_G)$, $\omega$, $L_{e}$ and $B_e(t)$ for $e \in \pmb{\mathrm{E}}_G$ and $t > t_{now}$}

\vspace{0.4em}
\KwOut{Forwarding tree (minimum weight Steiner Tree) $T_{R_{new}}$ and transmission schedule (traffic allocation) for $R_{new}$ for $t > t_{now}$}

\nonl\hrulefill

\vspace{0.4em}
To every edge $e \in \pmb{\mathrm{E}}_G$, assign weight $(L_{e} + \mathrm{V}_{R_{new}})$\;

\vspace{0.4em}
Find the minimum weight Steiner tree $T_{R_{new}}$ that connects $S_{R_{new}} \cup \pmb{\mathrm{D}}_{R_{new}}$. We used GreedyFLAC \cite{Watel2014, DSTAlgoEvaluation}\;


\vspace{0.4em}
$t^{\prime}$ ~$\gets$~ $t_{now}+1$ and $\mathcal{V}^{\prime}$ ~$\gets$~ $\mathcal{V}_{R_{new}}$ \;

\vspace{0.4em}
\While{$\mathcal{V}^{\prime} > 0$} {
 \vspace{0.4em}
 $B_{T_{R_{new}}}(t^{\prime})$ ~$\gets$~ $\min_{e \in \pmb{\mathrm{E}}_{T_{R_{new}}}}(B_e(t^{\prime}))$ \;
 
 \vspace{0.4em}
 Schedule $R_{new}$ on $T_{R_{new}}$ with rate $\min(B_{T_{R_{new}}}(t^{\prime}),
 \frac{\mathcal{V}^{\prime}}{\omega})$ at timeslot $t^{\prime}$ \;
 
 \vspace{0.4em}
 $t^{\prime}$ ~$\gets$~ $t^{\prime}+1$ and $\mathcal{V}^{\prime}$ ~$\gets$~ $\mathcal{V}^{\prime}-\min(B_{T_{R_{new}}}(t^{\prime}),
 \frac{\mathcal{V}^{\prime}}{\omega}) \times \omega$ \;
}

\vspace{0.4em}
\Return{\textnormal{$T_{R_{new}}$ and the transmission schedule of $R_{new}$}}\;

\caption{Allocate($R_{new}$)} \label{algop2mp_1}
\end{algorithm}

\subsection{Evaluation} \label{eval}
We evaluated DCCast using synthetic traffic. We assumed a total capacity of $1.0$ for each timeslot over every link. The arrival of requests followed a Poisson distribution with rate $\lambda_{P2MP} = 1$. Demand of every request was calculated using an exponential distribution with mean $20$ added to a constant value of $10$ (fixing the minimum demand to $10$). All simulations were performed over as many timeslots as needed to finish all requests with arrival time of last request set to be $500$ or less. Presented results are normalized by minimum values in each chart.

\begin{table}
    \begin{center}
    \caption{Schemes used for comparison.} \label{table_1}
    \vspace{0.5em}
        \begin{tabular}{ |l|p{11.5cm}| } 
        
\hline
\textbf{Scheme} & \textbf{Method} \\
\hline
\hline
MINMAX      & Selects forwarding trees to minimize maximum load on any link. Schedules traffic using FCFS policy \S \ref{dccast}. \\
\hline
RANDOM      & Selects random forwarding trees. Schedules traffic using FCFS policy \S \ref{dccast}. \\
\hline
BATCHING    & Batches (enqueues) new requests arriving in time windows of $T$. At the end of batching windows, jointly schedules all new requests according to Shortest Job First (SJF) policy and picks their forwarding trees using weight assignment of Algorithm \ref{algo_1}. \\
\hline
SRPT        & Upon arrival of a new request, jointly reschedules all existing requests and the new request according to SRPT policy \S \ref{dccast} and picks new forwarding trees for all requests using weight assignment of Algorithm \ref{algo_1}. \\
\hline
P2P-SRPT-LP & Views each P2MP request as multiple independent point-to-point (P2P) requests. Uses a Linear Programming (LP) model along with SRPT policy \S \ref{dccast} to (re)schedule each request over $K$-Shortest Paths between its source and destination upon arrival of new requests. \\
\hline
P2P-FCFS-LP & Similar to above while using FCFS policy \S \ref{dccast}. \\
\hline
            
        \end{tabular}
    \end{center}
\end{table}

We measure three different metrics: \textbf{total bandwidth used} as well as \textbf{mean and tail transfer completion times}. The total bandwidth used is the sum of all traffic over all timeslots and all links, i.e., the total network capacity consumed during simulation running time. The completion time of a transfer is defined as its arrival time to the time its last bit is delivered to the destination(s). We performed simulations using Google's GScale topology \cite{b4}, with $12$ nodes and $19$ edges, on a single machine (Intel Core i7-6700T and 24 GBs of RAM). All simulations were coded in Java and used Gurobi Optimizer \cite{gurobi} to solve linear programs for P2P schemes. We increased the destinations (copies) for each object from $1$ to $6$ picking recipients according to uniform distribution. Table \ref{table_1} shows list of considered schemes. In this table, the first $4$ approaches are P2MP schemes and last $2$ are P2P schemes that operate by breaking each P2MP transfer into multiple P2P transfers.

\begin{figure}
    \centering
    \includegraphics[width=0.7\textwidth]{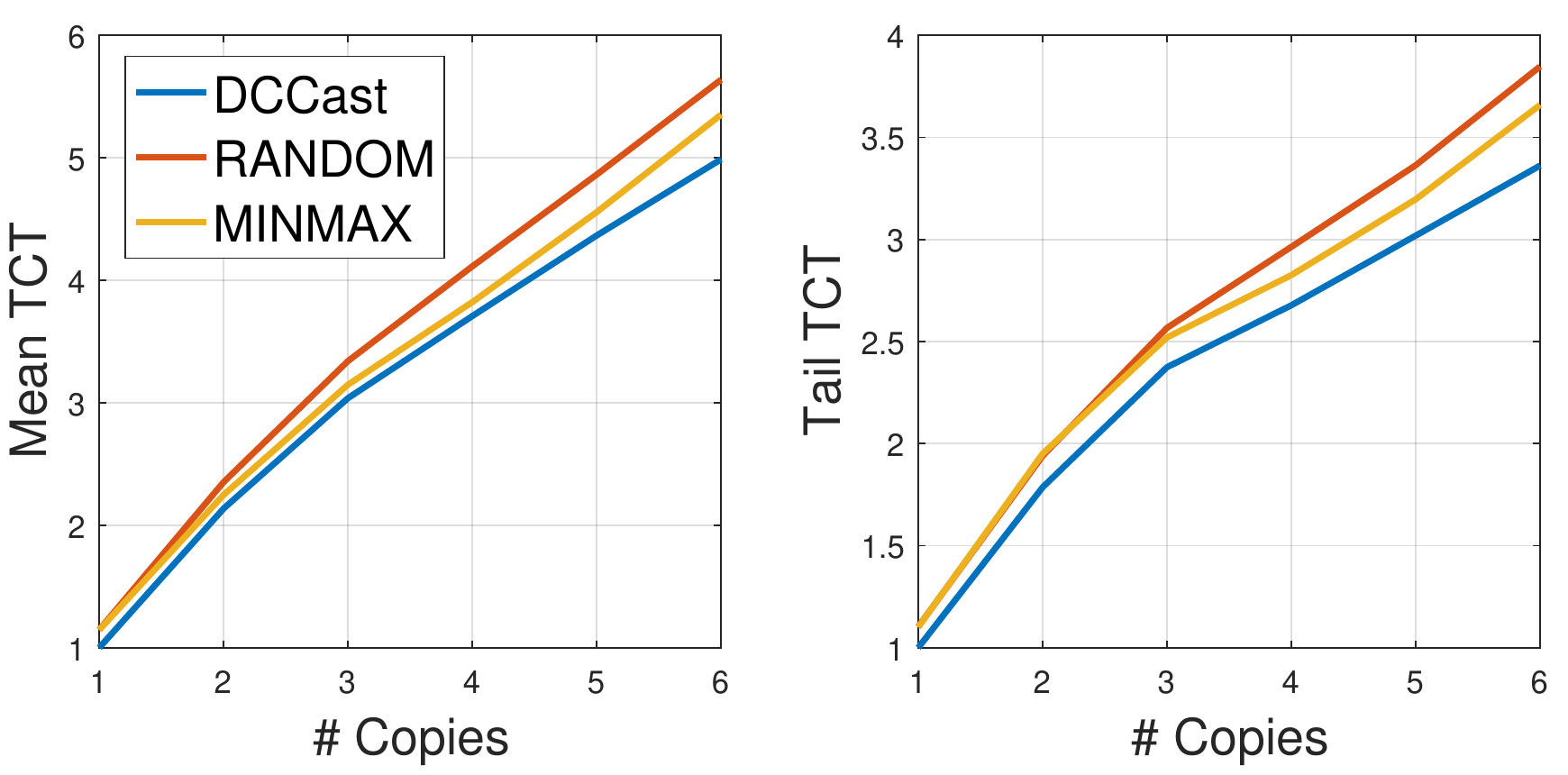}
    \caption{Tree Selection (GScale Topo)}
    \label{fig:p2mp_all_1}
\end{figure}

\begin{figure}
    \centering
    \includegraphics[width=0.7\textwidth]{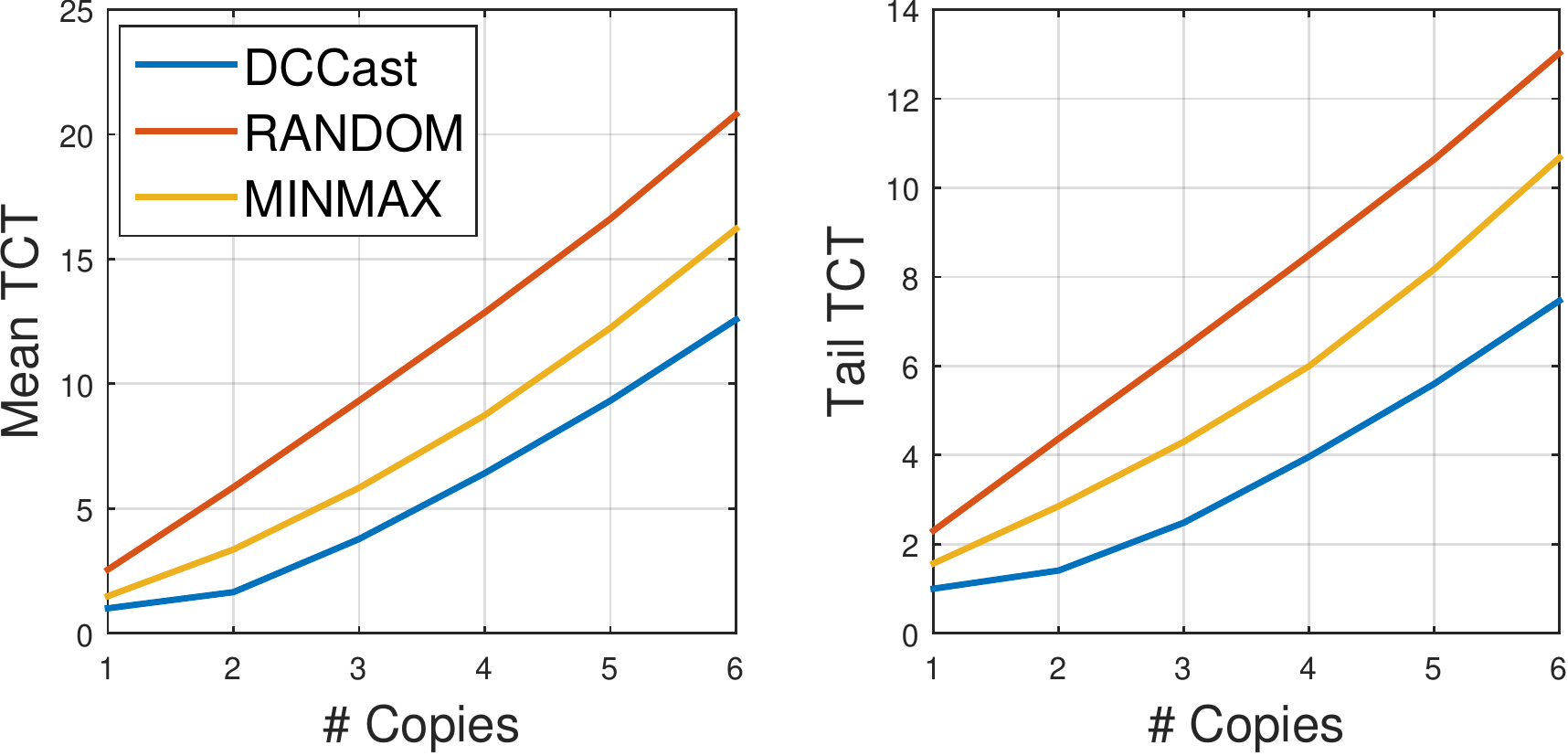}
    \caption{Tree Selection (Random topology, $\lvert \pmb{\mathrm{V}}_G \lvert=50$)}
    \label{fig:p2mp_all_1_2}
\end{figure}

We evaluated various forwarding tree selection criteria over both GScale topology and a larger random topology with $50$ nodes and $150$ edges as shown in Figures \ref{fig:p2mp_all_1} and \ref{fig:p2mp_all_1_2}, respectively. In case of GScale, DCCast performs slightly better than RANDOM and MINMAX in completion times while using equal overall bandwidth (not in figure). In case of larger random topologies, DCCast's dominance is more obvious regarding completion times while using same or less bandwidth (not in figure).

\begin{figure}
    \centering
    \includegraphics[width=0.7\textwidth]{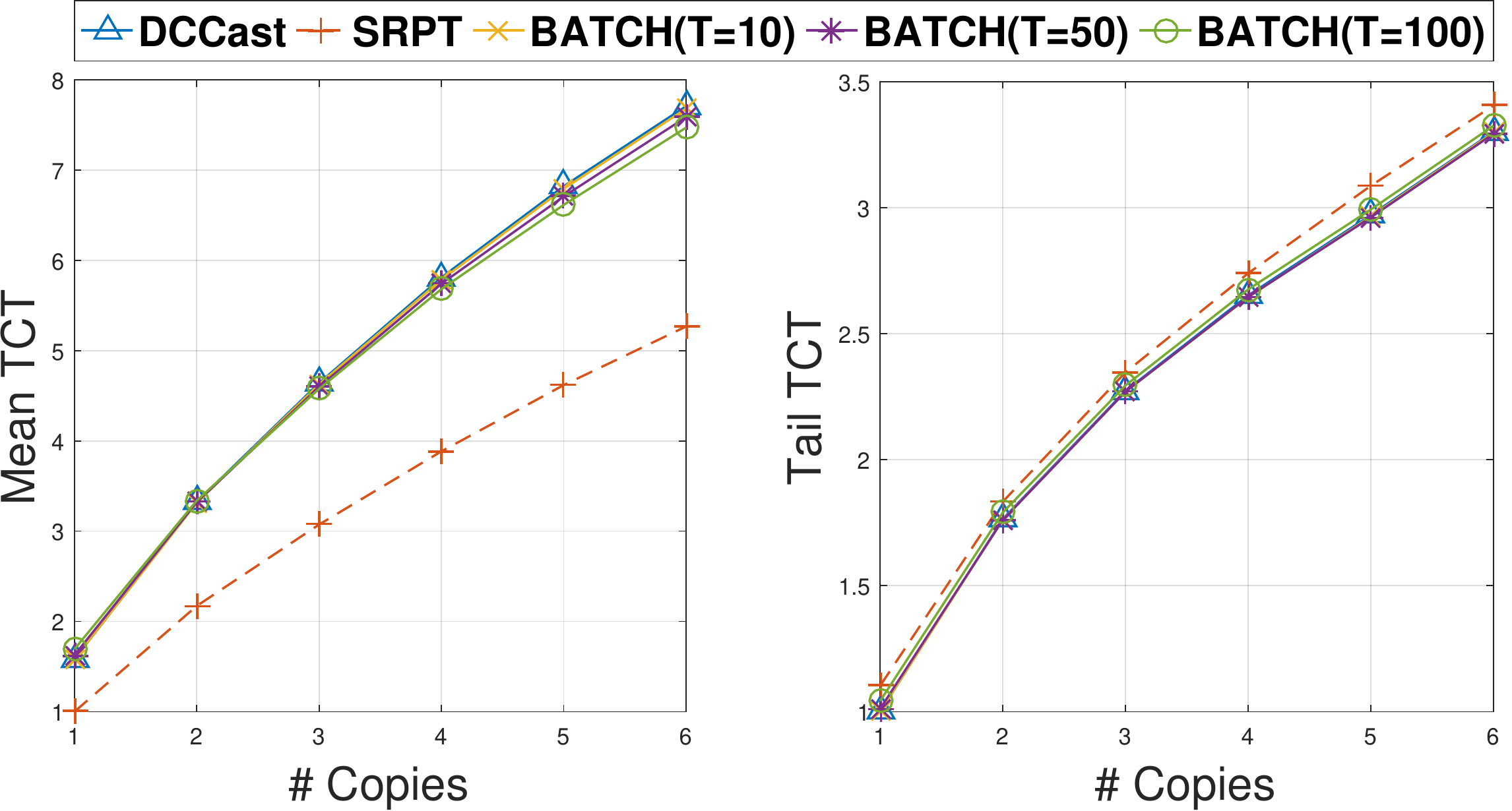}
    \caption{Various scheduling policies and the effect of batching.}
    \label{fig:p2mp_all_2}
\end{figure}

We also experimented various scheduling disciplines over forwarding trees as shown in Figure \ref{fig:p2mp_all_2}. The SRPT discipline performs considerably better with respect to mean completion times; it however may lead to starvation of larger transfers if smaller ones keep arriving. It has to compute and install new forwarding trees and recalculate the whole schedule, for all requests currently in the system with residual demands, upon arrival of every new request. This could impose significant rule installation overhead which is considered negligible in our evaluations. It might also lead to lots of packet loss and reordering. Batching improves performance marginally compared to DCCast and could be an alternate road to take. Generally, a smaller batch size results in a smaller initial scheduling latency while a larger batch size makes it possible to employ collective knowledge of many requests in a batch for optimized scheduling. Batching might be more effective for systems with bursty request arrival patterns. All schemes performed almost similarly regarding tail completion times and total bandwidth usage (not in figure).

\begin{figure}
    \centering
    \includegraphics[width=\textwidth]{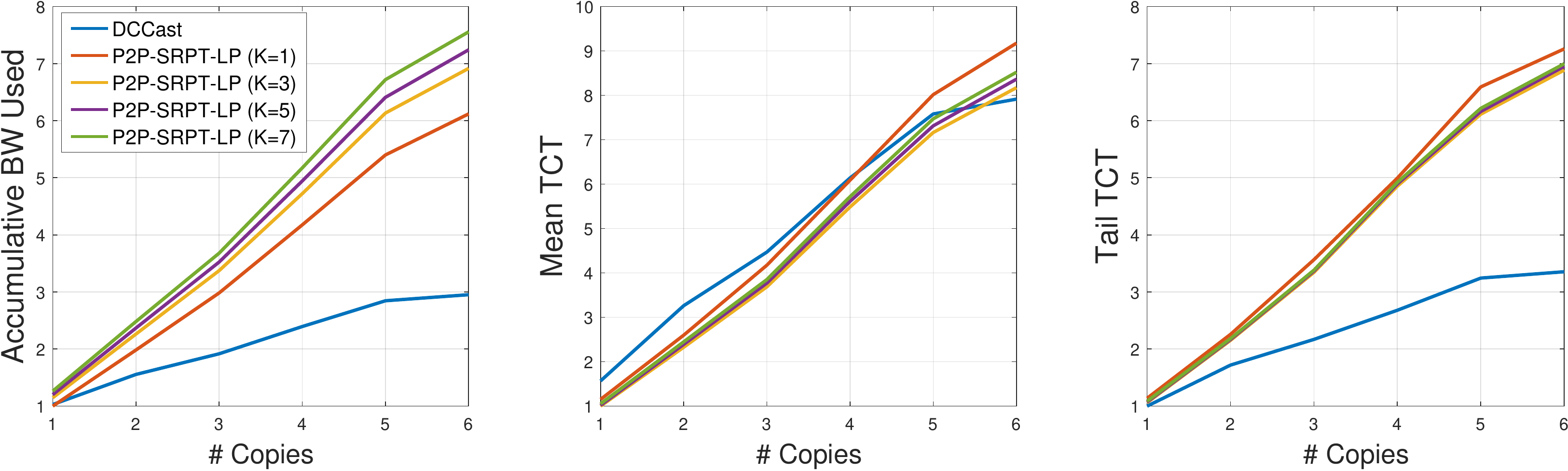}
    \caption{DCCast vs Point-To-Point (P2P-SRPT-LP).}
    \label{fig:p2mp_all_3}
\end{figure}

In Figure \ref{fig:p2mp_all_3}, we compare DCCast with a Point-To-Point scheme (P2P-SRPT-LP) using SRPT scheduling policy which uses various number of shortest paths (i.e., $K$ shortest paths) and delivers each copy independently. The total bandwidth usage is close for all schemes when there is only one destination per request. Both bandwidth usage and tail completion times of DCCast are up to $50\%$ less than that of P2P-SRPT-LP as the number of destinations per transfer increases. Although DCCast follows the FCFS policy, its mean completion time is close to that of P2P-SRPT-LP and surpasses it for $6$ copies due to bandwidth savings which leave more headroom for new transfers. In a different experiment, we compared DCCast with P2P-FCFS-LP and obtained somewhat similar results. DCCast again saved up to $50\%$ bandwidth and reduced tail completion times by up to almost $50\%$ while increasing the number of destinations per transfer.

\begin{figure}[t!]
    \centering
    \includegraphics[width=\textwidth]{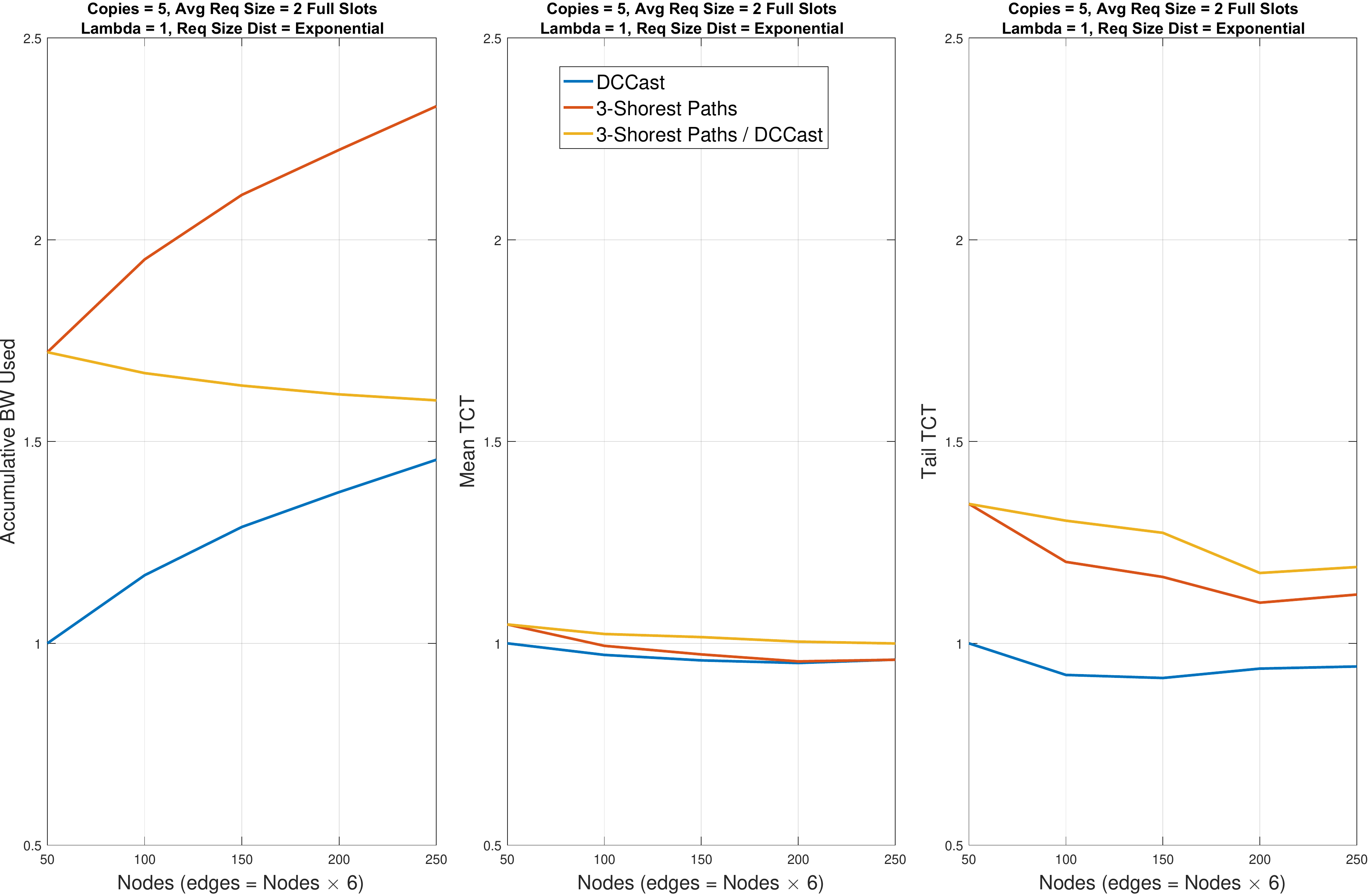}
    \caption{Performance of $3$-Shortest Paths (P2P) vs DCCast as network grows.}
    \label{fig:perf_by_size}
\end{figure}

\begin{figure}[t!]
    \centering
    \includegraphics[width=\textwidth]{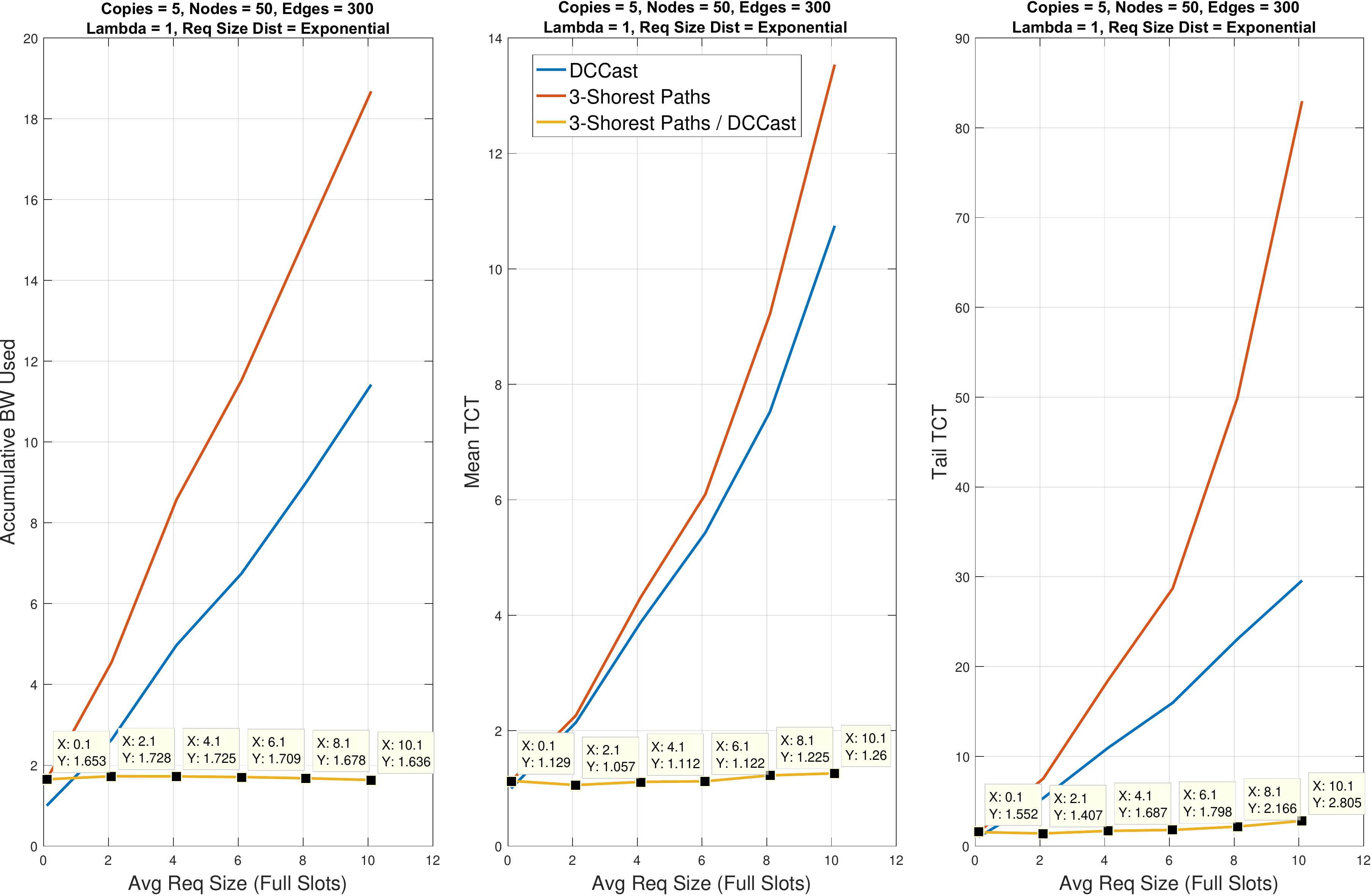}
    \caption{Performance of $3$-Shortest Paths (P2P) vs DCCast as incoming network load increases.}
    \label{fig:perf_by_load}
\end{figure}

Finally, we studied the effect of load and network size on DCCast comparing it with a P2P scheme that is based on $3$-Shortest Paths. Figure \ref{fig:perf_by_size} shows that when network grows in size, there is minor change in performance of P2MP routing. The total bandwidth usage increase obviously since paths become longer. However, the growth in bandwidth usage of P2P scheme considered is a little more than that of DCCast. Figure \ref{fig:perf_by_load} shows the effect of input load on performance of same schemes. As can be seen, all performance metrics grow much slower for DCCast compared to P2P shortest paths (lower values are better).

\begin{figure}[t!]
    \centering
    \includegraphics[width=0.5\textwidth]{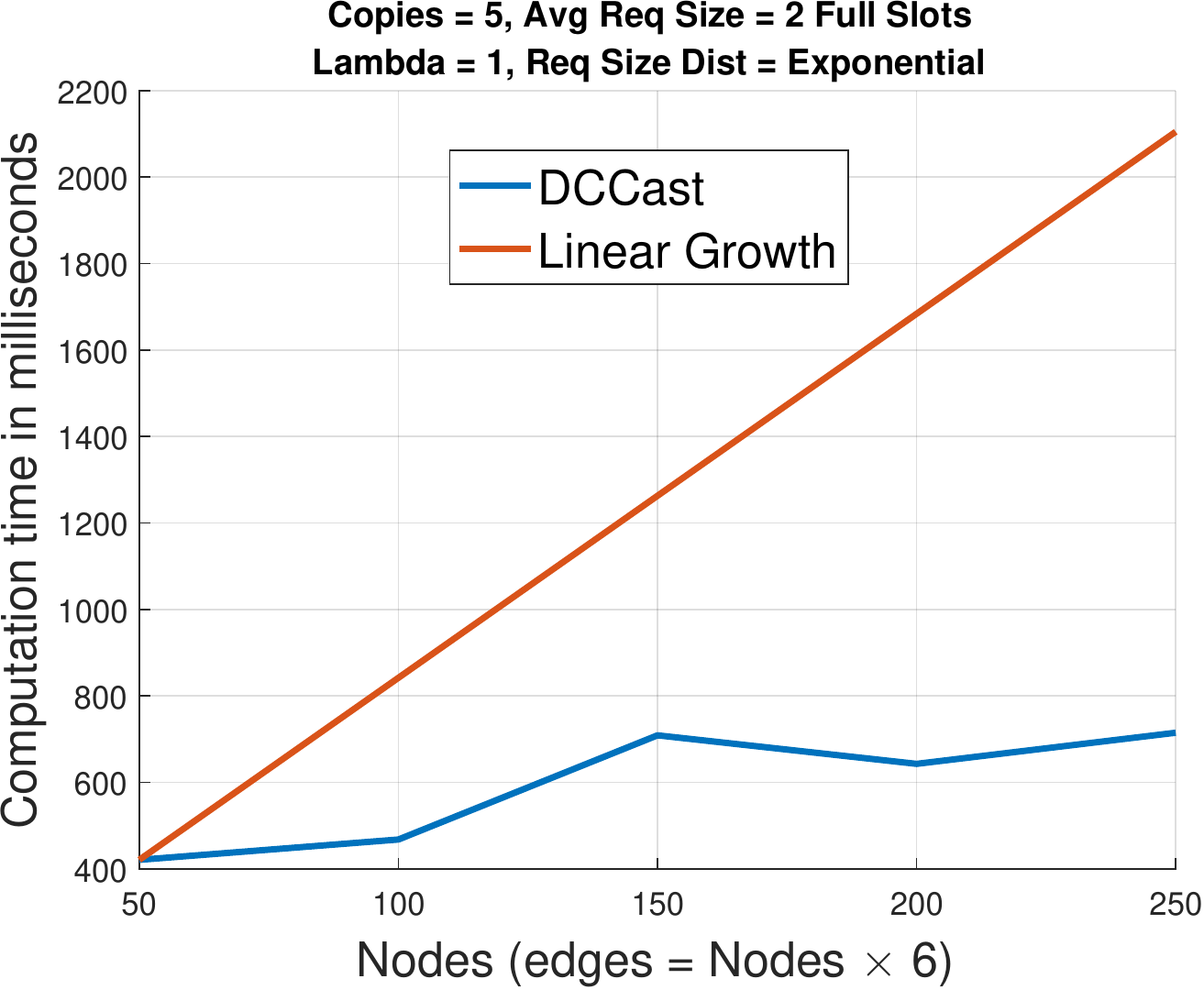}
    \caption{Computational overhead of DCCast as network size grows.}
    \label{fig:comp_network_size}
\end{figure}

\newpage
\noindent\textbf{Computational Overhead:} We used a large network with $50$ nodes and $300$ edges and considered P2MP transfers with $5$ destinations per transfer. Transfers were generated according to Poisson distribution with arrival times ranging from $0$ to $1000$ timeslots and the simulation ran until all transfers were completed. Mean processing time of a single \textit{timeslot} increased from $1.2ms$ to $50ms$ per timeslot while increasing $\lambda_{P2MP}$ from $1$ to $10$. Mean processing time of a single \textit{transfer} (which accounts for finding a tree and scheduling the transfer) was $1.2ms$ and $5ms$ per transfer for $\lambda_{P2MP}$ equal to $1$ and $10$, respectively. This is negligible compared to timeslot lengths of minutes in prior work \cite{amoeba}. We also looked at the computational overhead of DCCast as network size grows shown in Figure \ref{fig:comp_network_size}. As can be seen, the growth is sub-linear.

\newpage
\section{Fast Admission Control for Point to Multipoint Transfers with Deadlines}
Existing techniques to performing inter-DC transfers are either unable to guarantee the deadlines for inter-DC multicast transfers or can only do so by treating multicast transfers as separate P2P transfers. We present Deadline-aware DCCast (DDCCast), a quick yet effective deadline aware point to multipoint technique based on the ALAP traffic allocation policy. DDCCast performs careful admission control using temporal planning, rate-allocation, and rate-limiting to avoid congestion while sending traffic over forwarding trees that are adaptively selected to reduce network capacity consumption and maximize the number of admitted transfers. We perform experiments confirming DDCCast's potential to reduce total bandwidth usage by up to $45\%$ while admitting up to $25\%$ more traffic into the network compared to alternatives that guarantee deadlines.

\subsection{System Model}
We use the same notations expressed earlier in Table \ref{table_var_0} and Table \ref{table_var_dccast}. Similarly, to provide flexible bandwidth allocation, we consider a slotted timeline \cite{tempus, amoeba, dcroute} where the transmission rate of senders is constant during each slot, but can be updated from one slot to the next. This can be achieved using rate-limiting techniques at the end-points \cite{swan, bwe}.

A central scheduler is assumed that receives transfer requests from end-points, performs admission control to determine feasibility, calculates an initial temporal schedule, and informs the end-points of next timeslot's rate-allocation when the timeslot begins. The allocation for future slots can change as new requests are submitted, however, only the scheduler knows about schedules beyond the current timeslot and it can update such schedules as new requests are submitted. We focus on scheduling large transfers that can take minutes or more to complete \cite{tempus} and therefore, the time to submit a transfer request, calculate the routes, and install forwarding rules is considered negligible in comparison. We also assume equal link capacity for all links to simplify the problem. We consider an online scenario where requests may arrive at any time and go through an admission control process; if admitted, they are scheduled to be completed prior to their deadlines. To prevent thrashing, similar to previous works \cite{amoeba, dcroute}, we also assume that once a request is admitted, it cannot be evicted.

A transfer request $R_{i}$ is considered \textbf{active} if it has been admitted but not completed. At any moment, there may be $K \ge 0$ different active requests with various deadlines. We define \textbf{active window} as the range of time from $t_{now}+1$ (next timeslot) to $t_{end}$, the timeslot of the latest deadline, defined as $\max(t_{d_{R_{i_1}}},\dots,t_{d_{R_{i_K}}})$. At the end of each timeslot, all requests can be updated to reflect their remaining (residual) demands by deducting volume sent during a timeslot from their total demand at the beginning of a timeslot. To perform a P2MP transfer $R$, the source $S_R$ transmits traffic over a Steiner Tree \cite{steiner_tree_problem} that spans across all destinations $\pmb{\mathrm{D}}_{R_1}$ to $\pmb{\mathrm{D}}_{R_n}$ which we refer to as the P2MP request's forwarding tree. The transmission rate over a forwarding tree at every timeslot is the minimum of available bandwidth over all edges of the tree at that timeslot.

\subsection{Point to Multipoint Transfers with Deadlines}
We focus on the case when a P2MP transfer is only valuable if all of its destinations receive the associated object prior to the specified deadline, i.e., all receivers have the same deadline. As a result, a transfer should only be accepted if this requirement can be guaranteed given no failures or unexpected loss of capacity across the network.

\vspace{0.5em}
\noindent\textbf{P2MP Deadline Problem:} Determine feasibility of allocating transfer $R_{K+1}$ using any forwarding tree over the inter-DC network $G$, given $K$ existing requests $R_1$ to $R_K$ with residual demands $\mathcal{V}^{r}_{R_1}$ to $\mathcal{V}^{r}_{R_K}$ each with their own forwarding trees. If feasible, the transfer is admitted and the algorithm should determine the forwarding tree that minimizes overall bandwidth consumption. The objective is to maximize the total traffic admitted into the network.

\vspace{0.5em}
The most general approach to solving the \textit{P2MP Deadline Problem} is to form a Mixed Integer Linear Program (MILP) that considers capacity of links over various timeslots along with transfer deadlines and reschedules all active requests along with the new request. The solution would be a new schedule for every active transfer (over the same trees) and a new tree with a rate allocation schedule for the new request. Solving MILPs can be computationally intensive and may take a long time. This is especially problematic if MILPs have to be solved upon arrival of requests for admission control where admission control latency can lead to creation of backlogs. We discuss our fast heuristic next.

\subsection{Deadline-aware DCCast (DDCCast)}
\begin{figure}
    \centering
    \includegraphics[width=\textwidth]{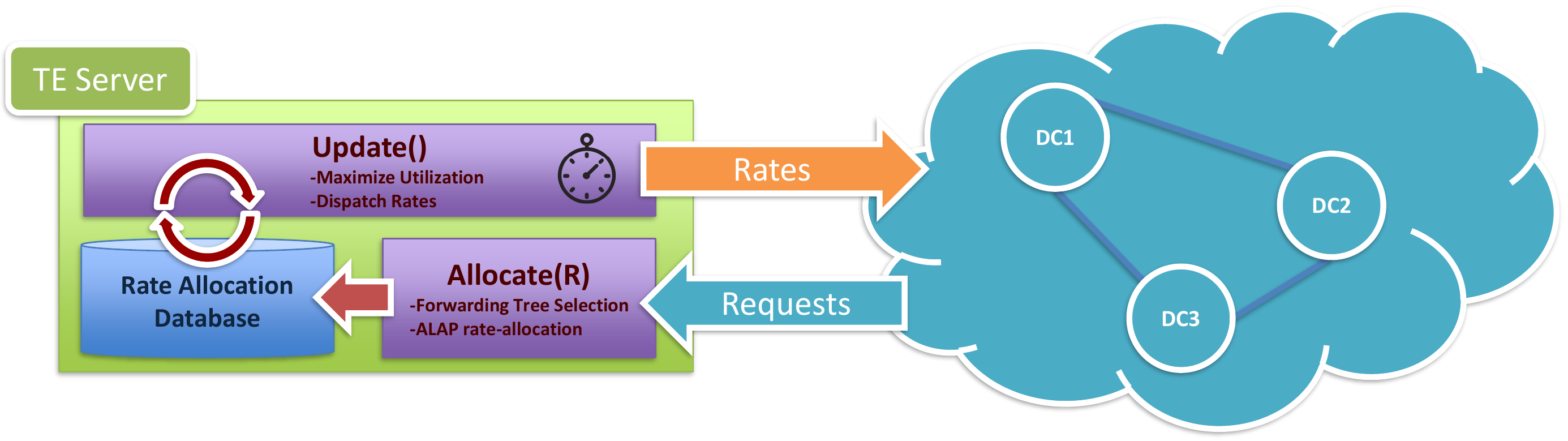}
    \caption{DDCCast (Deadline-Aware DCCast \ref{dccast}) architecture.}
    \label{fig:ddccast}
\end{figure}

The architecture of DDCCast (Deadline-Aware DCCast \cite{dccast, dccastgit}) is shown in Figure \ref{fig:ddccast}. There are two main procedures of \texttt{Update()} and \texttt{Allocate($R_{new}$)}. The former simply reads the rate-allocations from the database and dispatches them to all end-points at the beginning of every timeslot. The latter performs admission control, forwarding tree selection and rate-allocation according to the ALAP policy. The rates are then updated in a database. Also, at the beginning of every timeslot, if there is unused capacity, the \texttt{Update()} procedure moves back some of the future allocations, starting with the closest allocation to the current timeslot that can be moved back, to maximize utilization. Afterwards, to keep the allocation ALAP, it may sweep the timeline and further push any allocations that can be pushed forward closest to their deadlines. This technique is similar to the one used by DCRoute in Chapter \ref{chapter_admission_control} with the minor difference that it is applied over the edges of multicast trees. We discuss the main parts of DDCCast in the following.

\subsubsection{Forwarding Tree Selection}
For every new transfer, this procedure selects a forwarding tree that connects the sender to all receivers over the inter-DC network. This is done by assigning weights to edges of the inter-DC network and selecting a minimum weight Steiner Tree \cite{steiner_tree_problem}. Weight of a forwarding tree is sum of the weights of its edges. For every transfer $R_{new}$ with volume $\mathcal{V}_{R_{new}}$, we assign edge $e \in \pmb{\mathrm{E}}_G$ of the inter-DC network a weight of $(\mathcal{V}_{R_{new}} + L_{e}(t_{d_{R_{new}}}))$ where $L_{e}(t)$ is the total load on edge $e$ up to and including timeslot $t$. Running a minimum weight Steiner Tree heuristic gives us a forwarding tree $T_{R_{new}}$. This process is performed only once for every request upon their arrival. 

We explain the motivation behind our approach to tree selection. Ideally for routing, we seek a tree with minimum number of edges that connects the source datacenter to all destination datacenters (i.e., a minimum edge Steiner Tree), but such tree may not have enough capacity available on all edges to complete $R_{new}$ prior to $t_{d_{R_{new}}}$. Therefore, a different Steiner tree, which can be larger but offers more available bandwidth may be chosen. It is possible that larger trees provide higher available capacity by using longer paths through least loaded edges, but they consume more bandwidth since they send $\mathcal{V}_{R_{new}}$ over a larger number of edges. To model this behavior, we use a weight assignment that allows balancing two possibly conflicting objectives, i.e., finding the forwarding tree with highest available capacity by potentially taking longer paths (to balance load across the network), while minimizing the total network capacity used by minimizing the number of edges used. Our evaluations presented earlier in \ref{dccast} show that this cost assignment performs more effectively compared to minimizing the maximum utilization on the network which is a well-known policy that is frequently used for traffic engineering over wide area networks.

\subsubsection{Admission Control}
After finding a P2MP forwarding tree, we need to first verify if the new transfer can be accommodated over the tree. We perform admission control by calculating the available bandwidth over the tree (i.e., $\forall e \in \pmb{\mathrm{E}}_{T_{R_{new}}}$) for all timeslots of $t_{now}+1$ to $t_{d_{R_{new}}}$. We then sum the available bandwidth across these timeslots and admit the request if the total is not less than $\mathcal{V}_{R_{new}}$.

This admission control approach does not guarantee that a rejected request could not have been accommodated on $G$. It is possible that a request is rejected although it could have been accepted if a different forwarding tree had been chosen. In general, finding the tree with maximum available bandwidth prior to a deadline is a hard problem given that the maximum available rate over a tree is the minimum of what is available over its edges per timeslot. In addition, even if this problem could be optimally solved in polynomial time, it is unclear whether it would lead to an improved solution since this is an online resource packing problem with multiple capacity and demand constraints.

\subsubsection{Traffic Allocation and Adjustment}
Once admitted, the traffic allocation process places every new request according to ALAP policy which guarantees meeting deadlines while postponing the use of bandwidth until necessary. Adjustments are done in \texttt{Update()} procedure upon beginning of timeslots. To maximize utilization and use the network efficiently, we adjust the schedules when there is unused capacity. Upon the beginning of every timeslot, we pull traffic from closest timeslots in the future over each forwarding tree and send it in current timeslot, if there is available capacity across all edges of such a P2MP forwarding tree. For a network, it may not be possible to schedule traffic ALAP on all edges since allocations may need to span over multiple edges all of which may not have available bandwidth. Therefore, after maximizing the utilization of the upcoming timeslot (i.e., $t_{now}+1$), we sweep the timeline starting $t_{now}+2$ and push allocations forward as much as possible until no schedule can be pushed further toward its deadline.

\subsection{Evaluation}
We evaluated DDCCast using synthetic traffic generated in accordance with several related works \cite{amoeba, dcroute}. The arrival of requests followed a Poisson distribution with rate $\lambda$. The deadline $t_{d_{R_{new}}}$ of every request $R_{new}$ was generated using an exponential distribution with a mean value of $10$ timeslots. Demand of $R_{new}$ was then calculated using another exponential distribution with a mean of $\frac{t_{d_{R_{new}}}-t_{now}}{8}$. All simulations were performed over $500$ timeslots and each scenario was repeated $10$ times and the average measurements have been reported. We assumed a total capacity of $1.0$ for every timeslot over every link.

\vspace{0.5em}
\noindent\textbf{Setup:} We performed our simulations over Google's GScale topology \cite{b4} with $12$ datacenters and $19$ bidirectional edges. We assumed a machine attached to each datacenter generating traffic destined to other (multiple) datacenters. The simulations were performed on a single machine equipped with an Intel Core i7-6700T CPU and 24GBs of RAM. All simulations were coded in Java, and to solve linear programs for Amoeba, we used Gurobi \cite{gurobi}.

\vspace{0.5em}
\noindent\textbf{Performance Metrics:} We measured two metrics of \textbf{total bandwidth used} and \textbf{total traffic volume admitted}. Both parameters were calculated over the whole network and all timeslots. The first parameter is the sum of all traffic over all timeslots and all links. The second parameter determines what volume of offered load from all end-points was admitted into the network.

\vspace{0.5em}
\noindent\textbf{Schemes:} Following schemes were considered: DDCCast, DCRoute,\footnote{DCRoute was presented earlier in Chapter \ref{chapter_admission_control}.} and Amoeba \cite{amoeba} all of which aim to guarantee the deadlines, maximize total utilization, and perform admission control. DCRoute and Amoeba do not have the notion of point to multipoint forwarding trees. As a result, to perform the following simulations, each P2MP transfer with multiple destinations in DDCCast is broken into several independent P2P transfers from the source to each destination and then plugged into DCRoute and Amoeba. We only compare DDCCast with these two works since other works either do not support deadlines \cite{bwe, swan} or focus on different objectives.

\subsubsection{Effect of Number of Destinations}
\begin{figure}
    \centering
    \includegraphics[width=0.9\textwidth]{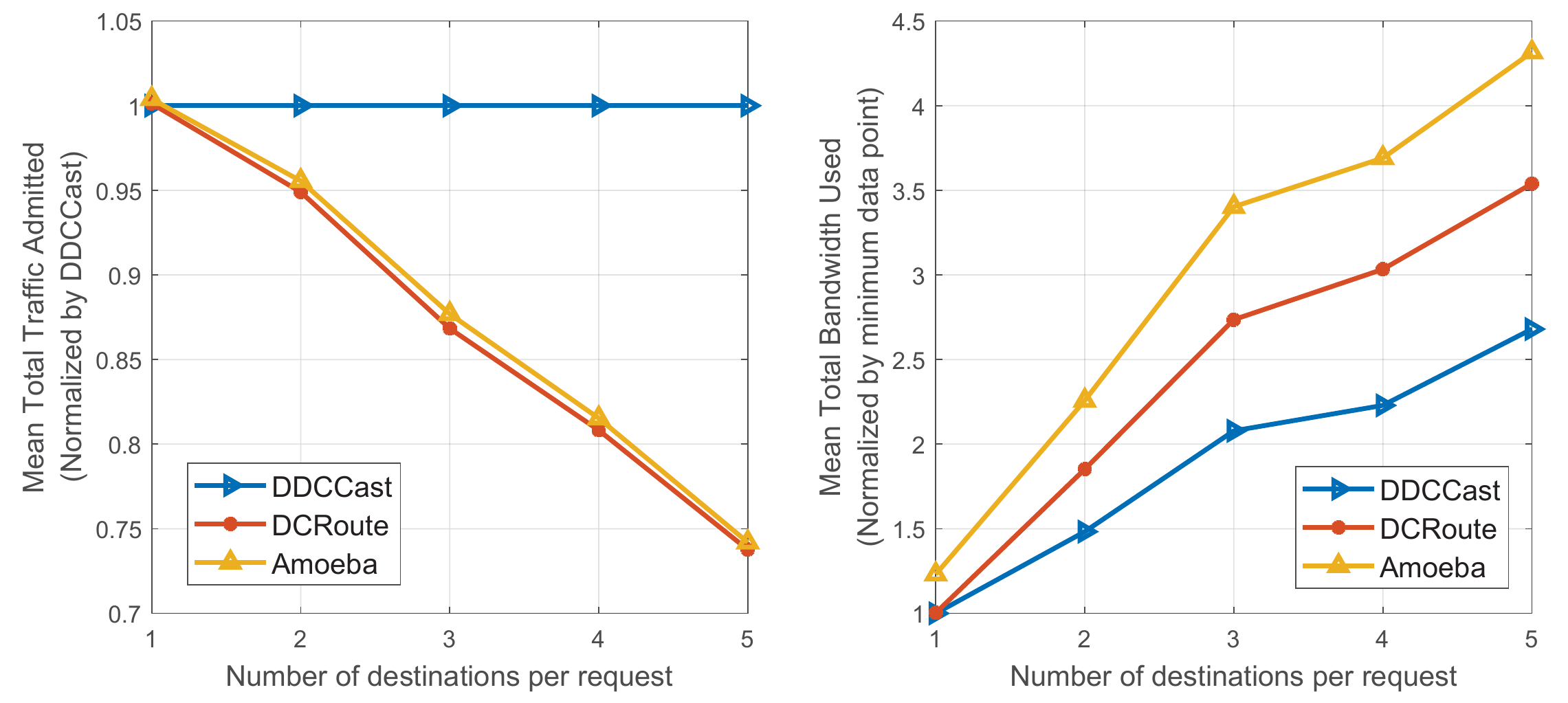}
    \caption{Capacity consumption and total admitted traffic by $\vert \pmb{\mathrm{D}}_{R} \vert$ (given $\lambda = 2$)}
    \label{fig:p2mp_copies}
\end{figure}

Figure \ref{fig:p2mp_copies} shows the results of this experiment. We increased the number of destinations for each transfer from $1$ to $5$ and picked random destinations for each transfer. The total volume of traffic used by Amoeba \cite{amoeba} is up to $1.8\times$ the volume used by DDCCast. Even in case of one destination Amoeba uses $1.2\times$ the bandwidth of DCCast and DCRoute. This occurs because Amoeba routes traffic across the $K$ static shortest paths, and as $K$ increases, some of these paths may not be as short as the shortest path. Therefore, even for a small incoming network load, a portion of traffic may traverse longer paths and increase total bandwidth usage. DDCCast saves bandwidth by using P2MP forwarding trees. DDCCast admits $25\%$ more traffic compared to Amoeba when sending objects to $5$ destinations while using $45\%$ less overall network capacity.

\subsubsection{Effect of Transfer Arrival Rate (i.e., Incoming Load)}
\begin{figure}
    \centering
    \includegraphics[width=0.9\textwidth]{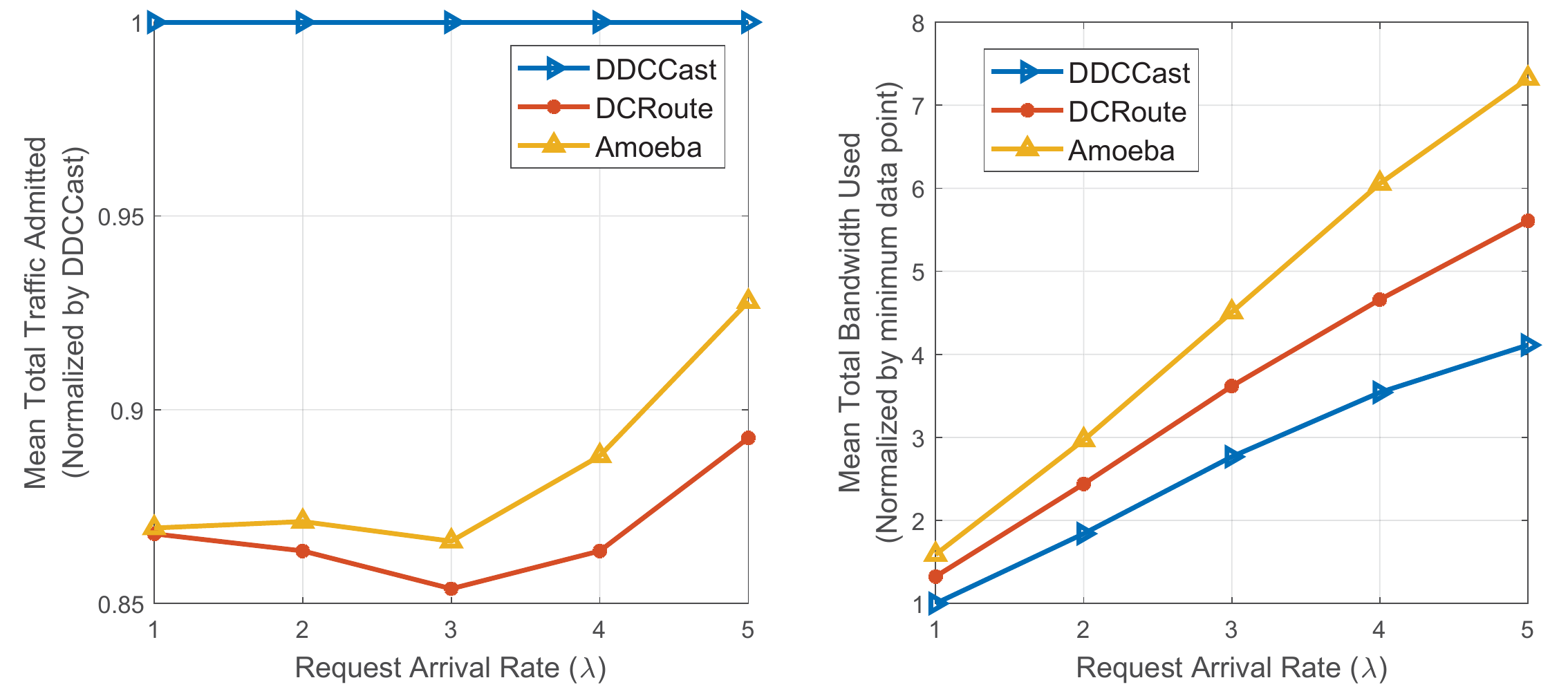}
    \caption{Capacity consumption and total admitted traffic by $\lambda$ (given $\vert \pmb{\mathrm{D}}_{R_i} \vert = 3, \forall i$)}
    \label{fig:p2mp_lambda}
\end{figure}

We investigate the effect of $\lambda$ while sending an object to three destinations. Results of this experiment have been shown in Figure \ref{fig:p2mp_lambda}. Volume of admitted traffic is about $10\%$ higher for DDCCast compared with other two schemes over all arrival rates. Also, similar to the previous experiment, DDCCast's total bandwidth usage is between $37\%$ to $45\%$ less than Amoeba \cite{amoeba} and $28\%$ less than DCRoute \cite{dcroute}.

\section{Conclusions}
In this chapter, we studied efficient inter-DC P2MP transfers which are multicast transfers with known source and set of receivers upon submission to the inter-DC network. We investigated an adaptive approach to selection of forwarding trees (i.e., multicast trees) which reduced total capacity consumption while balancing load across the network. It is possible to set up such trees using commodity hardware that support multicast forwarding, or SDN frameworks such as OpenFlow \cite{openflow} along with application of Group Tables \cite{openflow-1.3.1}.\footnote{See Appendix \ref{chapter_sdn_gt_all} for a discussion of switch support for group tables.} Such trees can be configured upon arrival of transfers and torn down upon their completion. Our evaluations show that by adaptively selecting forwarding trees according to edge load and transfer size, we can reduce the total network capacity consumption while either reducing completion times, or admitting more traffic given guaranteed deadlines.

\clearpage
\chapter{Speeding up P2MP Transfers using Receiver Set Partitioning} \label{chapter_p2mp_quickcast}
In the previous chapter, we discussed using atomically selected forwarding trees (i.e., multicast trees) to copy an object from one datacenter to multiple datacenters over an inter-DC network. This allowed us to save network capacity while reducing the time needed to cast objects to many locations. Although one can perform inter-DC P2MP transfers using a single multicast forwarding tree, that might lead to poor performance as the slowest receiver on each tree dictates the completion time for all receivers. In this chapter, we discuss using multiple trees per transfer, each connected to a subset of receivers, which alleviates this concern. The choice of multicast trees also determines the total bandwidth usage.

We approach this problem by breaking it into three sub-problems of partitioning, tree selection, and rate allocation. We present an algorithm, called QuickCast, which is computationally fast and allows us to significantly speed up multiple receivers per multicast transfer with control over extra bandwidth consumption. We evaluate QuickCast against a variety of synthetic and real traffic patterns as well as real WAN topologies. Compared to performing bulk multicast transfers as separate unicast transfers, QuickCast achieves up to $3.64\times$ reduction in mean completion times while at the same time using $0.71\times$ the bandwidth. Also, QuickCast allows the top $50\%$ of receivers to complete between $3\times$ to $35\times$ faster on average compared with when a single forwarding multicast tree is used for data delivery.

\section{Background and Related Work}
In general, it is not required that the receivers of a P2MP transfer complete data reception at the same time. For many applications, speeding up several receivers per P2MP transfer can translate to improved end-user quality of experience and increased availability. For example, faster replication of video content to regional datacenters enhances average user's experience in social media applications or making a newly trained model available at regional datacenters allows speedier access to new application features for millions of users. Several recent works focus on improving the performance of unicast transfers over dedicated inter-DC networks \cite{b4, swan, tempus, amoeba, owan}. However, performing bulk multicast transfers as many separate unicast transfers can lead to excessive bandwidth usage and will increase receiver completion times.

Although there exists extensive work on multicasting, it is not possible to apply those solutions to our problem as existing research has focused on different goals and considers different constraints. For example, earlier research in multicasting aims at dynamically building and pruning multicast trees as receivers join or leave \cite{ip_multicast}, building multicast overlays that reduce control traffic overhead and improve scalability \cite{nice}, or choosing multicast trees that satisfy a fixed available bandwidth across all edges as requested by applications \cite{online_multicast_bw_guarantees, sdn_multicast}, minimize congestion within datacenters \cite{avalanche, datacast}, reduce data recovery costs assuming some recovery nodes \cite{raera}, or maximize the throughput of a single multicast flow \cite{MPMC_2013, MPMC_2016}. To our knowledge, none of the related research efforts aimed at minimizing the mean completion times of receivers for concurrent bulk multicast transfers while considering the overall bandwidth usage, which is the focus of this chapter.

In this chapter, we break the bulk multicast transfer routing, and scheduling problem with the objective of minimizing mean completion times of receivers into three sub-problems of the receiver set partitioning, multicast forwarding tree selection per receiver partition, and rate allocation per forwarding tree. We briefly describe each problem as follows.

\vspace{0.5em}
\noindent\textbf{Receiver Set Partitioning:} As different receivers can have different completion times, a natural way to improve completion times is to partition receivers into multiple sets with each receiver set having a separate tree. This reduces the effect of slow receivers on faster ones. We employ a partitioning technique that groups receivers of every bulk multicast transfer into multiple partitions according to their mutual distance (in hops) on the inter-DC graph. With this approach, the partitioning of receivers into any $N > 1$ partitions consumes minimal additional bandwidth on average. We also offer a configuration parameter called the partitioning factor that is used to decide on the right number of partitions that create a balance between receiver completion times improvements and the total bandwidth consumption.

\vspace{0.5em}
\noindent\textbf{Forwarding Tree Selection:} To avoid heavily loaded routes, multicast trees should be chosen dynamically per partition according to the receivers in that partition and the distribution of traffic load across network edges. We utilize a computationally efficient approach for forwarding tree selection that connects a sender to a partition of its receivers by assigning weights to edges of the inter-DC graph, and using a minimum weight Steiner tree heuristic. We define a weight assignment according to the traffic load scheduled on edges and their capacity and empirically show that this weight assignment offers improved receiver completion times at minimal bandwidth consumption.

\vspace{0.5em}
\noindent\textbf{Rate Allocation:} Given the receiver partitions and their forwarding trees, formulating the rate allocation for minimizing mean completion times of receivers leads to a hard problem. We consider the popular scheduling policies of fair sharing, Shortest Remaining Processing Time (SRPT), and First Come First Serve (FCFS). We reason why fair sharing is preferred compared to policies that strictly prioritize transfers (i.e., SRPT, FCFS, etc.) for network throughput maximization when focusing on bulk multicast transfers especially ones with many receivers per transfer. We empirically show that using max-min fairness \cite{max-min-fairness}, which is a form of fair sharing, we can considerably improve the average network throughput which in turn reduces receiver completion times.

\subsubsection{Motivating Example}
\begin{figure}[t]
    \centering
    \includegraphics[width=0.8\columnwidth]{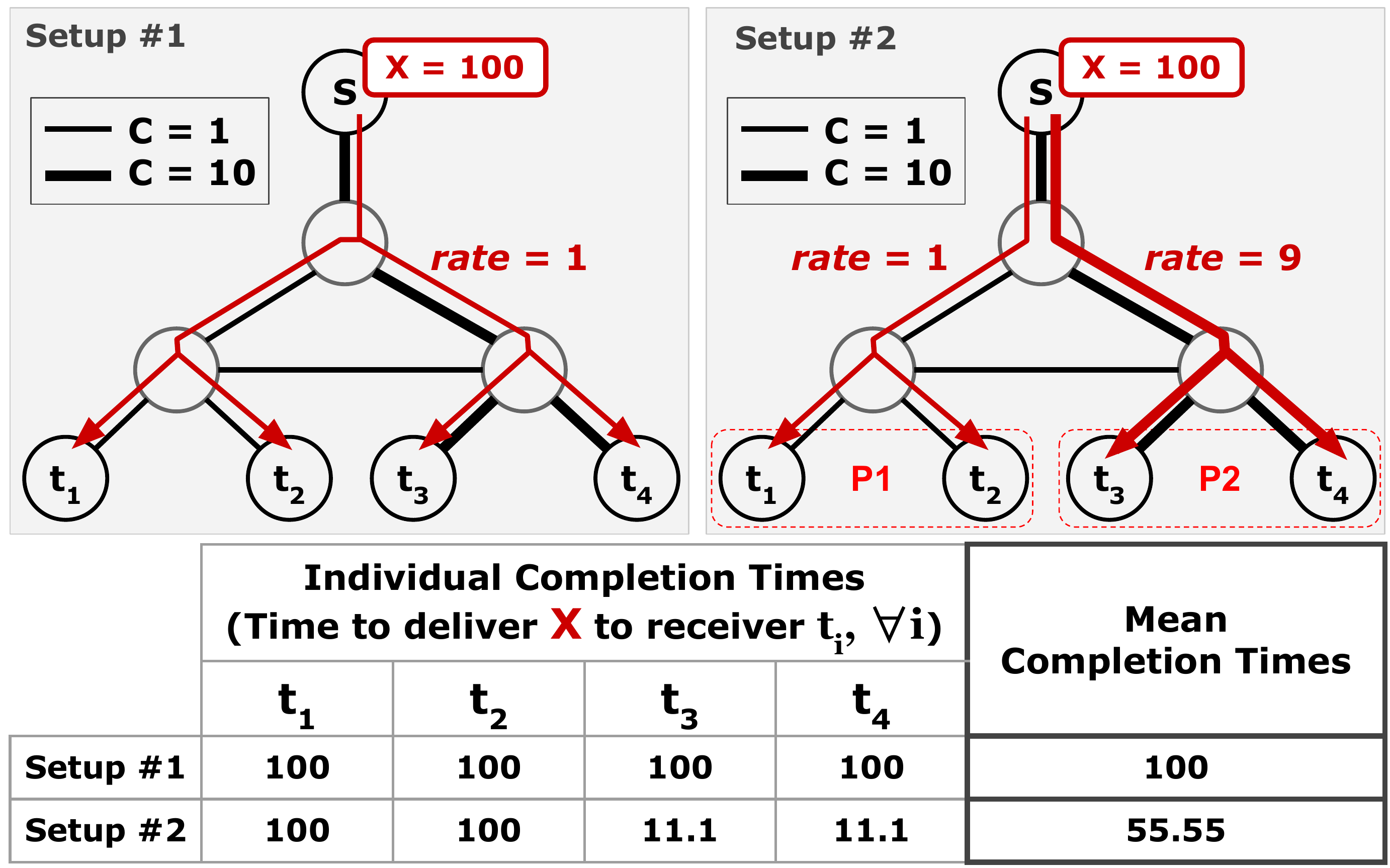}
    \caption{Using multiple smaller multicast trees we can improve the completion times of several receivers while marginally increasing total network capacity consumption.}
    \label{fig:motivating_example}
\end{figure}

Figure \ref{fig:motivating_example} shows an example of delivering a large object \textbf{X} from source $S$ to destinations $\{t_1,t_2,t_3,t_4\}$ which has a volume of 100 units. We have two types of links with capacities of 1 and 10 units of traffic per time unit. We can use a single multicast tree to connect the sender to all receivers which will allow us to transmit at the bottleneck rate of 1 to all receivers. However, one can group receivers into two partitions of $P1$ and $P2$ and attach each partition with a separate multicast tree. Then we can select transmission rates so that we minimize the mean completion times. In this case, assigning a rate of 1 to the tree attached to $P1$ and a rate of 9 to the tree attached to $P2$ will attain this goal while respecting link capacity over all links (the link attached to $S$ is the bottleneck). As another possibility, we could have assigned a rate of 10 to the tree attached to $P2$, allowing $\{t_3,t_4\}$ to finish in 10 units of time, while suspending the tree attached to $P1$ until time 11. As a result, the tree attached to $P1$ would have started at 11 allowing $\{t_1,t_2\}$ to finish at 110. In this dissertation, we aim to improve the speed of several receivers per bulk multicast transfer without hurting the completion times of the slow receivers. In computing the completion times, we ignore the propagation and queuing latencies as the focus of this dissertation is on delivering bulk objects for which the transmission time dominates the propagation or queuing latency along the trees.

\section{System Model}
We consider a scenario where bulk multicast transfers arrive at the inter-DC network in an online fashion. We will use the same notations as that of Table \ref{table_var_0}. We will also use some additional definitions as described in Table \ref{table_var_quickcast}. In general, synchronization is not required across receivers of a bulk multicast transfer and therefore, receivers are allowed to complete at different times as long as they all receive the multicast object completely. Incoming requests are processed as they come by a traffic engineering server that manages the forwarding state of the whole network in a logically centralized manner for installation and eviction of multicast trees. Upon arrival of a request, this server decides on the number of partitions and receivers that are grouped per partition and a multicast tree per partition.

Periodically, the TES computes the transmission rates for all multicast trees at the beginning of every timeslot and dispatches them to senders for rate limiting. This allows for a congestion free network since the rates are computed according to link capacity constraints and other ongoing transfers. To minimize control plane overhead, partitions and forwarding trees are fixed once they are established for an incoming transfer. In this context, the bulk multicast transfer routing and scheduling problem can be formally stated as follows.

\begin{table}
\begin{center}
\caption{Definition of variables used in this chapter besides those defined in Table \ref{table_var_0}.} \label{table_var_quickcast}
\vspace{0.5em}
\begin{tabular}{ |p{2cm}|p{11.5cm}| }
    \hline
    \textbf{Variable} & \textbf{Definition} \\
    \hline
    \hline
    $U_e$ & Edge $e$'s bandwidth utilization, $0 \le U_e \le 1$ \\
    \hline
    $T$ & A directed Steiner tree connected to a partition of receivers \\
    \hline
    $P$ & A partition of receivers of a request $R$ \\
    \hline
    $\pmb{\mathrm{P}}_R$ & Set$\langle\rangle$ of partitions of request $R$, $\vert \pmb{\mathrm{P}}_R \vert \le \vert \pmb{\mathrm{D}}_R \vert$ \\
    \hline
    $T_P$ & The forwarding tree (i.e., multicast tree) of partition $P$ \\
    \hline
    $r_{T_P}(t)$ & The transmission rate over $T_P$ of partition $P \in \pmb{\mathrm{P}}$ at timeslot $t$ \\
    \hline
    $\mathcal{V}_{P}^r$ & Residual volume of some partition $P \in \pmb{\mathrm{P}}$ \\
    \hline
    $L_e$ & Edge $e$'s total traffic load at time $t_{now}$, i.e., total outstanding bytes scaled by $e$'s inverse capacity \\
    \hline
    $p_f \ge 1$ & Configuration parameter; determines a partitioning cost threshold \\
    \hline
    $N_{max}$ & Configuration parameter; maximum number of partitions allowed per transfer \\
    \hline
\end{tabular}
\end{center}
\end{table}

\vspace{0.5em}
\noindent\textbf{Partitioning Problem:} Given an inter-DC network $G(\pmb{\mathrm{V}}_G,\pmb{\mathrm{E}}_G)$ with the edge capacity $C_e, \forall e \in \pmb{\mathrm{E}}_G$ and the set of all partitions $\{P \in \pmb{\mathrm{P}}_R ~\vert~ \forall R \in \pmb{\mathrm{R}}, \mathcal{V}^{r}_{P} > 0\}$, for a newly arriving bulk multicast transfer $R_{new}$, the traffic engineering server needs to compute a set of receiver partitions $\pmb{\mathrm{P}}_{R}$ each with one or more receivers, and select a forwarding tree $T_P, \forall P \in \pmb{\mathrm{P}}_{R}$.

\vspace{0.5em}
\noindent\textbf{Rate-allocation Problem:} Per timeslot $t$, the traffic engineering server needs to compute the rates $r_{T_P}(t), ~\{P \in \pmb{\mathrm{P}}_R ~\vert~ \forall R \in \pmb{\mathrm{R}}, \mathcal{V}^{r}_{P} > 0\}$. The objective is to minimize the average time for a receiver to complete data reception while keeping the total bandwidth consumption below a certain threshold compared to the minimum possible, i.e., a minimum edge Steiner tree per transfer.

\vspace{0.5em}
Both the number of ways to partition receivers into subsets and the number of candidate forwarding trees per subset grow exponentially with the problem size. It is, in general, not clear how partitioning and selection of forwarding trees correlate with both receiver completion times and total bandwidth usage. Even the simple objective of minimizing the total bandwidth usage is a hard problem. Also, assuming known forwarding trees, selecting transmission rates per timeslot per tree for minimization of mean receiver completion times is a hard problem. Finally, this is an online problem with unknown future arrivals which adds to the complexity.

\section{Optimizing Receiver Completion Times with Minimum Bandwidth Usage}
As stated earlier, we need to address the three sub-problems of receiver set partitioning, tree selection, and rate allocation. Since the partitioning sub-problem uses the tree selection sub-problem, we first discuss tree selection in the following. As the last problem, we will address rate allocation. Since the total bandwidth usage is a function of transfer properties, i.e., number of receivers, transfer volume, and the location of sender and receivers, and the network topology, it is highly sophisticated to design a solution that guarantees a limit on the total bandwidth usage. Instead, we aim to reduce the receiver completion times while minimally increasing bandwidth usage.

\subsection{Forwarding Tree Selection}
The tree selection problem states that given a network topology with link capacity knowledge, how to choose a Steiner tree that connects a sender to all of its receivers. The objective is to minimize the completion times of receivers\footnote{All receivers on a tree complete at the same time.} while minimally increasing the total bandwidth usage. Since the total bandwidth usage is directly proportional to the number of edges on selected trees, we would want to keep trees as small as possible. Reduction in completion times can be achieved by avoiding edges that have a large outstanding traffic load. For this purpose, we use an approach similar to the one used in Chapter \ref{chapter_p2mp_dccast} which worked by assigning proper weights to the edges of the inter-DC graph and choosing a minimum weight Steiner tree. The weight assignment we use next also takes into account the variable link capacities over the topology.

\vspace{0.5em}
\noindent\textbf{Weight Assignment:} We use the metric of link load $L_{e}, \forall e \in \pmb{\mathrm{E}}_G$ that is defined in Table \ref{table_var_quickcast} and can be computed as $L_e = \frac{1}{C_e} \sum_{P \in \pmb{\mathrm{P}}_{R_{new}}, \forall R ~\vert~ e \in \pmb{\mathrm{E}}_{T_P}} \mathcal{V}_{P}^r$. Note that this is different from what we used in Chapter \ref{chapter_p2mp_dccast} in that we divide the total outstanding volume of traffic allocated on a link by its capacity.

\vspace{0.5em}
We can compute a link's load since we know the remaining volume of current transfers and the edges that they use. A link's load is a measure of how busy it is expected to be in the next few timeslots. It increases as new transfers are scheduled on a link, and diminishes as traffic flows through it. To select a forwarding tree from a source to a set of receivers, we use an edge weight of $L_{e} + \frac{\mathcal{V}_{R_{new}}}{C_{e}}$ and select a minimum weight Steiner tree. The selected tree will most likely exclude any links that are expected to be highly busy. Addition of the second element in the weight (new request's volume divided by capacity) helps select smaller trees in case there is not much load on most edges.

Algorithm \ref{algo_dccast} applies the weight assignment approach mentioned above to select a forwarding tree that balances the traffic load across available trees and finds a minimum weight Steiner tree using the GreedyFLAC heuristic \cite{Watel2014}. In \S \ref{evaluations}, we explore a variety of weights for forwarding tree selection as shown in Table \ref{table_cost} and see that this weight assignment provides consistently close to minimum values for the three performance metrics of mean and tail receiver completion times as well as total bandwidth usage.

\SetAlgoVlined
\begin{algorithm}[t]
\caption{Forwarding Tree Selection Algorithm} \label{algo_dccast}
{\small
\SetKw{KwBy}{by}
\SetKwProg{ComputeTree}{ComputeTree}{}{}

\vspace{0.4em}
\KwIn{Request $R_{new}$, partition $P \in \pmb{\mathrm{P}}_{R_{new}}$, $G(\pmb{\mathrm{V}}_G,\pmb{\mathrm{E}}_G)$, and $L_{e}, \forall e \in \pmb{\mathrm{E}}_G$ }

\vspace{0.4em}
\KwOut{A forwarding tree (set of edges)}

\nonl\hrulefill
    
\vspace{0.4em}
\ComputeTree{$\mathrm{(}P,R_{new}\mathrm{)}$}{
    
    \vspace{0.4em}
    Assign a weight of $(L_{e} + \frac{\mathcal{V}_{R_{new}}}{C_e})$ to every edge $e$, $\forall e \in \pmb{\mathrm{E}}_G$\;
    
    \vspace{0.4em}
    Find a minimum weight Steiner tree $T_P$ which connects the nodes $\{S_{R_{new}} \cup P$\}\;
    
    \vspace{0.4em}
    $L_e \gets L_e + \frac{\mathcal{V}_{R_{new}}}{C_e},~\forall e \in \pmb{\mathrm{E}}_{T_P}$\;
    
    \vspace{0.4em}
    \Return{$T_P$}\;
}
}
\end{algorithm}

\vspace{0.5em}
\noindent\textbf{Worst-case Complexity:} Algorithm \ref{algo_dccast} computes one minimum weight Steiner tree. For a request $R_{new}$, the worst-case complexity of Algorithm \ref{algo_dccast} is $O(\lvert \pmb{\mathrm{V}}_G \rvert^3 \lvert \pmb{\mathrm{D}}_{R_{new}} \rvert^2+\lvert \pmb{\mathrm{E}}_G \rvert)$ given the complexity of GreedyFLAC \cite{Watel2014}.

\subsection{Receiver Set Partitioning} \label{partitioning}
The maximum transmission rate on a tree is that of the link with minimum capacity. To improve bandwidth utilization of inter-DC backbone, we can replace a large forwarding tree with multiple smaller trees each connecting the source to a subset of receivers. By partitioning, we isolate some receivers from the bottlenecks allowing them to receive data at a higher rate. We aim to find a set of partitions each with at least one receiver that allows for reducing the average receiver completion times while minimally increasing the bandwidth usage. Bottlenecks may appear either due to competing transfers or differences in link capacity. In the former case, some edges may be shared by multiple trees which lead to lower available bandwidth per tree. Such conditions may arise more frequently under heavy load. In the latter case, differences in link capacity can increase completion times especially in large networks and with many receivers per transfer.

Receiver set partitioning to minimize the impact of bottlenecks and reduce completion times is a sophisticated open problem. It is best if partitions are selected in a way that no additional bottlenecks are created. Also, increasing the number of partitions may in general increase bandwidth consumption (multiple smaller trees may have more edges in total compared to one large tree). Therefore, we need to come up with the right number of partitions and receivers that are grouped per partition. We propose a partitioning approach, called the hierarchical partitioning, that is computationally efficient and uses a partitioning factor to decide on the number of partitions and receivers that are grouped in those partitions.

\subsubsection{Number of Partitions} Transfers may have a highly varying number of receivers. Generally, the number of partitions should be computed based on the number of receivers, where they are located in the network, and the network topology. Also, using more partitions can lead to the creation of unnecessary bottlenecks due to shared links. We compute the number of partitions per transfer according to the total traffic load on network edges and considering a threshold that limits the cost of additional bandwidth consumption.

\subsubsection{Limitations of Partitioning} Partitioning, in general, cannot improve tail completion times of transfers as tail is usually driven by physical resource constraints, i.e., low capacity links or links with high contention.

\subsubsection{Hierarchical Partitioning}
We group receivers into partitions according to their mutual distance which is defined as the number of hops on the shortest hop path that connects any two receivers. Hierarchical clustering \cite{clustering_methods} approaches such as agglomerative clustering can be used to compute the groups by initially assuming that every receiver has its partition and then by merging the two closest partitions at each step which generates a hierarchy of partitioning solutions. Each layer of the hierarchy then gives us one possible solution with a given number of partitions.

With this approach, the partitioning of receivers into any $N > 1$ partitions consumes minimal additional bandwidth on average compared to any other partitioning with $N$ partitions. That is because assigning a receiver to any other partition will likely increase the total number of edges needed to connect the source to all receivers; otherwise, that receiver would not have been grouped with the other receivers in its current partition in the first place. There is, however, no guarantee since hierarchical clustering works based on a greedy heuristic.

After building a partitioning hierarchy, the algorithm selects the layer with the maximum number of partitions whose total sum of tree weights stays below a threshold that can be configured as a system parameter. Choosing the maximum partitions allows us to minimize the effect of slow receivers given the threshold, which is a multiple of the weight of a single tree that would connect the sender to all receivers and can be looked at as a bandwidth budget. We call the multiplication coefficient the partitioning factor $p_f$. Algorithm \ref{algo_quick} shows this process in detail. The partitioning factor $p_f$ plays a vital role in the operation of QuickCast as it determines the extra cost we are willing to pay in bandwidth for improved completion times. In general, a $p_f$ greater than one but close to it should allow partitioning to separate very slow receivers from several other nodes. A $p_f$ that is considerably larger than one may generate too many partitions and potentially create many shared links which reduce throughput and additional edges that increase bandwidth usage. If $p_f$ is less than one, a single partition will be used.

\SetAlgoVlined
\begin{algorithm}[t]
\caption{Compute Partitions and Trees} \label{algo_quick}
{\small
\SetKw{KwBy}{by}
\SetKwProg{ComputePartitionsAndTrees}{ComputePartitionsAndTrees}{}{}

\vspace{0.4em}
\KwIn{Request $R_{new}$, $G(\pmb{\mathrm{V}}_G,\pmb{\mathrm{E}}_G)$, and $L_{e}, \forall e \in \pmb{\mathrm{E}}_G$}

\vspace{0.4em}
\KwOut{Pairs of (partition, forwarding tree)}

\nonl\hrulefill

\vspace{0.4em}
\ComputePartitionsAndTrees{$\mathrm{(}R_{new},N_{max}\mathrm{)}$}{

    \vspace{0.4em}
    Assign a weight of $(L_{e} + \frac{\mathcal{V}_{R_{new}}}{C_e})$ to $e$, $\forall e \in \pmb{\mathrm{E}}_G$\;
    
    \vspace{0.4em}
    Find the minimum weight Steiner tree $T_{R_{new}}$ which connects the nodes $\{S_{R_{new}} \cup \pmb{\mathrm{D}}_{R_{new}}\}$ and its total weight $W_{T_{R_{new}}}$\;
    
    \vspace{0.4em}
    \ForEach{$\{\alpha,\beta\}, \alpha \in \pmb{\mathrm{D}}_{R_{new}},\beta \in \pmb{\mathrm{D}}_{R_{new}},~\alpha \neq \beta$}{
        \vspace{0.4em}
        $\mathrm{DIST}_{\alpha,\beta} \gets$ number of edges on the minimum hop path from $\alpha$ to $\beta$\;
    }
    
    \vspace{0.4em}
    Compute the agglomerative clustering hierarchy for $\pmb{\mathrm{D}}_{R_{new}}$ using average linkage and distance $\mathrm{DIST}_{i,j}$ which will have $l$ clusters at layer $1 \le l \le \lvert \pmb{\mathrm{D}}_{R_{new}} \rvert$\;
    
    \vspace{0.4em}
    \For{$l = \min(N_{max}, \lvert \pmb{\mathrm{D}}_{R_{new}} \rvert)$ \KwTo $2$ \KwBy $-1$}{

        \vspace{0.4em}
        $\pmb{\mathrm{P}}_{l} \gets$ set of clusters at layer $l$ of agglomerative hierarchy, each cluster forms a partition\;

        \vspace{0.4em}
        \ForEach{$P \in \pmb{\mathrm{P}}_{l}$}{
            \vspace{0.4em}
            Find the minimum weight Steiner tree $T_{P}$ which connects the nodes $\{S_{R_{new}} \cup P\}$\;
        }
        
        \vspace{0.4em}
        \If{$\sum_{P \in \pmb{\mathrm{P}}_{l}} W_{T_{P}} \le p_f \times W_{T_{R_{new}}}$}{
            
            \vspace{0.4em}
            \ForEach{$P \in \pmb{\mathrm{P}}_{l}$}{
                \vspace{0.4em}
                $T_{P} \gets$ \textbf{ComputeTree}~($P$,$R_{new}$)\;
            }
            
            \vspace{0.4em}
            \Return{$(P,~T_{P}),~\forall P \in \pmb{\mathrm{P}}_{l}$}\;
        }
    }
    
    \vspace{0.4em}
    $L_e \gets L_e + \frac{\mathcal{V}_{R_{new}}}{C_e},~\forall e \in T_{R_{new}}$\;
    
    \vspace{0.4em}
    \Return{$(\pmb{\mathrm{D}}_{R_{new}},~T_{R_{new}})$}\;
}
}
\end{algorithm}

\vspace{0.5em}
\noindent\textbf{Worst-case Complexity:} Algorithm \ref{algo_quick} performs multiple calls to the GreedyFLAC \cite{Watel2014}. It uses the hierarchical clustering with average linkage which has a worst-case complexity of $O(\lvert \pmb{\mathrm{D}}_{R_{new}} \rvert^3)$. To compute the pairwise distances of receivers, we use breadth first search with has a complexity of $O(\lvert \pmb{\mathrm{V}}_G \rvert+\lvert \pmb{\mathrm{E}}_G \rvert)$. Worst-case complexity of Algorithm \ref{algo_quick} is $O((\lvert \pmb{\mathrm{V}}_G \rvert^3+\lvert \pmb{\mathrm{E}}_G \rvert) \lvert \pmb{\mathrm{D}}_{R_{new}} \rvert^2+\lvert \pmb{\mathrm{D}}_{R_{new}} \rvert^3)$.

\subsection{Rate Allocation} \label{rate-allocation}
To compute the transmission rates per tree per timeslot, one can formulate an optimization problem with the capacity and demand constraints, and consider minimizing the mean receiver completion times as the objective. This is, however, a hard problem and can be modeled using mixed-integer programming by assuming a binary variable per timeslot per tree that shows whether that tree has completed by that timeslot. One can come up with approximation algorithms to this problem which is considered part of the future work.

We consider the three popular scheduling policies of FCFS, SRPT, and fair sharing according to max-min fairness \cite{max-min-fairness} which have been extensively used for network scheduling. These policies can be applied independently of partitioning and forwarding tree selection techniques. Each one of these three policies has its unique features. FCFS and SRPT both prioritize transfers; the former according to arrival times and the latter according to transfer volumes and so obtain a meager fairness score \cite{fairness_theory}. SRPT has been extensively used for minimizing flow completion times within datacenters \cite{pfabric, pias, epn}. Strictly prioritizing transfers over forwarding trees (as done by SRPT and FCFS), however, can lead to low overall link utilization and increased completion times, especially when trees are large. This might happen due to bandwidth contention on shared edges which can prevent some transfers from making progress. Fair sharing allows all transfers to make progress which mitigates such contention enabling concurrent multicast transfers to all make progress. In \S \ref{eval-rate-alloc}, we empirically compare the performance of these scheduling policies and show that fair sharing based on max-min fairness can significantly outperform both FCFS and SRPT in average network throughput especially with a larger number of receivers per tree. As a result, we will use QuickCast along with the fair sharing policy based on max-min fairness.

The TES periodically computes the transmission rates per multicast tree every timeslot to maximize utilization and cope with inaccurate inter-DC link capacity measurements, imprecise rate limiting, and dropped packets due to corruption. To account for inaccurate rate limiting, dropped packets and link capacity estimation errors, which all can lead to a difference between the actual volume of data delivered and the number of bytes transmitted, we propose that senders keep track of actual data delivered to their receivers per forwarding tree. At the end of every timeslot, every sender reports to the traffic engineering server how much data it was able to deliver allowing it to compute rates accordingly for the timeslot that follows. Newly arriving transfers will be assigned rates starting the next timeslot.

\section{Evaluation} \label{evaluations}
We considered various topologies and transfer size distributions as shown in Tables \ref{table_topology} and \ref{table_traffic}. Also, for Algorithm \ref{algo_quick}, unless otherwise stated, we used $p_f = 1.1$ which limits the overall bandwidth usage while offering significant gains. In the following sections, we first evaluated a variety of weight assignments for multicast tree selection considering receiver completion times and bandwidth usage. We showed that the weight proposed in Algorithm \ref{algo_dccast} offers close to minimum completion times with minimal extra bandwidth consumption. Next, we evaluated the proposed partitioning technique and considered two cases of $N_{max}=2$,\footnote{Two partitions is the minimum needed to separate several receivers from the slowest receiver per P2MP transfer.} and $N_{max}=\lvert \pmb{\mathrm{D}}_{R_{new}} \rvert$.

We measured the performance of QuickCast while varying the number of receivers and showed that it offers consistent gains. We also measured the speedup observed by different receivers ranked by their speed per multicast transfer, and the effect of partitioning factor $p_f$ on the gains in completion times as well as bandwidth usage. In addition, we evaluated the effect of different scheduling policies on average network throughput and showed that with increasing number of multicast receivers, fair sharing offers higher throughput compared to both FCFS and SRPT. Finally, we showed that QuickCast is computationally fast by measuring its running time and that the maximum number of group table forwarding entries it uses across all switches is only a fraction of what is usually available in a physical switch across the several considered scenarios.

\vspace{0.5em}
\noindent\textbf{Network Topologies:} Table \ref{table_topology} shows the list of topologies we considered. These topologies provide capacity information for all links which range from 45 Mbps to 10 Gbps. We normalized all link capacities dividing them by the maximum link capacity. We also assumed all bidirectional links with equal capacity in either direction.

\begin{table}
\begin{center}
\caption{Various topologies used in evaluation.} \label{table_topology}
\vspace{0.5em}
\begin{tabular}{ |p{3cm}|p{10.5cm}| }
    \hline
    \textbf{Name} & \textbf{Description} \\
    \hline
    \hline
    ANS \cite{ans} & A backbone and transit network that spans across the United States with $18$ nodes and $25$ links. All links have equal capacity of 45 Mbps. \\
    \hline
    GEANT \cite{geant} & A backbone and transit network that spans across the Europe with $34$ nodes and $52$ links. Link capacity ranges from 45 Mbps to 10 Gbps. \\
    \hline
    UNINETT \cite{uninett} & A large-sized backbone that spans across Norway with $69$ nodes and $98$ links. Most links have a capacity of 1, 2.5 or 10 Gbps. \\
    \hline
\end{tabular}
\end{center}
\end{table}

\vspace{0.5em}
\noindent\textbf{Traffic Patterns:}
Table \ref{table_traffic} shows the considered distributions for transfer volumes. Transfer arrival followed a Poisson distribution with rate $\lambda$. We considered no units for time or bandwidth. For all simulations, we assumed a timeslot length of $\delta = 1.0$. For Pareto distribution, we considered a minimum transfer volume equal to that of $2$ full timeslots and limited maximum transfer volume to that of $2000$ full timeslots. Unless otherwise stated, we considered an average demand equal to volume of $20$ full timeslots per transfer for all traffic distributions (we fixed the mean values of all distributions to the same value). Per simulation instance, we assumed equal number of transfers per sender and for every transfer, we selected the receivers from all existing nodes according to the uniform distribution (with equal probability from all nodes).

\vspace{0.2em}
\noindent\textbf{Assumptions:} We focused on computing gains and assumed accurate knowledge of inter-DC link capacity, and precise rate control at the end-points which together lead to a congestion free network. We also assumed no dropped packets due to corruption or errors, and no link failures.

\vspace{0.2em}
\noindent\textbf{Simulation Setup:} We developed a simulator in Java (JDK 8). We performed all simulations on one machine (Core i7-6700 and 24 GB of RAM). We used the Java implementation of GreedyFLAC \cite{DSTAlgoEvaluation} for minimum weight Steiner trees.

\begin{table}
\begin{center}
\caption{Transfer size distributions (parameters in \S \ref{evaluations}).} \label{table_traffic}
\vspace{0.5em}
\begin{tabular}{ |p{5cm}|p{8.5cm}| }
    \hline
    \textbf{Name} & \textbf{Description} \\
    \hline
    \hline
    Light-tailed & Based on Exponential distribution. \\
    \hline
    Heavy-tailed & Based on Pareto distribution. \\
    \hline
    Cache-Follower (Facebook) & Generated by cache applications over Facebook inter-DC WAN \cite{social_inside}. \\
    \hline
    Hadoop (Facebook) & Generated by geo-distributed analytics over Facebook inter-DC WAN \cite{social_inside}. \\
    \hline
\end{tabular}
\end{center}
\end{table}

\subsection{Weight Assignment Techniques for Tree Selection}
We empirically evaluate and analyze several weights for selection of forwarding trees. Table \ref{table_cost} lists the weight assignment approaches considered for tree selection (please see Table \ref{table_var_quickcast} for definition of variables). We considered three edge weight metrics of utilization (i.e., the fraction of a link's bandwidth currently in use), load (i.e., the total volume of traffic that an edge will carry starting current time), and load plus the volume of the newly arriving transfer request.

We also considered the weight of a tree to be either the weight of its edge with maximum weight or the sum of weights of its edges. An exponential weight is used to approximate selection of trees with minimum highest weight, similar to the approach used in \cite{tempus}. The benefit of the weight \#6 over \#5 is that in case there is no load or minimal load on some edges, selecting the minimum weight tree will lead to minimum edge trees that reduce bandwidth usage. Also, with this approach, we tend to avoid large trees for large transfers which helps further reduce bandwidth usage.

\begin{table}[t]
\begin{center}
\caption{Various weights for tree selection for incoming request $R_{new}$.} \label{table_cost}
\vspace{0.5em}
\renewcommand{\arraystretch}{1.2}
\begin{tabular}{ |p{0.5cm}|p{5cm}|p{8cm}| }
    \hline
    \textbf{\#} & Weight of edge $e, \forall e \in \pmb{\mathrm{E}}_G$ & \textbf{Properties of Selected Trees} \\
    \hline
    \hline
    1 & $1.0$ & A fixed minimum edge Steiner tree \\
    \hline
    2 & $\exp(U_e)$ & Minimum highest utilization over edges \\
    \hline
    3 & $\exp(L_e)$ & Minimum highest load over edges \\
    \hline
    4 & $U_e$ & Minimum sum of utilization over edges \\
    \hline
    5 & $L_e$ & Minimum sum of load over edges \\
    \hline
    6 & $L_e+\frac{\mathcal{V}_{R_{new}}}{C_e}$ & Minimum final sum of load over edges \\[0.5mm]
    \hline
    7 & $1.0+\frac{\exp(U_e)} {\sum_{e \in \pmb{\mathrm{E}}_G} \exp(U_e)}$ & Minimum edges, min-max utilization  \\[1mm]
    \hline
    8 & $1.0+\frac{\exp(L_e)} {\sum_{e \in \pmb{\mathrm{E}}_G} \exp(L_e)}$ & Minimum edges, min-max load \\[1mm]
    \hline
    9 & $1.0+\frac{U_e} {\sum_{e \in \pmb{\mathrm{E}}_G} U_e}$ & Minimum edges, min-sum of utilization \\[1mm]
    \hline
    10 & $1.0+\frac{L_e} {\sum_{e \in \pmb{\mathrm{E}}_G} L_e}$ & Minimum edges, min-sum of load \\[1mm]
    \hline
\end{tabular}
\end{center}
\end{table}

\begin{figure}[p]
\centering
\resizebox{0.7\columnwidth}{!}{
\begin{tabular}{c|c|c|c|c|c|c|c|c|c|c|c|c|}
\cline{2-13}
\multicolumn{1}{l|}{} & \multicolumn{12}{c|}{Mean Receiver Completion Times} \\ \cline{2-13}
\multicolumn{1}{l|}{} & \multicolumn{6}{c|}{ANS} & \multicolumn{6}{c|}{GEANT} \\ \cline{2-13} 
\multicolumn{1}{l|}{} & \multicolumn{3}{c|}{Light-tailed} & \multicolumn{3}{c|}{Heavy-tailed} & \multicolumn{3}{c|}{Light-tailed} & \multicolumn{3}{c|}{Heavy-tailed} \\ \hline 
\multicolumn{1}{|c|}{\cellcolor[HTML]{EFEFEF}\#} & $\mathcal{F}$ & $\mathcal{S}$ & $\mathcal{M}$ & $\mathcal{F}$ & $\mathcal{S}$ & $\mathcal{M}$ & $\mathcal{F}$ & $\mathcal{S}$ & $\mathcal{M}$ & $\mathcal{F}$ & $\mathcal{S}$ & $\mathcal{M}$ \\ \hline
\multicolumn{1}{|c|}{\cellcolor[HTML]{EFEFEF}1}  & \cellcolor[HTML]{34FF34}10- & \cellcolor[HTML]{34FF34}10- & \cellcolor[HTML]{F8FF00}20- & \cellcolor[HTML]{F8FF00}20- & \cellcolor[HTML]{34FF34}10- & \cellcolor[HTML]{F8FF00}20- & \cellcolor[HTML]{F56B00}50- & \cellcolor[HTML]{F8A102}40- & \cellcolor[HTML]{FE0000}50+ & \cellcolor[HTML]{FE0000}50+ & \cellcolor[HTML]{F8A102}40- & \cellcolor[HTML]{F8A102}40- \\ \hline
\multicolumn{1}{|c|}{\cellcolor[HTML]{EFEFEF}2}  & \cellcolor[HTML]{F8FF00}20- & \cellcolor[HTML]{F8FF00}20- & \cellcolor[HTML]{34FF34}10- & \cellcolor[HTML]{F8FF00}20- & \cellcolor[HTML]{FFCC67}30- & \cellcolor[HTML]{34FF34}10- & \cellcolor[HTML]{34FF34}10- & \cellcolor[HTML]{F8FF00}20- & \cellcolor[HTML]{34FF34}10- & \cellcolor[HTML]{F8FF00}20- & \cellcolor[HTML]{34FF34}10- & \cellcolor[HTML]{34FF34}10- \\ \hline
\multicolumn{1}{|c|}{\cellcolor[HTML]{EFEFEF}3}  & \cellcolor[HTML]{F8FF00}20- & \cellcolor[HTML]{F8FF00}20- & \cellcolor[HTML]{34FF34}10- & \cellcolor[HTML]{F8FF00}20- & \cellcolor[HTML]{F56B00}50- & \cellcolor[HTML]{FFCC67}30- & \cellcolor[HTML]{34FF34}10- & \cellcolor[HTML]{34FF34}10- & \cellcolor[HTML]{34FF34}10- & \cellcolor[HTML]{34FF34}10- & \cellcolor[HTML]{34FF34}10- & \cellcolor[HTML]{34FF34}10- \\ \hline
\multicolumn{1}{|c|}{\cellcolor[HTML]{EFEFEF}4}  & \cellcolor[HTML]{F8A102}40- & \cellcolor[HTML]{F8A102}40- & \cellcolor[HTML]{34FF34}10- & \cellcolor[HTML]{F8A102}40- & \cellcolor[HTML]{F8A102}40- & \cellcolor[HTML]{34FF34}10- & \cellcolor[HTML]{F8FF00}20- & \cellcolor[HTML]{FFCC67}30- & \cellcolor[HTML]{34FF34}10- & \cellcolor[HTML]{F8FF00}20- & \cellcolor[HTML]{34FF34}10- & \cellcolor[HTML]{34FF34}10- \\ \hline
\multicolumn{1}{|c|}{\cellcolor[HTML]{EFEFEF}5}  & \cellcolor[HTML]{34FF34}10- & \cellcolor[HTML]{34FF34}10- & \cellcolor[HTML]{34FF34}10- & \cellcolor[HTML]{34FF34}10- & \cellcolor[HTML]{F8FF00}20- & \cellcolor[HTML]{F8FF00}20- & \cellcolor[HTML]{34FF34}10- & \cellcolor[HTML]{34FF34}10- & \cellcolor[HTML]{34FF34}10- & \cellcolor[HTML]{34FF34}10- & \cellcolor[HTML]{34FF34}10- & \cellcolor[HTML]{34FF34}10- \\ \hline
\multicolumn{1}{|c|}{\cellcolor[HTML]{EFEFEF}\textbf{6}}  & \cellcolor[HTML]{34FF34}10- & \cellcolor[HTML]{34FF34}10- & \cellcolor[HTML]{34FF34}10- & \cellcolor[HTML]{34FF34}10- & \cellcolor[HTML]{F8FF00}20- & \cellcolor[HTML]{F8FF00}20- & \cellcolor[HTML]{34FF34}10- & \cellcolor[HTML]{34FF34}10- & \cellcolor[HTML]{34FF34}10- & \cellcolor[HTML]{34FF34}10- & \cellcolor[HTML]{34FF34}10- & \cellcolor[HTML]{34FF34}10- \\ \hline
\multicolumn{1}{|c|}{\cellcolor[HTML]{EFEFEF}7}  & \cellcolor[HTML]{34FF34}10- & \cellcolor[HTML]{34FF34}10- & \cellcolor[HTML]{34FF34}10- & \cellcolor[HTML]{34FF34}10- & \cellcolor[HTML]{34FF34}10- & \cellcolor[HTML]{34FF34}10- & \cellcolor[HTML]{F8A102}40- & \cellcolor[HTML]{FFCC67}30- & \cellcolor[HTML]{FFCC67}30- & \cellcolor[HTML]{F8A102}40- & \cellcolor[HTML]{FFCC67}30- & \cellcolor[HTML]{F8FF00}20- \\ \hline
\multicolumn{1}{|c|}{\cellcolor[HTML]{EFEFEF}8}  & \cellcolor[HTML]{34FF34}10- & \cellcolor[HTML]{34FF34}10- & \cellcolor[HTML]{34FF34}10- & \cellcolor[HTML]{34FF34}10- & \cellcolor[HTML]{34FF34}10- & \cellcolor[HTML]{34FF34}10- & \cellcolor[HTML]{F56B00}50- & \cellcolor[HTML]{F8A102}40- & \cellcolor[HTML]{FE0000}50+ & \cellcolor[HTML]{FE0000}50+ & \cellcolor[HTML]{F8A102}40- & \cellcolor[HTML]{F8A102}40- \\ \hline
\multicolumn{1}{|c|}{\cellcolor[HTML]{EFEFEF}9}  & \cellcolor[HTML]{34FF34}10- & \cellcolor[HTML]{34FF34}10- & \cellcolor[HTML]{34FF34}10- & \cellcolor[HTML]{34FF34}10- & \cellcolor[HTML]{34FF34}10- & \cellcolor[HTML]{34FF34}10- & \cellcolor[HTML]{F8A102}40- & \cellcolor[HTML]{F8A102}40- & \cellcolor[HTML]{FFCC67}30- & \cellcolor[HTML]{F8A102}40- & \cellcolor[HTML]{FFCC67}30- & \cellcolor[HTML]{FFCC67}30- \\ \hline
\multicolumn{1}{|c|}{\cellcolor[HTML]{EFEFEF}10} & \cellcolor[HTML]{34FF34}10- & \cellcolor[HTML]{34FF34}10- & \cellcolor[HTML]{34FF34}10- & \cellcolor[HTML]{34FF34}10- & \cellcolor[HTML]{34FF34}10- & \cellcolor[HTML]{34FF34}10- & \cellcolor[HTML]{FE0000}50+ & \cellcolor[HTML]{FE0000}50+ & \cellcolor[HTML]{FE0000}50+ & \cellcolor[HTML]{FE0000}50+ & \cellcolor[HTML]{FE0000}50+ & \cellcolor[HTML]{F56B00}50- \\ \hline
\end{tabular}
}\vspace{0.2em}
\\
\vspace{0.5em}
\resizebox{0.7\columnwidth}{!}{
\begin{tabular}{c|c|c|c|c|c|c|c|c|c|c|c|c|}
\cline{2-13}
\multicolumn{1}{l|}{} & \multicolumn{12}{c|}{Tail Receiver Completion Times} \\ \cline{2-13}
\multicolumn{1}{l|}{} & \multicolumn{6}{c|}{ANS} & \multicolumn{6}{c|}{GEANT} \\ \cline{2-13}
\multicolumn{1}{l|}{} & \multicolumn{3}{c|}{Light-tailed} & \multicolumn{3}{c|}{Heavy-tailed} & \multicolumn{3}{c|}{Light-tailed} & \multicolumn{3}{c|}{Heavy-tailed} \\ \hline
\multicolumn{1}{|c|}{\cellcolor[HTML]{EFEFEF}\#} & $\mathcal{F}$ & $\mathcal{S}$ & $\mathcal{M}$ & $\mathcal{F}$ & $\mathcal{S}$ & $\mathcal{M}$ & $\mathcal{F}$ & $\mathcal{S}$ & $\mathcal{M}$ & $\mathcal{F}$ & $\mathcal{S}$ & $\mathcal{M}$ \\ \hline
\multicolumn{1}{|c|}{\cellcolor[HTML]{EFEFEF}1}  & \cellcolor[HTML]{F8FF00}20- & \cellcolor[HTML]{F8FF00}20- & \cellcolor[HTML]{FFCC67}30- & \cellcolor[HTML]{F8FF00}20- & \cellcolor[HTML]{34FF34}10- & \cellcolor[HTML]{F8FF00}20- & \cellcolor[HTML]{F56B00}50- & \cellcolor[HTML]{F56B00}50- & \cellcolor[HTML]{FE0000}50+ & \cellcolor[HTML]{FE0000}50+ & \cellcolor[HTML]{FE0000}50+ & \cellcolor[HTML]{F56B00}50- \\ \hline
\multicolumn{1}{|c|}{\cellcolor[HTML]{EFEFEF}2}  & \cellcolor[HTML]{FFCC67}30- & \cellcolor[HTML]{F8FF00}20- & \cellcolor[HTML]{F8FF00}20- & \cellcolor[HTML]{FFCC67}30- & \cellcolor[HTML]{FFCC67}30- & \cellcolor[HTML]{F8FF00}20- & \cellcolor[HTML]{F8FF00}20- & \cellcolor[HTML]{FFCC67}30- & \cellcolor[HTML]{F8FF00}20- & \cellcolor[HTML]{FFCC67}30- & \cellcolor[HTML]{F8FF00}20- & \cellcolor[HTML]{34FF34}10- \\ \hline
\multicolumn{1}{|c|}{\cellcolor[HTML]{EFEFEF}3}  & \cellcolor[HTML]{F8FF00}20- & \cellcolor[HTML]{F8FF00}20- & \cellcolor[HTML]{34FF34}10- & \cellcolor[HTML]{34FF34}10- & \cellcolor[HTML]{34FF34}10- & \cellcolor[HTML]{34FF34}10- & \cellcolor[HTML]{34FF34}10- & \cellcolor[HTML]{34FF34}10- & \cellcolor[HTML]{34FF34}10- & \cellcolor[HTML]{34FF34}10- & \cellcolor[HTML]{34FF34}10- & \cellcolor[HTML]{34FF34}10- \\ \hline
\multicolumn{1}{|c|}{\cellcolor[HTML]{EFEFEF}4}  & \cellcolor[HTML]{F8A102}40- & \cellcolor[HTML]{F8A102}40- & \cellcolor[HTML]{34FF34}10- & \cellcolor[HTML]{FFCC67}30- & \cellcolor[HTML]{FFCC67}30- & \cellcolor[HTML]{34FF34}10- & \cellcolor[HTML]{FFCC67}30- & \cellcolor[HTML]{FFCC67}30- & \cellcolor[HTML]{F8FF00}20- & \cellcolor[HTML]{F8FF00}20- & \cellcolor[HTML]{F8FF00}20- & \cellcolor[HTML]{34FF34}10- \\ \hline
\multicolumn{1}{|c|}{\cellcolor[HTML]{EFEFEF}5}  & \cellcolor[HTML]{34FF34}10- & \cellcolor[HTML]{34FF34}10- & \cellcolor[HTML]{34FF34}10- & \cellcolor[HTML]{34FF34}10- & \cellcolor[HTML]{34FF34}10- & \cellcolor[HTML]{34FF34}10- & \cellcolor[HTML]{34FF34}10- & \cellcolor[HTML]{34FF34}10- & \cellcolor[HTML]{34FF34}10- & \cellcolor[HTML]{34FF34}10- & \cellcolor[HTML]{34FF34}10- & \cellcolor[HTML]{34FF34}10- \\ \hline
\multicolumn{1}{|c|}{\cellcolor[HTML]{EFEFEF}\textbf{6}}  & \cellcolor[HTML]{34FF34}10- & \cellcolor[HTML]{34FF34}10- & \cellcolor[HTML]{34FF34}10- & \cellcolor[HTML]{34FF34}10- & \cellcolor[HTML]{34FF34}10- & \cellcolor[HTML]{34FF34}10- & \cellcolor[HTML]{34FF34}10- & \cellcolor[HTML]{34FF34}10- & \cellcolor[HTML]{34FF34}10- & \cellcolor[HTML]{34FF34}10- & \cellcolor[HTML]{34FF34}10- & \cellcolor[HTML]{34FF34}10- \\ \hline
\multicolumn{1}{|c|}{\cellcolor[HTML]{EFEFEF}7}  & \cellcolor[HTML]{F8FF00}20- & \cellcolor[HTML]{34FF34}10- & \cellcolor[HTML]{F8FF00}20- & \cellcolor[HTML]{F8FF00}20- & \cellcolor[HTML]{34FF34}10- & \cellcolor[HTML]{34FF34}10- & \cellcolor[HTML]{F8A102}40- & \cellcolor[HTML]{FFCC67}30- & \cellcolor[HTML]{F8A102}40- & \cellcolor[HTML]{F56B00}50- & \cellcolor[HTML]{F8A102}40- & \cellcolor[HTML]{FE0000}50+ \\ \hline
\multicolumn{1}{|c|}{\cellcolor[HTML]{EFEFEF}8}  & \cellcolor[HTML]{34FF34}10- & \cellcolor[HTML]{34FF34}10- & \cellcolor[HTML]{34FF34}10- & \cellcolor[HTML]{34FF34}10- & \cellcolor[HTML]{34FF34}10- & \cellcolor[HTML]{34FF34}10- & \cellcolor[HTML]{F56B00}50- & \cellcolor[HTML]{F56B00}50- & \cellcolor[HTML]{FE0000}50+ & \cellcolor[HTML]{FE0000}50+ & \cellcolor[HTML]{FE0000}50+ & \cellcolor[HTML]{F56B00}50- \\ \hline
\multicolumn{1}{|c|}{\cellcolor[HTML]{EFEFEF}9}  & \cellcolor[HTML]{34FF34}10- & \cellcolor[HTML]{F8FF00}20- & \cellcolor[HTML]{F8FF00}20- & \cellcolor[HTML]{F8FF00}20- & \cellcolor[HTML]{34FF34}10- & \cellcolor[HTML]{34FF34}10- & \cellcolor[HTML]{FFCC67}30- & \cellcolor[HTML]{FFCC67}30- & \cellcolor[HTML]{F8A102}40- & \cellcolor[HTML]{F8A102}40- & \cellcolor[HTML]{FFCC67}30- & \cellcolor[HTML]{FE0000}50+ \\ \hline
\multicolumn{1}{|c|}{\cellcolor[HTML]{EFEFEF}10} & \cellcolor[HTML]{34FF34}10- & \cellcolor[HTML]{34FF34}10- & \cellcolor[HTML]{34FF34}10- & \cellcolor[HTML]{34FF34}10- & \cellcolor[HTML]{34FF34}10- & \cellcolor[HTML]{34FF34}10- & \cellcolor[HTML]{F8A102}40- & \cellcolor[HTML]{FE0000}50+ & \cellcolor[HTML]{F56B00}50- & \cellcolor[HTML]{F8A102}40- & \cellcolor[HTML]{FE0000}50+ & \cellcolor[HTML]{F56B00}50- \\ \hline
\end{tabular}
}\vspace{0.2em}
\\
\vspace{0.5em}
\resizebox{0.7\columnwidth}{!}{
\begin{tabular}{c|c|c|c|c|c|c|c|c|c|c|c|c|}
\cline{2-13}
\multicolumn{1}{l|}{} & \multicolumn{12}{c|}{Total Bandwidth Used} \\ \cline{2-13}
\multicolumn{1}{l|}{} & \multicolumn{6}{c|}{ANS} & \multicolumn{6}{c|}{GEANT} \\ \cline{2-13}
\multicolumn{1}{l|}{} & \multicolumn{3}{c|}{Light-tailed} & \multicolumn{3}{c|}{Heavy-tailed} & \multicolumn{3}{c|}{Light-tailed} & \multicolumn{3}{c|}{Heavy-tailed} \\ \hline
\multicolumn{1}{|c|}{\cellcolor[HTML]{EFEFEF}\#} & $\mathcal{F}$ & $\mathcal{S}$ & $\mathcal{M}$ & $\mathcal{F}$ & $\mathcal{S}$ & $\mathcal{M}$ & $\mathcal{F}$ & $\mathcal{S}$ & $\mathcal{M}$ & $\mathcal{F}$ & $\mathcal{S}$ & $\mathcal{M}$ \\ \hline
\multicolumn{1}{|c|}{\cellcolor[HTML]{EFEFEF}1}  & \cellcolor[HTML]{34FF34}10- & \cellcolor[HTML]{34FF34}10- & \cellcolor[HTML]{34FF34}10- & \cellcolor[HTML]{34FF34}10- & \cellcolor[HTML]{34FF34}10- & \cellcolor[HTML]{34FF34}10- & \cellcolor[HTML]{34FF34}10- & \cellcolor[HTML]{34FF34}10- & \cellcolor[HTML]{34FF34}10- & \cellcolor[HTML]{34FF34}10- & \cellcolor[HTML]{34FF34}10- & \cellcolor[HTML]{34FF34}10- \\ \hline
\multicolumn{1}{|c|}{\cellcolor[HTML]{EFEFEF}2}  & \cellcolor[HTML]{F8FF00}20- & \cellcolor[HTML]{F8FF00}20- & \cellcolor[HTML]{F8FF00}20- & \cellcolor[HTML]{F8FF00}20- & \cellcolor[HTML]{F8FF00}20- & \cellcolor[HTML]{F8FF00}20- & \cellcolor[HTML]{F8A102}40- & \cellcolor[HTML]{F8A102}40- & \cellcolor[HTML]{F8A102}40- & \cellcolor[HTML]{F8A102}40- & \cellcolor[HTML]{F56B00}50- & \cellcolor[HTML]{F8A102}40- \\ \hline
\multicolumn{1}{|c|}{\cellcolor[HTML]{EFEFEF}3}  & \cellcolor[HTML]{F8FF00}20- & \cellcolor[HTML]{F8FF00}20- & \cellcolor[HTML]{F8FF00}20- & \cellcolor[HTML]{F8FF00}20- & \cellcolor[HTML]{F8FF00}20- & \cellcolor[HTML]{F8FF00}20- & \cellcolor[HTML]{34FF34}10- & \cellcolor[HTML]{34FF34}10- & \cellcolor[HTML]{34FF34}10- & \cellcolor[HTML]{34FF34}10- & \cellcolor[HTML]{34FF34}10- & \cellcolor[HTML]{34FF34}10- \\ \hline
\multicolumn{1}{|c|}{\cellcolor[HTML]{EFEFEF}4}  & \cellcolor[HTML]{FFCC67}30- & \cellcolor[HTML]{FFCC67}30- & \cellcolor[HTML]{34FF34}10- & \cellcolor[HTML]{FFCC67}30- & \cellcolor[HTML]{FFCC67}30- & \cellcolor[HTML]{34FF34}10- & \cellcolor[HTML]{F8FF00}20- & \cellcolor[HTML]{FFCC67}30- & \cellcolor[HTML]{34FF34}10- & \cellcolor[HTML]{F8FF00}20- & \cellcolor[HTML]{FFCC67}30- & \cellcolor[HTML]{F8FF00}20- \\ \hline
\multicolumn{1}{|c|}{\cellcolor[HTML]{EFEFEF}5}  & \cellcolor[HTML]{34FF34}10- & \cellcolor[HTML]{34FF34}10- & \cellcolor[HTML]{34FF34}10- & \cellcolor[HTML]{34FF34}10- & \cellcolor[HTML]{34FF34}10- & \cellcolor[HTML]{34FF34}10- & \cellcolor[HTML]{34FF34}10- & \cellcolor[HTML]{34FF34}10- & \cellcolor[HTML]{34FF34}10- & \cellcolor[HTML]{34FF34}10- & \cellcolor[HTML]{34FF34}10- & \cellcolor[HTML]{F8FF00}20- \\ \hline
\multicolumn{1}{|c|}{\cellcolor[HTML]{EFEFEF}\textbf{6}}  & \cellcolor[HTML]{34FF34}10- & \cellcolor[HTML]{34FF34}10- & \cellcolor[HTML]{34FF34}10- & \cellcolor[HTML]{34FF34}10- & \cellcolor[HTML]{34FF34}10- & \cellcolor[HTML]{34FF34}10- & \cellcolor[HTML]{34FF34}10- & \cellcolor[HTML]{34FF34}10- & \cellcolor[HTML]{34FF34}10- & \cellcolor[HTML]{34FF34}10- & \cellcolor[HTML]{34FF34}10- & \cellcolor[HTML]{F8FF00}20- \\ \hline
\multicolumn{1}{|c|}{\cellcolor[HTML]{EFEFEF}7}  & \cellcolor[HTML]{34FF34}10- & \cellcolor[HTML]{34FF34}10- & \cellcolor[HTML]{34FF34}10- & \cellcolor[HTML]{34FF34}10- & \cellcolor[HTML]{34FF34}10- & \cellcolor[HTML]{34FF34}10- & \cellcolor[HTML]{34FF34}10- & \cellcolor[HTML]{34FF34}10- & \cellcolor[HTML]{34FF34}10- & \cellcolor[HTML]{34FF34}10- & \cellcolor[HTML]{34FF34}10- & \cellcolor[HTML]{34FF34}10- \\ \hline
\multicolumn{1}{|c|}{\cellcolor[HTML]{EFEFEF}8}  & \cellcolor[HTML]{34FF34}10- & \cellcolor[HTML]{34FF34}10- & \cellcolor[HTML]{34FF34}10- & \cellcolor[HTML]{34FF34}10- & \cellcolor[HTML]{34FF34}10- & \cellcolor[HTML]{34FF34}10- & \cellcolor[HTML]{34FF34}10- & \cellcolor[HTML]{34FF34}10- & \cellcolor[HTML]{34FF34}10- & \cellcolor[HTML]{34FF34}10- & \cellcolor[HTML]{34FF34}10- & \cellcolor[HTML]{34FF34}10- \\ \hline
\multicolumn{1}{|c|}{\cellcolor[HTML]{EFEFEF}9}  & \cellcolor[HTML]{34FF34}10- & \cellcolor[HTML]{34FF34}10- & \cellcolor[HTML]{34FF34}10- & \cellcolor[HTML]{34FF34}10- & \cellcolor[HTML]{34FF34}10- & \cellcolor[HTML]{34FF34}10- & \cellcolor[HTML]{34FF34}10- & \cellcolor[HTML]{34FF34}10- & \cellcolor[HTML]{34FF34}10- & \cellcolor[HTML]{34FF34}10- & \cellcolor[HTML]{34FF34}10- & \cellcolor[HTML]{34FF34}10- \\ \hline
\multicolumn{1}{|c|}{\cellcolor[HTML]{EFEFEF}10} & \cellcolor[HTML]{34FF34}10- & \cellcolor[HTML]{34FF34}10- & \cellcolor[HTML]{34FF34}10- & \cellcolor[HTML]{34FF34}10- & \cellcolor[HTML]{34FF34}10- & \cellcolor[HTML]{34FF34}10- & \cellcolor[HTML]{34FF34}10- & \cellcolor[HTML]{34FF34}10- & \cellcolor[HTML]{34FF34}10- & \cellcolor[HTML]{34FF34}10- & \cellcolor[HTML]{34FF34}10- & \cellcolor[HTML]{34FF34}10- \\ \hline
\end{tabular}
}%
\vspace{0.4em}
\resizebox{0.7\columnwidth}{!}{
\begin{tabular}{|l|
>{\columncolor[HTML]{34FF34}}l |l|l|
>{\columncolor[HTML]{F8FF00}}l |l|l|
>{\columncolor[HTML]{FFCC67}}l |}
\cline{1-2} \cline{4-5} \cline{7-8}
$< 10\%$ from min & 10- & & $< 20\%$ from min & 20- & & $< 30\%$ from min & 30- \\ \cline{1-2} \cline{4-5} \cline{7-8}
\end{tabular}
}%
\vspace{0.1em}
\resizebox{0.7\columnwidth}{!}{
\begin{tabular}{|l|
>{\columncolor[HTML]{F8A102}}l |l|l|
>{\columncolor[HTML]{F56B00}}l |l|l|
>{\columncolor[HTML]{FE0000}}l |}
\cline{1-2} \cline{4-5} \cline{7-8}
$< 40\%$ from min & 40- & & $< 50\%$ from min & 50- & & $\ge 50\%$ from min & 50+ \\ \cline{1-2} \cline{4-5} \cline{7-8}
\end{tabular}
}%
\vspace{0.2em}
\caption{Evaluation of various weights for tree selection ($\mathcal{F}$, $\mathcal{S}$ and $\mathcal{M}$ refer to scheduling policies FCFS, SRPT and Fair Sharing, respectively).} \label{fig:weights.assignment}
\end{figure}

Figure \ref{fig:weights.assignment} shows our simulation results of receiver completion times for bulk multicast transfers with $10$ receivers for a fixed arrival rate of $\lambda=1$. We considered both light-tailed and heavy-tailed transfer volume distributions. Techniques \#1, \#7, \#8, \#9 and \#10 all used minimal edge Steiner trees, and so offer minimum bandwidth usage. However, this comes at the cost of increasing completion times especially when edges have a non-homogeneous capacity.

Techniques \#2 and \#4 use utilization as criteria for load balancing. Minimizing maximum link utilization has long been a popular objective for traffic engineering over WAN. As can be seen, they have the highest bandwidth usage compared to other techniques (up to $40\%$ above the minimum) for almost all scenarios while their completion times are at least $20\%$ worse than the minimum for several scenarios.

Techniques \#3, \#5, and \#6 operate based on link load (i.e., total outstanding volume of traffic per edge) among which technique \#3 (minimizing maximum load) has the highest variation between best and worst case performance (up to $40\%$ worse than the minimum in mean completion times). 

Techniques \#5 and \#6 (minimizing the sum of load including and excluding the new multicast request) on the other hand offer consistently good performance that is up to $13\%$ above the minimum (for all performance metrics) across all scheduling policies, topologies, and traffic patterns. These techniques offer lower completion times for the GEANT topology with non-uniform link capacity. Technique \#6 also provides slightly better bandwidth usage and better completion times compared to \#5 for the majority of scenarios (not shown). Our proposals rely on technique \#6 for selection of load-aware forwarding trees, as shown in Algorithm \ref{algo_dccast}.

\subsection{Receiver Set Partitioning}
Receiver set partitioning allows separation of faster receivers from the slowest (or slower ones). This is essential to improve network utilization and speed up transfers when there are competing transfers or physical bottlenecks. For example, both GEANT and UNINETT have edges that vary by at least a factor of $10\times$ in capacity. We evaluate QuickCast over a variety of scenarios.

\subsubsection{Effect of Number of Receivers}
\begin{figure}
    \centering
    \subfigure[$\lambda = 1$ (heavy load)]
    {
        \includegraphics[width=\textwidth]{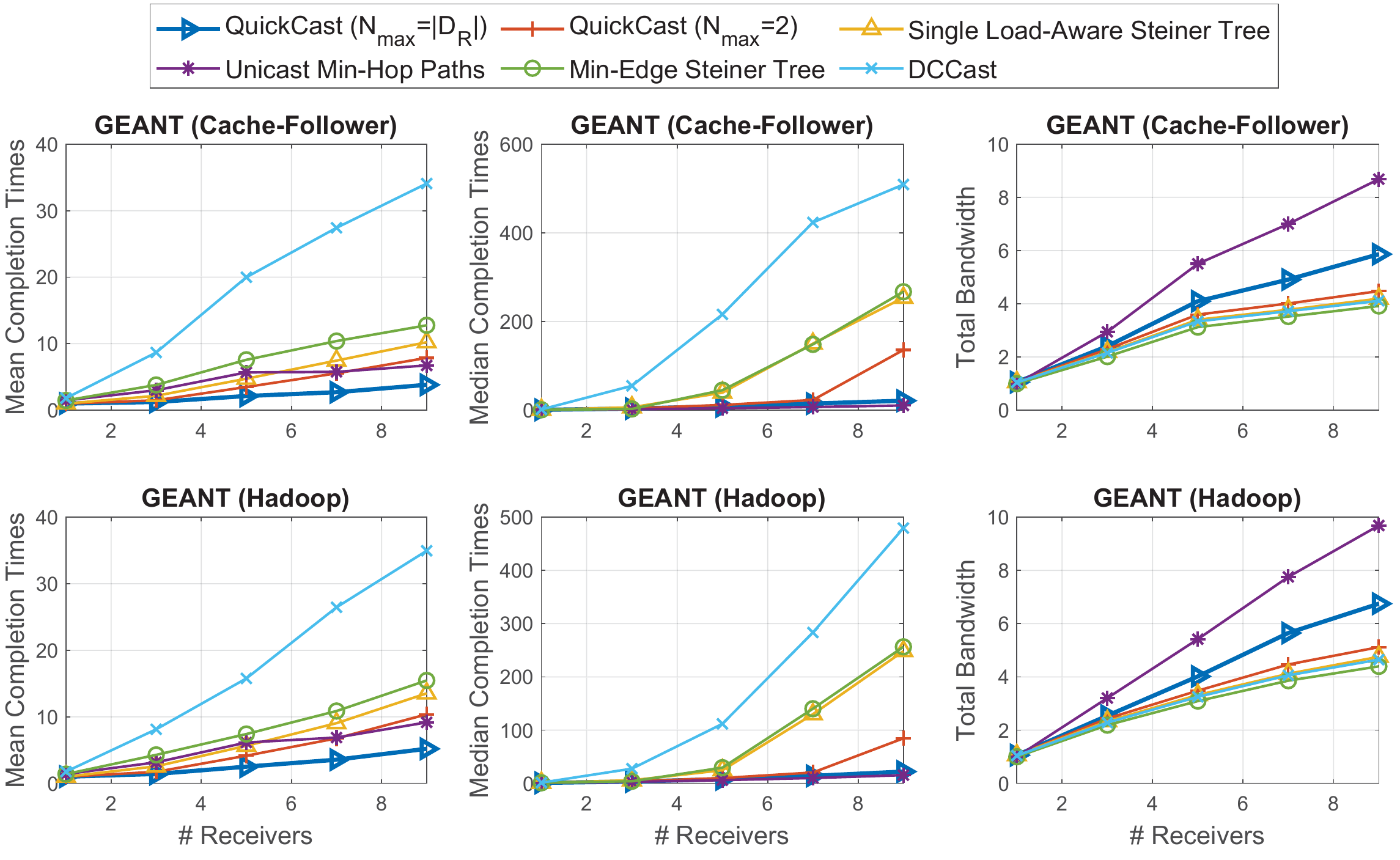}
    }
    \\
    \subfigure[$\lambda = 0.001$ (light load)]
    {
        \includegraphics[width=\textwidth]{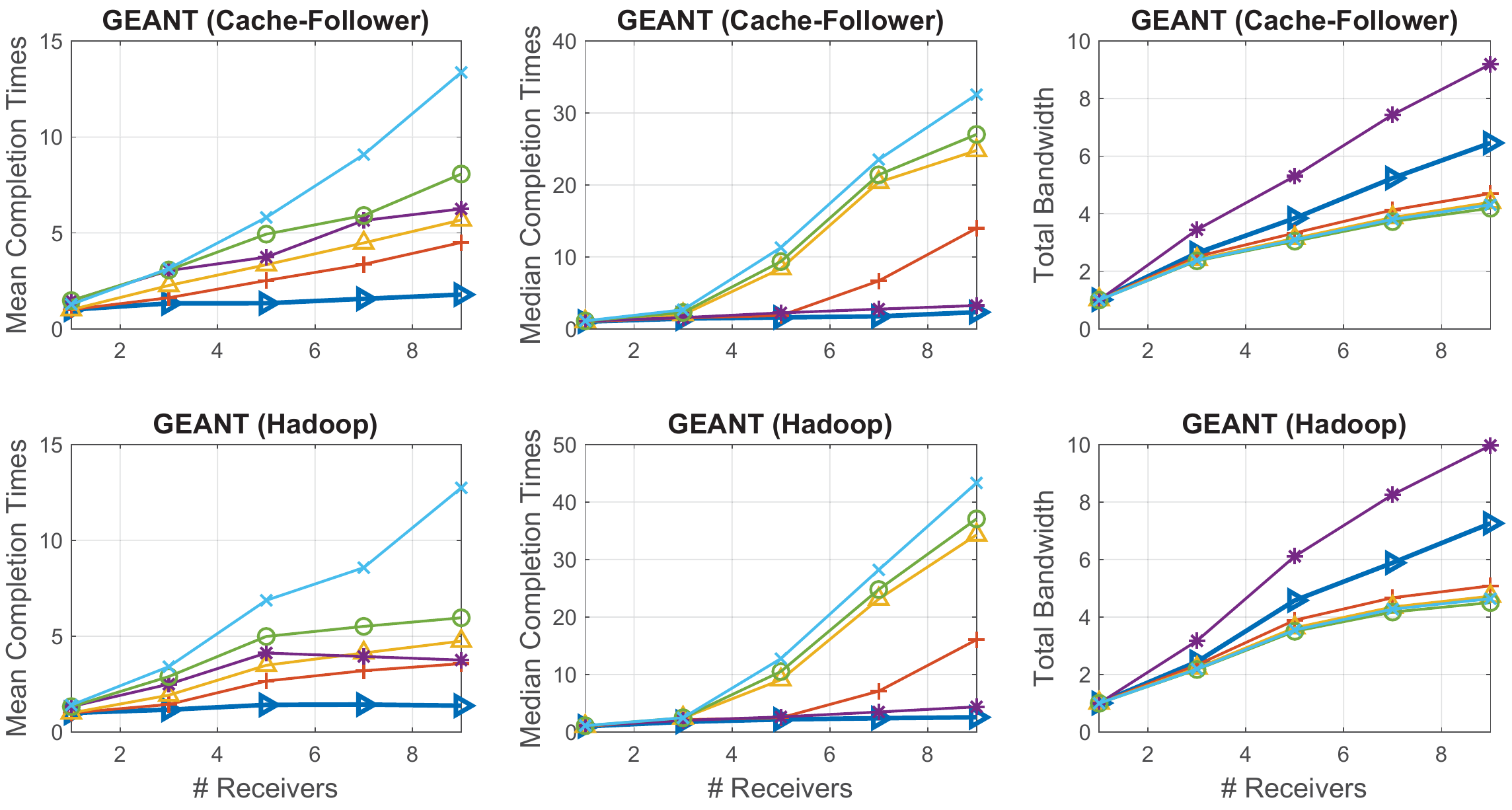}
    }
    \caption{Various schemes for bulk multicast transfers. All schemes use max-min fair rates except for DCCast which uses FCFS and are performed on GEANT topology. Plots are normalized by minimum (lower is better). We used Cache-Follower and Hadoop traffic patterns in Table \ref{table_traffic}.}
    \label{fig:overall}
\end{figure}

We provide an overall comparison of several schemes (QuickCast, Single Load-Aware Steiner Tree, and DCCast \cite{dccast}) along with two basic solutions of using a minimum edge Steiner tree and unicast minimum hop path routing as shown in Figure \ref{fig:overall}. We also considered both light and heavy load regimes. We used real inter-DC traffic patterns reported by Facebook for two applications of Cache-Follower and Hadoop \cite{social_inside}. Also, all schemes use the fair sharing rate allocation based on max-min fairness except DCCast which uses the FCFS policy.

The minimum edge Steiner tree leads to the minimum bandwidth consumption. The unicast minimum hop path routing approach separates all receivers per bulk multicast transfer. It, however, uses a significantly larger volume of traffic and also does not offer the best mean completion times for the following reasons. First, it exhausts network capacity quickly which increases tail completion times by a significant factor (not shown here). Second, it can lead to many additional shared links that increase contention across flows and reduce throughput per receiver. The significant increase in completion times of higher percentiles increases the average completion times of the unicast approach.

With $N_{max}=\lvert \pmb{\mathrm{D}}_{R_{new}} \rvert$, we see that QuickCast offers the best mean and median completion times, i.e., up to $2.84\times$ less compared to QuickCast with $N_{max}=2$, up to $3.64\times$ less compared to unicast minimum hop routing, and up to $3.33\times$ less than single load-aware Steiner tree. To achieve this gain, QuickCast with $N_{max}=\lvert \pmb{\mathrm{D}}_{R_{new}} \rvert$ uses at most $1.49\times$ more bandwidth compared to using minimum edge Steiner trees which is still $1.4\times$ less than bandwidth usage of unicast minimum hop routing. We also see that while increasing the number of receivers, QuickCast with $N_{max}=\lvert \pmb{\mathrm{D}}_{R_{new}} \rvert$ offers consistently small median completion times by separating fast and slow receivers since the number of partitions are not limited. Overall, we see a higher gain under light load as there is more capacity available to utilize. We also recognize that QuickCast with either $N_{max}=2$ or $N_{max}=\lvert \pmb{\mathrm{D}}_{R_{new}} \rvert$ performs almost always better than unicast minimum hop routing in mean completion times.

\subsubsection{Speedup by Receiver Rank}
\begin{figure}
    \centering
    \includegraphics[width=\textwidth]{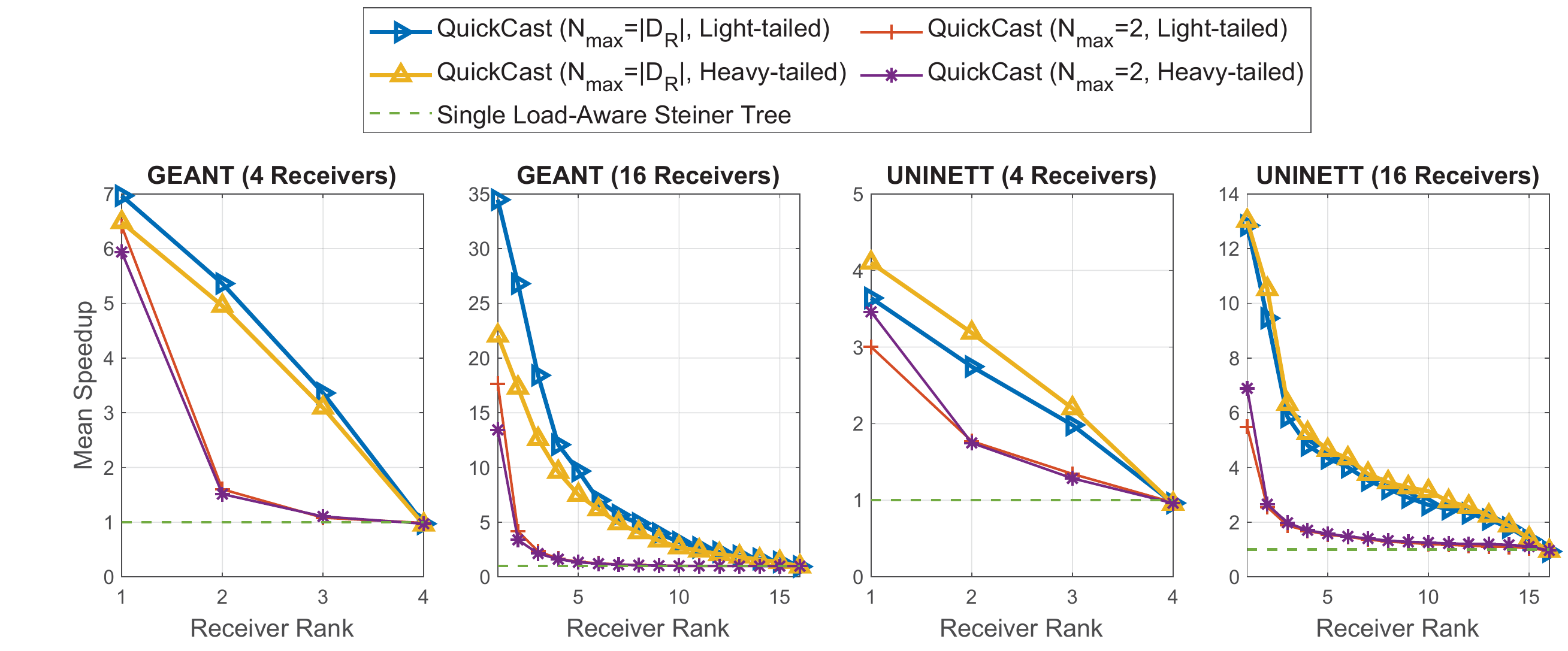}
    \caption{Mean receiver completion time speedup (larger is better) of receivers compared to single load-aware Steiner tree (Algorithm \ref{algo_dccast}) by their rank (receivers sorted by their speed from fastest to slowest per transfer), receivers selected according to uniform distribution from all nodes, we considered $\lambda = 1$.}
    \label{fig:speedup_1}
\end{figure}

Figure \ref{fig:speedup_1} shows how QuickCast can speed up multiple receivers per transfer by separating them from the slower receivers. The gains are normalized by when a single partition is used per bulk multicast transfer. In case the number of partitions is limited to two similar to \cite{quickcast}, the highest gain is usually obtained by the first two to three receivers while allowing more partitions, we can get considerably higher gain for a significant fraction of receivers. Also, by not limiting the partitions to two, we see higher gains for all receiver ranks that is above $2\times$ for multiple receiver ranks. This comes at the cost of higher bandwidth consumption which we saw earlier in the previous experiment.

\subsubsection{Partitioning Factor}
\begin{figure}
    \centering
    \includegraphics[width=\textwidth]{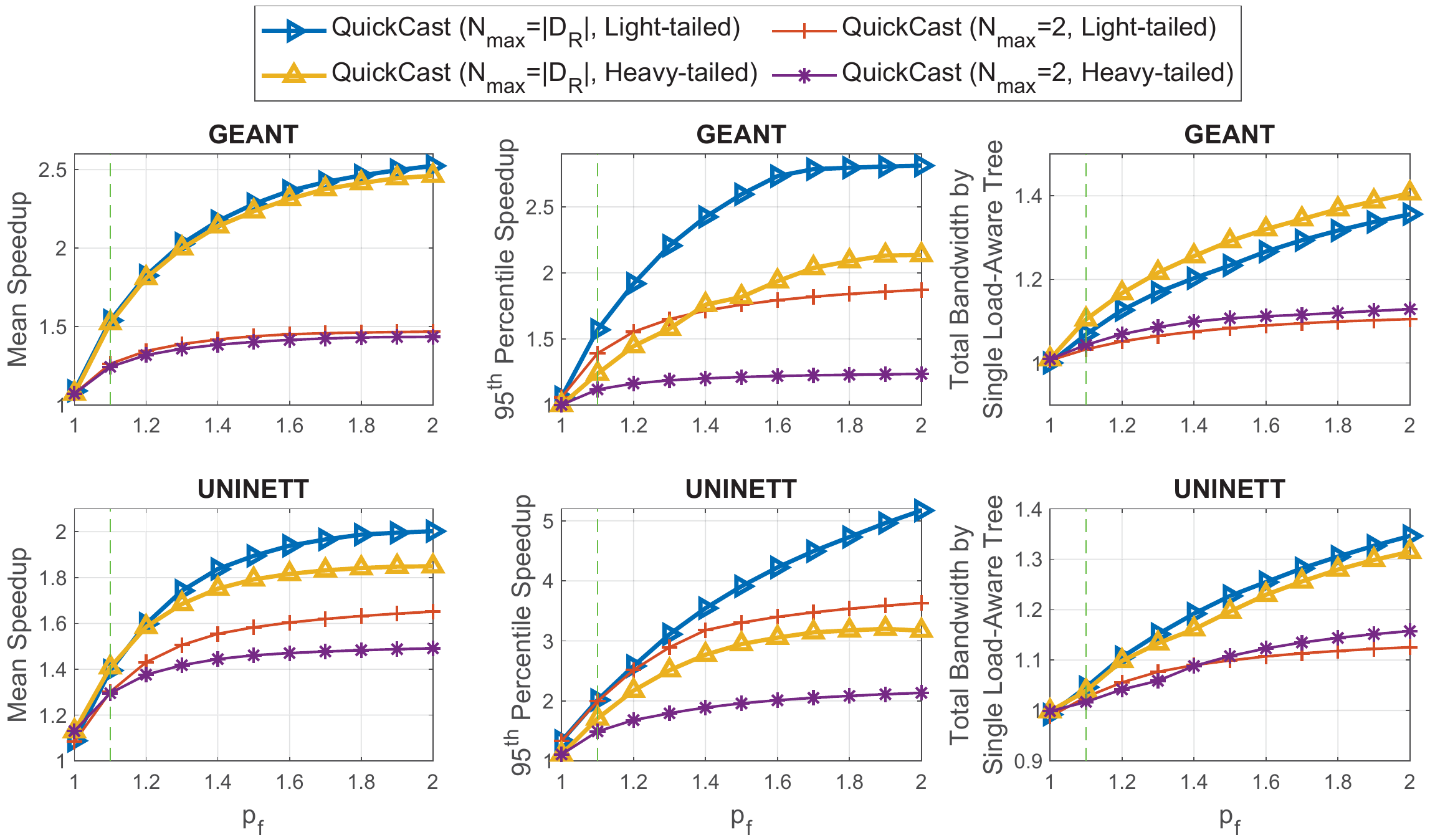}
    \caption{Performance of QuickCast as a function of partitioning factor $p_f$. We assumed 4 receivers and an arrival rate of $\lambda = 1$.}
    \label{fig:pf}
\end{figure}

The performance of QuickCast as a function of the partitioning factor (i.e., $p_f$) has been shown in Figure \ref{fig:pf} where gains are normalized by single load-aware Steiner tree which uses a single partition per bulk multicast transfer. We computed per receiver mean and 95\textsuperscript{th} percentile completion times as well as bandwidth usage.

As can be seen, bandwidth consumption increases with partitioning factor as more requests' receivers are partitioned into two or more groups. The gains in completion times keep increasing if $N_{max}$ is not limited as we increase $p_f$. That, however, can ultimately lead to unicast delivery to all receivers (i.e., every receiver as a separate partition) and excessive bandwidth usage. We see a diminishing return type of curve as $p_f$ is increased with the highest returns coming when we increase $p_f$ from 1 to 1.1 (marked with a green dashed lined). That is because using too many partitions can saturate network capacity while not improving the separation of fast and slow nodes considerably.

At $p_f=1.1$, we see up to 10\% additional bandwidth usage compared to single load-aware Steiner tree while mean completion times improve by between 40\% to 50\%. According to other experiments not shown here, with large $p_f$, it is possible even to see reductions in gain that come from excessive bandwidth consumption and increased contention over capacity. Note that this experiment was performed considering four receivers per bulk multicast transfer. Using more receivers can lead to more bandwidth usage with the same $p_f$, an increased slope at values of $p_f$ close to 1, and faster saturation of network capacity as we increase $p_f$. Therefore, using smaller $p_f$ is preferred with more receivers per transfer.

\begin{figure}
    \centering
    \includegraphics[width=\textwidth]{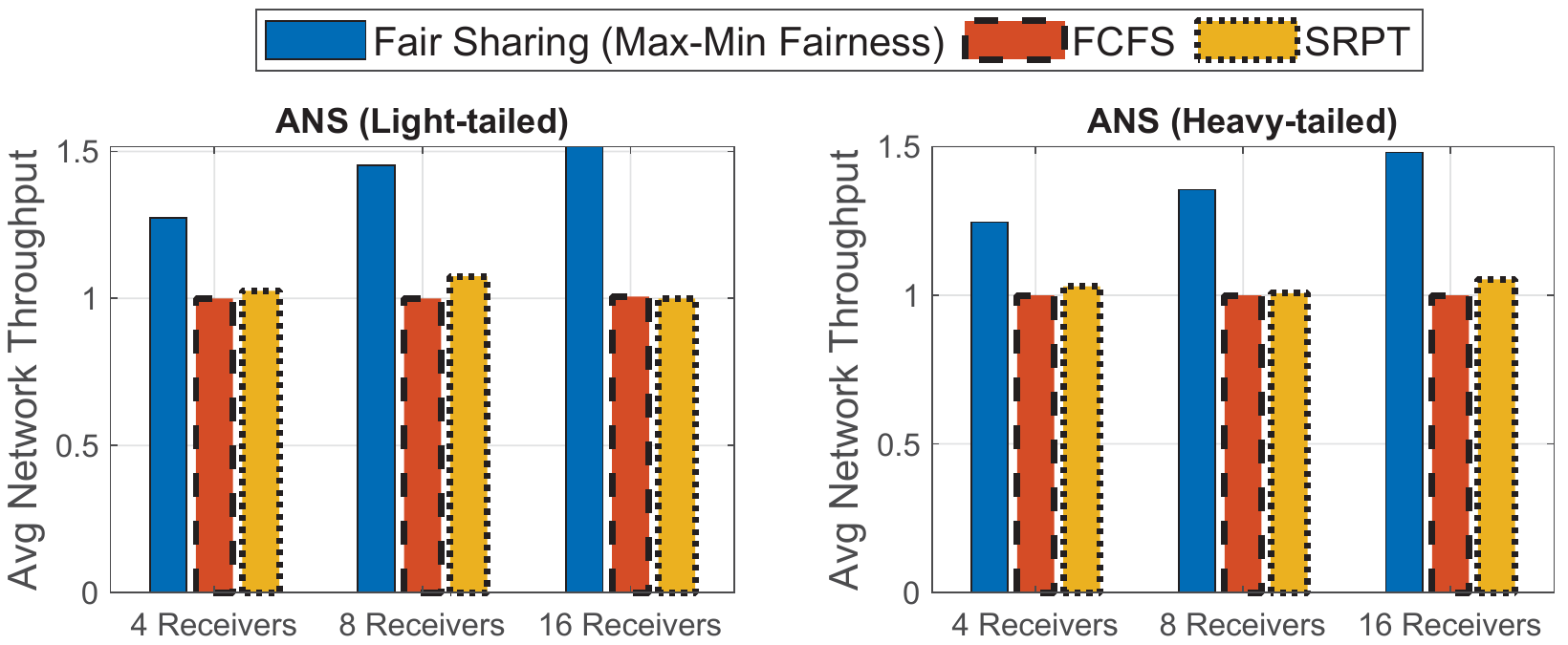}
    \caption{Average throughput of bulk multicast transfers obtained by running different scheduling policies. We started 100 transfers at the time zero, senders and receivers were selected according to the uniform distribution. Each group of bars is normalized by the minimum in that group.}
    \label{fig:thr}
\end{figure}

\subsection{Effect of Rate Allocation Policies} \label{eval-rate-alloc}
As explained earlier in \S \ref{rate-allocation}, when scheduling traffic over large forwarding trees, fair sharing can sometimes offer significantly higher throughput and hence better completion times. We performed an experiment over the ANS topology and with both light-tailed and heavy-tailed traffic distributions. ANS topology has uniform link capacity across all edges which helps us rule out the effect of capacity variations on throughput obtained via different scheduling policies. We also considered an increasing number of receivers from 4 to 8 and 16. Figure \ref{fig:thr} shows the results. We see that fair sharing offers a higher average throughput across all ongoing transfers compared to FCFS and SRPT and that with more receivers, the benefit of using fair sharing increases to up to $1.5\times$ with 16 receivers per transfer.

\subsection{Running Time}
To ensure scalability of proposed algorithms, we measured the running time of our algorithms over various topologies (with different sizes) and with varying rates of arrival. We assumed two arrival rates of $\lambda=0.001$ and $\lambda=1$ which account for light and heavy load regimes. We also considered eight receivers per transfer and all the three topologies of ANS, GEANT, and UNINETT. We saw that the running time of Algorithm \ref{algo_dccast}, and \ref{algo_quick} remained below one millisecond and 20 milliseconds, respectively, across all of these scenarios. These numbers are less than the propagation latency between the majority of senders and receivers over considered topologies (a simple TCP handshake would take at least twice the propagation latency). More efficient realization of these algorithms can further reduce their running time (e.g., implementation in C/C++ instead of Java).

\subsection{Forwarding Plane Resource Usage}
QuickCast can be realized using software-defined networking and OpenFlow compatible switches. To forward packets to multiple outgoing ports on switches where trees branch out to numerous edges, we can use group tables which have been supported by OpenFlow since early versions. Besides, an increasing number of physical switch makers have added support for group tables. To allow forwarding to multiple outgoing ports, the group table entries should be of type ``ALL", i.e., \texttt{OFPGT\_ALL} in the OpenFlow specifications. Group tables are highly scarce (compared to TCAM entries) and so should be used with care. Some new switches support 512 or 1024 entries per switch. Another critical parameter is the maximum number of action buckets per entry which primarily determines the maximum possible branching degree for trees. Across the switches we looked at, we found that the minimum supported value was 8 action buckets which should be enough for WAN topologies as most of such do not have any nodes with this connectivity degree.

In general, reasoning about the number of group table entries needed to realize different schemes is hard since it depends on how the trees are formed which is highly intertwined with edge weights that depend on the distribution of load. For example, consider a complete binary tree with 8 receivers as leaves and the sender at the root. This will require 6 group table entries to transmit to all receivers with two action buckets per each intermediate node on the tree (branching at the sender does not need a group table entry). If instead, we used an intermediate node to connect to all receivers with a branching degree of 8, we would only need one group table entry with eight action buckets.

We measured the number of group table entries needed to realize QuickCast. We computed the average of the maximum, and maximum of the maximum number of entries used per switch during the simulation for the topologies of ANS, GEANT, and UNINETT, with arrival rates of $\lambda=0.001$ and $\lambda=1$, considering both light-tailed and heavy-tailed traffic patterns and assuming that each bulk multicast transfer had eight receivers. The experiment was terminated when 200 transfers arrived. Looking at the maximum helps us see whether there are enough entries at all times to handle all concurrent transfers. Interestingly, we saw that by using multiple trees per transfer, both the average and maximum of the maximum number of group table entries used were less than when a single tree was used per transfer. One reason is that using a single tree slows down faster receivers which may lead to more concurrent receivers that increase the number of group entries. Also, by partitioning receivers, we make subsequent trees smaller and allow them to branch out closer to their receivers which balances the use of group table entries usage across the switches reducing the maximum. Finally, by using more partitions, the maximum number of times a tree needs to branch to reach all of its receivers decreases. Across all the scenarios considered above, the maximum of maximum group table entries at any timeslot was 123, and the average of the maximum was at most 68 for QuickCast. Furthermore, by setting $N_{max}=\lvert \pmb{\mathrm{D}}_{R_{new}} \rvert$ which allows for more partitions, the maximum of maximum group table entries decreased by up to 17\% across all scenarios.

\section{Conclusions}
Many P2MP transfers do not require that all receivers finish reception at the same time. Moreover, attaching all receivers of a P2MP transfer to the sender using a single forwarding trees limits the speed of all receivers to that of the slowest one. We introduced the bulk multicast routing and scheduling problem to minimize mean completion times of receivers and split it into three sub-problems of receiver set partitioning, tree selection, and rate allocation. We then presented QuickCast which applies three heuristic techniques to offer approximate solutions to these three hard sub-problems. We performed extensive evaluations to validate the effectiveness of QuickCast. In general, the gains are a function of network connectivity, link capacities, and transfer properties. Considering multiple network topologies and transfer size distributions, we found that QuickCast offers significant speedups for multiple receivers per P2MP transfer while negligibly increasing the total bandwidth consumption. Interestingly, we also found that the number of forwarding rules at network switches needed to realize QuickCast can be considerably less than when a single forwarding tree is used per P2MP transfer which makes it more practical.

\clearpage
\chapter{Mixed Completion Time Objectives for P2MP Transfers over Inter-DC Networks} \label{chapter_iris}
Bulk transfers from one to multiple datacenters can have many different completion time objectives ranging from quickly replicating some $k$ copies to minimizing the time by which the last destination receives a full replica. We design an SDN-style wide-area traffic scheduler that optimizes different completion time objectives for various requests. The scheduler builds, for each bulk transfer, one or more multicast forwarding trees which preferentially use lightly loaded network links. Multiple multicast trees are used per bulk transfer to insulate destinations that have higher available bandwidth and can hence finish quickly from congested destinations. 


When receivers of a bulk multicast transfer have very different network bandwidth available on paths from the sender, the slowest receiver dictates the completion time for all receivers. As discussed in Chapter \ref{chapter_p2mp_quickcast}, using multiple multicast trees to separate the faster receivers which will improve the average receiver's completion time. However, each additional tree consumes more network bandwidth and at the extremum, this idea devolves to one tree per receiver. We aim to answer the following questions:

\begin{enumerate}
    \item What is the right number of trees per transfer?
    \item Which receivers should be grouped in each tree?
\end{enumerate}

We analyze a relaxed version of this partitioning problem where each partition is a subset of receivers attached to the sender with a separate forwarding tree. We first propose a partitioning technique that reduces the average receiver completion times of receivers by isolating slow and fast receivers. We study this approach in the relaxed setting of having a congestion-free network core, i.e., links in/out of the datacenters are the capacity bottlenecks, and considering max-min fair rate allocation from the underlying network. We then develop a partitioning technique for real-world inter-datacenter networks, without relaxations, and inspired by the findings from studying the relaxed scenario. The partitioning technique operates by building a hierarchy of valid partitioning solutions and selecting the one that offers the best average receiver completion times. Our evaluation of this partitioning technique on real-world topologies, including ones with bottlenecks in the network core, show that the technique yields completion times that are close to a lower bound and hence nearly optimal.

Moreover, we incorporate binary objective vectors which allow applications to indicate transfer-specific objectives for receivers' completion times. Using the application-provided objective vectors, we can optimize for mixed completion time objectives based on the trade-off between total network capacity consumption and the receivers' average completion times.

We present the {\nameiris} heuristic, which computes a partitioning of receivers for every transfer given a binary objective vector. {\nameiris} aims to minimize the completion time of receivers whose rank is indicated by applications/users with a one in the objective vector while saving as much bandwidth as possible by grouping receivers whose ranks are indicated with consecutive zeros in the objective vector.

{\nameiris} operates in a logically centralized manner, receives bulk multicast transfer requests from end-points, and computes receiver partitions along with their multicast forwarding trees. We create forwarding trees using group tables~\cite{openflow-1.1.0}. {\nameiris} uses a RESTful API to communicate with the end-points allowing them to specify their transfer properties and requirements (i.e., objective vectors) using which it computes and installs the required rules in the forwarding plane. We believe our techniques are easily applicable in today's inter-DC networks~\cite{b4, swan-backbone, facebook-express-backbone}.

We perform extensive simulations and Mininet emulations with {\nameiris} using synthetic and real-world Facebook inter-DC traffic patterns over large WAN topologies. Simulation results show that {\nameiris} speeds up transfers to a small number of receivers~(e.g., $\geq 8$ receivers) by $\ge 2\times$ on the average completion time while the bandwidth used is $\leq 1.13\times$ compared to state-of-the-art. Transfers with more receivers receive larger benefits. For transfers to at least $16$ receivers, $75\%$ of the receivers complete at least $5\times$ faster and the fastest receiver completes $2.5\times$ faster compared to state-of-the-art. Compared to performing multicast as multiple unicast transfers with shortest path routing, {\nameiris} reduces mean completion times by about $2\times$ while using $0.66\times$ of the bandwidth. Finally, Mininet emulations show that {\nameiris} reduces the maximum group table entries needed by up to $3\times$.

\vspace{0.5em}
\noindent\textbf{Motivating Examples:}
Back-end geo-distributed applications running on datacenters can have different requirements on how their objects are replicated to other datacenters. Hence, inter-DC traffic is usually a mix of transfers with various completion time objectives. For example, while reproducing $n$ copies of an object to $n$ different datacenters/locations, one application may want to transfer $k$ copies quickly to any among $n$ given receivers, and another application may want to minimize the time when the last copy finishes. In the former case, grouping the slower $n-k$ receivers into one partition consumes less bandwidth and this spare bandwidth could be used to speed up the other transfers. In the latter case, by grouping all receivers except the slowest receiver together (i.e., into one tree), we can isolate them from the slowest receiver with minimal bandwidth consumption. Minimizing the completion times of all receivers is another possible objective. Our technique takes as input a binary objective vector whose $i$\textsuperscript{th} element expresses interest in the completion time of the $i$\textsuperscript{th} fastest receiver; it aims to minimize the completion times of receivers whose rank is set to one in this objective vector. It is easy to see that following values of the objective vector achieve the goals discussed so far; when $k=1$, $n=3$, $\{1, 0, 0\}$, $\{0, 0, 1\}$ and $\{1, 1, 1\}$ aim to minimize the completion time of the fastest $k$ out of $n$ receivers, the slowest receiver, and all receivers, respectively.

\section{System Model} \label{system_model_iris}
Similar to previous chapters, a TES runs our algorithms in a logically centralized manner to decide how traffic is forwarded in-network. P2MP transfers are processed in an online fashion as they arrive with the main objective of optimizing completion times. Also, forwarding entries, which are installed for every transfer upon arrival, are fixed until the transfers' completion may only be updated in case of failures.

We consider max-min fair \cite{max-min-fairness} rate allocation across multicast forwarding trees. Traffic is transmitted with the same rate from the source to all the receivers attached to a forwarding tree. To reach max-min fair rates, such rates can either be computed centrally over specific time periods, i.e., timeslots, and then be used for end-point traffic shaping or end-points can gradually converge to such rates in a distributed fashion in a way similar to TCP \cite{mctcp} (fairness is considered across trees). In our evaluations, we will consider the former approach for increased network utilization. Using a fair sharing policy addresses the starvation problem (such as in SRPT policy) and prevents larger transfers from blocking edges (such as in FCFS policy).

We use the notion of \textbf{objective vectors} to allow applications to define transfer-specific requirements which in general can improve overall system performance and reduce bandwidth consumption. An objective vector for a transfer is a vector of zeros and ones which is the same size as the number of receivers of that transfer. From left to right, the binary digit $i$ in this vector is associated with the $i\textsuperscript{th}$ fastest receiver. A one in the objective vector indicates that we are specifically interested in the completion time of the receiver associated with that rank in the vector. By assigning zeros and ones to different receiver ranks, it is possible to respect different applications' preferences or requirements while allowing the system to optimize bandwidth consumption further. The application/user, however, needs not be aware of the mapping between the downlink speeds (rank in the objective vector) and the receivers themselves.

Table \ref{table_omega} offers several examples. For instance, an objective vector of $\{0,0,0,1,0,0,0,0\}$ indicates the application's interest in the fourth fastest receiver. To respect the application's objective, we initially isolate the fourth receiver and do not group it with any other receiver. The first three fastest receivers can be grouped into a partition to save bandwidth. The same goes for the four slowest receivers. However, we do not group all receivers indicated with zeros into one partition initially (i.e., the top three receivers and the bottom four) to avoid slowing some of them down unnecessarily (in this case, the top three receivers). This forms the basis for the partitioning technique proposed in \S \ref{partitioning} that operates by building a hierarchy with multiple layers, where each layer is a valid partitioning solution, and selects the layer that gives the smallest average receiver completion times.

\begin{table}[t]
\centering
\caption{Behavior of several objective vectors.} \label{table_omega}
\vspace{0.5em}
\begin{tabular}{ |p{4.5cm}|p{9cm}| }
    \hline
    \textbf{Objective Vector ($\omega$)} & \textbf{Outcome (given $n$ receivers)} \\
    \hline
    \hline
    \raisebox{-0.6\totalheight}{\includegraphics[height=3em]{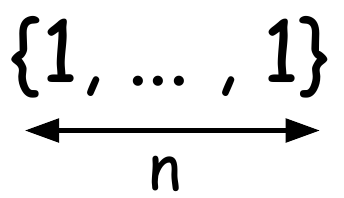}} & Interested in completion times of all individual receivers \\
    \hline
    \raisebox{-0.9\totalheight}{\includegraphics[height=3em]{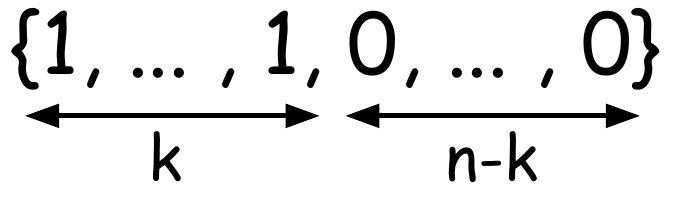}} & Interested in completion times of the top $k$ receivers (groups the bottom $n-k$ receivers to save bandwidth) \\
    \hline
    \raisebox{-0.9\totalheight}{\includegraphics[height=3em]{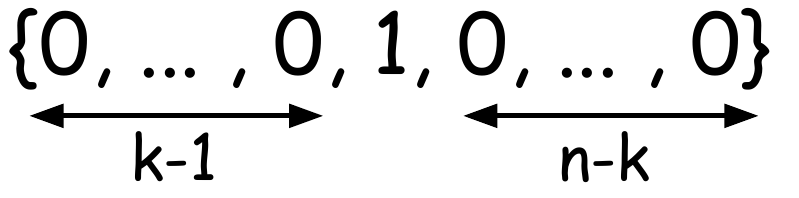}} & Interested in the completion time of the $k$\textsuperscript{th} receiver (groups the top $k-1$ receivers into a fast partition, and the bottom $n-k$ receivers into a slow one to save bandwidth) \\
    \hline
    \raisebox{-0.9\totalheight}{\includegraphics[height=3em]{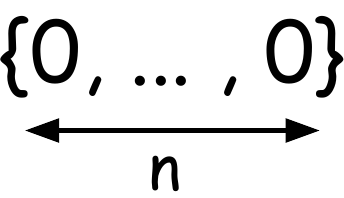}} & Not interested in the completion time of any specific receiver (all receivers form a single partition) \\
    \hline
\end{tabular}
\end{table}

\vspace{0.5em}
\noindent\textbf{Problem Statement:}
Given an inter-DC topology with known available bandwidth per link, the traffic engineering server is responsible for \textit{partitioning receivers} and \textit{selecting a forwarding tree per partition} for every incoming bulk multicast transfer. A bulk multicast transfer is specified by its source, set of receivers and volume of data to be delivered. The primary objective is minimizing average receiver completion times. In case an objective vector is specified, we want to minimize average completion times of receivers whose ranks are indicated with a $1$ in the vector as well as receivers indicated with consecutive $0$'s in the vector together as groups (receivers noted with consecutive $0$'s use the same forwarding tree and will have the same completion times). Minimizing bandwidth consumption, which is directly proportional to the size of selected forwarding trees, is considered a secondary objective.

\subsection{Online Greedy Optimization Model} \label{greedy_online}
The online bulk multicast partitioning and forwarding tree selection problem can be formulated using Eq. \ref{opt}-\ref{const_cap} added the constraint that our rate allocation is max-min fair across forwarding trees for any selection of the partitions and the trees. We will use the notation defined in Table \ref{table_var_0} as well as those in Table \ref{table_var_iris}.

\begin{table}[t]
\centering
\caption{Definition of variables used in this chapter besides those defined in Table \ref{table_var_0}.} \label{table_var_iris}
\vspace{0.5em}
\begin{tabular}{ |p{1.5cm}|p{12cm}| }
    \hline
    \textbf{Variable} & \textbf{Definition} \\
    \hline
    \hline
    $T$ & A directed Steiner tree \\
    \hline
    $r_T(t)$ & The transmission rate over tree $T$ at timeslot $t$ \\
    \hline
    $P$ & A receiver partition of some request \\
    \hline
    $\pmb{\mathrm{P}}_{R}$ & Set$\langle\rangle$ of partitions of some request $R$ \\
    \hline
    $T_{P}$ & The forwarding tree of partition $P$ \\
    \hline
    $\mathcal{V}^{r}_{P}$ & Current residual volume of partition $P$ of request $R$ \\
    \hline
    $\kappa_P$ & Estimated minimum completion time of partition $P$ \\
    \hline
    $L_e$ & Edge $e$'s total load (see \S \ref{forwardingtree}) \\
    \hline
    $\pi_R$ & Objective vector assigned to request $R$ \\
    \hline
    $\pi_R^{\star}$ & Weighted completion time vector computed from $\pi_R$ by replacing the last zero in a pack of consecutive zeros with the number of consecutive zeros in that pack (e.g., $\pi_R=\{0,0,0,1,0,0\} \rightarrow \pi_R^{\star}=\{0,0,3,1,0,2\}$) \\
    \hline
    $\tau_{R}$ & Vector of completion times of receivers of request $R$ sorted from fastest to slowest \\
    \hline
\end{tabular}
\end{table}

The set $\pmb{\mathrm{R}}$ includes both the new transfer $R_{new}$ and all the ones already in the system for which we already have the partitions and forwarding trees. The optimization objective of Eq. \ref{opt} is to minimize the weighted sum of completion times of receivers of all requests $R \in \pmb{\mathrm{R}}$ according to their objective vectors, and the total bandwidth consumption of $R_{new}$ by partitioning its receivers and selecting their forwarding trees (indicated by the term $\sum_{P \in \pmb{\mathrm{P}}_{R_{new}}} \mathcal{V}_{P} \lvert T_{P} \rvert$). Operators can choose the non-negative coefficient $\epsilon$ according to the overall system objective to give a higher weight to minimizing the weighted completion time of receivers than reducing bandwidth consumption. Eq. \ref{const_dem} shows the demand constraints which state that the total sum of transmission rates over every tree for future timeslots is equal to the remaining volume of data per partition (each partition uses one tree). Eq. \ref{const_cap} presents the capacity constraints which state that the total sum of transmission rates per timeslot for all trees that share a common edge has to not go beyond its available bandwidth.

\begin{align}
    \min~~~&\sum_{R \in \pmb{\mathrm{R}}} \Big( \tau_{R} \cdot \pi_R^{\star} \Big) + \epsilon \sum_{P \in \pmb{\mathrm{P}}_{R_{new}}} \mathcal{V}_{P} \lvert T_{P} \rvert \label{opt}\\
    \textrm{Subject to}~~~&\sum_{t} r_{T_P}(t) = \mathcal{V}^{r}_{P} & \forall P \in \pmb{\mathrm{P}}_{R}, R \in \pmb{\mathrm{R}} \label{const_dem} \\
    &\sum_{\{P \in \pmb{\mathrm{P}}_{R}, R \in \pmb{\mathrm{R}} ~\vert~ e \in T_P\}} r_{T_P}(t) \le B_e(t) \qquad & \forall t, e \label{const_cap}
\end{align}

This online discrete optimization problem is highly complex as it is unclear how receivers should be partitioned into multiple subsets to reduce completion times and there is an exponential number of possibilities. Selection of forwarding trees to minimize completion times is also a hard problem. In \S \ref{Iris}, we will present a heuristic that aims to approximate a solution to this optimization problem inspired by the findings in \S \ref{partitioning_model}.

\section{Partitioning of Receivers on a Relaxed Topology} \label{partitioning_model}
\begin{figure}[b]
    \centering
    \includegraphics[width=0.5\columnwidth]{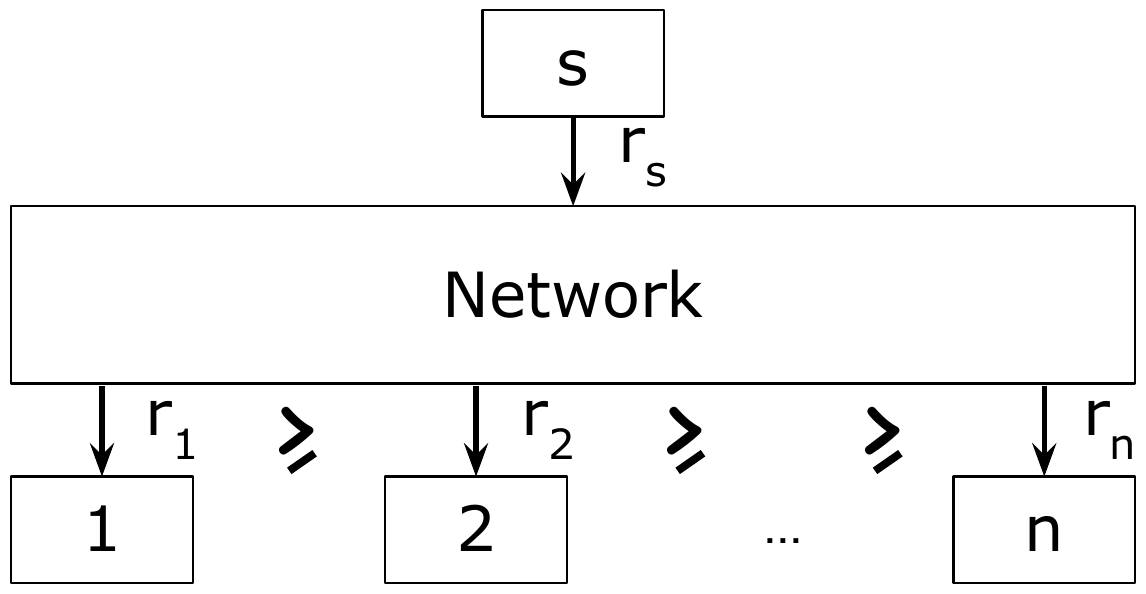}
    \caption{A relaxed topology with infinite core capacity, and uplink and downlink capacities of $r_s$ and $r_1 \ge \dots \ge r_n$.}
    \label{fig:problem_formulation}
\end{figure}

Due to the high complexity of the partitioning problem as a result of physical topology, we first study a relaxed topology where every datacenter is attached with a single uplink/downlink to a network with infinite core capacity (and so the network core cannot become a bottleneck). As shown in Figure \ref{fig:problem_formulation}, the sender has a maximum uplink rate of $r_s$ and transmits to a set of $n$ receivers with different maximum downlink rates of $r_i, \forall i \in \{1,\dots,n\}$. In \S \ref{forwardingtree}, we discuss a load-balancing forwarding tree selection approach that aims to distribute load across the network to minimize the effect of bottlenecks within the network core. Also, inspired by the findings in this section, we will develop an effective partitioning heuristic in \S \ref{partitioning}. Without loss of generality, let us also assume that the receivers in Figure \ref{fig:problem_formulation} are sorted by their downlink rates in descending order. The sender can initiate multicast flows to any partition (i.e., a subset of receivers) given that every receiver appears in exactly one partition. All receivers in a partition will have the same multicast rate that is the rate of the slowest receiver in the partition. To compute rates at the uplink, we consider the max-min fair rate allocation policy (see \S \ref{system_model_iris}). In this context, we would like to compute the number of partitions as well as the receivers that should be grouped per partition to minimize mean completion times.

\vspace{0.5em}
\textbf{Theorem 1.} Given receivers sorted by their downlink rates, partitioning that groups consecutive receivers is pareto-optimal with regards to minimizing completion times.

\vspace{0.5em}
\textbf{\textit{Proof.}} We use proof by contradiction. Let us assume a partitioning where non-consecutive receivers are grouped together, that is, there exist two partitions $P_1$ and $P_2$ where part of partition $P_1$ falls in between receivers of $P_2$ or the other way around. Let us call the slowest receivers of $P_1$ and $P_2$ as $j_1$ and $j_2$, respectively. Across $j_1$ and $j_2$, let us pick the fastest and call it $f(j_1,j_2)$. If $f(j_1,j_2) = j_1$ (i.e., in the non-decreasing order of downlink speed from left to right, $P_2$ appears before $P_1$ as in $P_2\{\dots\}~P_1\{\dots,j_1\}~P_2\{\dots,j_2\}~\dots$), then by swapping the fastest receiver in $P_2$ and $j_1$, we can improve the rate of $P_1$ while keeping the rate of $P_2$ the same. If $f(j_1,j_2) = j_2$, then by swapping the fastest receiver in $P_1$ and $j_2$, we can improve the rate of $P_2$ while keeping the rate of $P_1$ the same. This can be done in both cases without changing the number of partitions, or number of receivers per partition across all partitions. Since the new partitioning has a higher or equal achievable rate for one of the partitions, the total average completion times will be less than or equal to that of original partitioning, which means the original partitioning could not have been optimal.

\subsection{Our Partitioning Approach}
\begin{figure}
    \centering
    \includegraphics[width=\columnwidth]{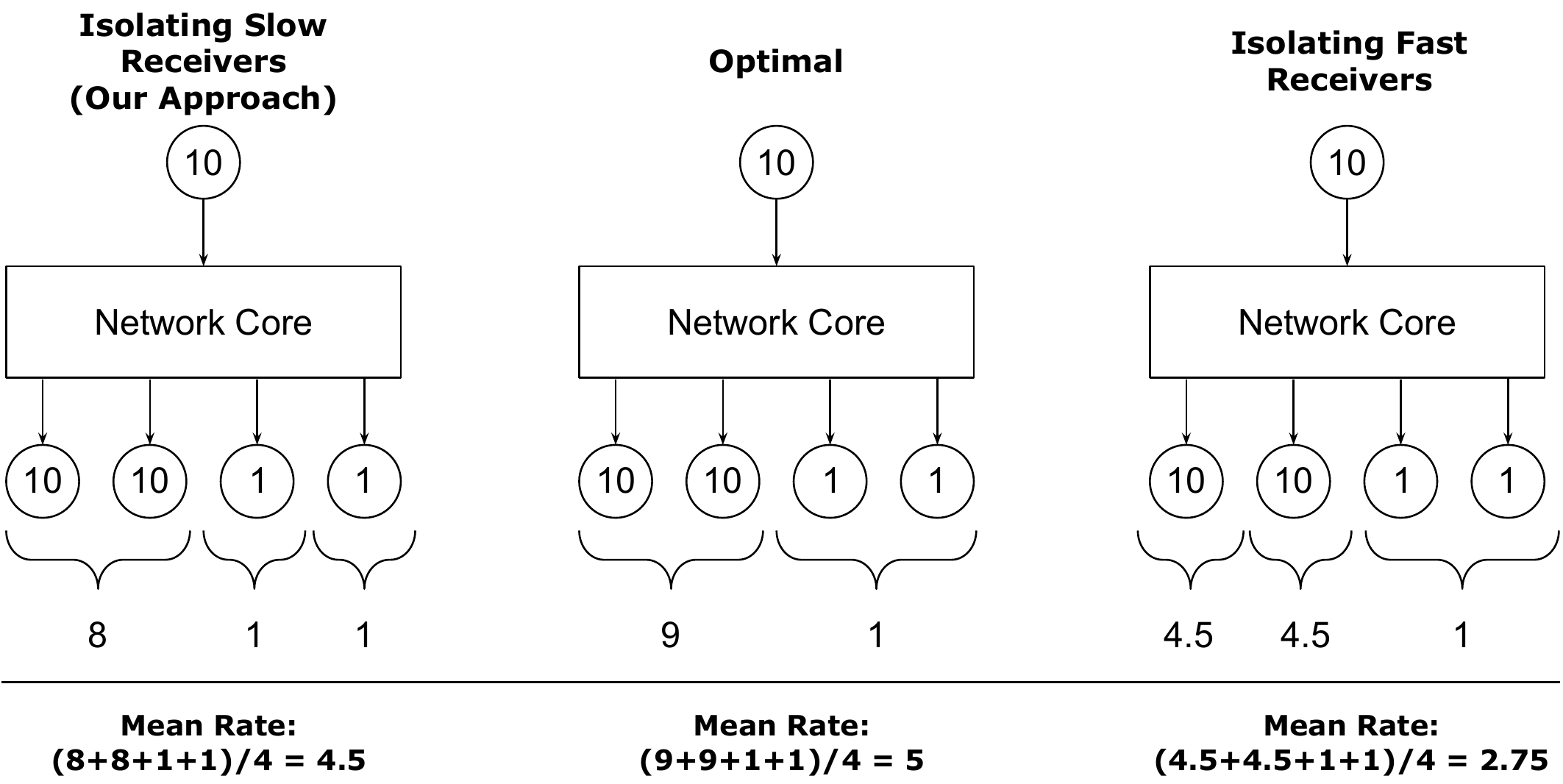}
    \caption{Various partitioning solutions for a scenario with four receivers. Numbers show the downlink and uplink speeds of nodes and curly brackets indicate the partitions where all nodes in a partition receive data at the same rate. The objective is to maximize the average rate of receivers given the max-min fairness policy.}
    \label{fig:partitioning_example_1}
\end{figure}

Based on Theorem 1, the number of possible partitioning scenarios that can be considered for minimum average completion times is the number of compositions of integer $n$, that is, $2^{n-1}$ ways which can be a large space to search. To reduce complexity, we isolate slow receivers from the rest of receivers to minimize their effect. In other words, given an integer $1 \le M \le n$, we group the first $n - M + 1$ fastest receivers into one partition and the rest of the receivers as separate $1$-receiver partitions ($M - 1$ in total). Since we do not know the value of integer $M$, we will try all possible values, that is, $n$ in total which will help us find the right threshold for the separation of fast and slow receivers. In particular, we compute the total average downlink rate of all receivers for the given transfer for every value of $M$ and select the $M$ that maximizes the average rate.\footnote{Or alternatively minimizes the average completion times of receivers.} As shown in Figure \ref{fig:problem_formulation}, the uplink at the sender has a rate of $r_s$ which will be divided across all the multicast flows that deliver data to the receivers. Isolating a slow receiver only takes a small fraction of the sender's uplink which is why this technique is effective as we will later see in evaluations. An example of this approach and how it compares with the optimal solution is shown in Figure \ref{fig:partitioning_example_1} where our solution selects $M=3$ partitions isolating the two slow receivers.

\begin{figure}
    \centering
    \includegraphics[width=0.8\columnwidth]{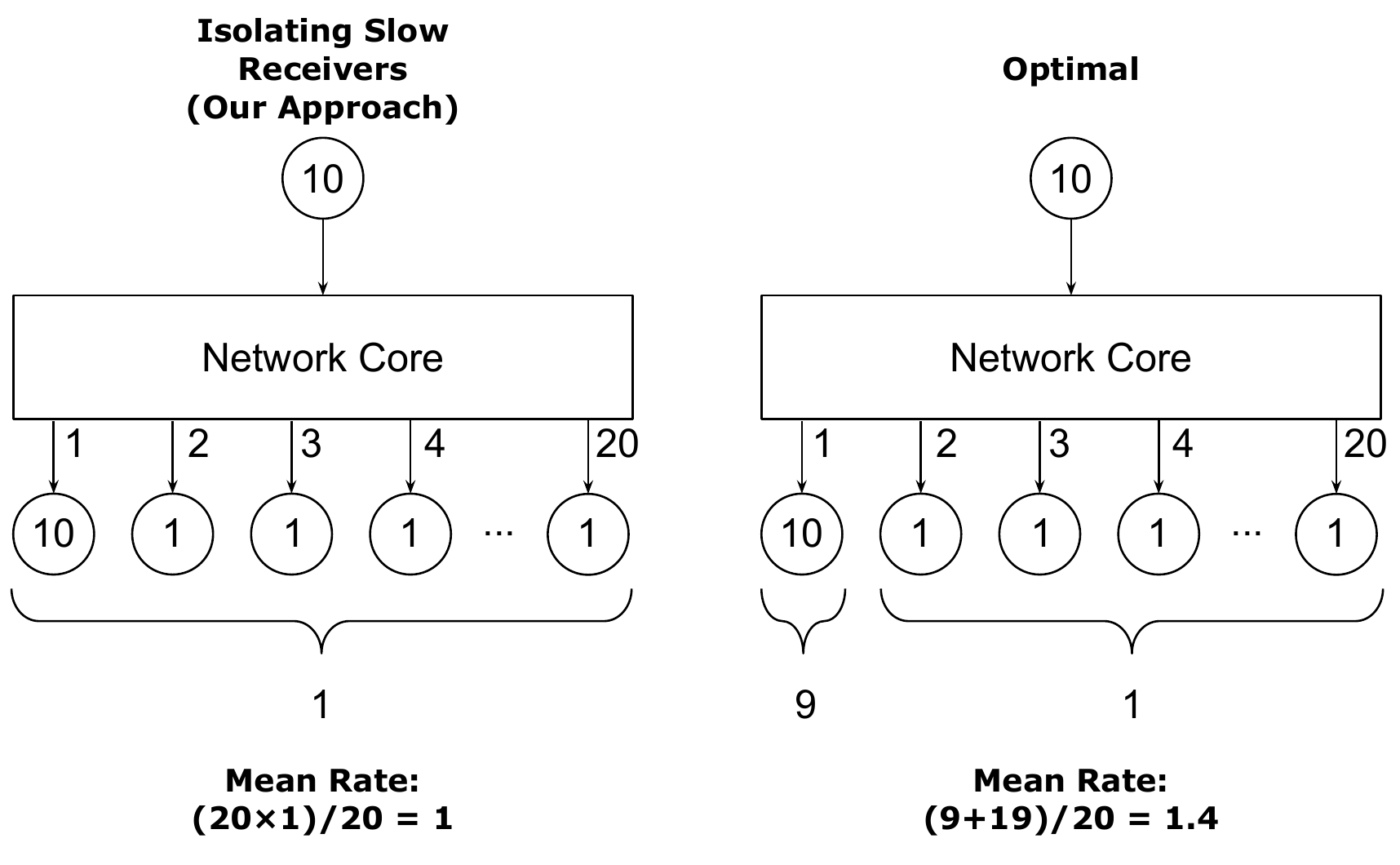}
    \caption{A worst-case scenario for the proposed partitioning scenario. Numbers within the nodes show the downlink and uplink speeds of nodes and curly brackets indicate the partitions where all nodes in a partition receive data at the same rate. The objective is to maximize the average rate of receivers given the max-min fairness policy.}
    \label{fig:partitioning_example_2}
\end{figure}

A main determining factor in the effectiveness of this approach is how $r_s$ compares with $\sum_{1 \le i \le n} r_i$. If $r_s$ is larger, then simply using $n$ partitions will offer the maximum total rate to the receivers. The opposite is when $r_s \ll \sum_{1 \le i \le n} r_i$ in which case using a single partition offers the highest total rate. In other cases, given the partitioning approach mentioned above, the worst-case scenario happens when there are many slow receivers and only a handful of fast receivers. An example has been shown in Figure \ref{fig:partitioning_example_2}. In the scenario on the left, our approach groups all the receivers into one partition where they all receive data at the rate of one. That is because by isolating slow receivers we can either get a rate of one or less than one if we isolate more than nine slow receivers, which means using one partition is enough. The optimal case, however, groups all the slow receivers into one partition. In general, scenarios like this rarely happen as the number of slow receivers over inter-datacenter networks is usually small, i.e., most datacenters are connected using high capacity links with large available bandwidth. In general, since we consider all values of $M$ from $1$ to $n$ partitions, the solution obtained from our partitioning approach cannot be worse than the two baseline approaches of using a single multicast tree for all receivers and unicasting to all receivers using separate paths.

\subsection{Incorporating Objective Vectors}
\begin{figure}
    \centering
    \includegraphics[width=0.8\columnwidth]{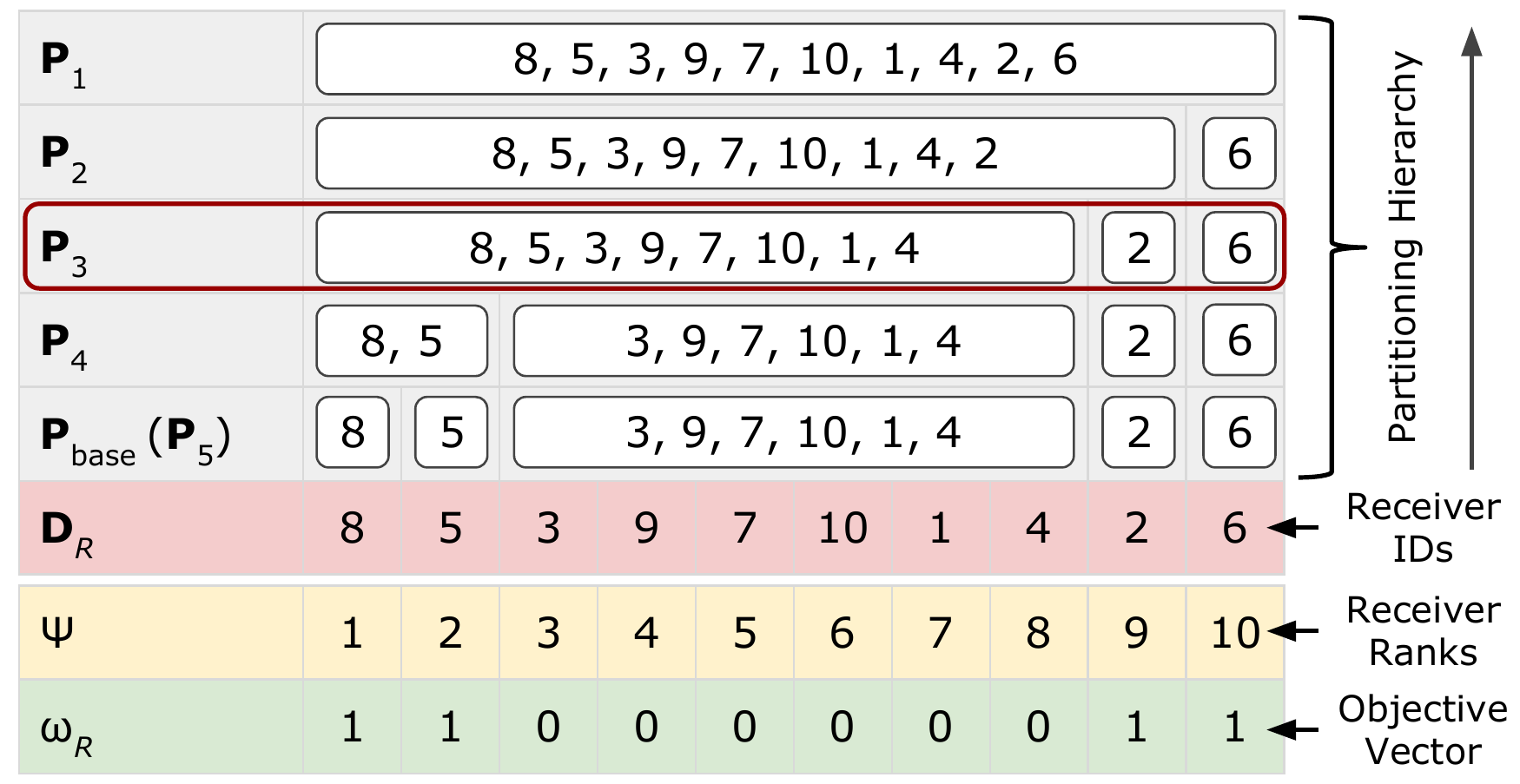}
    \caption{Example of a partitioning hierarchy for a transfer with 10 receivers (the topology not shown).}
    \label{fig:wancast_clustering_example}
\end{figure}

We allow users to supply an objective vector along with their multicast transfers to better optimize the network performance, that is, total network capacity consumption and receiver completion times. We incorporate the objective vectors by grouping receivers with consecutive ranks that are indicated with zeros in the objective vector and treating them as one partition in the whole process. That is because the users have indicated no interest in the completion times of those receivers, so we might as well reduce the network capacity usage by grouping them from the beginning. Figure \ref{fig:wancast_clustering_example} shows an example of building possible solutions by isolating slow receivers and incorporating the user-supplied objective vector, which we refer to as the partitioning hierarchy. Please note that this hierarchy moves in the reverse direction, that is, instead of isolating slow receivers, it merges fast receivers from bottom to the top.

Each layer in this hierarchy, labeled as $\bm{\mathrm{P}}_{i}, 1 \le i \le 5$, represents a valid partitioning solution.\footnote{The associated network topology is not shown.} We see that receivers indicated with consecutive zeros in $\omega_{R}$ are merged into one big partition at the base layer or $\bm{\mathrm{P}}_{5}$. Also, we see that as we move up, the two fastest partitions at each layer are merged, which reduces total bandwidth consumption. For each layer, we compute the average completion time of receivers and then select the layer that offers the least value, in this case, $\bm{\mathrm{P}}_{3}$ was chosen.

\section{Iris} \label{Iris}
We apply the partitioning technique discussed in the previous chapter to real-world inter-datacenter networks. We develop a heuristic for partitioning receivers on real-world topologies without relaxations of \S \ref{partitioning_model}. We will generate multiple valid partitioning solutions in the form of a hierarchy where layers of the hierarchy present feasible partitioning solutions and each layer is formed by merging the two fastest partitions of the layer below.\footnote{In general, it is not possible to offer optimality guarantees due to the highly varying factors of network topology, transfer arrivals, and the distribution of transfer volumes. However, our extensive simulations in \S \ref{evaluations} show that our approach can offer significant improvement on other approaches over various topologies and traffic patterns. Also, as a result of building a hierarchy of partitioning options and selecting the best one, our solution will be at least as good as either using a single multicast tree or using unicasting to all receivers.}

\begin{figure}
    \centering
    \includegraphics[width=\columnwidth]{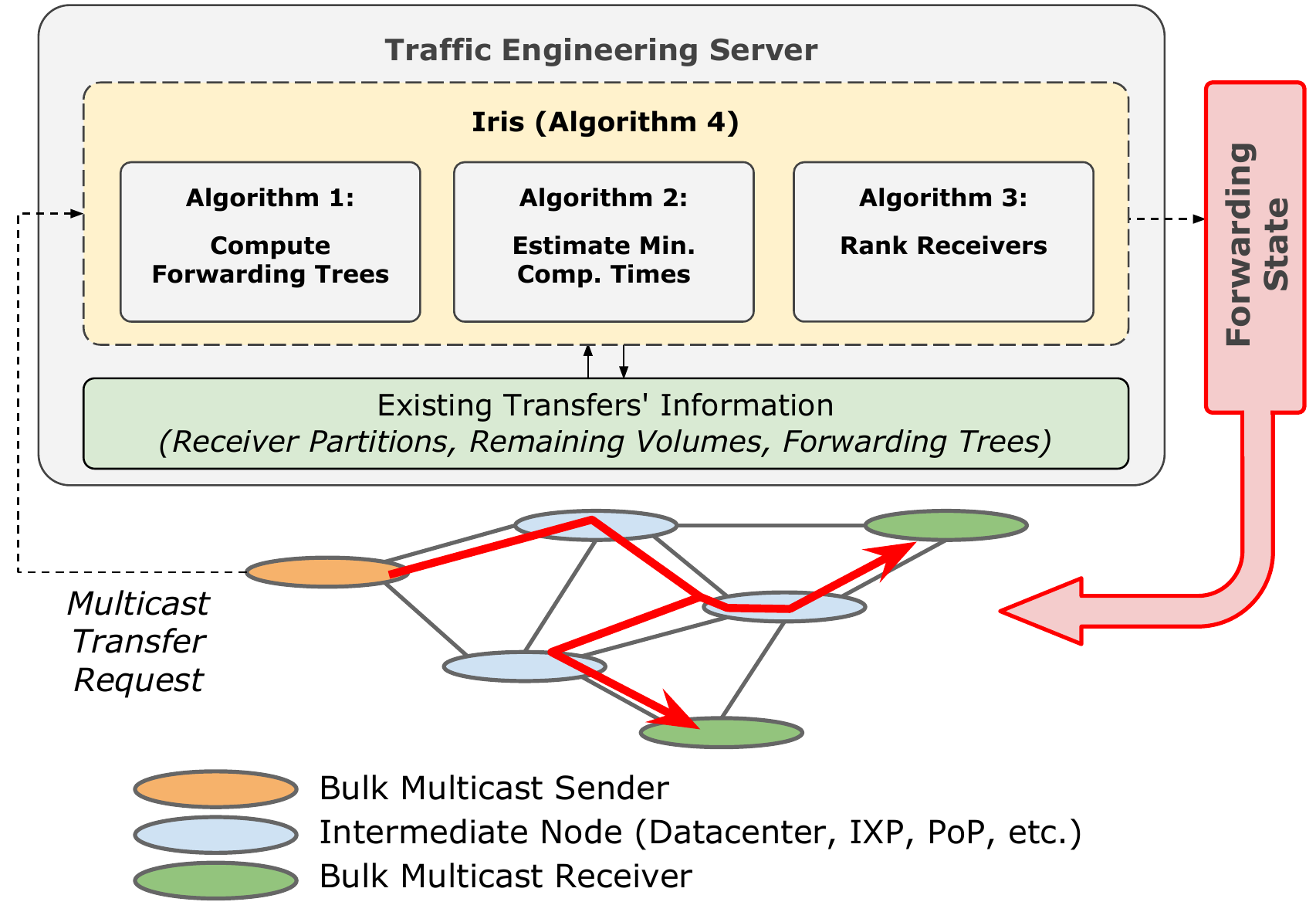}
    \caption{Pipeline of {\nameiris}.}
    \label{fig:iris_pipeline}
\end{figure}

We present {\nameiris}, a heuristic that runs on the traffic engineering server to manage bulk multicast transfers.\footnote{Unicast transfers are a special case with a single receiver.} When a bulk multicast transfer arrives at an end-point, it will communicate the request to the traffic engineering server which will then invoke {\nameiris}. It uses the knowledge of physical layer topology, available bandwidth on edges after deducting the share of high priority user traffic and other running transfers to compute partitions and forwarding trees. The traffic engineering server pulls end-points' actual progress periodically to determine their exact remaining volume across transfers to compute the total outstanding load per edge for all edges. {\nameiris} consists of four modules as shown in Figure \ref{fig:iris_pipeline} which we discuss in the following subsections. {\nameiris} aims to find an approximate solution to the optimization problem of Eq. \ref{opt} assuming $\epsilon \ll 1$ to prioritize minimizing completion times over minimizing bandwidth consumption. We will empirically evaluate {\nameiris} by comparing it to recent work and a lower bound in \S \ref{evaluations}.

\subsection{Choosing Forwarding Trees} \label{forwardingtree}
Load aware forwarding trees are selected given the link capacity information on the topology and according to other ongoing bulk multicast transfers across the network to reduce the completion times by mitigating the effect of bottlenecks. Tree selection should also aim to keep bandwidth consumption low by minimizing the number of edges per tree where an edge could refer to any of the links on the physical topology. To select a forwarding tree, a general approach that can capture a wide range of selection policies is to assign weights to edges of the inter-DC graph $G$ and select a minimum weight Steiner tree \cite{steiner_tree_problem}. Per edge $e \in \pmb{\mathrm{E}}_G$, we assume a virtual queue that increases by volume of every transfer scheduled on that edge and decreases as traffic flows through it. Since edges differ in capacity, completing the same virtual queue size may need significantly different times for different links. We define a metric called load as $L_e = \frac{1}{B_e} \sum_{\{P \in \pmb{\mathrm{P}}_{R}, \forall R \in \pmb{\mathrm{R}} ~\vert~ e \in T_{P}\}} \mathcal{V}^{r}_{P}$. This equation sums up the remaining volumes of all trees that use a specific edge (total virtual queue size) and divides that by the average available bandwidth on that edge to compute the minimum possible time it takes for all ongoing transfers on that edge to complete. To keep completion times low, we need to avoid edges for which this value is large.

With this metric available, to select a forwarding tree given a sender and several receivers, we will first assign an edge weight of $L_e + \frac{\mathcal{V}_{R_{new}}}{B_e}$ to all edges $e \in \pmb{\mathrm{E}}_G$ and then select a minimum weight Steiner tree as shown in Algorithm \ref{tree_algorithm}. With this edge weight, compared to edge utilization which has been extensively used in literature for traffic engineering, we achieve a more stable measure of how busy a link is expected to be in the near future on average. We considered the second term in edge weight to reduce total bandwidth use when there are multiple trees with the same weight. It also leads to the selection of smaller trees for larger transfers which decreases the total bandwidth consumption of {\nameiris} further in the long run.

\SetAlgoVlined
\begin{algorithm}[t]
\caption{Compute A Forwarding Tree} \label{tree_algorithm}
\SetKw{KwBy}{by}

\small
\vspace{0.4em}
\KwIn{Steiner tree terminal nodes $\pmb{\mathrm{\Gamma}} \subset \{S_{R_{new}} \cup \pmb{\mathrm{D}}_{R_{new}}\}$, request $R_{new}$}

\vspace{0.4em}
\KwOut{A Steiner tree}

\nonl\hrulefill

\SetKwProg{CompForwardingTree}{CompForwardingTree}{}{}

\vspace{0.4em}
\CompForwardingTree{$\mathrm{(}\pmb{\mathrm{\Gamma}},R_{new}\mathrm{)}$}{
    
    \vspace{0.4em}
    To every edge $e \in \pmb{\mathrm{E}}_G$, assign a weight of $(L_{e} + \frac{\mathcal{V}_{R_{new}}}{B_e})$\;
    
    \vspace{0.4em}
    \Return{A minimum weight Steiner tree that connects the nodes in set $\pmb{\mathrm{\Gamma}}$ (we used a hueristic \cite{DSTAlgoEvaluation})}\;
}
\end{algorithm}

\subsection{Estimating Minimum Completion Times} \label{ct_estimation}
The purpose of this procedure is to estimate the minimum completion time of different partitions of a given transfer considering available bandwidth over the edges and applying max-min fair rate allocation when there are shared links across forwarding trees.  Algorithms \ref{rank_algorithm} and \ref{wancast_algorithm} then use the minimum completion time per partition to rank the receivers (i.e., faster receivers have an earlier completion time) and then decide which partitions to merge. Computing the minimum completion times is done by assuming that the new transfer request has access to all the available bandwidth and compared to computing the exact completion times is much faster. Besides, calculating the exact completion times is not particularly more effective due to the continuously changing state of the system as new transfer requests arrive. Since available bandwidth over future timeslots is not precisely known, we can use estimate values similar to other work \cite{netstitcher, tempus, amoeba}. Algorithm \ref{ct_calc_algorithm} shows how the minimum completion times are computed.

\SetAlgoVlined
\begin{algorithm}[t]
\caption{Computing Minimum Completion Times} \label{ct_calc_algorithm}
\SetKw{KwBy}{by}

\small
\vspace{0.4em}
\KwIn{Request $R_{new}$, a set of partitions $\pmb{\mathrm{P}}$ where $P \subset \pmb{\mathrm{D}}_{R_{new}}, \forall P \in \pmb{\mathrm{P}}$}

\vspace{0.4em}
\KwOut{The minimum completion time of every partition in $\pmb{\mathrm{P}}$}

\nonl\hrulefill

\SetKwProg{MinimumCompletionTimes}{MinimumCompletionTimes}{}{}

\vspace{0.4em}
\MinimumCompletionTimes{$\mathrm{(}\pmb{\mathrm{P}},R_{new}\mathrm{)}$}{

    \vspace{0.4em}
    $\pmb{\mathrm{f}} \gets \emptyset$, $t \gets t_{now}+1$\;

    \vspace{0.4em}
    $\gamma_P \gets \mathcal{V}_R,~\forall P \in \pmb{\mathrm{P}}$\;

    \vspace{0.4em}
    $T_P \gets$ \texttt{CompForwardingTree(}$P,R_{new}$\texttt{)}, $\forall P \in \pmb{\mathrm{P}}$\;

    \vspace{0.4em}
    \While{$\lvert \pmb{\mathrm{f}} \rvert < \lvert \mathrm{\pmb{\mathrm{P}}} \rvert$}{

        \vspace{0.4em}
        Compute $r_{T_P}(t),\forall P \in \{\pmb{\mathrm{P}} - \pmb{\mathrm{f}}\}$, max-min fair rate \cite{max-min-fairness} allocated to tree $T_P$ at timeslot $t$ given available bandwidth of $B_e(t)$ on every edge $e \in \pmb{\mathrm{E}}_G$\;

        \vspace{0.4em}
        $\gamma_P \gets \gamma_P - \omega ~r_{T_P}(t),~\forall P \in \pmb{\mathrm{P}}$\;

        \vspace{0.4em}
        \ForEach{$P \in \{\pmb{\mathrm{P}} - \pmb{\mathrm{f}}\}$}{
            \vspace{0.4em}
            \If{$\gamma_P = 0$}{
                \vspace{0.4em}
                $\kappa_P \gets t$, $\pmb{\mathrm{f}} \gets \{\pmb{\mathrm{f}} \cup P\}$\;
            }
        }
        
        \vspace{0.4em}
        $t \gets t+1$\;
    
    }
    
    \vspace{0.4em}
    \Return{$\kappa_P,\forall P \in \pmb{\mathrm{P}}$}
}
\end{algorithm}

\subsection{Assigning Ranks to Receivers} \label{receiver_rank_assignment}
Algorithm \ref{rank_algorithm} assigns ranks to individual receivers according to their minimum completion times taking into account available bandwidth over edges as well as edges' load in the path selection process. This ranking is used along with the provided objective vector later to partition receivers.

\SetAlgoVlined
\begin{algorithm}[t]
\caption{Assign Receiver Ranks} \label{rank_algorithm}
\SetKw{KwBy}{by}

\small
\vspace{0.4em}
\KwIn{Request $R_{new}$}

\vspace{0.4em}
\KwOut{$\psi_r$, i.e., the rank of receiver $r \in \pmb{\mathrm{D}}_{R_{new}}$}

\nonl\hrulefill

\SetKwProg{AssignReceiverRanks}{AssignReceiverRanks}{}{}

\vspace{0.4em}
\AssignReceiverRanks{$\mathrm{(}R_{new}\mathrm{)}$}{
    
    \vspace{0.4em}
    {\color{gray}/* Every receiver is treated as a separate partition */}
    
    \vspace{0.4em}
    $\{\kappa_r,~\forall r \in \pmb{\mathrm{D}}_{R_{new}}\} \gets $ \texttt{MinimumCompletionTimes(}$\pmb{\mathrm{D}}_{R_{new}}, R_{new}$\texttt{)}\;
    
    \vspace{0.4em}
    $\psi_r \gets$ Position of receiver $r$ in the list of all receivers sorted by their estimated minimum completion times (fastest receiver is assigned a rank of $1$), $\forall r \in \pmb{\mathrm{D}}_{R_{new}}$\;
    
    \vspace{0.4em}
    \Return{$\psi_r,\forall r \in \pmb{\mathrm{D}}_{R_{new}}$}\;
}
\end{algorithm}

\SetAlgoVlined
\begin{algorithm}[p]
\caption{Compute Receiver Partitions and Trees ({\nameiris})} \label{wancast_algorithm}
\SetKw{KwBy}{by}

\small
\vspace{0.4em}
\KwIn{Request $R_{new}$, binary objective vector $\pi_{R_{new}}$}

\vspace{0.4em}
\KwOut{Partitions of request $R_{new}$ and their forwarding trees}

\nonl\hrulefill

\SetKwProg{CompPartitionsAndTrees}{CompPartitionsAndTrees}{}{}

\vspace{0.4em}
\CompPartitionsAndTrees{$\mathrm{(}R_{new},\pi_{R_{new}}\mathrm{)}$}{

    \vspace{0.4em}
    {\color{gray}/* Initial partitioning using the objective vector $\pi_{R_{new}}$ */}

    \vspace{0.4em}
    $\{\psi_r, \forall r \in \pmb{\mathrm{D}}_{R_{new}}\} \gets$ \texttt{AssignReceiverRanks(}$R_{new}$\texttt{)}\;

    \vspace{0.4em}
    $\pmb{\mathrm{D}}_{R_{new}}^{s} \gets$ Receivers $r$ sorted by $\psi_r, \forall r \in \pmb{\mathrm{D}}_{R_{new}}$ ascending\;

    \vspace{0.4em}
    $\pmb{\mathrm{P}}_{base} \gets \{$Any receiver $r \in \pmb{\mathrm{D}}_{R_{new}}$ for which $\pi_{R_{new}}<\psi_r>$ is $1$ as a separate partition$\} \cup \{$Group receivers that appear consecutively on $\pmb{\mathrm{D}}_{R_{new}}^{s}$ for which $\pi_{R_{new}}<\psi_r>$ is $0$, each group forms a separate partition$\}$\;
    
    \vspace{0.4em}
    {\color{gray}/* Building the partitioning hierarchy for $\pmb{\mathrm{P}}_{base}$ */}
    
    \vspace{0.4em}
    $\pmb{\mathrm{P}}_{\lvert \pmb{\mathrm{P}}_{base} \rvert} \gets$ $\pmb{\mathrm{P}}_{base}$\;
    
    \vspace{0.4em}
    \For{$l = \lvert \pmb{\mathrm{P}}_{base} \rvert$ \KwTo $l = 1$ \KwBy $-1$}{
        \vspace{0.4em}
        $\{\kappa_P,~\forall P \in \pmb{\mathrm{P}}_l\} \gets $ \texttt{MinimumCompletionTimes(}$\mathrm{\pmb{\mathrm{P}}_l,R_{new}}$\texttt{)}\;
        
        \vspace{0.4em}
        $\kappa_l \gets \sum_{P \in \pmb{\mathrm{P}}_l} (\lvert P \rvert~\kappa_P)$\;
        
        \vspace{0.4em}
        Assuming receivers are sorted from left to right by increasing order of rank, merge the two partitions on the left, $P$ and $Q$, to form $PQ$\;
        
        \vspace{0.4em}
        $\pmb{\mathrm{P}}_{l-1} \gets \{PQ\} \cup \{\pmb{\mathrm{P}}_{l}-\{P,Q\}\}$\;
    }

    \vspace{0.4em}
    Find $l_{min}$ for which $\kappa_{l_{min}} \le \min_{1 \le l \le \lvert \pmb{\mathrm{P}}_{base} \rvert} \kappa_l$, if multiple layers have the same $\kappa_l$, choose the layer with minimum total weight over all of its forwarding trees, i.e., select $l_{min}$ to optimize $\min(\sum_{P \in \pmb{\mathrm{P}}_{l_{min}}} (\sum_{e \in T_P} (L_e + \frac{\mathcal{V}_{R_{new}}}{B_e})))$\;

    \vspace{0.4em}
    \ForEach{$P \in \pmb{\mathrm{P}}_{l_{min}}$}{
        \vspace{0.4em}
        $T_{P} \gets$ \texttt{CompForwardingTree(}$P,R_{new}$\texttt{)}\;
        
        \vspace{0.4em}
        \ForEach{$e \in T_{P}$}{
            \vspace{0.4em}
            $L_e \gets L_e + \frac{\mathcal{V}_{R_{new}}}{B_e}$, $W_e \gets W_e + \frac{\mathcal{V}_{R_{new}}}{B_e}$\;
        }
    }
    
    \vspace{0.4em}
    \Return{$(P,~T_P),~\forall P \in \pmb{\mathrm{P}}_{l_{min}}$}\;
}
\end{algorithm}

\subsection{The {\nameiris} Algorithm} \label{partitioning_iris}
The {\nameiris} algorithm computes receiver partitions using hierarchical partitioning and assigns each partition a multicast forwarding tree. The partitioning problem is solved per transfer and determines the number of partitions and the receivers that are grouped per partition. {\nameiris} uses a partitioning technique inspired by the findings of \S \ref{partitioning_model} that is computationally fast, significantly improves receiver completion times, and operates only relying on network topology and available bandwidth per edge (i.e., after deducting the quota for higher priority user traffic). Algorithm \ref{wancast_algorithm} illustrates how {\nameiris} partitions receivers with an objective vector. Given that each node in a real-world topology may have multiple interfaces, we cannot directly compute the right number of partitions using Theorem 2. As a result, we build a partitioning hierarchy with numerous layers and examine the various number of partitions from bottom to the top of the hierarchy while looking at the average of minimum completion times. By building a hierarchy, we consider the discrete nature of forwarding tree selection on the physical network topology. The process consists of two steps as follows.

We first use the receiver ranks from Algorithm \ref{rank_algorithm} and the objective vector to create the base of partitioning hierarchy, $\pmb{\mathrm{P}}_{base}$. We first sort the receivers by their ranks from fastest to slowest and then group them according to the weights in the objective vector. For any receiver whose rank in the objective vector has a value of $1$, we consider a separate partition (single node partition) which allows the receiver to complete as fast as possible by not attaching it to any other receiver. Next, we group receivers with consecutive ranks that are assigned a value of $0$ in the objective vector into partitions with potentially more than one receiver, which allows us to save as much bandwidth as possible since the user has not indicated interest in their completion times.

Now that we have a set of base partitions $\pmb{\mathrm{P}}_{base}$, a heuristic creates a hierarchy of partitioning solutions with $\lvert \pmb{\mathrm{P}}_{base} \rvert$ layers where every layer $1 \le l \le \lvert \pmb{\mathrm{P}}_{base} \rvert$ is made up of a set of partitions $\pmb{\mathrm{P}}_{l}$. Each layer is created by merging two partitions from the layer below going from the bottom to the top of hierarchy. At the bottom of the hierarchy, we have the base partitions. Also, at any layer, any partition $P$ is attached to the sender using a separate forwarding tree $T_{P}$. We first compute the average of minimum completion times of all receivers at the bottom of the hierarchy. We continue by merging the two partitions that hold receivers with highest ranks. When merging two partitions, the faster partition is slowed down to the speed of slower partition. A new forwarding tree is computed for the resulting partition using the forwarding tree selection heuristic of Algorithm \ref{tree_algorithm} to all receivers in that partition, and the average of minimum completion times for all receivers are recomputed. This process continues until we reach a single partition that holds all receivers. In the end, we select the layer at which the average of minimum completion times across all receivers is minimum, which gives us the number of partitions, the receivers that are grouped per partition, and their associated forwarding trees. If there are multiple layers with the minimum average completion times, the one with minimum total forwarding tree weight across its forwarding trees is chosen which on average leads to better load distribution.

\begin{table}
\centering
\caption{Various topologies and traffic patterns used in evaluation. One unit of traffic is equal to what can be transmitted at the rate of the fastest link over a given topology per timeslot.} \label{table_evaluations}
\vspace{0.5em}
\begin{tabular}{p{2cm}|p{2.5cm}|p{8.5cm}|}
\cline{2-3}
 & \textbf{Name} & \textbf{Description} \\ \hline
\multicolumn{1}{|l|}{\multirow{2}{*}{Topology}} & GEANT & Backbone and transit network across Europe with 34 nodes and 52 links. Link capacity from 45 Mbps to 10 Gbps. \\ \cline{2-3} 
\multicolumn{1}{|l|}{} & UNINETT & Backbone network across Norway with 69 nodes and 98 links. Most links have a capacity of 1, 2.5 or 10 Gbps. \\ \hline \hline
\multicolumn{1}{|l|}{\multirow{4}{*}{Traffic Pattern}} & Light-tailed & Based on Exponential distribution with a mean of $20$ units per transfer. \\ \cline{2-3} 
\multicolumn{1}{|l|}{} & Heavy-tailed & Based on Pareto distribution with the minimum of $2$ units, the mean of $20$ units, and the maximum capped at $2000$ units per transfer. \\ \cline{2-3} 
\multicolumn{1}{|l|}{} & Hadoop & Generated by geo-distributed data analytics over Facebook's inter-DC WAN (distribution mean of $20$ units per transfer). \\ \cline{2-3} 
\multicolumn{1}{|l|}{} & Cache-follower & Generated by geo-distributed cache applications over Facebook's inter-DC WAN (distribution mean of $20$ units per transfer). \\ \hline
\end{tabular}
\end{table}

\section{Evaluation} \label{evaluations_iris}
We considered various topologies and transfer size distributions as shown in Table \ref{table_evaluations}. We selected two research topologies with given capacity information on edges from the Internet Topology Zoo \cite{zoo}. We could not use other commercial topologies as the exact connectivity and link capacity information were not publicly disclosed. We also considered multiple transfer volume distributions including synthetic (light-tailed and heavy-tailed) and real-world Facebook inter-DC traffic patterns (Hadoop and Cache-follower) \cite{social_inside}. Transfer arrival pattern was according to Poisson distribution with a rate of $\lambda$ per timeslot. For simplicity, we assumed an equal number of receivers for all bulk multicast transfers per experiment. We performed simulations and Mininet emulations to evaluate {\nameiris}. We compare {\nameiris} with multiple baseline techniques and QuickCast, presented in Chapter \ref{chapter_p2mp_quickcast}, which also focuses on partitioning receivers into groups for improved completion times.

\subsection{Computing a Lower Bound} \label{aggregate_topo}
\begin{figure}[t]
    \centering
    \includegraphics[width=0.7\textwidth]{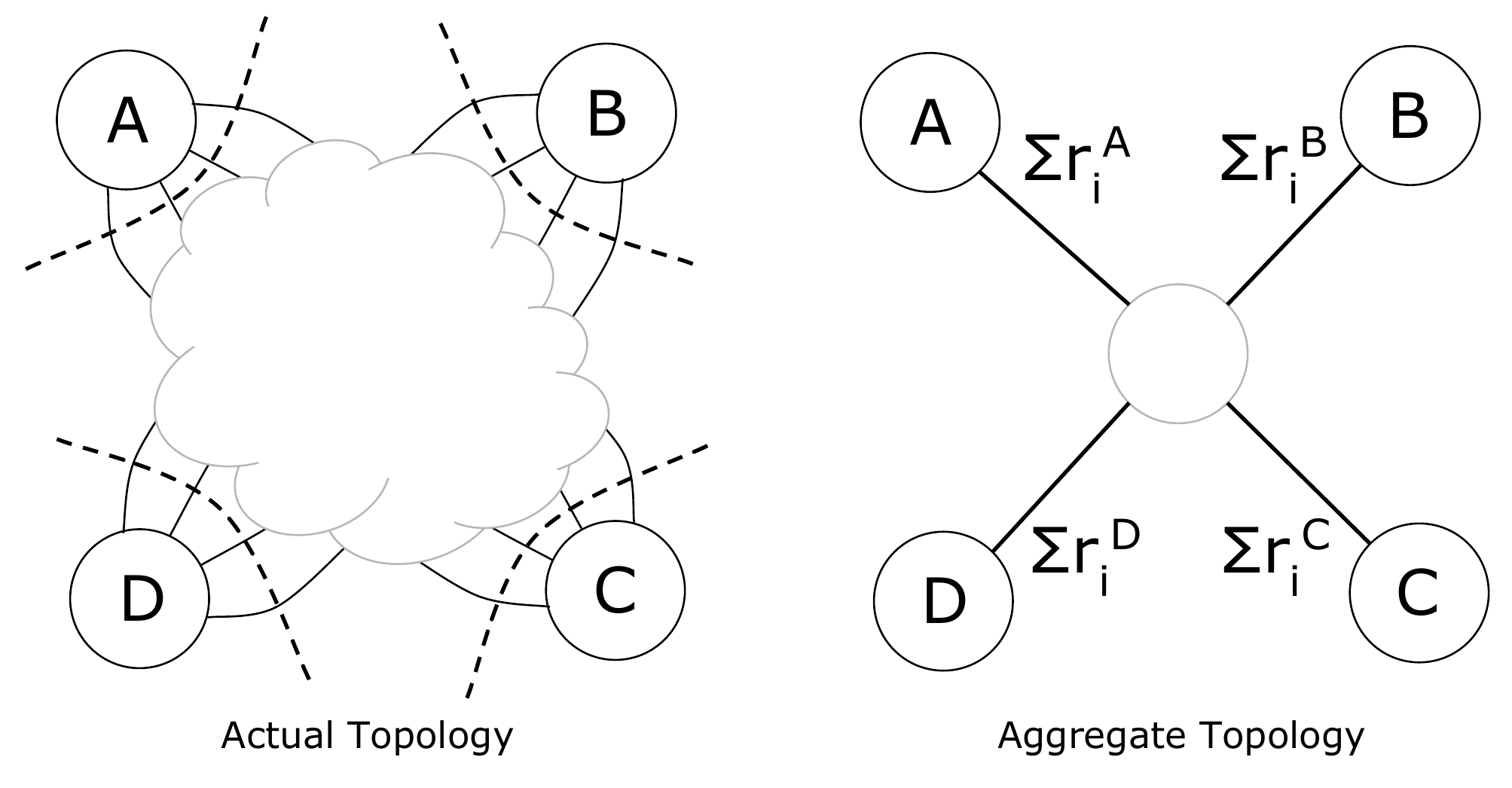}
    \caption{The physical topology, and the aggregate topology to compute a lower bound on receiver completion times. The aggregate topology is not part of how \nameiris~operates and is only used for evaluation in this section.}
    \label{fig:aggregate_topology}
\end{figure}

We develop a technique to compute a lower bound on receiver completion times by creating an aggregate topology from the actual topology. As shown in Figure \ref{fig:aggregate_topology}, to create the aggregate topology, we combine all downlinks and uplinks with rates $r_i^{[node]}$ for all interfaces $i$ per node to a single uplink and downlink with their rates set to the sum of rates of physical links. Also, the aggregate topology connects all nodes in a star topology using their uplinks and downlinks and so assumes no bottlenecks within the network. Since this topology is a relaxed version of the physical topology, any solution that is valid for the physical topology is valid on this topology as well. Therefore, the solution to the aggregate topology is a lower bound that can be computed efficiently but may be inapplicable to the actual physical topology. We will use this approach in \S \ref{min_avg_comp} for evaluation of {\nameiris}.

\subsection{Simulations}
In simulations, we focus on computing gains and therefore assume no dropped packets and accurate max-min fair rates. We normalized link capacities by maximum link rate per topology and fixed the timeslot length to $\omega = 1.0$.

\vspace{0.5em}
\noindent\textbf{Accounting for the Effect of User Traffic:} We account for the effect of higher priority user traffic in the simulations. The amount of available bandwidth per edge per timeslot, i.e., $B_e(t)$, is computed by deducting the rate of user traffic from the link capacity $C_e$. Recent work has shown that this rate can be safely estimated \cite{tempus, netstitcher}. For evaluations, we assume that user traffic can take up to $30\%$ of a link's capacity with a minimum of $5\%$ and that its rate follows a periodic pattern going from low to high and to low again. Per link, we consider a random period in the range of $10$ to $100$ timeslots that is generated and assigned per experiment instance.

\subsubsection{Minimizing Average Completion Times} \label{min_avg_comp}
This is when the objective vector is made of all ones. The partitioning hierarchy then begins with all receivers forming their $1$-receiver partitions. This is a highly general objective and can be considered as the default approach when the application/user does not specify an objective vector. We discuss multiple simulation experiments.

\begin{figure}
    \centering
    \includegraphics[width=\textwidth]{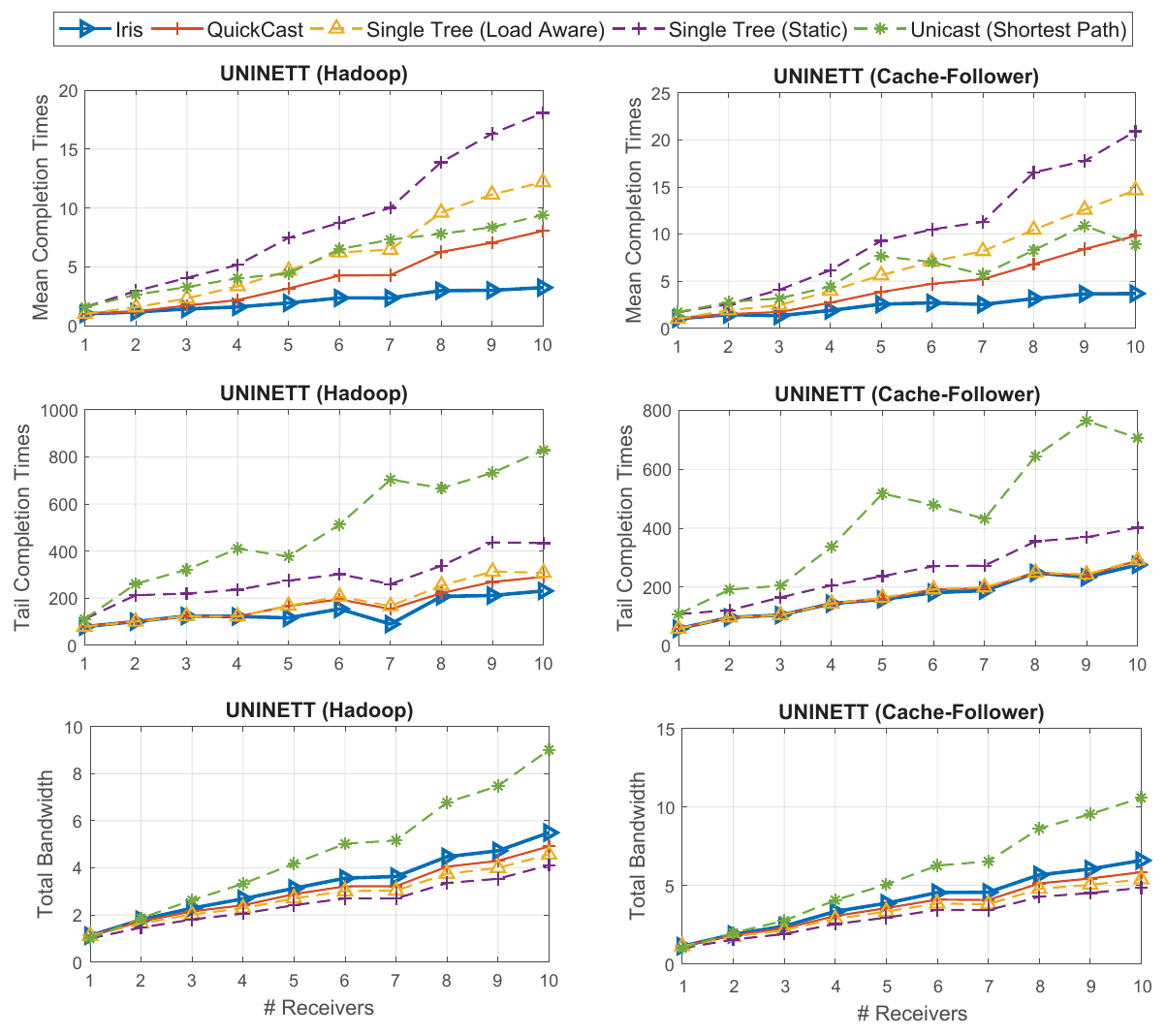}
    \caption{Comparison of various techniques by number of multicast receivers. Plots are normalized by the minimum data point (mean and tail charts are normalized by the same minimum), $\lambda = 1$, and lower values are better.}
    \label{fig:overall_iris}
\end{figure}

In Figure \ref{fig:overall_iris}, we measure the completion times (mean and tail) as well as bandwidth consumption by the number of receivers (tail is 99.9\textsuperscript{th} percentile). We consider two baseline cases: unicast shortest path and static single tree (i.e., minimum edge Steiner tree) routing. The shortest path routing is the unicast scenario that uses minimum bandwidth possible. The minimum edge Steiner tree routing uses minimum bandwidth possible while connecting all receivers with a single tree. The first observation is that using unicast, although leads to highest separation of fast and slow receivers, does not lead to the fastest completion as it can lead to many shared bottlenecks and that is why we see long tail times. {\nameiris} offers the minimum completion times (mean and tail) across all scenarios. Also, its completion times grow much slower compared to others as the number of receivers (and so overall network load) increases. This is while {\nameiris} uses only up to $35\%$ additional bandwidth compared to the static single tree (unicast shortest path routing uses up to $2.25\times$). Compared to QuickCast, {\nameiris} offers up to $26\%$ lower tail times and up to $2.72\times$ better mean times while using up to $13\%$ extra bandwidth.

\begin{figure}
    \centering
    \includegraphics[width=0.8\textwidth]{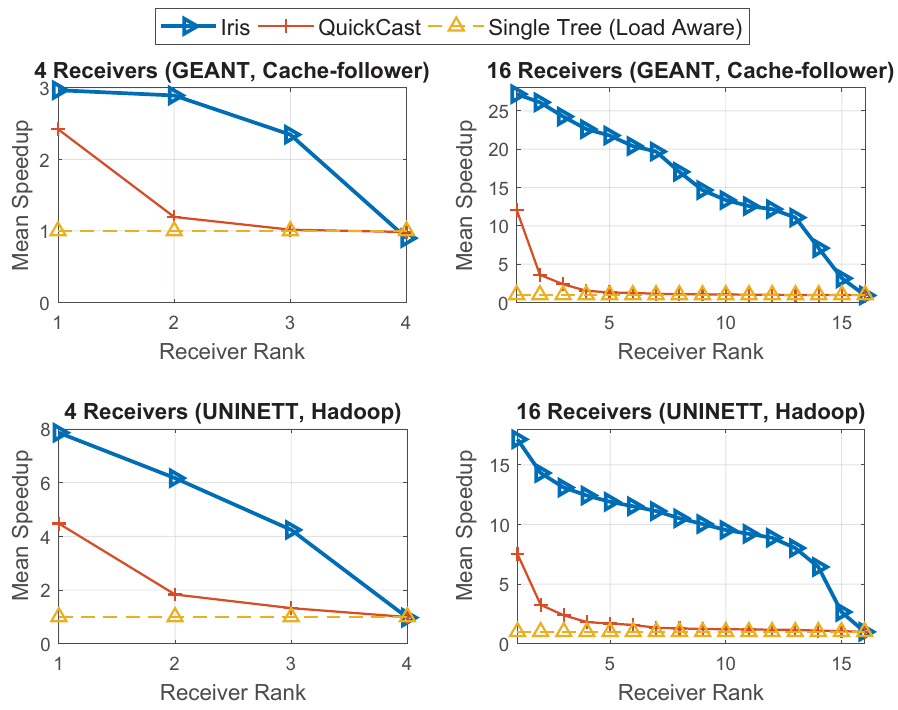}
    \caption{Mean completion time speedup (larger is better) of receivers normalized by no partitioning (load aware single tree) case given their rank from fastest to slowest, every node initiates equal number of transfers, receivers were selected according to uniform distribution from all nodes, and we considered $\lambda$ of 1.}
    \label{fig:speedup_1_iris}
\end{figure}

In Figure \ref{fig:speedup_1_iris}, we show the completion times speedup of receivers by their rank. As seen, gains depend on the topology, traffic pattern, and receiver's rank. The dashed line is the baseline, i.e., no-partitioning case. Compared to QuickCast \cite{quickcast}, the fastest node always completes faster and up to $2.25\times$ faster with {\nameiris}. Also, the majority of receivers complete significantly faster. In case of four receivers, the top $75\%$ receivers complete between $2\times$ to $4\times$ faster than baseline and with sixteen receivers, the top $75\%$ receivers complete at least $8\times$ faster than baseline. This is when QuickCast's gain drops quickly to one after the top $25\%$ of receivers.

\begin{figure}
    \centering
    \includegraphics[width=0.8\textwidth]{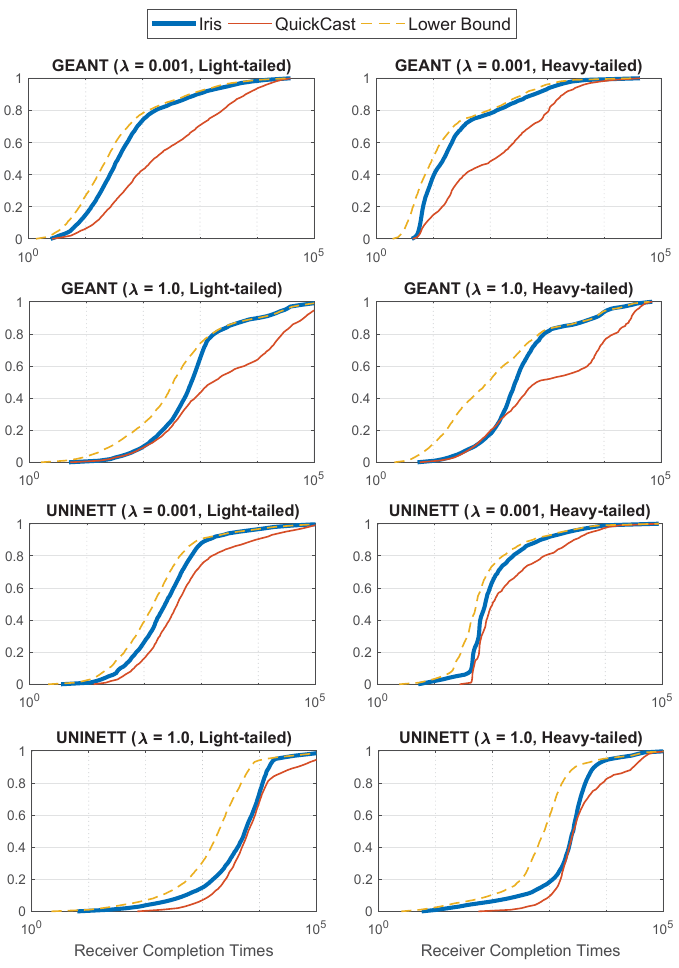}
    \caption{CDF of receiver completion times. Every transfer has 8 receivers selected uniformly across all nodes. ``Lower Bound'' is computed by finding the aggregate topology and applying Theorem 2.}
    \label{fig:speedup_2}
\end{figure}

In Figure \ref{fig:speedup_2}, we measure the CDF of completion times for all receivers. As seen, tail completion times are two to three orders of magnitude longer than median completion times which is due to variable link capacity and transfer volumes. We evaluate the completion times of QuickCast and {\nameiris} and compare them with a lower bound which considers the aggregate topology (see \S \ref{aggregate_topo}) and applies Theorem 2 directly. It is likely that no feasible solution exists that achieves this lower bound. Under low arrival rate (light load), we see that {\nameiris} tracks the lower bound nicely with a marginal difference. Under high arrival rate (heavy load), {\nameiris} stays close to the lower bound for lower and higher percentiles while not far from it for others.

\subsubsection{Other Objective Vectors}
\begin{figure}
    \centering
    \includegraphics[width=0.8\textwidth]{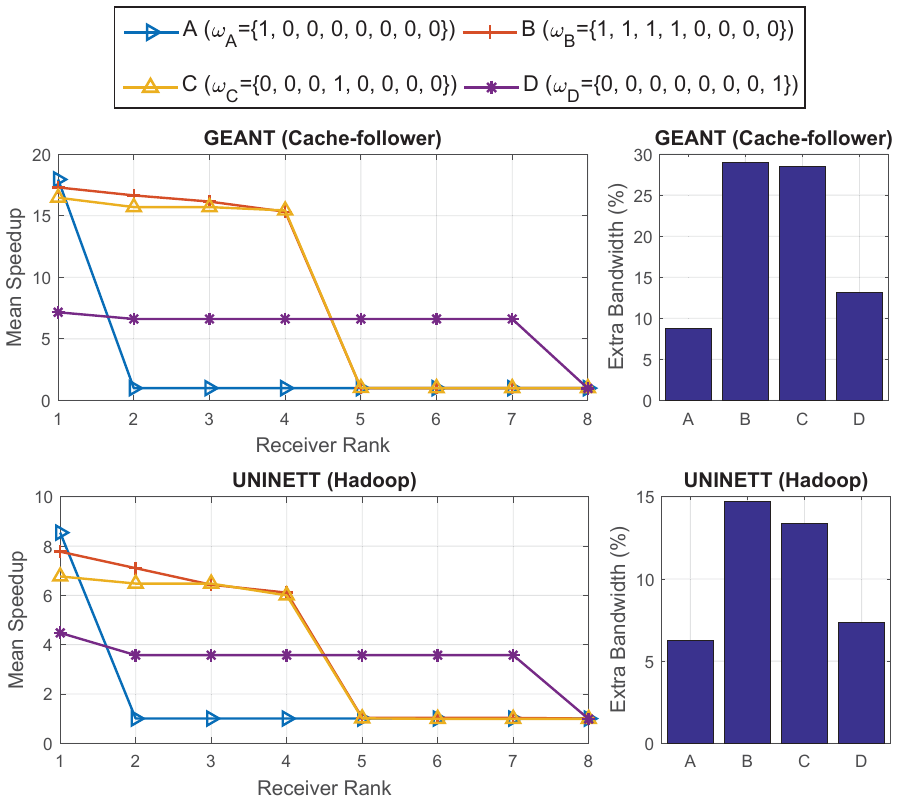}
    \caption{Gain by rank for different receivers per transfer averaged over all transfers for four different objective vectors. We set $\lambda = 0.1$ and there are $8$ receivers.}
    \label{fig:vectors}
\end{figure}

We discuss four different objective vectors of $A$, $B$, $C$ and $D$ as shown in Figure \ref{fig:vectors}. This figure shows the mean speedup of receivers given their ranks, and the bandwidth consumption associated with each vector. In $A$, we aim to finish one copy quickly while not being concerned with completion times of other receivers. We see a gain of between $9\times$ to $18\times$ across the two topologies considered for the first receiver. We also see that this approach uses much less extra bandwidth compared to when we have a vector with more ones (e.g., case $B$). In $B$, we aim to speed up the first four receivers (we care about each one) while in $C$, we want to speed up the fourth receiver not directly concerning ourselves with the top three receivers. As can be seen, $B$ offers increasing speedups for the top three receivers while $C$'s speedup is flatter. Also, $C$ uses less bandwidth compared to $B$ by grouping the top three receivers into one partition at the base of the hierarchy. Finally, $D$'s vector specifies that the application/user only cares about the completion time of the last receiver which means that receiver will be put in a separate partition at the base of the hierarchy while other receivers will be grouped into one partition. Since the slowest receiver is usually limited by its downlink speed, this cannot improve its completion time. However, with minimum extra bandwidth, this speeds up all receivers except the slowest by as much as possible. Except for the slowest, all receivers observe a speedup of between $3\times$ to $6\times$ while using $8\%$ to $16\%$ less bandwidth compared to $B$. A tradeoff is observed, that is, $D$ offers lower speedup but consistent gain for more receivers with less bandwidth use compared to $B$.

\subsection{Mininet Emulations}
\begin{figure}
    \centering
    \includegraphics[width=0.64\textwidth]{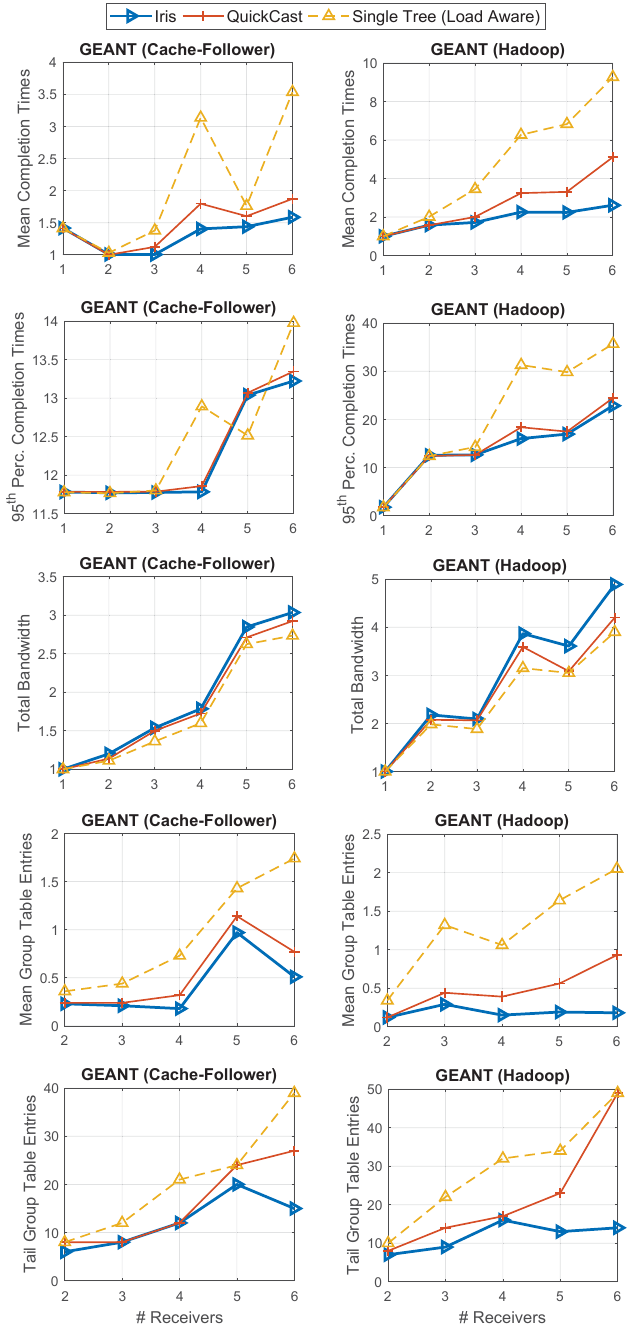}
    \caption{Mininet Emulation Results}
    \label{fig:mininet}
\end{figure}

We used Mininet to build and test a prototype of {\nameiris} and compare it with QuickCast and set up the testbed on CloudLab \cite{cloudlab}. We used OpenvSwitch (OVS) 2.9 in the OpenFlow 1.3 compatibility mode along with the Floodlight controller 1.2 connecting them to a control network. We assumed fixed available bandwidth over edges according to GEANT topology \cite{geant} while scaling downlinks' capacity so that the maximum is 500 Mbps. We did this to reduce the CPU overhead of traffic shaping over TCLink Mininet modules. Our traffic engineering program communicated with end-points through a RESTful API. We used NORM \cite{norm_navy} for multicast session management along with its rate-control module. To increase efficiency, we computed max-min fair rates centrally at the traffic engineering program and let the end-points shape their traffic using NORM's rate control module. The experiment was performed using twelve trace files generated according to Facebook traffic patterns (concerning transfer volume) \cite{social_inside}, and each trace file had 200 requests in total with an arrival rate of one request per timeslot based on Poisson distribution. We also considered timeslots of one second, a minimum transfer volume of 5 MBs and limited the maximum transfer volume to 500 MBs (which also match the distribution of YouTube video sizes \cite{you_tube}). We considered three schemes of {\nameiris}, QuickCast and a single tree approach (no partitioning). The total emulation time was about 24 hours. Figure \ref{fig:mininet} shows our emulation results. To allow comparison between the tail (95\textsuperscript{th} percentile) and mean values, we have normalized both plots by the same minimum in each row. Also, the group table usage plots are not normalized and show the actual average and actual maximum across all switches. The reason why data points jump up and down is the randomness of generated traces that comes from transfers (volume, source, receivers, arrival pattern, etc).

\vspace{0.5em}
\noindent\textbf{Completion Times and Bandwidth:} {\nameiris} can improve on QuickCast by speeding up mean receiver completion times by up to $2.5\times$. It also offers up to $4\times$ better mean completion times compared to using a single forwarding tree per transfer. We also see that compared to using one multicast tree, {\nameiris} consumes at most $25\%$ extra bandwidth.

\vspace{0.5em}
\noindent\textbf{Forwarding Plane:} We see that {\nameiris} uses up to about $4\times$ less group table entries at the switches where the maximum number of entries were exhausted which allows more parallel transfers across the same network. {\nameiris} achieves this by allowing a larger number of partitions per transfer whenever it does not hurt the completion times. By allowing more partitions, each tree will branch less times on average reducing the number of group table entries.

\vspace{0.5em}
\noindent\textbf{Running Time:} Across all experiments, the computation time needed to run {\nameiris} to calculate partitions and forwarding trees stayed below 5 ms per request.

\subsection{Practical Concerns}
New challenges, such as increased communication latency across network elements and failures, may arise while deploying {\nameiris} on a real-world geographically distributed network. Communication latency may not affect the performance considerably as we focus on long-running internal transfers that are notably more resilient to latency overhead of scheduling and routing compared to interactive user traffic. Failures may affect physical links or the TES. Loss of a physical link can be addressed by rerouting the affected transfers reactively either by the network controller or by using the SDN fast failover mechanisms. End-points may be equipped with distributed congestion control, such as the one presented in \cite{mctcp}, which they can fall back to in case the centralized traffic engineering fails.

\section{Conclusions}
In this chapter, we presented the problem of grouping receivers into multiple partitions per P2MP transfer to minimize the effect of receiver downlink speed discrepancy on completion times of receivers. We analyzed a relaxed version of this problem and came up with a partitioning that minimizes mean completion times given max-min fair rates. We also set forth the idea of applications/users expressing their requirements in the form of binary objective vectors which allows us to optimize resource consumption and performance further. We then described {\nameiris}, a system that computes partitions and forwarding trees for incoming bulk multicast transfers as they arrive given objective vectors. We showed that {\nameiris} could significantly reduce mean completion times with a small increase in bandwidth consumption and can fulfill the requirements expressed using objective vectors while saving bandwidth whenever possible. It is worth noting that performance of any partitioning and forwarding tree selection algorithm rests profoundly on the network topology and transfer properties.

\clearpage
\chapter{Speeding up P2MP Transfers using Parallel Steiner Trees} \label{chapter_p2mp_parallel}
In Chapters \ref{chapter_p2mp_dccast} to \ref{chapter_iris}, we discussed different ways of managing Point to Multipoint (P2MP) inter-DC transfers via using dynamically selected forwarding trees to balance load across the network and reduce network capacity consumption. In all past efforts, we attached each receiver to the sender using a single forwarding tree.\footnote{In case of partitioning, every receiver belonged to exactly one partition and so was connected to the sender using a single forwarding tree.} In general, however, it may be possible to increase receivers' download speeds by using multiple parallel trees that connect the sender to all receivers\footnote{In case of partitioning, all receivers in every partition are attached using one forwarding tree to the sender as in Chapters \ref{chapter_p2mp_quickcast} and \ref{chapter_iris}.} which is what we will explore in this chapter. We will show that by using two forwarding trees per receiver, we can reduce the completion times of receivers by up to $40\%$ while only increasing the total network capacity usage by up to $10\%$. We also find that using up to more than two parallel trees offers a negligible benefit or even hurts the performance due to excessive bandwidth usage and creation of unnecessary bottlenecks.

\section{Motivating Example}
\begin{figure}
    \centering
    \includegraphics[width=\textwidth]{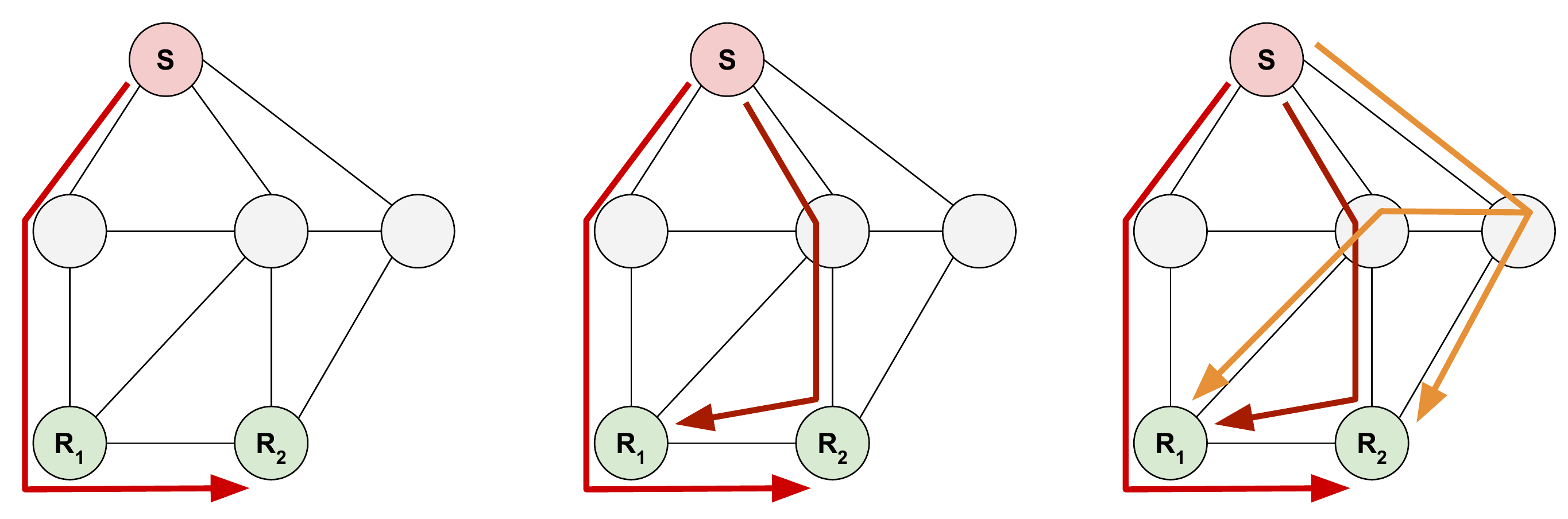}
    \caption{Using parallel forwarding trees we can increase the overall network throughput to all receivers. We may have to pay some extra bandwidth cost as we add more trees.}
    \label{fig:parallel_tree_example}
\end{figure}

By using parallel trees, we can substantially increase the multicast forwarding throughput possibly at little extra network capacity cost. Figure \ref{fig:parallel_tree_example} shows how adding more trees can improve the overall receiver throughput. Assuming equal link capacity of $1$ for all edges, the single tree case on the left offers a total rate of $1$. Adding one more tree in an edge-disjoint manner will double the rate. If we consider an equal division of traffic across the two trees, the total network bandwidth usage is not increase compared to the single tree case. Now find the network on the right. We see three trees that will give us a total rate of $3$. However, the last tree has four edges. Assuming equal division of traffic across all three trees, we see that this will increase the total bandwidth usage by $1.11\times$. Also, we see that adding more trees will not help us improve completion times due to the creation of bottlenecks (since trees will not be edge-disjoint anymore).

\section{System Model}
We adopt the same system model as presented in \S \ref{dccast_sys_model}. Namely, we consider a slotted timeline, and a centralized traffic engineering mechanism that determines what trees will be used by an arriving transfer. The central controller also computes the rates at which senders transmit traffic on each tree. We also focus on bulk and internal data transfers that are not in the critical path of user experience and so are resilient to some degree of latency. We assume heterogeneous link capacities as presented by real WAN topologies. We will use the same notation as that in Tables \ref{table_var_0} and \ref{table_var_dccast}.

\section{Application of Parallel Forwarding Trees}
We discuss how to dynamically select parallel edge-disjoint forwarding trees according to network load across different edges and then discuss various rate-allocation (i.e., traffic scheduling) policies.

\subsection{Adaptive Edge-disjoint Parallel Forwarding Tree Selection}
Although using a single forwarding tree for every transfer minimizes packet reordering and total network capacity consumption, it can considerably limit the overall achievable network throughput for P2MP transfers. Under light load, this will lead to inefficient use of network capacity artificially increasing the completion times of P2MP transfers. We discuss our approach to selection of multiple forwarding trees.

To perform a P2MP transfer $R_{new}$ with volume $\mathcal{V}_{R_{new}}$, the source $S_{R_{new}}$ transmits traffic over edge-disjoint Steiner trees that span across $\pmb{\mathrm{D}}_{R_{new}}$. In this chapter, we do not discuss receiver set partitioning as that subject can be applied orthogonal to the parallel tree selection approach by treating each partition as a separate P2MP request. At any timeslot, traffic for any transfer flows with the same rate over all links of a forwarding tree to reach all the destinations at the same time. The problem of scheduling a P2MP transfer then translates to finding multiple forwarding trees and a transmission schedule over every tree for every arriving transfer in an online manner. A relevant problem is the minimum weight Steiner tree \cite{steiner_tree_problem} that can help minimize total bandwidth usage with proper weight assignment. Although it is a hard problem, heuristic algorithms exist that often provide near optimal solutions \cite{robins2005tighter, Watel2014}.

To select multiple Steiner trees, we use the metric load $L_e$ that is defined for every edge $e$ as the total remaining volume of traffic for all the trees that include that edge. We first assign every edge a weight of $W_e = \frac{L_e + \mathcal{V}_{R_{new}}}{C_e}$ which is the minimum time it would take for all the transfers that share that edge to complete (if $R_{new}$ were to be placed on that edge). The algorithm starts by first selecting a minimum weight Steiner tree using a heuristic algorithm. We then mark all of the edges of this tree as deleted and run the minimum weight Steiner tree selection algorithm again. This process is repeated until either no more trees can be found (i.e., some receivers are disconnected) or we reach a maximum of $K$ trees set by the operators as a configuration parameter.

This approach offers several benefits. Since trees are selected dynamically as load changes on edges, they tend to avoid highly busy links. Also, as trees assigned to a transfer are edge-disjoint, this approach avoids creating additional bottlenecks that cause competition across trees of the same transfer. Finally, by limiting the maximum number of trees, operators can choose between speeding up the transfers (using more trees) or minimizing total bandwidth consumption (using fewer trees) by changing the value of $K$. This value could be chosen as a function of network load, i.e., under heavier load operators can reduce $K$ and increase it as load decreases.

\vspace{0.5em}
\noindent\textbf{Updating $L_e$:} While using multiple forwarding trees, after selection of such trees, the load on their edges needs to be increased according to $\mathcal{V}_{R_{new}}$. Since we do not know, originally, how much of the traffic will be sent over each tree, it is unclear how to increase the load on the edges of different trees. This is because according to the scheduling policy used to send traffic and the future transfers that arrive, the volume of traffic sent over different trees per transfer can change. For example, if a transfer has two trees and one of them has to compete with a future transfer, the volume of traffic sent over the other tree will automatically increase as a result as soon as the future transfer arrives. To address this, we use a heuristic technique as follows. We assume that at any time, the remaining volume of a transfer is equally divided across all its trees. If one tree sends a lot of traffic, that reduction in load will be equally divided and deducted from all of the trees for that transfer. Although the exact load on every edge will potentially not be accurate, this approach offers an efficient approximation of load which helps us to quickly select future forwarding trees.

\subsection{Scheduling Policies}
Similar to previous chapters, we consider well-known scheduling policies of First Come First Serve (FCFS), Shortest Remaining Processing Time (SRPT), and fair sharing based on Max-Min Fairness (MMF). These scheduling policies have different properties. Fair sharing is the most widely used policy as it allows many users to fairly access the network bandwidth over network bottlenecks. SRPT allows more internal data transfers to be completed in any given period of time. FCFS can also be used to offer more accurate guarantees to applications on when their transfers will complete.

\section{Evaluation} \label{evaluations_parallel}
We considered various topologies and transfer size distributions. In the following, we perform experiments to measure the effectiveness of using parallel forwarding trees on multiple toplogies and using multiple transfer size distributions.

\vspace{0.5em}
\noindent\textbf{Network Topologies:}
We use the same topologies discussed in Chapter \ref{chapter_p2mp_quickcast}. These topologies provide capacity information for all links which range from 45 Mbps to 10 Gbps. We normalized all link capacities dividing them by the maximum link capacity. We also assumed all bidirectional links with equal capacity in either direction.


\vspace{0.5em}
\noindent\textbf{Traffic Patterns:}
We use the same transfer size distributions discussed in Chapter \ref{chapter_p2mp_quickcast}. Transfer arrival followed a Poisson distribution with rate $\lambda$. We considered no units for time or bandwidth. For all simulations, we assumed a timeslot length of $\omega = 1.0$. For Pareto distribution, we considered a minimum transfer volume equal to that of $2$ full timeslots and limited maximum transfer volume to that of $2000$ full timeslots. Unless otherwise stated, we considered an average demand equal to volume of $20$ full timeslots per transfer for all traffic distributions (we fixed the mean values of all distributions to the same value). Per simulation instance, we assumed equal number of transfers per sender and for every transfer, we selected the receivers from all existing nodes according to the uniform distribution (with equal probability from all nodes).

\vspace{0.5em}
\noindent\textbf{Assumptions:} We focused on computing gains and assumed accurate knowledge of inter-DC link capacity, and precise rate control at the end-points which together lead to a congestion free network. We also assumed no dropped packets due to corruption or errors, and no link failures.

\vspace{0.5em}
\noindent\textbf{Simulation Setup:} We developed a simulator in Java (JDK 8). We performed all simulations on one machine (Core i7-6700 and 24 GB of RAM). We used the Java implementation of GreedyFLAC \cite{DSTAlgoEvaluation} for minimum weight Steiner trees.


\subsection{Effect of Number of Parallel Trees}
\begin{figure}
    \centering
    \includegraphics[width=\textwidth]{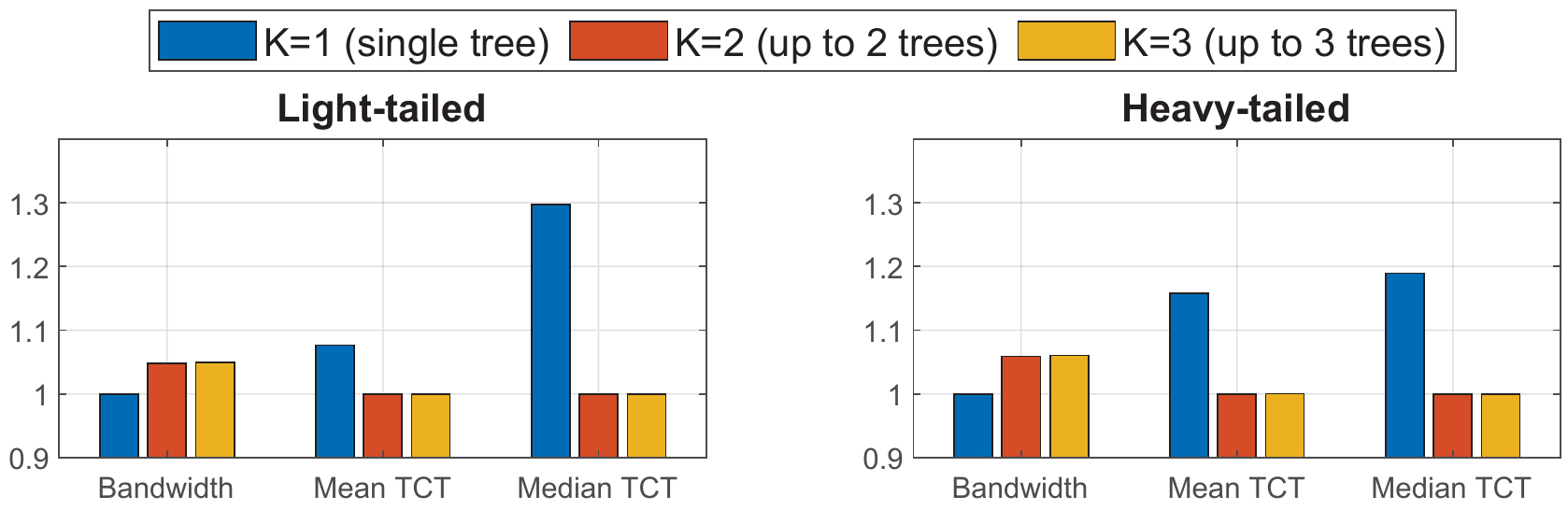}
    \caption{The effect of number of parallel trees on total bandwidth consumption and transfer completion times (TCTs). Other experiment parameters are $\lambda = 0.01$, max-min fair rate computation, and GEANT \cite{geant} topology.}
    \label{fig:exp_1_parallel_trees}
\end{figure}

Figure \ref{fig:exp_1_parallel_trees} shows the effect of maximum number of trees per transfer (i.e., $K$). We see that almost all the gain is obtained with $2$ parallel trees and increasing it further does not improve the completion times. Adding more trees, however, increases the total network bandwidth usage. We see that while the network bandwidth consumption increases by about $7\%$ in the settings of this experiment, the mean completion times improve by up to $17\%$ and the median completion times improve by up to $30\%$. The gain in mean completion times is less than that of median as a result of the tail completion times which are usually much higher than median since the transfer size distribution is skewed. Also, it is worth noting that having parallel trees cannot improve tail completion times as the tail is restricted by physical constraints such as low capacity links.

\subsection{Effect of Number of Copies}
\begin{figure}
    \centering
    \includegraphics[width=\textwidth]{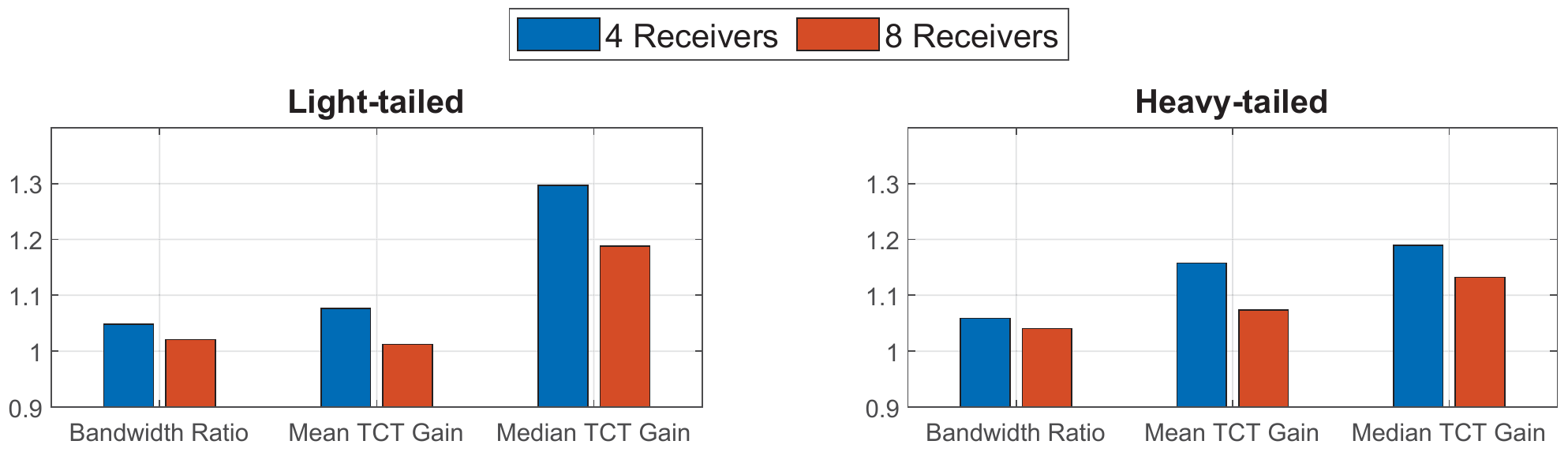}
    \caption{The effect of number of receivers on total bandwidth consumption ratio (i.e., $\frac{\textnormal{Bandwidth of }K=2}{\textnormal{Bandwidth of }K=1}$) and transfer completion times (TCTs) gain (i.e., $\frac{\textnormal{completion time of }K=1}{\textnormal{completion time of }K=2}$). Other experiment parameters are $\lambda = 0.01$, max-min fair rate computation, and GEANT \cite{geant} topology.}
    \label{fig:exp_2_parallel_trees}
\end{figure}

In Figure \ref{fig:exp_2_parallel_trees} we explore the effect of number of receivers per transfer. With more receivers, we will have larger trees which make it harder in general to find edge-disjoint parallel trees. As a result, we see that the gain in mean and median completion times drops with more receivers. One way to increase the effect of parallel trees in scenarios with many receivers per transfer is to reduce the receivers per tree by partitioning receivers using some technique, for example those discussed in Chapters \ref{chapter_p2mp_quickcast} and \ref{chapter_iris}.

\subsection{Effect of Transfer Size Distribution}
\begin{figure}[p]
    \centering
    \includegraphics[width=\textwidth]{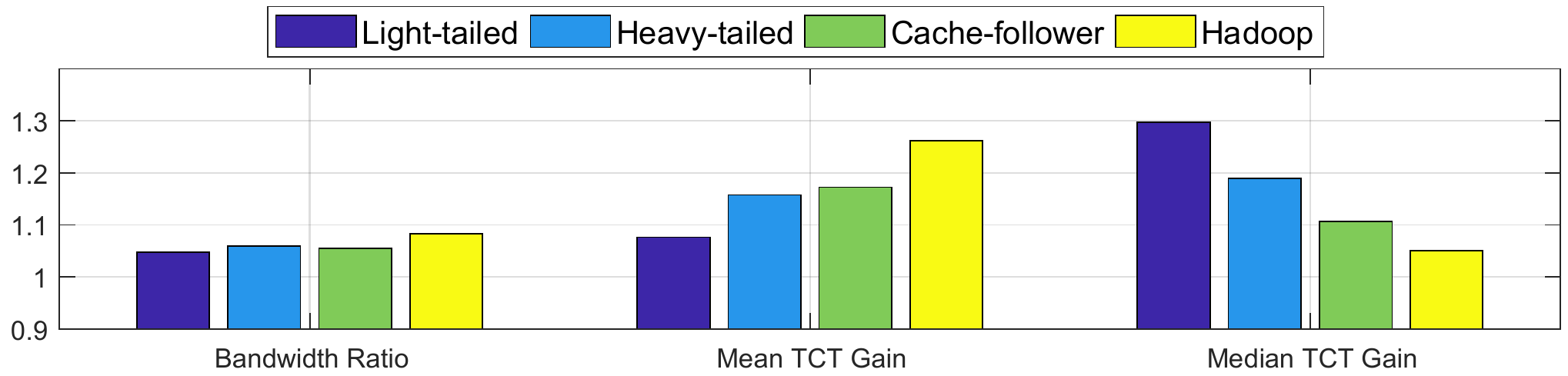}
    \caption{The effect of transfer size distribution on total bandwidth consumption ratio (i.e., $\frac{\textnormal{Bandwidth of }K=2}{\textnormal{Bandwidth of }K=1}$) and transfer completion times (TCTs) gain (i.e., $\frac{\textnormal{completion time of }K=1}{\textnormal{completion time of }K=2}$). Other experiment parameters are $\lambda = 0.01$, $4$ receivers per transfer, max-min fair rate computation, and GEANT \cite{geant} topology.}
    \label{fig:exp_3_parallel_trees}
\end{figure}

Figure \ref{fig:exp_3_parallel_trees} shows the effect of different transfer size distributions which include both synthetic and real distributions. Since trees are selected dynamically, we see that the total bandwidth consumption also changes with the traffic pattern. We also see that the gain in mean and median completion times depend highly on the traffic distribution ranging from $5\%$ to $30\%$. Interestingly, we also see that the gain in mean completion times has an inverse relationship with that of median completion times. We believe this behavior is a result of how distributions affect the tail completion times. For example, with the synthetic light-tailed and heavy-tailed distributions, the tail grows larger with $K=2$, while for the real traffic patterns of Cache-follower and Hadoop we see a decrease in tail completion times (not shown in the figure). The common result is that regardless of the traffic patterns, we always obtain considerable gains in either mean or median completion times with up to $10\%$ increase in bandwidth usage.

\subsection{Effect of Topology}
\begin{figure}[p]
    \centering
    \includegraphics[width=\textwidth]{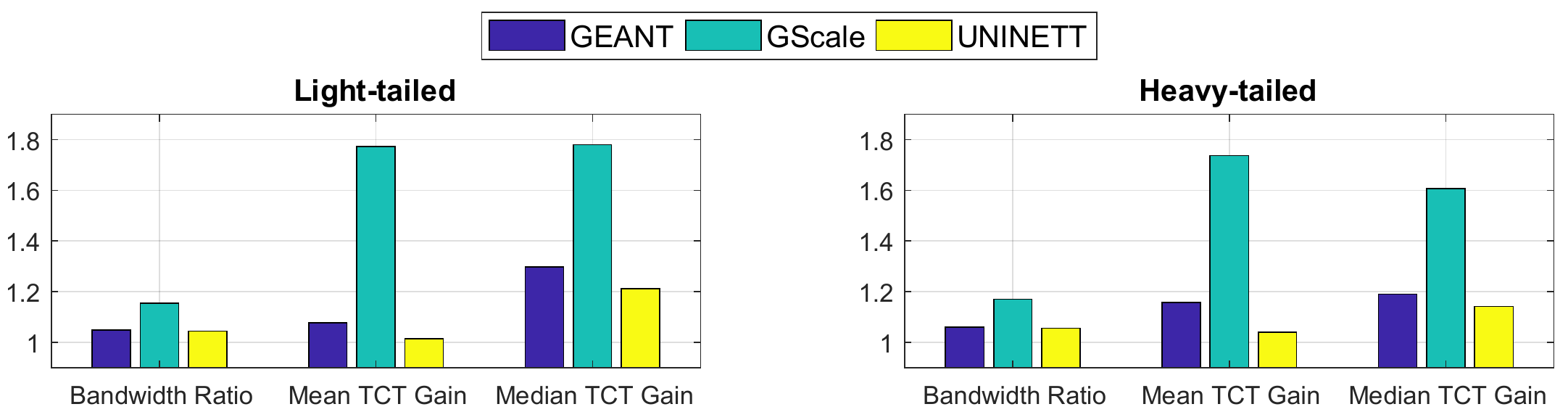}
    \caption{The effect of topology on total bandwidth consumption ratio (i.e., $\frac{\textnormal{Bandwidth of }K=2}{\textnormal{Bandwidth of }K=1}$) and transfer completion times (TCTs) gain (i.e., $\frac{\textnormal{completion time of }K=1}{\textnormal{completion time of }K=2}$). Other experiment parameters are $\lambda = 0.01$, $4$ receivers per transfer, and max-min fair rate computation.}
    \label{fig:exp_4_parallel_trees}
\end{figure}

We explore the effect of different topologies as shown in Figure \ref{fig:exp_4_parallel_trees}. We see that GScale offers significantly higher gains in completion times compared to the other two topologies. That is because we assumed a uniform capacity of $1$ across the edges of GScale while GEANT and UNINETT have many low capacity edges which negatively affect the gains. We also see a higher bandwidth usage over GScale that is up to $18\%$ which is due to the ability of the routing algorithm to use parallel trees for more transfers. GScale is a smaller topology and is better connected compared to UNINETT and GEANT which is why we can build more parallel trees on average.

\subsection{Effect of Scheduling Policies}
\begin{figure}[p]
    \centering
    \includegraphics[width=\textwidth]{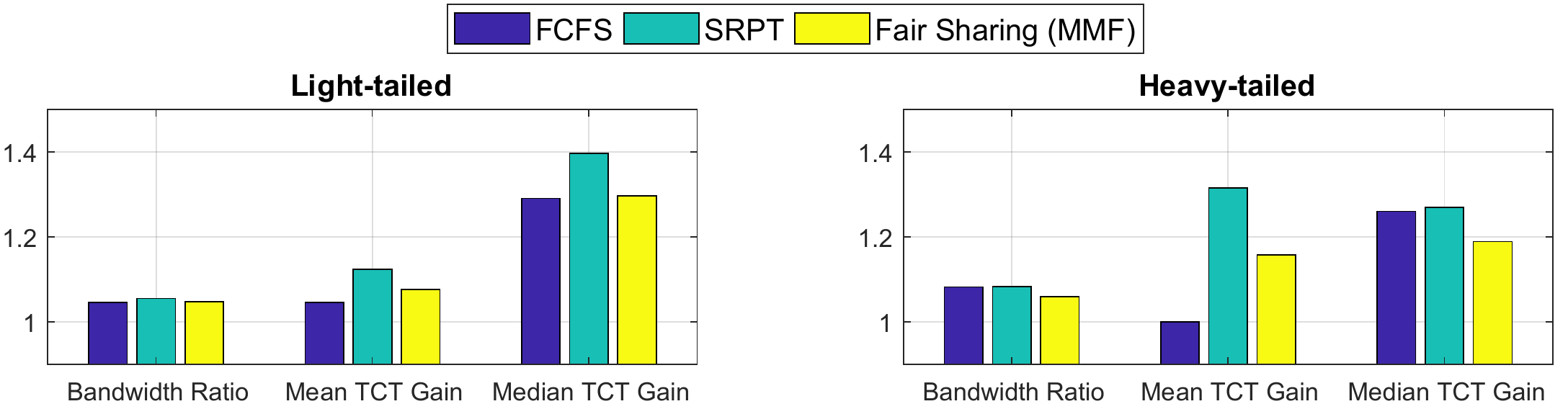}
    \caption{The effect of traffic scheduling policy on total bandwidth consumption ratio (i.e., $\frac{\textnormal{Bandwidth of }K=2}{\textnormal{Bandwidth of }K=1}$) and transfer completion times (TCTs) gain (i.e., $\frac{\textnormal{completion time of }K=1}{\textnormal{completion time of }K=2}$). Other experiment parameters are $\lambda = 0.01$, $4$ receivers per transfer, and GEANT \cite{geant} topology.}
    \label{fig:exp_5_parallel_trees}
\end{figure}

Figure \ref{fig:exp_5_parallel_trees} shows the effect of scheduling policies on the flow completion times gain and the total bandwidth use. We see that using parallel trees offers the most gain when applying the SRPT policy. This is because small transfers obtain much higher throughput as soon as they arrive since the policy preempts any other larger transfers. Fair sharing and FCFS both offer considerable gains in median completion times with fair sharing offering a higher average gain that is due to better tail completion times. In other words, with FCFS, few large transfers can fully block some links and slow down all other transfers whose trees use those edges. Overall, we see that using parallel trees marginally increases bandwidth use while considerably improving completion times regardless of the scheduling policy.

\subsection{Effect of Network Load}
\begin{figure}
    \centering
    \includegraphics[width=\textwidth]{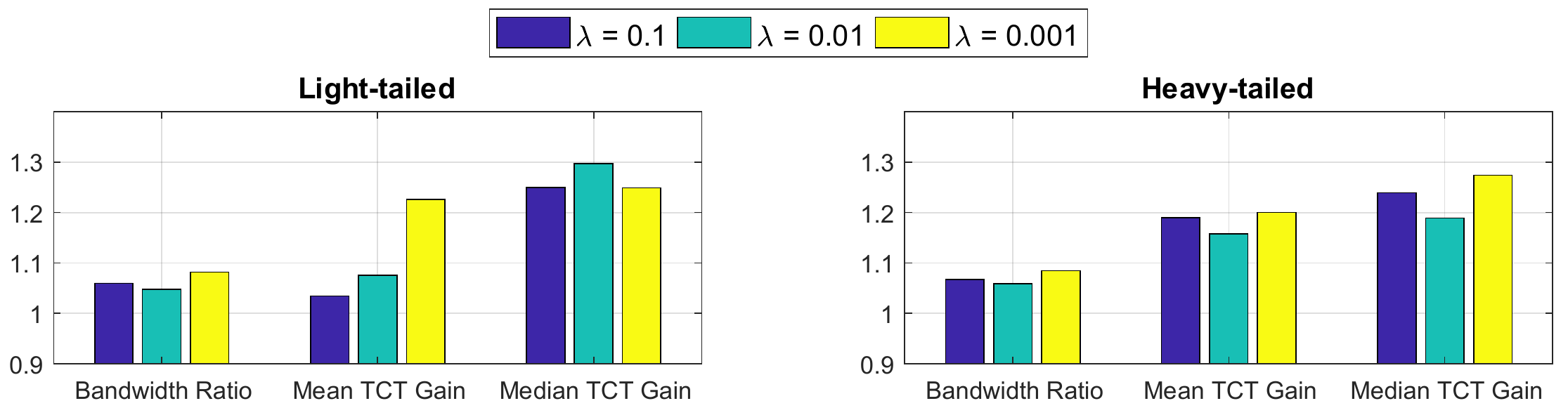}
    \caption{The effect of transfer arrival rate (i.e., network load) on total bandwidth consumption ratio (i.e., $\frac{\textnormal{Bandwidth of }K=2}{\textnormal{Bandwidth of }K=1}$) and transfer completion times (TCTs) gain (i.e., $\frac{\textnormal{completion time of }K=1}{\textnormal{completion time of }K=2}$). Other experiment parameters are $4$ receivers per transfer, max-min fair rate computation, and GEANT \cite{geant} topology.}
    \label{fig:exp_6_parallel_trees}
\end{figure}

In Figure \ref{fig:exp_6_parallel_trees} we evaluate the effect of network load. Overall, it appears that with lower network load, we observer higher gains in completion times and slightly higher bandwidth consumption. Under light load, most network edges are not loaded and so using parallel trees allows us to increase throughput for ongoing transfers with minimal interference. As load increases, we expect higher contention across competing transfers for access to network capacity which reduces the gains of having parallel trees.

\newpage
\section{Conclusions}
In this chapter, we evaluated the benefits of parallel forwarding trees for inter-DC P2MP transfers. The approach is to use edge-disjoint forwarding trees to reduce the interference across the trees of one transfer while maximizing throughput. We used a load-adaptive approach for selection for forwarding trees that selects up to $K$ such trees that balance load across the network. We also discussed a weight assignment technique for updating load weights over the edges of trees for efficient computation of weights. According to our evaluations with different traffic patterns, topologies, network load, number of parallel trees, and scheduling policies, we find that using up to two parallel trees per transfer can considerably improve the completion times of transfers while slightly increasing the total network bandwidth use. We also find that for better-connected networks with fewer bottlenecks, using parallel trees offer higher gains in completion times.

\clearpage
\chapter{Summary and Future Directions} \label{chapter_summary}
As organizations continue to build more datacenters around the globe, communication across these datacenters becomes more and more important for highly distributed applications with globally distributed users. Increasingly, companies use private dedicated high speed networks to connect datacenters to offer high quality infrastructure for distributed applications. For such costly networks to be profitable, it is necessary to maximize performance and efficiency.

In this dissertation, we made the case for coordinated control of routing over inter-DC networks and traffic transmission at the end-points. Since inter-DC networks are relatively small with tens to hundreds of nodes, such coordination is possible and is currently used by multiple organizations. A traffic engineering sever that is logically centralized receives traffic demands from end-points as well as network status updates from the network. Combined with the topology information, the traffic engineering server can then compute the routes over which traffic is forwarded over inter-DC networks and the rate at which traffic is transmitted from end-points.

We focused on multiple research domains concerned with traffic engineering over inter-DC networks. First, we noticed that a large portion of inter-DC traffic is formed by large inter-DC flows which we refer to as transfers. We realized that current adaptive routing techniques based on link utilization or static topology information are insufficient for minimizing the completion times of such transfers. We then developed Best Worst-case Routing (BWR), which is a routing heuristic that aims to route new transfers to minimize their worst-case completion times. We showed that this technique can improve completion times regardless of the scheduling policy used for transmission of traffic.

We then discussed the deadline requirement of many inter-DC transfers and studied admission control for large transfers. Admission control helps prevent over committing existing resources and makes sure that admitted transfers meet the deadlines they aimed for. Our major contribution has been to make such admission control as fast as possible to handle large number of transfers as they arrive. The admission control considers both routing of traffic and transmission control. We considered both cases of single path routing and multipath routing and showed that using up to two parallel paths offers considerable gains in admitted traffic. For fast admission control, we applied a new traffic allocation strategy that pushes traffic for every transfer as close as possible to their deadlines which we call the As Late As Possible (ALAP) scheduling policy. With this allocation strategy, we can quickly determine if a new transfer can meet its deadline and compute a feasible allocation without formulating complex optimization problems.

Next, we considered the problem of delivering objects from one location to multiple locations while paying attention to performance metrics such as completion times and deadlines. This problem has the one-to-many transmission property in common with the traditional multicasting problem, but has the added property that all the receivers of a transfer are known at the arrival time which allows us to select a multicast tree unon its arrival. We called such transfers Point to Multipoint (P2MP) transfers. We used Steiner trees to minimize bandwidth usage while selecting them in a way that distributes load by shifting traffic across various trees to exercise all available capacity. This approach allowed us to reduce tail completion times while handling more traffic. We also discussed the same problem given deadlines for P2MP transfers and showed that using the same adaptive tree selection technique combined with the ALAP scheduling policy, we can admit more traffic to the network and guarantee deadlines as well.

We then explored ways of further reducing the completion times of some receivers for P2MP transfers given that not all the receivers have to receive complete data at the same time. We observed that a single slow receiver, can slow down all receivers attached to the sender on a tree and proposed to break receivers into multiple partitions. Each partition is then connected to the sender using a separate tree. By grouping receivers according to their download speeds or by according to their proximity we can then improve their overall reception rate. We presented algorithms for performing such partitioning and showed that it is effective. We also showed that the effectiveness of these techniques is a function of network topology and link capacity distribution as well as distribution of transfer volumes.

Finally, we aimed to further improve the completion times of a P2MP transfer by using parallel forwarding trees. We explored the application of edge-disjoint forwarding trees that are selected adaptively according to network load. We realized that parallel trees can considerably improve completion times while minimally increase bandwidth consumption. We also found that selecting more than two parallel trees does not offer any benefits in most cases but increases bandwidth consumption.

\section{Future Directions}
We propose a few research directions for interested researchers to explore. We categorize these ideas according to the part of dissertation they target.

\subsection{Adaptive Routing over Inter-DC Networks} \label{adaptive_routing}
\begin{figure}
    \centering
    \includegraphics[width=0.6\textwidth]{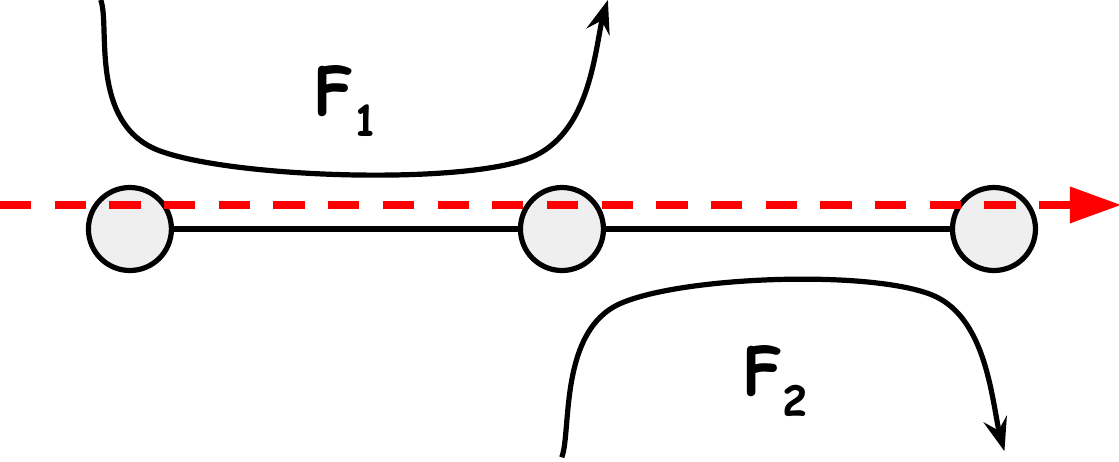}
    \caption{Example scenario used in \S \ref{adaptive_routing}}
    \label{fig:future_work_example}
\end{figure}

We presented BWR as an effective routing technique that improves transfer completion times regardless of the scheduling policy used for traffic. The method we developed to compute the worst-case completion times of transfers per path can be further improved. Our current implementation is merely summing up the remaining volumes of all the flows that intersect a path can be in general not tight enough and is too conservative for worst-case. For example, consider a path with two hops as shown in Figure \ref{fig:future_work_example}. On the first hop, we intersect $F_1$ with $4$ remaining packets, and on the second hop, we intersect $F_2$ with $5$ remaining data units. The current method will simply use $4 + 5$ as the worst-case start time on the path, but if the whole topology has only these two links, $F_1$ and $F_2$ can transmit in parallel, which means the worst-case will be $\max(F_1, F_2)$ that is $5$. Knowing this, an algorithm should prefer this path over another with a worst-case of $8$ (not shown in the figure). Computing the exact worst-case may be possible by using a dependency graph which can be computationally expensive. Furthermore, since optimizing one arriving flow does not necessarily provide benefits to future arrivals, we did not explore finding the exact worst-case. Also, extending to multipath BWR routing, and the effect of inaccurate flow size information on routing performance can be other directions to explore.

\subsection{Deadline-aware Point to Multipoint Transfers}
The approach we presented in this dissertation for deadline-based admission control requires that all destinations can be reached before the deadline for that transfer. This, however, may be too restrictive for many applications: we might prefer to maximize the number of receivers that complete before specified deadlines per transfer while considering a minimum number of replicas that need be made before a given deadline. This objective is more practical in the sense that minimum replicas represent some degree of reliability (which means we guarantee a required reliability degree) while allowing more transfers to be admitted which increases network utilization and efficiency.

\subsection{Receiver Completion Times of Point to Multipoint Transfers}
Due to varying load on edges as a result of time zone differences, the total bandwidth per tree may not be significant as trees span across many regions. To address this issue, store-and-forward can be used along with parallel trees to utilize the capacity of wide area networks further. With store-and-forward, one can build large scale overlay networks across datacenters and use intermediate nodes as large temporary buffers that store data in case the incoming rate is higher than the outgoing rate per transfer. As time passes bandwidth increases on outgoing edges of such nodes, the temporary buffer used will drain to next hop overlay nodes. An overlay node can consist of multiple servers in every datacenter with enough capacity to store data over highly loaded hours and consume later. With this approach, overlay nodes will use simple point to point connections but on a per-hop basis to build a multicast overlay network \cite{mc_icc_overlay}.

\subsection{Large-scale Implementation and Evaluation of Algorithms for Fast and Efficient Point to Multipoint Transfers}
In this dissertation, we developed various algorithms for fast and efficient P2MP transfers and evaluated them through simulations. Large-scale evaluation of our techniques and algorithms over real inter-DC networks and using practical inter-DC applications is another direction for future research. For example, forwarding trees can be realized using SDN Group Tables \cite{openflow-1.1.0}, Bit Index Explicit Replication (BIER) \cite{bier}, and via standard multicast tables at the inter-DC switches. These approaches offer various trade-offs concerning the latency of installing a forwarding tree, the number of forwarding trees that can be set up at any given time, and the maximum rate at which traffic can be forwarded over forwarding trees. Comparison and analysis of how various ways of implementing forwarding trees can affect the efficiency and speed of inter-DC transfers is an exciting and valuable topic for future research.

\clearpage
\appendix
\renewcommand\chaptername{Appendix}
\chapter{NP-Hardness Proof for Best Worst-case Routing} \label{chapter_bwr_hardness}
From Chapter \ref{chapter_adaptive_routing}, we recall that over the inter-DC graph $G$, each edge $e$ was associated with a set of ongoing flows $\pmb{\mathrm{F}}_{e}$ and each flow $F_i$ had a remaining volume $\mathcal{V}^{r}_{i}$. The Best Worst-case Routing (BWR) problem was to select the path with minimum total weight between two vertices $s$ and $t$ with the weight computed as sum of the remaining volumes of all flows that have at least one common edge with the path. In other words, given a set of flows $\pmb{\mathrm{F}}$, to find a best worst-case path, we are looking for a subset of flows $\gamma \subseteq \pmb{\mathrm{F}}$ with minimum total sum of remaining data units where there exists a path from $s$ to $t$ removing all edges that have a flow in $\pmb{\mathrm{F}}-\gamma$. In the following, we show that one instance of this problem where the remaining volumes of all flows is set to $1$ is NP-Hard, and so BWR must be as well.

\begin{figure}
    \centering
    \includegraphics[width=\textwidth]{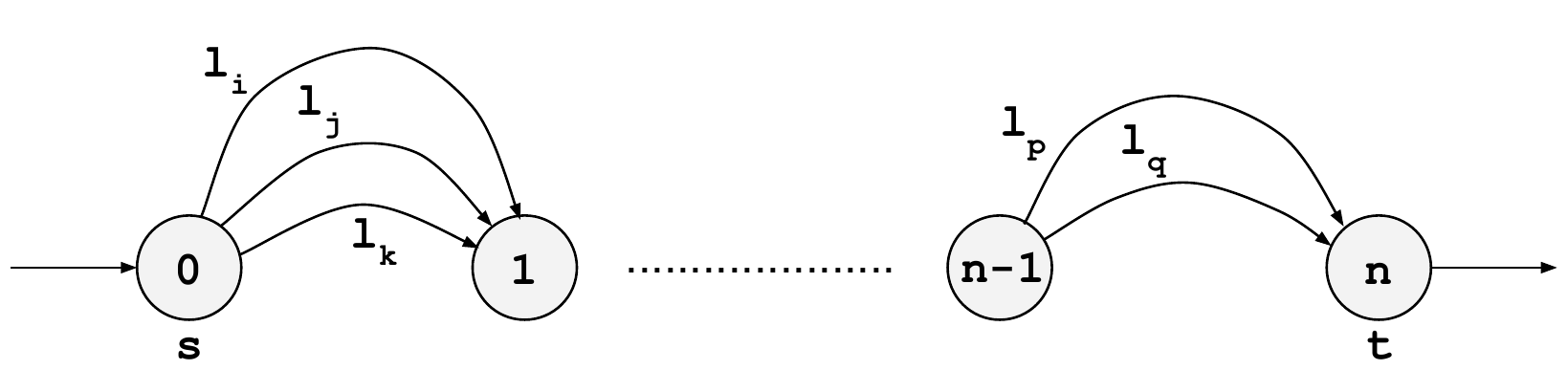}
    \caption{Network used in Problem 1.}
    \label{fig:np-hardness}
\end{figure}

\vspace{0.5em}
\textbf{Problem 1.} Consider the multi-graph of Figure \ref{fig:np-hardness} and a set of labels $\pmb{\mathrm{L}} = \{l_{1}, \dots, l_{m}\}$. Between any two vertices $i$ and $i+1$, we have at least one edge and each edge is associated with exactly one label. Also, there are no edges between vertices with non-consecutive numbers. We want to find a path $P$ from $s$ to $t$ so that the total number of distinct labels on the edges associated with $P$ is minimized.

\vspace{0.5em}
\textbf{Proposition 1.} Problem 1 is NP-Hard.

\vspace{0.5em}
\textbf{\textit{Proof.}} We will reduce the well-known Set Cover problem to Problem 1. Consider an arbitrary instance of Set Cover with the universal set $U = \{1, 2, \dots, n\}$ and collection of subsets $S = \{S_1, S_2, ..., S_m\}$. Construct a multi-graph $G$ with $n+1$ vertices labeled $0, 1, ..., n$. For every $S_j$ that contains element $i \in U$, add an edge ($i-1$, $i$) with label $l_j$ to $G$ corresponding to that subset. Now, any set cover using $k$ subsets corresponds to a path from node $0$ to node $n$ that uses $k$ labels. Conversely, any path on $G$ from node $0$ to node $n$ which uses $k$ labels, covers all the elements of $U$ and so corresponds to a set cover with $k$ subsets. Therefore, finding a path in $G$ with minimum total distinct labels corresponds to finding a minimum set cover on $S$, which is NP-hard.\footnote{Please note that the multi-graph created can be converted to a simple graph by replacing every edge with two edges labeled with the same label and a node in between.}


\vspace{0.5em}
\textbf{Problem 2.} Consider some graph $G(V, E)$ where each edge is associated with a set of labels $\pmb{\mathrm{L}} = \{l_{1}, \dots, l_{m}\}$. We want to find a path $P$ from $s$ to $t$ so that the total number of distinct labels on the edges associated with $P$ is minimized.

\vspace{0.5em}
\textbf{Proposition 2.} Problem 2 is NP-Hard.

\vspace{0.5em}
\textbf{\textit{Proof.}} Problem 1 is an instance of Problem 2 where each edge is associated with exactly one label and there are edges only between consecutive nodes. Therefore, Problem 2 must be NP-Hard.

\vspace{0.5em}
The flow routing problem has one more constraint, that is, all edges of a flow appear in consecutive order. In other words, all edges associated with a specific label appear on the graph $G(V, E)$ in consecutive order from the source of the flow to its destination.

\vspace{0.5em}
\textbf{Proposition 3.} Assuming a flow size of $1$ for all flows, BWR is NP-Hard.

\vspace{0.5em}
\textbf{\textit{Proof.}} We will reduce Problem 2 into an instance of Problem 3. Let us take an instance of Problem 2. We associate every label from the set of all labels $\pmb{\mathrm{L}}$ with exactly one flow. For any label $l$, if all the edges on $G$ are connected consecutively, we do not make any changes. Otherwise, we add dummy edges labeled $l$ to $G$ so that all edges with label $l$ appear consecutively in the new graph $G^{1}$. We repeat this for all such labels $l$ arriving at graph $G^{k}$ where $k$ is the number of labels for which edges do not appear consecutively on $G$. Next, to any dummy edge in $G^{k}$, we add $m + 1$ dummy new and distinct labels which will extend the set of all labels to $\pmb{\mathrm{L}}^\prime$. The graph $G^{k}$ with the set of labels $\pmb{\mathrm{L}}^\prime$ together form a valid instance of Problem 3. Any solution found for this instance is also a solution to Problem 2 and vice versa. That is because solutions found over the new graph with new labels will not include any of the dummy edges as such solutions will include at least $m + 1$ labels. This concludes our proof that BWR is NP-Hard.

\chapter{SDN Switches that Support Group Table ALL} \label{chapter_sdn_gt_all}
We list some Software Defined Networking (SDN) \cite{sdn} products that support the Group Table ALL feature which can be used to forward incoming packets to multiple outgoing ports via packet replication. When using Group Tables,\footnote{Group Tables are a feature supported by OpenFlow \cite{openflow} that allow complex group operations on incoming packets for purposes such as fast failover, load balancing, and multicasting \cite{of-1.1}.} each group entry can have multiple action buckets each programmed with a set of actions. The ``ALL" feature means all action buckets of a group entry will be executed and each bucket will be supplied with a copy of incoming packet that matches the group entry predicates. This feature has been in OpenFlow standard since version $1.1$ and was added for the purpose of flooding, broadcasting or multicasting \cite{of-1.1}. 

Despite being part of the OpenFlow specifications, this feature has not been widely supported by switch vendors. As of 2016, physical switches on the market have started providing support for this feature. We merely list several products that currently support this feature and cite related documents which contain detailed information on how these features are actually supported (e.g., maximum number of entries, maximum action buckets per entry, and whether group chaining is allowed). Table \ref{table:switches} provides a list of several products that can be used for building multicast forwarding trees.


\begin{table}[b]
\centering
\caption{SDN products with support for \texttt{OFPGT\_ALL}.} \label{table:switches}
{
\begin{tabular}{|p{4cm}|p{10cm}|}
    \hline
     \textbf{Vendor} & \textbf{Product} \\
    \hline
    \hline
     HP \cite{hp-1} & HP 5920 \& 5900 Switch Series \\
    \hline
     HP \cite{hp-2} & HP 5130 EI Switch Series \\
    \hline
     HP \cite{hp-3} & HP Switch 2920 series, HP Switch 3500 series, HP Switch 3800 series, HP Switch 5400 series, v1 and v2 modules, HP Switch 5406R series, HP Switch 5412A series, HP Switch 6200 series, HP Switch 6600 series, HP Switch 8200 series, v1 and v2 modules \\
    \hline
     Juniper Networks \cite{juniper-1} & MX Series, EX9200, QFX5100 and EX4600 \\
    \hline
     Alcatel-Lucent \cite{alcatel-1} & OmniSwitch 10K, OmniSwitch 9900, OmniSwitch 6900, OmniSwitch 6860, and OmniSwitch 6865 \\
    \hline
     IBM \cite{ibm-1} & IBM System Networking RackSwitch G8264 \\
    \hline
     Brocade \cite{brocade-1} & Brocade VDX 2741, Brocade VDX 6740, Brocade VDX 6940 and Brocade VDX 8770 \\
    \hline
\end{tabular}
}
\end{table}

\clearpage
{\footnotesize \bibliographystyle{unsrt}
\bibliography{citations}}

\end{document}